# From Cosmic Web to Supernova Remnants: Modeling FRB DM to Trace Baryons across Multiple Scales

by

Zhao Zhang

A thesis submitted in partial fulfillment for the degree of Doctor of Philosophy

in the

Department of Earth and Space Science

Graduate School of Science

Osaka University

June 2026

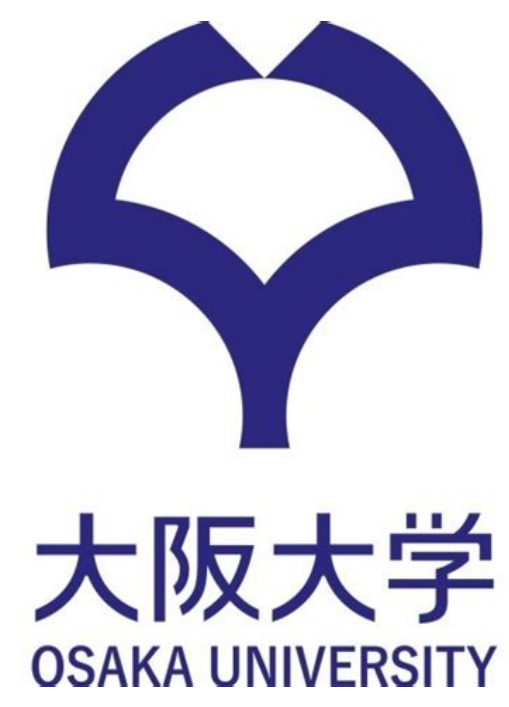

# Declaration of Authorship

I, Zhao Zhang, declare that this thesis titled, 'From Cosmic Web to Supernova Remnants: Modeling FRB Dispersion Measures for Baryon Tracing across Scales' and the work presented in it are my own. I confirm that:

- This work was done wholly or mainly while in candidature for a research degree at this University.
- Where any part of this thesis has previously been submitted for a degree or any other qualification at this University or any other institution, this has been clearly stated.
- Where I have consulted the published work of others, this is always clearly attributed.
- Where I have quoted from the work of others, the source is always given. With the exception of such quotations, this thesis is entirely my own work.
- I have acknowledged all main sources of help.
- Where the thesis is based on work done by myself jointly with others, I have made clear exactly what was done by others and what I have contributed myself.

Signed: ________________________________________

Date: ________________________________________

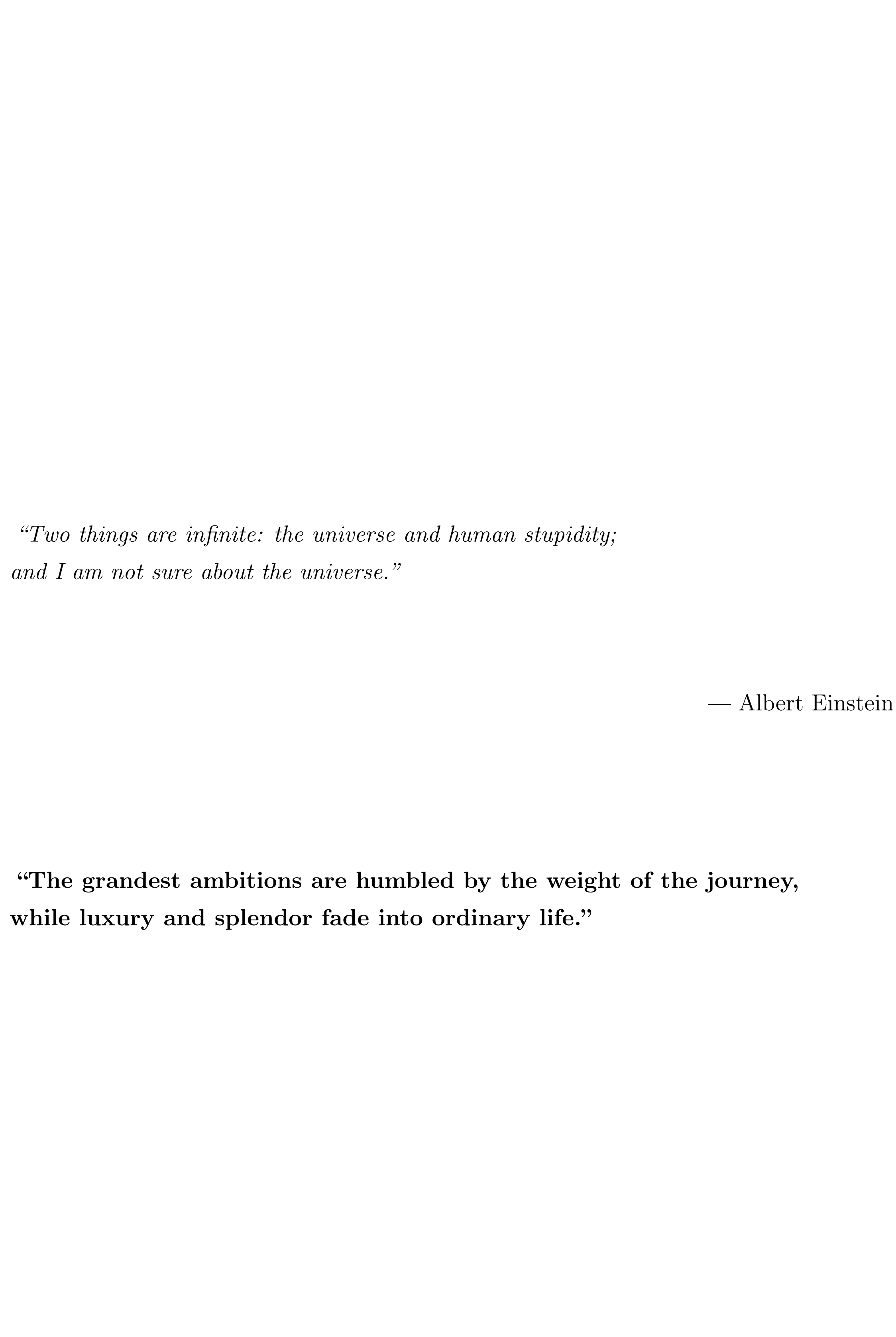

*"Two things are infinite: the universe and human stupidity;
and I am not sure about the universe."*

— Albert Einstein

**"The grandest ambitions are humbled by the weight of the journey,
while luxury and splendor fade into ordinary life."**

# *Abstract*

**Background:** Fast radio bursts (FRBs) provide a powerful probe of ionized baryons through their dispersion measures (DMs), but the observed signal is intrinsically multi-scale: it contains contributions from the intergalactic medium (IGM), circumgalactic gas and foreground halos, host galaxies, and source-local plasma. This thesis addresses the FRB DM problem from cosmic-web to source-local scales, with the dual goal of constraining the distribution of the missing baryons and determining how local environments can bias or enrich FRB-based cosmological inferences.

**Method:** I combine cosmological and zoom-in hydrodynamical simulations with detailed local-environment forward modeling. On large scales, I use the CROCODILE framework based on the GADGET3/4-OSAKA smoothed particle hydrodynamics code, including star formation, supernova feedback, and active galactic nucleus (AGN) feedback, to construct light cones, gas density profiles, and FRB DM statistics. These calculations are used to quantify the DM–$z$ relation, the partition of diffuse baryons between the IGM and halos/CGM, and the host-galaxy contribution across dwarf, Milky Way-like, and cluster environments. On local scales, I model a young magnetar embedded in supernova ejecta using time-dependent one-dimensional hydrodynamical simulations with non-equilibrium ionization and radiative cooling, exploring both single-star and binary-stripped progenitors to follow shock structure, ionization evolution, radio transparency, and the time dependence of $\mathrm{DM}_{\mathrm{source}}$ and $\mathrm{RM}_{\mathrm{source}}$.

**Results:** The cosmological analysis shows that AGN feedback lowers central halo gas densities, reshapes the CGM–IGM transition, and significantly modulates foreground-halo DM along FRB sight lines. From the DM–$z$ relation integrated to $z = 1$, I constrain the diffuse baryon fraction to $f_{\mathrm{diff}} = 0.865^{+0.101}_{-0.165}$ for the fiducial model and $0.856^{+0.101}_{-0.162}$ for the NoBH model, where $f_{\mathrm{diff}} \equiv f_{\mathrm{IGM}} + f_{\mathrm{halos}}$. I further quantify the redshift evolution of $f_{\mathrm{CGM}}$, $f_{\mathrm{IGM}}$, and $\langle f_{\mathrm{diff,obs}} \rangle$, showing that FRB-inferred baryon fractions depend on both physical gas partition and line-of-sight selection effects. Zoom-in simulations demonstrate that host contributions are strongly environment dependent, remaining typically below $100\,\mathrm{pc\,cm^{-3}}$ for central dwarf-galaxy cases but exceeding $1300\,\mathrm{pc\,cm^{-3}}$ in cluster environments. The local-environment study shows that the dominant time-variable DM component generally arises from the unshocked ejecta, whereas the shocked region contributes only $\lesssim 10\,\mathrm{pc\,cm^{-3}}$. At early times the ejecta contribution follows $\mathrm{DM} \propto t^{-\alpha}$ with $\alpha \simeq 1.8$–$1.9$. Binary-stripped progenitors generally yield lower DMs than single-star models at fixed initial mass, while composition-dependent mean molecular weights introduce non-monotonic mass trends. Matching the observed $\mathrm{dDM}/\mathrm{d}t$ of FRB 20190520B and the late declining stage of FRB 20121102 implies source-local SNR contributions of tens to hundreds of $\mathrm{pc\,cm^{-3}}$. Most models become transparent to GHz emission by $t_{\rm esc} \lesssim 70$ yr, and within the shock-only RM framework only the $11\,M_{\odot}$ single-star model reproduces the RM evolution of FRB 20121102.

**Conclusions:** This thesis shows that FRB dispersion measures must be interpreted as layered signals rather than as a single cosmological observable. Robust FRB cosmology requires explicit modeling of foreground halos, host environments, and source-local plasma, while physically consistent local-environment models are essential for separating cosmological and astrophysical DM components. FRBs are therefore best understood not only as distance indicators, but as multi-scale baryon probes linking the cosmic web, gaseous halos, host galaxies, and young compact-object environments within a unified framework.

# *Acknowledgements*

Completing this thesis marks the end of a long chapter of my student life. Looking back on this journey, I feel deeply grateful to the many people who have supported, guided, and inspired me along the way.

First and foremost, I would like to express my sincere gratitude to my supervisor, Prof. Kentaro Nagamine, for his continuous guidance, support, and mentorship throughout my four years of doctoral study. Under his supervision, I gradually learned how to independently develop and carry out scientific projects, how to approach scientific questions with rigor and critical thinking, and how to search for the underlying physical truth behind complicated phenomena. When I first entered this field, I knew almost nothing about cosmology. Through his guidance, I was able to grow into a researcher capable of independently exploring and completing scientific projects. He also provided me with many valuable opportunities for international collaboration and academic exchange, allowing me to interact with outstanding researchers from diverse backgrounds. Through these experiences, I witnessed my own growth and gradually developed confidence in my scientific abilities, which has had a profound impact on me both academically and personally.

I would also like to sincerely thank all of my thesis referees — Prof. Hironori Matsumoto, Prof. Kentaro Terada, Prof. Kento Masuda, and Prof. Hirokazu Odaka — for their careful reading of my thesis and for their many valuable comments and constructive suggestions. Their feedback greatly improved the scientific quality and completeness of this thesis.

I am deeply grateful to all the professors, collaborators, colleagues, and friends who have supported me through scientific discussions, collaborations, and technical assistance throughout my research journey. In particular, I would like to thank Prof. Bing Zhang, Prof. K. G. Lee, Prof. Daisuke Nagai, Prof. Fayin Wang, Kunihito Ioka-san, Yoshiyuki Inoue-san, Daisuke Toyouchi-san, Yuri Oku, Ke Xu, Nicolas Ledos, Keita Fukushima, Umer Rehman, Kenji Kihara, Ellis Owen, Kazuki Tomaru, Yuto Kuwayama, Suchetha Coorey, Pablo Granizo, Mingyong Jung, Pallas, Shotaro Yamasaki, Ruinan Li, Ruichong Hu, Jiaming Zhuge, Haoyu Yuan, Shengyu Yan, Wei Deng, Yining Wei, Tianyue Li and many others for their insightful discussions, constructive suggestions, and inspiring exchanges of ideas. The scientific interactions and collaborations with these talented researchers have been an invaluable part of my PhD journey.

I would also like to express my gratitude to my undergraduate and master's advisors, Prof. Houjun Lü and Prof. Bin-bin Zhang, for introducing me to scientific research and guiding me during the early stages of my academic journey. Their encouragement and support laid the foundation for my later scientific development during my doctoral studies. Even while I was studying abroad, they continued to care about both my research and my personal life.

I am also sincerely grateful to the members of the Toyonaka Rotary Club for their generous scholarship support and their continuous care for both my academic work and daily life throughout my PhD study. Their support greatly relieved many practical burdens and allowed me to focus more fully on my research. In particular, I would like to express my heartfelt gratitude to Miyata-san (emeritus professor of U. Osaka and a Rotarian), who introduced me to the Rotary Club as a scholarship student. Throughout my PhD journey, he consistently showed great care for my studies, research, and daily life, and generously offered me encouragement, support, and guidance. His kindness and consideration have meant a great deal to me during my time in Japan.

I would also like to express my deepest gratitude to my parents for their unconditional love, support, and sacrifice throughout my life. They not only taught me how to grow as a person, but also nurtured my curiosity and passion for physics and mathematics from an early age. It is because of their encouragement and influence that I was able to embark on this scientific journey — a path that continues to fill me with excitement, passion, and a sense of wonder about the Universe.

Finally, and most importantly, I would like to thank my wife, Feng Gao. During our six years together, we spent nearly five years living apart while each pursuing our own studies and dreams. Despite the distance, we continued to support and encourage each other through difficult moments, uncertainty, and self-doubt. Her constant care and support in all aspects of daily life allowed me to devote myself more fully to research and to continue pursuing my scientific aspirations without distraction.

As I complete this thesis, we have also just begun a new chapter of our lives through marriage. The coincidence of these two important moments fills me with profound happiness and gratitude. For this reason, this thesis will remain not only a record of my academic journey, but also an unforgettable milestone in my life.

# Contents

# List of Figures

# List of Tables

# Abbreviations

| | |
|---|---|
| **FRB** | **F**ast **R**adio **B**urst |
| **SN/SNe** | **S**uper**N**ova(**E**) |
| **CCSN** | **C**ore-**C**ollapse **S**uper**N**ova |
| **SLSN** | **S**uper**L**uminous **S**uper**N**ova |
| **Ic-BL** | Type Ic broad-line supernova |
| **HN** | **H**yper**N**ova |
| **FBOT** | **F**ast **B**lue **O**ptical **T**ransient |
| **DM** | **D**ispersion **M**easure |
| **RM** | **R**otation **M**easure |
| **IGM** | **I**nter**g**alactic **M**edium |
| **CGM** | **C**ircum**g**alactic **M**edium |
| **HG** | **H**ost **G**alaxy |
| **HH** | **H**ost **H**alo |
| **FH** | **F**oreground **H**alo |
| **ISM** | **I**nter**s**tellar **M**edium |
| **SNR** | **S**uper**n**ova **R**emnant |
| **BS** | **B**inary-stripped **S**tar |
| **SS** | **S**ingle **S**tar |
| **NEI** | **N**on-**E**quilibrium **I**onization |
| **AGN** | **A**ctive **G**alactic **N**ucleus |
| **SPH** | **S**moothed **P**article **H**ydrodynamics |
| **CDM** | **C**old **D**ark **M**atter |
| **CMB** | **C**osmic **M**icrowave **B**ackground |
| **FLRW** | Friedmann–Lemaître–Robertson–Walker |
| **MW** | Milky Way |
| **WHIM** | Warm–Hot Intergalactic Medium |
| **CSM** | Circumstellar Medium |
| **BH** | Black Hole |
| **HMF** | Halo Mass Function |
| **NFW** | Navarro–Frenk–White |
| **IMF** | Initial Mass Function |
| **ZAMS** | Zero-Age Main Sequence |
| **RSG** | Red Supergiant |
| **WR** | Wolf–Rayet |
| **BSG** | Blue Supergiant |
| **OB** | OB star |
| **PWN** | Pulsar Wind Nebula |
| **MWN** | Magnetar Wind Nebula |
| **BNS** | Binary Neutron Star |
| **DSA** | Diffusive Shock Acceleration |
| **NLDSA** | Nonlinear Diffusive Shock Acceleration |
| **CCO** | Central Compact Object |
| **RLOF** | Roche-Lobe Overflow |
| **UVB** | Ultraviolet Background |
| **CHIME/FRB** | Canadian Hydrogen Intensity Mapping Experiment/Fast Radio Burst Project |
| **ASKAP** | Australian Square Kilometre Array Pathfinder |
| **NE2001** | Cordes–Lazio Galactic electron-density model |
| **YMW16** | Yao–Manchester–Wang 2016 Galactic electron-density model |

# Physical Constants

| | | | |
|---|---|---|---|
| Speed of Light | $c$ | = | $2.997\ 924\ 58 \times 10^{8}$ m s$^{-1}$ (exact) |
| Boltzmann Constant | $k_{\mathrm{B}}$ | = | $1.380\ 649 \times 10^{-23}$ J K$^{-1}$ (exact) |
| Gravitational Constant | $G$ | = | $6.67430 \times 10^{-11}$ m$^{3}$ kg$^{-1}$ s$^{-2}$ |
| Elementary Charge | $e$ | = | $1.602\ 176\ 634 \times 10^{-19}$ C (exact) |
| Electron Mass | $m_{\mathrm{e}}$ | = | $9.109\ 383\ 7015 \times 10^{-31}$ kg |
| Proton Mass | $m_{\mathrm{p}}$ | = | $1.672\ 621\ 923\ 69 \times 10^{-27}$ kg |
| Thomson Cross Section | $\sigma_{\mathrm{T}}$ | = | $6.652\ 458\ 732 \times 10^{-29}$ m$^{2}$ |
| Parsec | pc | = | $3.085\ 677\ 581 \times 10^{16}$ m |
| Solar Mass | $M_{\odot}$ | = | $1.988\ 47 \times 10^{30}$ kg |

# Symbols

| | | |
|---|---|---|
| $z$ | redshift | — |
| $\mathrm{DM}_{\mathrm{FRB}}$ | total observed FRB dispersion measure | $\mathrm{pc\,cm^{-3}}$ |
| $\mathrm{DM}_{\mathrm{MW}}$ | Milky Way dispersion measure | $\mathrm{pc\,cm^{-3}}$ |
| DM | dispersion measure | $\mathrm{pc\,cm^{-3}}$ |
| $\mathrm{DM}_{\mathrm{diff}}$ | diffuse-medium dispersion measure | $\mathrm{pc\,cm^{-3}}$ |
| $\mathrm{DM}_{\mathrm{IGM}}$ | intergalactic-medium dispersion measure | $\mathrm{pc\,cm^{-3}}$ |
| $\mathrm{DM}_{\mathrm{halos}}$ | foreground-halo dispersion measure | $\mathrm{pc\,cm^{-3}}$ |
| $\mathrm{DM}_{\mathrm{host}}$ | host-galaxy/halo dispersion measure | $\mathrm{pc\,cm^{-3}}$ |
| $\mathrm{DM}_{\mathrm{source}}$ | source-local dispersion measure | $\mathrm{pc\,cm^{-3}}$ |
| $\mathrm{DM}_{\mathrm{nonloc}}$ | non-local dispersion measure | $\mathrm{pc\,cm^{-3}}$ |
| $\mathrm{RM}_{\mathrm{source}}$ | source-local rotation measure | $\mathrm{rad\,m^{-2}}$ |
| $n_e$ | free electron number density | $\mathrm{cm^{-3}}$ |
| $\chi_e$ | electron fraction | — |
| $H_0$ | Hubble constant (simulation cosmology) | $\mathrm{km\,s^{-1}\,Mpc^{-1}}$ |
| $E(z)$ | dimensionless Hubble function | — |
| $\rho_{c,0}$ | present-day critical density | $\mathrm{kg\,m^{-3}}$ |
| $\Omega_m$ | matter density parameter (simulation cosmology) | — |
| $\Omega_\Lambda$ | dark-energy density parameter (simulation cosmology) | — |
| $\Omega_b$ | baryon density parameter (simulation cosmology) | — |
| $f_{\mathrm{IGM}}$ | intergalactic baryon fraction | — |
| $f_{\mathrm{CGM}}$ | circumgalactic baryon fraction | — |
| $f_{\mathrm{halos}}$ | halo baryon fraction contributing to FRB DM | — |
| $f_{\mathrm{diff}}$ | diffuse baryon fraction ($f_{\mathrm{IGM}} + f_{\mathrm{halos}}$) | — |
| $\langle f_{\mathrm{diff,obs}} \rangle$ | sightline-averaged observed diffuse baryon fraction | — |
| $R_{200}$ | halo radius enclosing mean density $200\rho_c$ | kpc |
| $M_{200}$ | halo mass within $R_{200}$ | $M_\odot$ |
| $b$ | halo impact parameter | kpc |
| $M_{BH}$ | black-hole mass | $M_\odot$ |
| $\dot{M}$ | mass-loss rate | $M_\odot\,yr^{-1}$ |
| $v_w$ | stellar-wind velocity | $km\,s^{-1}$ |
| $M_{ej}$ | ejecta mass | $M_\odot$ |
| $E_{SN}$ | supernova explosion energy | erg |
| $n_0$ | ambient ISM number density | $cm^{-3}$ |
| $\mu_e$ | mean molecular weight per electron | — |
| $\mu_m$ | mean molecular weight | — |
| $\mu_a$ | mean atomic weight | — |
| $K$ | wind-density normalization, $\dot{M}/(4\pi v_w)$ | $g\,cm^{-1}$ |
| $K_{13}$ | normalized wind-density parameter | — |
| $R_{ch}$ | characteristic SNR radius | pc |
| $t_{ch}$ | characteristic SNR timescale | yr |
| $t_{esc}$ | radio-transparent escape time | yr |
| $t_{occur}$ | occurrence time inferred from observed $dDM/dt$ | yr |

# Chapter 1

# Introduction

## 1.1 Cosmological Background

### 1.1.1 Standard cosmology and large-scale structure formation

Modern cosmology is commonly described within the framework of the Friedmann–Lemaître–Robertson–Walker (FLRW) metric, which assumes that the Universe is homogeneous and isotropic on sufficiently large scales. In comoving coordinates, the spacetime line element can be written as (Peebles & Ratra, 2003)

$$ds^2 = -c^2dt^2 + a^2(t)\left[\frac{dr^2}{1-Kr^2} + r^2(d\theta^2 + \sin^2\theta\, d\phi^2)\right], \tag{1.1}$$

where $a(t)$ is the cosmic scale factor and $K$ describes the spatial curvature. In this description, the expansion of the Universe is fully encoded in the time evolution of $a(t)$.

The Hubble parameter is defined as

$$H(t) \equiv \frac{\dot{a}}{a}, \tag{1.2}$$

and the cosmological redshift is related to the scale factor through

$$1 + z = \frac{a_0}{a(t)}, \tag{1.3}$$

where $a_0 \equiv a(t_0)$ is the present-day scale factor, usually set to unity. These relations form the basis for connecting cosmological observables to the expansion history of the Universe (Peebles & Ratra, 2003, Planck Collaboration et al., 2020). The expansion history is governed by the Friedmann equation,

$$H^2(z) = H_0^2 E^2(z), \tag{1.4}$$

with

$$E^2(z) = \Omega_{\rm m}(1+z)^3 + \Omega_{\rm r}(1+z)^4 + \Omega_\Lambda + \Omega_K(1+z)^2, \tag{1.5}$$

where $\Omega_{\rm m}$, $\Omega_{\rm r}$, $\Omega_\Lambda$, and $\Omega_K$ denote the present-day density parameters of non-relativistic matter, radiation, dark energy, and curvature, respectively, satisfying

$$\Omega_{\rm m} + \Omega_{\rm r} + \Omega_\Lambda + \Omega_K = 1. \tag{1.6}$$

Thus, non-relativistic matter scales as $(1+z)^3$, radiation as $(1+z)^4$, while a cosmological constant remains independent of redshift (Peebles & Ratra, 2003, Planck Collaboration et al., 2020).

Within the $\Lambda$ cold dark matter ($\Lambda$CDM) paradigm, small primordial density fluctuations grow through gravitational instability. Dark matter collapses first to form halos, and baryons subsequently fall into these potential wells, giving rise to galaxies, galaxy groups and clusters, and the filamentary cosmic web (Blumenthal et al., 1984, White & Rees, 1978, Bond et al., 1996, Springel, 2005, Springel et al., 2006).

The statistical abundance of collapsed halos is classically described by the Press–Schechter formalism, which connects the initial Gaussian density field to the halo mass function and provides a foundational framework for hierarchical structure formation (Press & Schechter, 1974).

Although dark matter dominates the gravitating matter budget, baryons are the directly observable component and are distributed across multiple phases, including the interstellar medium (ISM), the circumgalactic medium (CGM), and the intergalactic medium (IGM). Cosmological simulations and observations suggest that a substantial fraction of baryons at low redshift reside in diffuse ionized gas, especially in the IGM and CGM, motivating the long-standing "missing baryon problem" (Cen & Ostriker, 1999, Danforth & Shull, 2005, Cen, 2012, Shull et al., 2012, Tumlinson et al., 2017, Bregman, 2007).

These basic cosmological relations are also directly relevant for FRB studies. In particular, the function $E(z)$ determines the cosmological line element,

$$dl = \frac{c\,dz}{(1+z)H_0 E(z)}, \tag{1.7}$$

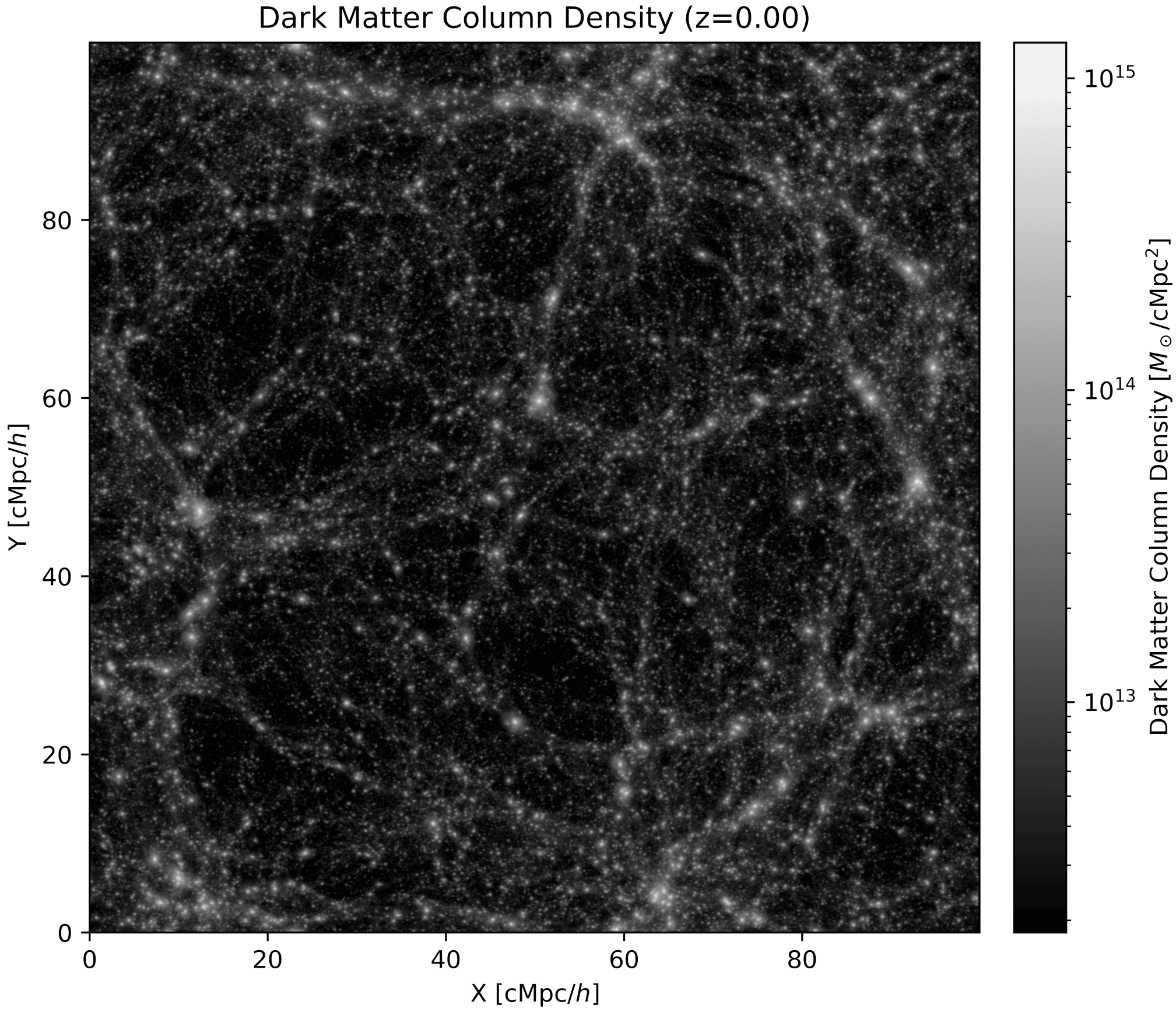

FIGURE 1.1: Dark-matter density projection at redshift $z = 0$, illustrating the filamentary cosmic web in our CROCODILE simulation. Figure adapted from Oku & Nagamine (2024); see also the classical large-scale-structure morphology discussed by Springel et al. (2006).

which enters the line-of-sight integral of the free-electron column density and therefore the cosmological contribution to the observed FRB dispersion measure (DM) (Peebles & Ratra, 2003, Zhang et al., 2025).

### 1.1.2 Baryon census and the missing baryon problem

Measurements of the cosmic microwave background (CMB) and Big Bang nucleosynthesis indicate that baryons constitute approximately $\Omega_b \approx 0.048$ of the total energy density of the Universe. However, observational inventories of baryons in the local Universe have historically accounted for only a fraction of this predicted baryonic mass (Danforth & Shull, 2005, 2008).

This discrepancy is commonly referred to as the "missing baryon problem". Observations suggest that a significant portion of baryons reside in diffuse and highly ionized gas distributed across the intergalactic medium (IGM) and the circumgalactic medium (CGM) of galaxies. Because this gas is tenuous and hot, it is difficult to detect directly through traditional emission or absorption-line observations (Tripp et al., 2006, Richter et al., 2004, 2006, Tilton et al., 2012, Prochaska et al., 2011). Complementary mass inventories in galaxies and clusters further highlight the census tension (Salucci & Persic, 1999, Reiprich & Böhringer, 2002, Zwaan et al., 2003).

Observations using X-ray (Nicastro et al., 2005a,b, Yao et al., 2012, Ren et al., 2014, Nicastro et al., 2018, Kovács et al., 2019, Zhang et al., 2024), Ly$\alpha$ (Penton et al., 2004, Danforth & Shull, 2008, Tejos et al., 2016, Kovács et al., 2019, Hu et al., 2024), and the Sunyaev–Zel'dovich (SZ) effect (de Graaff et al., 2019, Singari et al., 2020, Erciyes et al., 2023, Tanimura & Aghanim, 2023, Hadzhiyska et al., 2025) have placed important constraints on the WHIM. However, these methods remain limited by observational uncertainties, motivating complementary approaches for tracing diffuse ionized baryons. Beyond direct observations, cosmological simulations (Martizzi et al., 2019, Gheller & Vazza, 2019, Medlock & Cen, 2021) have also been extensively used to characterize WHIM baryon reservoirs and assess their role in alleviating the missing-baryon tension.

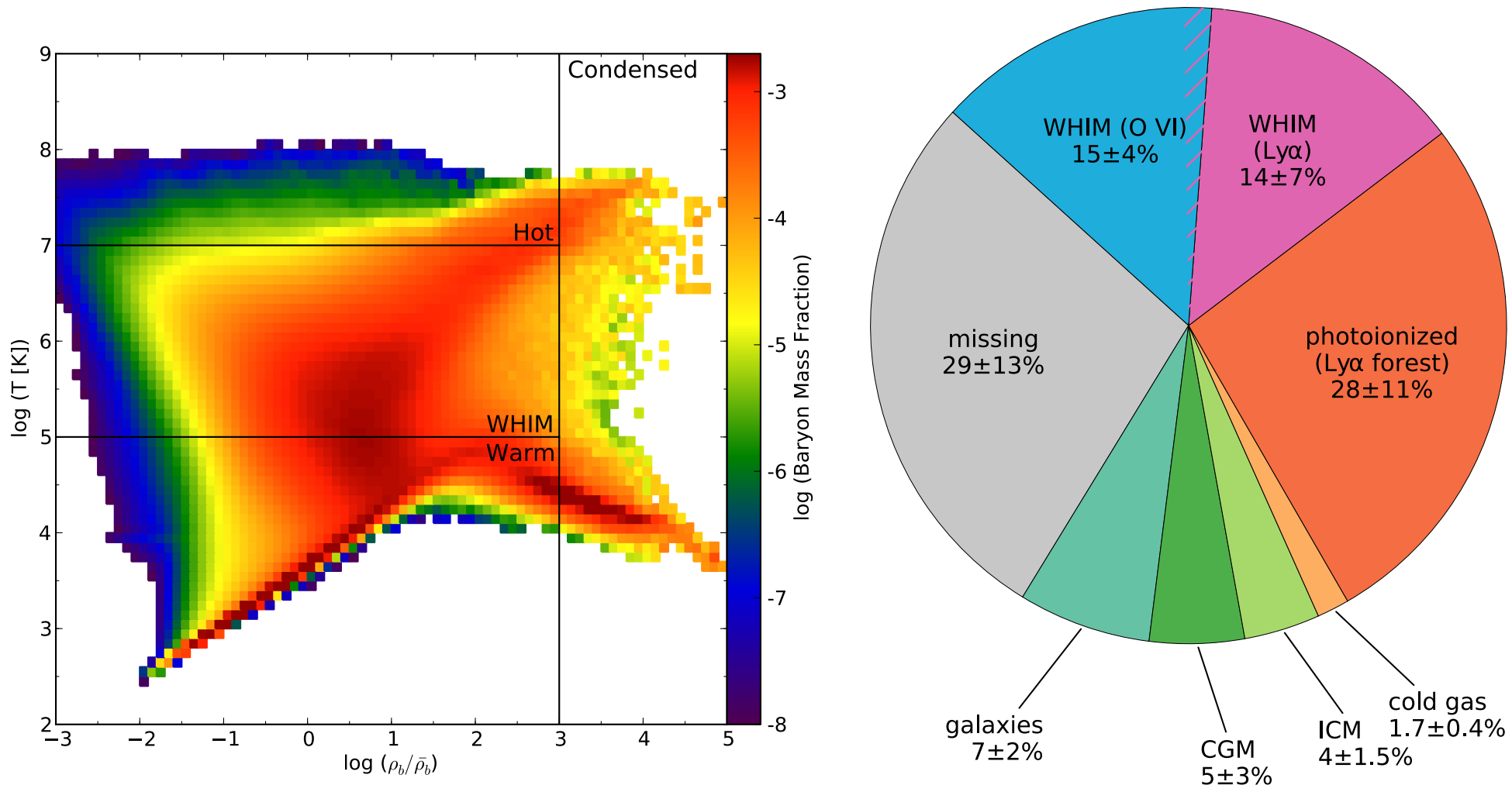

Figure 1.2: Two diagnostic views of the low-redshift baryon census. Left: phase-space distribution illustrating that a large baryon fraction is expected in diffuse and warm-hot ionized gas. Right: baryon-fraction budget showing that observed baryons in galaxies and hot intracluster gas do not fully account for the cosmological baryon density, highlighting the missing-baryon problem. Figure adapted from Shull et al. (2012).

As summarized in Figure 1.2, our work is initially motivated by this baryon-census tension, with a focus on developing FRB-based methods to recover the ionized baryon reservoir across cosmic-web, halo/CGM, and host-environment scales.

## 1.2 Fast Radio Bursts as Cosmological Probes

### 1.2.1 FRB phenomenology and observables

Fast radio bursts (FRBs) are millisecond-duration radio transients that are now known to be predominantly extragalactic (Lorimer et al., 2007, Thornton et al., 2013, FRB Collaboration et al., 2026). Their key observables include the arrival time, fluence, pulse width, scattering, polarization, rotation measure (RM), and, most importantly for this thesis, DM. Because DM traces the integrated free-electron column density along the line of sight, FRBs provide a powerful probe of diffuse ionized baryons across cosmological distances.

In addition to DM, RM provides a key and complementary observable for probing cosmic magnetism. The rotation measure is defined as

$$\mathrm{RM} = \frac{e^3}{2\pi m_e^2 c^4} \int n_e B_\parallel \, dl, \tag{1.8}$$

where $n_e$ is the free-electron number density and $B_\parallel$ is the line-of-sight magnetic-field component. Jointly modeling DM and RM therefore helps diagnose magnetic-field strength, field reversals, and magneto-ionic environments from source-local to intergalactic scales.

Observed FRB phenomenology spans multiple dimensions. In time domain, burst durations are typically at the millisecond level, with sub-ms structures in some events and repetition timescales ranging from seconds to years. In frequency domain, FRBs are currently detected from a few hundred MHz to several GHz, and many bursts show narrow-band or patchy spectral structure rather than simple broadband power-law behavior. In polarization domain, bursts can exhibit strong linear polarization and diverse RM amplitudes, indicating heterogeneous plasma and magnetic environments.

A practical classification is commonly made along two axes: repeating versus apparently non-repeating sources, and morphology-based classes such as one-off single-component bursts versus complex multi-component bursts with sub-burst drift. This phenomenological classification is observationally useful even though it does not yet map uniquely to progenitor physics.

The current observational landscape is driven by wide-field and high-cadence radio facilities, including CHIME/FRB, ASKAP, FAST, MeerKAT, the Deep Synoptic Array, UTMOST, and the VLA follow-up/localization network. Recently, the CHIME/FRB Collaboration released the second CHIME/FRB Catalog (FRB Collaboration et al., 2026), containing DMs for more than 3,000 unique FRB sources, and bringing the known FRB sample into the several-thousand regime.

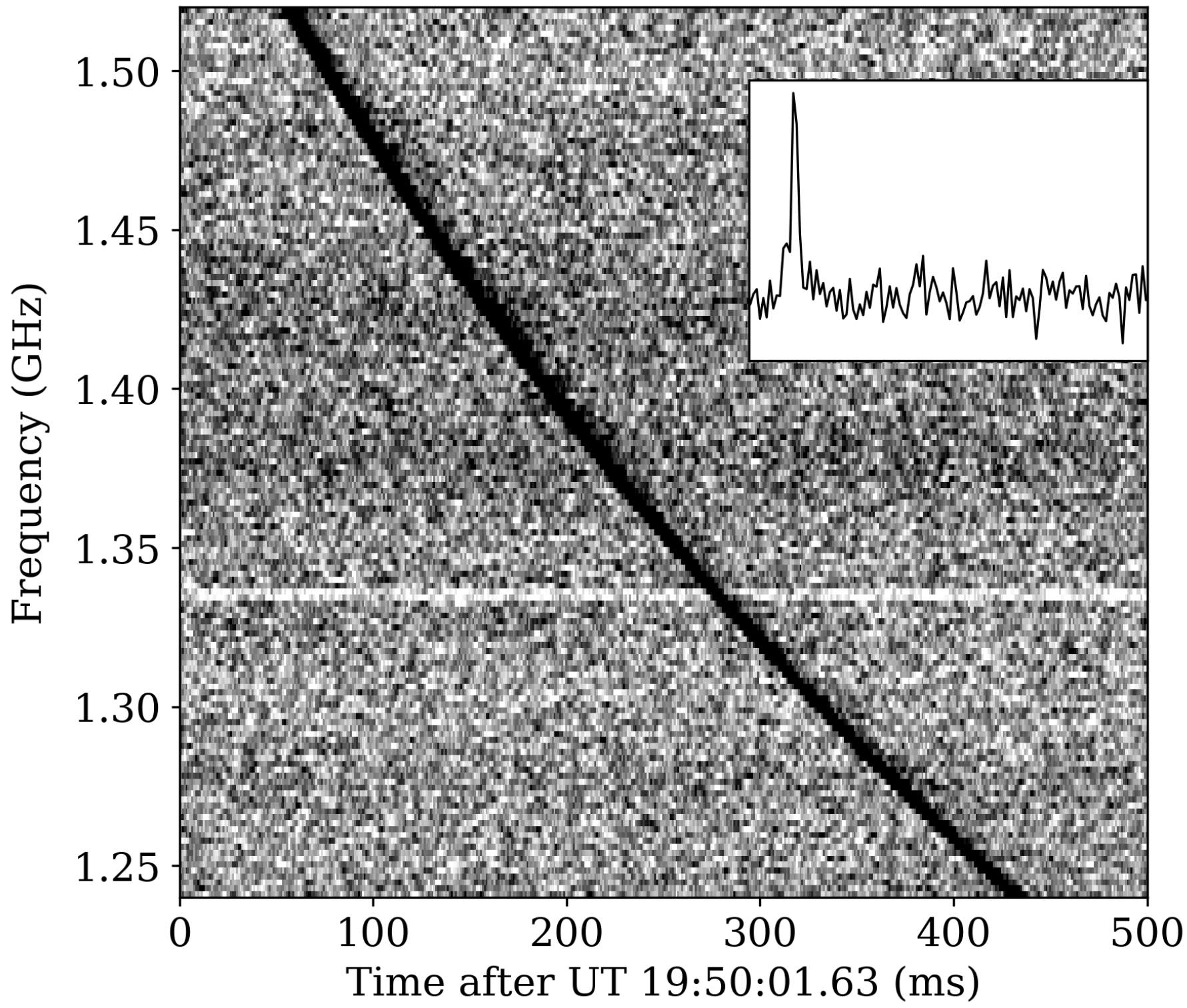

FIGURE 1.3: Simulated dynamic spectrum of a dispersed radio burst, illustrating the characteristic $\nu^{-2}$ frequency-dependent delay expected from plasma dispersion (e.g., Lorimer et al. (2007)).

### 1.2.2 FRB origins and magnetar formation channels

The formation mechanism of FRBs remains an active area of research. Magnetars are currently the leading progenitor channel, strongly supported by the Galactic magnetar SGR 1935+2154 and its FRB-like burst FRB 200428 (Bochenek et al., 2020). Other channels may contribute to a subpopulation of events. One proposed alternative is binary neutron star (BNS) mergers (Zhang, 2020, Wang et al., 2020); however, their event rate is too low to account for the bulk FRB population (Lipunov & Pruzhinskaya, 2014), and population studies suggest only a minor fraction of repeaters may arise from BNS-formed magnetars (Zhang et al., 2020a). Although unlikely to dominate, BNS channels may contribute to offset FRB sources through natal kicks, which can displace progenitor systems from host-galaxy centers and provide a complementary observational signature of progenitor diversity.

Recent population-synthesis studies further support a mixed-channel picture in which both single-star and binary evolutionary pathways may feed the observed magnetar population (Hu & Zhang, 2026). This is physically well motivated because a large fraction of massive stars (often quoted as $\sim 70\%$–$80\%$) are born in interacting binary systems (Sana et al., 2012). In this framework, magnetars can be produced through "classical" CCSN channels from single progenitors, as well as binary-mediated routes (including stripped progenitors and post-interaction systems) that modify pre-SN core structure and envelope properties (Laplace et al., 2021, Farmer et al., 2023). This context is directly relevant for our later FRB–SNR discussion, because different CCSN and binary-stripped pathways are expected to imprint different circum-source densities, ejecta masses, and ionization histories, and hence different source-local DM contributions, denoted as $\mathrm{DM}_{\mathrm{source}}$ (the source-environment component of the total FRB DM), and their temporal evolution.

In particular, Hu & Zhang (2026) found that the observed magnetar population is dominated by apparently isolated systems, but with strong imprints of binary evolution (see 1.4). Therefore, "no present-day companion" does not imply "no binary-interaction history," which is important when interpreting FRB-, SLSN-, Ic-BL-, and FBOT-associated magnetar progenitors. In the same framework, BNS-merger pathways are subdominant at the population level and generally negligible for the overall FRB event-rate budget.

The delay-time distribution in Hu & Zhang (2026) is also highly relevant for FRB host-galaxy studies. Single-star channels are typically prompt (tracking massive-star lifetimes), whereas binary channels broaden the delay-time distribution, with short delays possible through rapid tidal spin-up and long-delay tails produced by channels such as AIC and BWD mergers. This implies that magnetar-powered FRB sources need not appear exclusively in the youngest star-forming environments, and that a broader host demographic is physically expected.

### 1.2.3 Dispersion measure and cosmological interpretation

The basic idea of using dispersion from distant transients to study the ionized intergalactic medium predates FRB discoveries and was discussed in the context of gamma-ray bursts by Ioka (2003) and Inoue (2004). FRBs provide a much larger and cleaner statistical sample for this purpose, enabling precision studies of baryon distribution in the low-redshift Universe.

FRBs are brief but intense bursts of radio waves, typically reaching $10^{38}$–$10^{42}\,\mathrm{erg\,s^{-1}}$. Their DM, first highlighted in the discovery of FRB 010724 by Lorimer et al. (2007), arises from frequency-dependent propagation delays in ionized plasma.

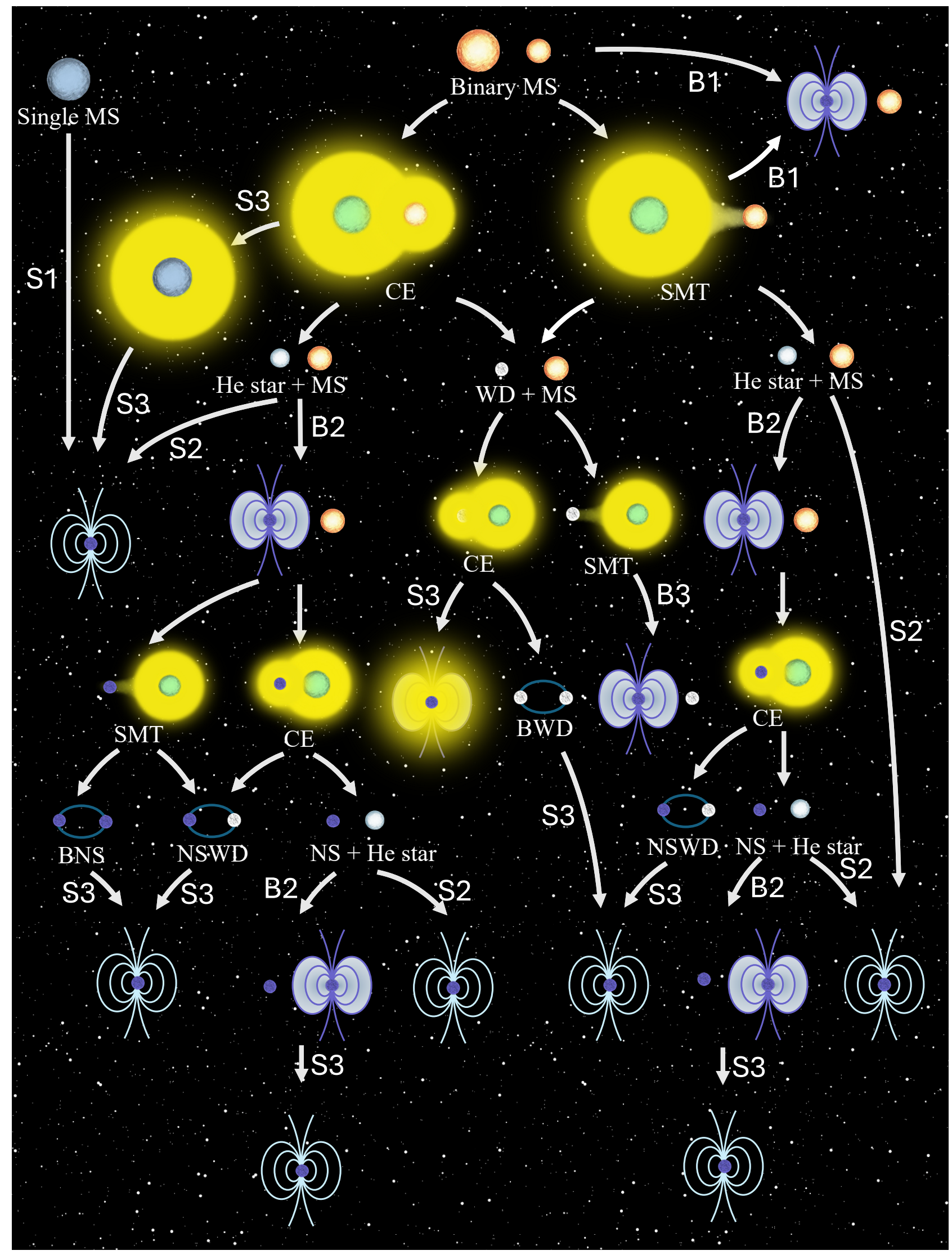

FIGURE 1.4: Schematic overview of magnetar formation channels relevant to FRB progenitors, including single-star and binary-evolution pathways. Figure adapted from Hu & Zhang (2026).

The DM quantifies this dispersive delay relative to vacuum propagation (Deng & Zhang, 2014, Zheng et al., 2014):

$$\Delta t_{\text{obs}} = \int \frac{\nu_p^2}{\nu_c^2} \frac{dl}{c} = \frac{e^2}{2\pi m_e c \nu_c^2} \text{DM}, \tag{1.9}$$

where $\nu_p = (n_{e,z} e^2/\pi m_e)^{1/2}$ is the plasma frequency, $\nu_c$ is the comoving frequency, and $e$, $m_e$, and $n_e$ are the electron charge, mass, and number density, respectively.

For plasma at redshift $z$, the rest-frame delay is

$$\Delta t_z = \frac{e^2}{2\pi m_e c \nu_z^2} \int_0^L n_{e,z} \, dl_z = \frac{e^2}{2\pi m_e c \nu_z^2} \text{DM}_z, \tag{1.10}$$

where $L$ is the path length through plasma, $\nu_z$ is the rest-frame frequency, and $\mathrm{DM}_z \equiv \int_0^L n_{e,z}\, dl_z$.

Using $\Delta t_z = \Delta t_{\mathrm{obs}}/(1+z)$ and $\nu_z = \nu_{\mathrm{c}}(1+z)$, Eq. (1.10) becomes

$$\Delta t_{\mathrm{obs}} = \frac{e^2}{2\pi m_e c\, \nu_{\mathrm{c}}^2} \int_0^L \frac{n_{e,z}}{1+z}\, dl_z, \tag{1.11}$$

and comparison with Eq. (1.9) gives the observer-frame DM:

$$\mathrm{DM} = \int_0^L \frac{n_{e,z}(l)}{1+z}\, dl_z. \tag{1.12}$$

For cosmological calculations, the proper-path element is given by Eq. (1.7), i.e., $dl_z \equiv dl = c\, dz/[(1+z)H_0 E(z)]$. In this chapter we adopt the flat ΛCDM form $E(z) = \sqrt{\Omega_m(1+z)^3 + \Omega_\Lambda}$. Using comoving electron density $n_{e,\mathrm{c}} = n_{e,z}/(1+z)^3$, Eq. (1.12) becomes

$$\mathrm{DM} = \frac{c}{H_0} \int_0^z \frac{n_{e,\mathrm{c}}(1+z')}{E(z')}\, dz'. \tag{1.13}$$

The cosmological use of FRB dispersion was established through both theory and simulations. McQuinn (2014) demonstrated that FRBs can constrain baryon distributions and galaxy-halo physics, while Deng & Zhang (2014) and Zheng et al. (2014) developed early FRB DM–$z$ frameworks for cosmological inference. Observationally, the empirical FRB DM–$z$ relation was first established using a small sample of well-localized bursts by Macquart et al. (2020). With the rapid increase in localized FRBs in recent years (e.g., Wang et al. 2025c), the enlarged sample allows a more direct assessment of this relation. As shown in Figure 1.5, while a broad increasing trend of DM with redshift is still present, the updated sample suggests a somewhat steeper relation compared to the original Macquart result, along with significant scatter across different sightlines. The shaded region

Building on this, Li et al. (2019a) proposed a cosmology-independent strategy to constrain the diffuse baryon fraction using joint measurements of FRB DM and luminosity distance. Subsequently, simulation-based studies have expanded the methodology to include large-scale-structure variance, halo contributions, and redshift-dependent systematics (Zhang et al., 2021, Medlock & Cen, 2021, Zhu & Feng, 2021, Lee et al., 2022, Khrykin et al., 2024b, Medlock et al., 2024, Connor et al., 2024, Zhang et al., 2025, Konietzka et al., 2025). With rapidly growing FRB samples from current surveys, FRBs are now emerging as a precision probe of the cosmic baryon distribution (Bhandari & Flynn, 2021, Zhang, 2023, Glowacki & Lee, 2024, Wu & Wang, 2024).

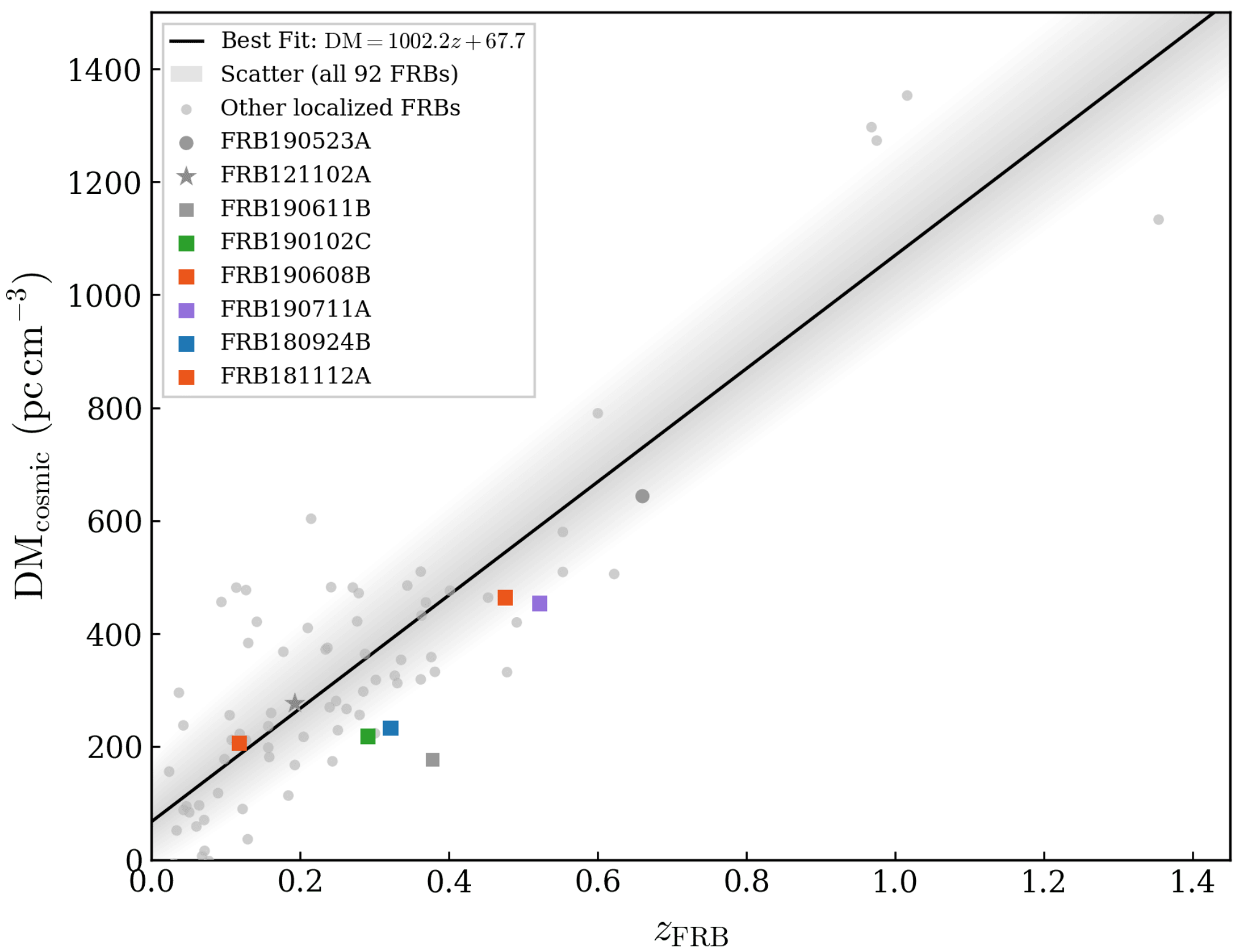

FIGURE 1.5: Updated DM–$z$ relation for 92 well-localized FRBs compiled from the literature, including the original sample of Macquart et al. (2020) and later updates such as Wang et al. (2025c). Gray circles show the full sample, while colored squares highlight the original Macquart benchmark sources. Three historically important but non-gold-sample FRBs (FRB 121102, FRB 190523, and FRB 190611) are marked with distinct gray symbols. $\mathrm{DM_{cosmic}}$ is computed as $\mathrm{DM_{obs}} - \mathrm{DM_{MW,ISM}}(\mathrm{NE2001}) - \mathrm{DM_{MW,halo}} - \mathrm{DM_{host}}$, with $\mathrm{DM_{MW,halo}} = 50\,\mathrm{pc\,cm^{-3}}$ and $\mathrm{DM_{host}} = 50(1+z)^{-1}\,\mathrm{pc\,cm^{-3}}$. The solid line shows the best-fit relation to all FRBs, and the shaded region indicates the scatter.

### 1.2.4 Multi-component decomposition of DM

For a given FRB, the total observed DM can be decomposed as

$$\mathrm{DM_{FRB}} = \mathrm{DM_{MW}} + \mathrm{DM_{halos}} + \mathrm{DM_{IGM}} + \frac{\mathrm{DM_{host}} + \mathrm{DM_{source}}}{1+z}, \tag{1.14}$$

where $\mathrm{DM_{MW}}$ is the Milky Way term (including both ISM and Galactic-halo contributions), $\mathrm{DM_{halos}}$ is the cumulative foreground-halo contribution, $\mathrm{DM_{IGM}}$ is the intergalactic contribution, $\mathrm{DM_{host}}$ is the host-galaxy (or host-halo) contribution, and $\mathrm{DM_{source}}$ is the local source-environment contribution.

The foreground-halo (FH) term is

$$\mathrm{DM_{halos}} = \sum_i \frac{\mathrm{DM}_{\mathrm{halo},i}}{1+z_i}, \tag{1.15}$$

where $i$ runs over all intervening halos intersected by the FRB sight line.

Following the non-local/local decomposition, we define

$$\mathrm{DM}_{\mathrm{diff}} \equiv \mathrm{DM}_{\mathrm{IGM}} + \mathrm{DM}_{\mathrm{halos}}, \tag{1.16}$$

$$\mathrm{DM}_{\mathrm{nonloc}} \equiv \mathrm{DM}_{\mathrm{MW}} + \mathrm{DM}_{\mathrm{diff}} + \frac{\mathrm{DM}_{\mathrm{host}}}{1+z}, \tag{1.17}$$

so that

$$\mathrm{DM}_{\mathrm{FRB}} = \mathrm{DM}_{\mathrm{nonloc}} + \frac{\mathrm{DM}_{\mathrm{source}}}{1+z}. \tag{1.18}$$

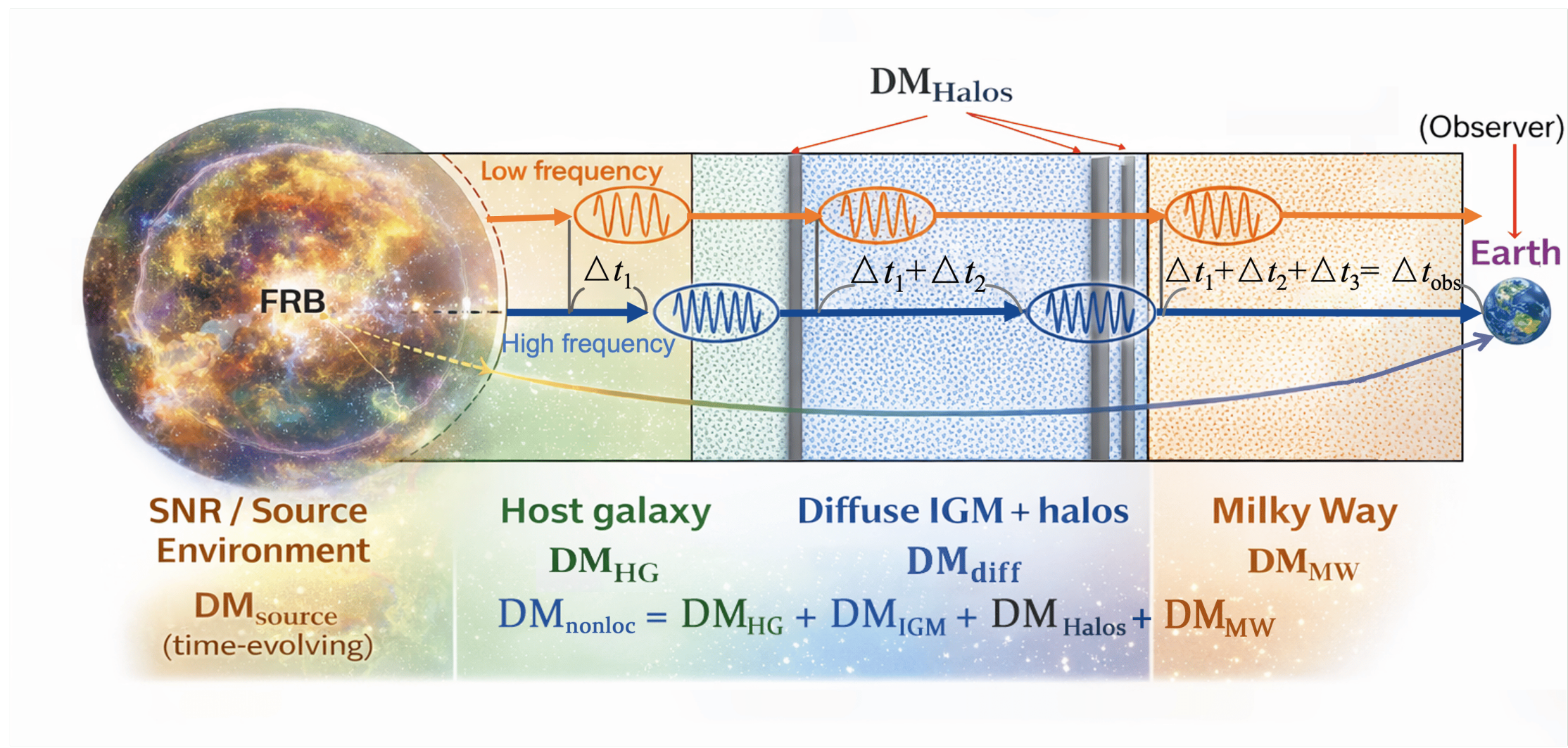

FIGURE 1.6: Schematic summary of the multi-component decomposition used for FRB dispersion measure analysis.

## 1.3 Multi-scale Nature of FRB Dispersion Measures

### 1.3.1 Cosmic web contribution

The dominant non-local contribution to FRB dispersion at cosmological distances is the diffuse baryonic component associated with large-scale structure (LSS), including low-density filaments, sheets, and mildly overdense intergalactic environments. In the idealized limit, this component is often summarized as an IGM term, but in practice it reflects a continuous cosmic-web distribution shaped by hierarchical structure formation and baryonic feedback processes. Before FRBs were established as precision DM probes, the missing-baryon problem was already studied with complementary tracers — e.g., O VI absorption, Ly$\alpha$ absorbers, and the Ly$\alpha$ forest (see the motivation overview in Figure 1.2). As discussed in the previous baryon-census section, FRBs now provide a highly complementary and increasingly powerful avenue in this broader multi-probe framework. McQuinn (2014) was the first to use simulations to demonstrate how FRBs can be utilized to probe cosmology and galaxy physics, laying the theoretical foundation for using FRBs as cosmic probes. They also predicted that, at $z \sim 1$, the sightline-to-sightline DM distribution is approximately Gaussian.[1] With localized FRBs, the observed DM–$z$ relation now provides a practical bridge from the classical baryon-census framework (Section 1.2.3) to simulation-based multi-scale DM decomposition (Macquart et al., 2020).

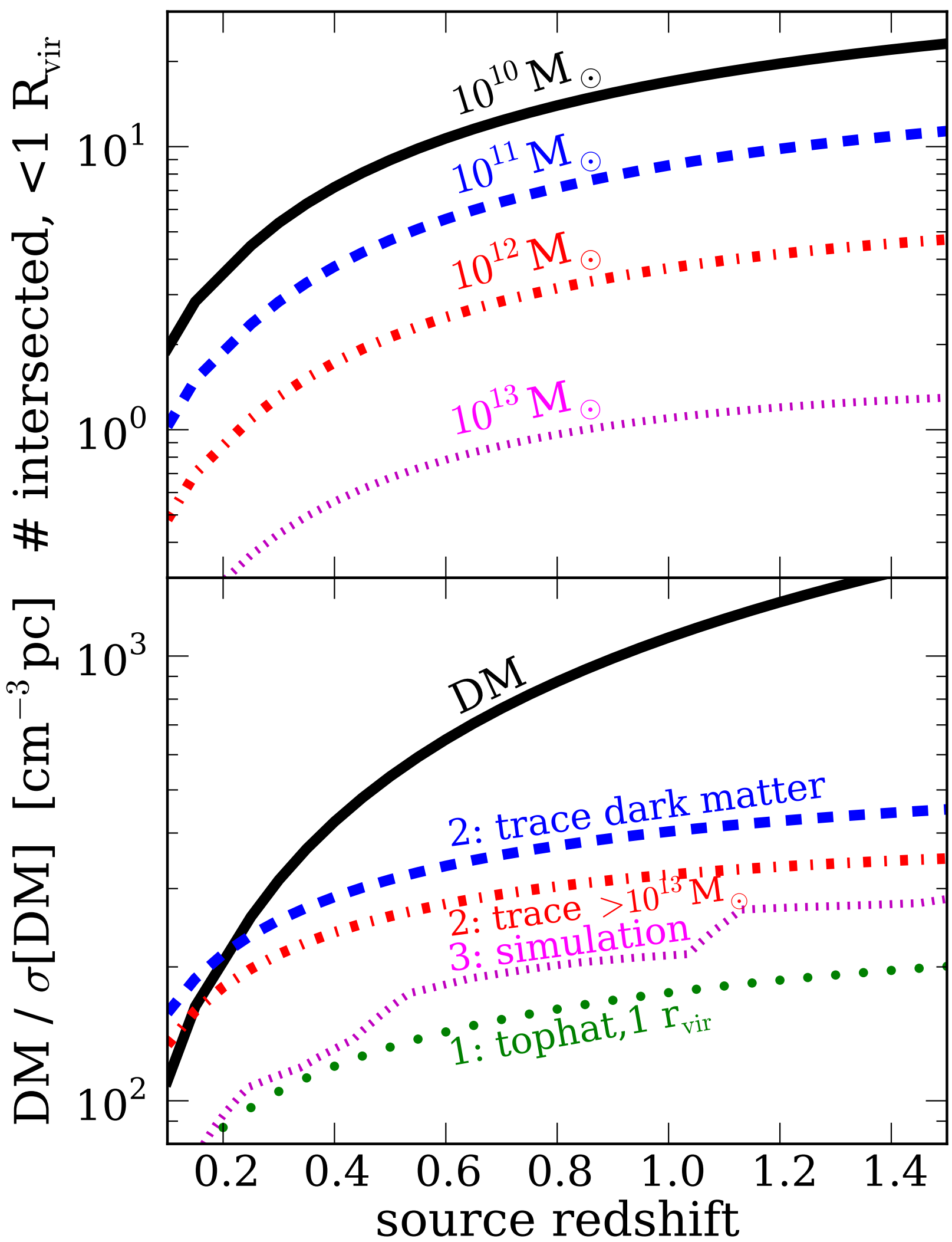

Figure 1.7: Simulated FRB DM$-z$ relation (adapted from McQuinn 2014). The top panel shows the number of halos above specified mass thresholds intersected by an average sightline within $1\,r_{\rm vir}$. The bottom panel shows the estimated $\sigma$[DM] for three baryonic profile models.

[1]Subsequent work with increasingly sophisticated simulations and larger FRB samples suggests that the distribution is often better described as log-normal, or as a skewed near-Gaussian form (Connor et al., 2024, Zhang et al., 2025).

Recent analyses using growing localized FRB samples have tightened observational constraints on cosmic baryons from FRB cosmology. In particular, Yang et al. (2022) and Wang & Wei (2023) used small samples of well-localized FRBs and the observed DM–$z$ relation to constrain $f_{\rm IGM}$ and cosmological parameters, reporting values around $f_{\rm IGM} \approx 0.86$ and $f_{\rm IGM} \approx 0.83$, respectively. In these two studies, the non-local DM contribution is modeled as a single effective IGM term, i.e., contributions along the line of sight are not explicitly decomposed into diffuse IGM and FH components. To compare later foreground-aware analyses, we define $f_{\rm diff} \equiv f_{\rm IGM} + f_{\rm halos}$. With this decomposition, Khrykin et al. (2024b) report $f_{\rm IGM} = 0.63$ and $f_{\rm halos} = 0.16$ (thus $f_{\rm diff} = 0.79$), whereas Connor et al. (2024) find $f_{\rm IGM} \sim 0.80$ and $f_{\rm halos} \sim 0.13$ (thus $f_{\rm diff} \sim 0.93$). Methodologically, Khrykin et al. (2024b) reconstruct foregrounds using an ARGO-based framework with mNFW halo profiles and halo-position information, while Connor et al. (2024) use the enlarged DSA-110 localized FRB sample plus earlier events for population-level constraints.

Cosmological numerical simulations provide the physical context for these observational constraints. LSS simulations consistently show that stellar and AGN feedback redistribute baryons across phases and environments, altering both mean DM and line-of-sight scatter (Sorini et al., 2022, Khrykin et al., 2024a, Zhu & Feng, 2021, Mo et al., 2025, Medlock et al., 2024). These models further demonstrate that FRB dispersion is not solely a smooth IGM integral: it encodes stochastic intersections with overdense structures and the integrated imprint of feedback-driven baryon cycling.

### 1.3.2 Halo and CGM contribution

As illustrated by Figure 1.8, baryons are continuously exchanged between galaxies, the CGM, and the IGM through inflow, outflow, feedback, and baryon recycling. These processes strongly modulate the baryon partition between halo gas and the diffuse IGM, making CGM physics central to any FRB-based baryon census.

Before FRB cosmology, CGM structure was primarily constrained with multi-ion absorption/emission diagnostics at different impact parameters (Figure 1.9; Tumlinson et al. 2017). In the FRB context, however, a low-redshift sightline that intersects only one foreground halo can provide a more direct integral constraint. If Milky Way and host contributions are accurately subtracted (or negligible, e.g., for sources at galaxy outskirts), the reconstructed FRB DM can be used to infer the CGM electron density at a given projected radius and, with a physical profile model, recover the full halo gas density profile.

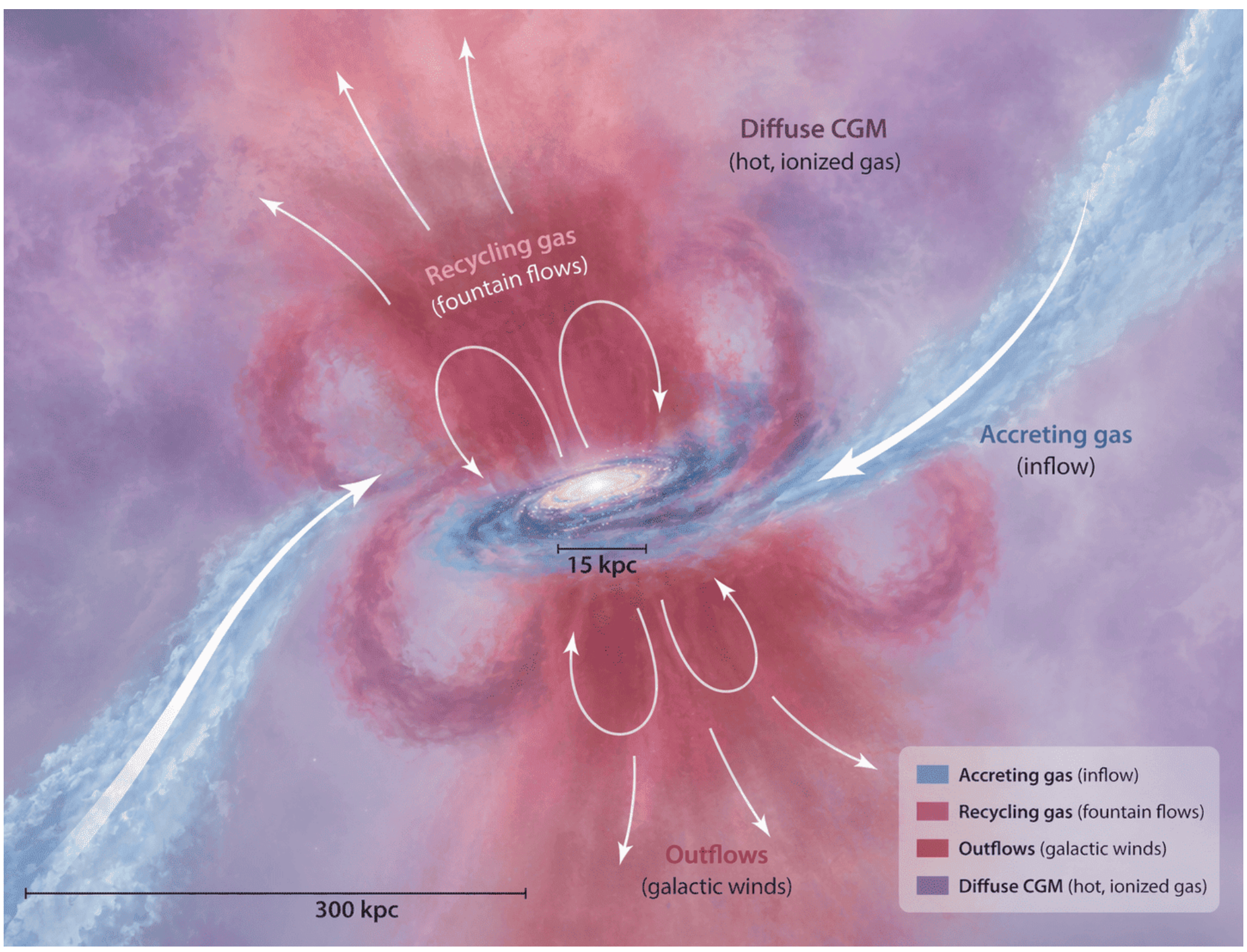

FIGURE 1.8: Conceptual overview of baryon cycling between galaxies, the CGM, and the IGM (inflow, outflow, feedback, and recycling). Figure adapted from Tumlinson et al. (2017).

Intervening galaxy halos and their circumgalactic media (CGM) provide a distinct and non-negligible contribution to FRB DM, especially along sight lines passing near massive systems. Early observational and conceptual work emphasized that foreground galactic environments can measurably enhance dispersion and bias cosmological interpretations if not modeled explicitly (Prochaska et al., 2019, Li et al., 2019b). More recent analyses using CHIME/FRB and localized samples have quantified this halo-linked component through cross-correlation with galaxy distributions and halo reconstructions (Connor & Ravi, 2022, Wu & McQuinn, 2023, Lee et al., 2022, 2023, Simha et al., 2023).

In Simha et al. (2023), four FRB sightlines were analyzed by identifying visible foreground galaxies from imaging/spectroscopic surveys and estimating their DM contributions with an mNFW-based halo model. This framework successfully explains the DM excess for two FRBs (top panels of Figure 1.10), but not for the other two. The residual excess may indicate (i) undetected low-mass foreground systems (e.g., dwarfs), motivating deeper wide-field searches with facilities such as LSST and JWST, or (ii) larger host-galaxy contributions, as discussed in Section 1.3.3.

As discussed above, recent FRB constraints indicate a non-negligible tension in baryon partition between diffuse IGM and halo/CGM reservoirs (Khrykin et al., 2024b, Connor et al., 2024). Here we emphasize that this tension is likely driven by both methodology and physics — including halo modeling choices, sample selection/foreground incompleteness, and feedback microphysics — which motivates the simulation-based interpretation in the next paragraph.

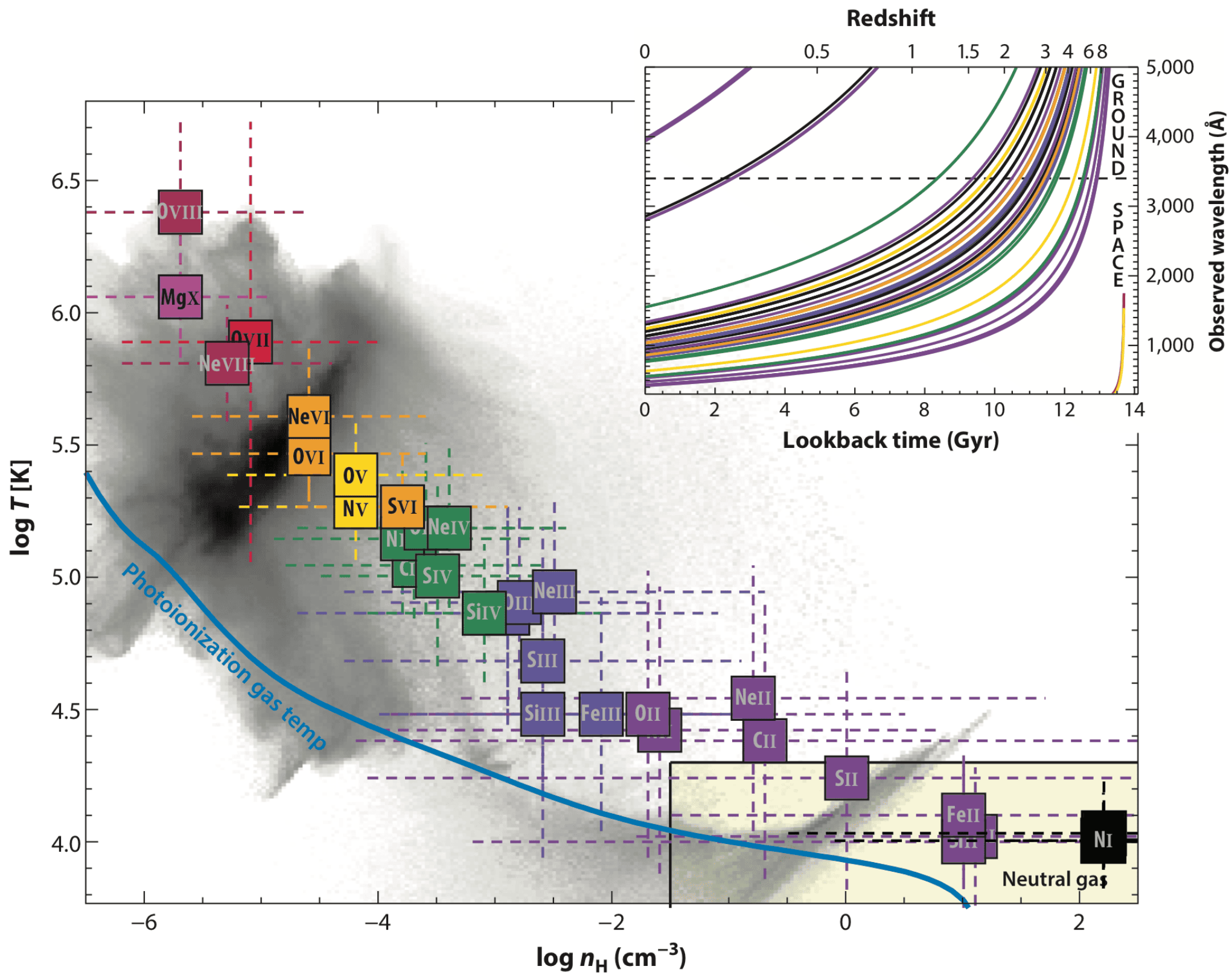

FIGURE 1.9: Observed metal absorption lines in the CGM (from low to high ionization species) as a function of impact parameter, highlighting how different ions trace distinct density/temperature phases from cool gas to warm–hot halo plasma. Figure adapted from Tumlinson et al. (2017).

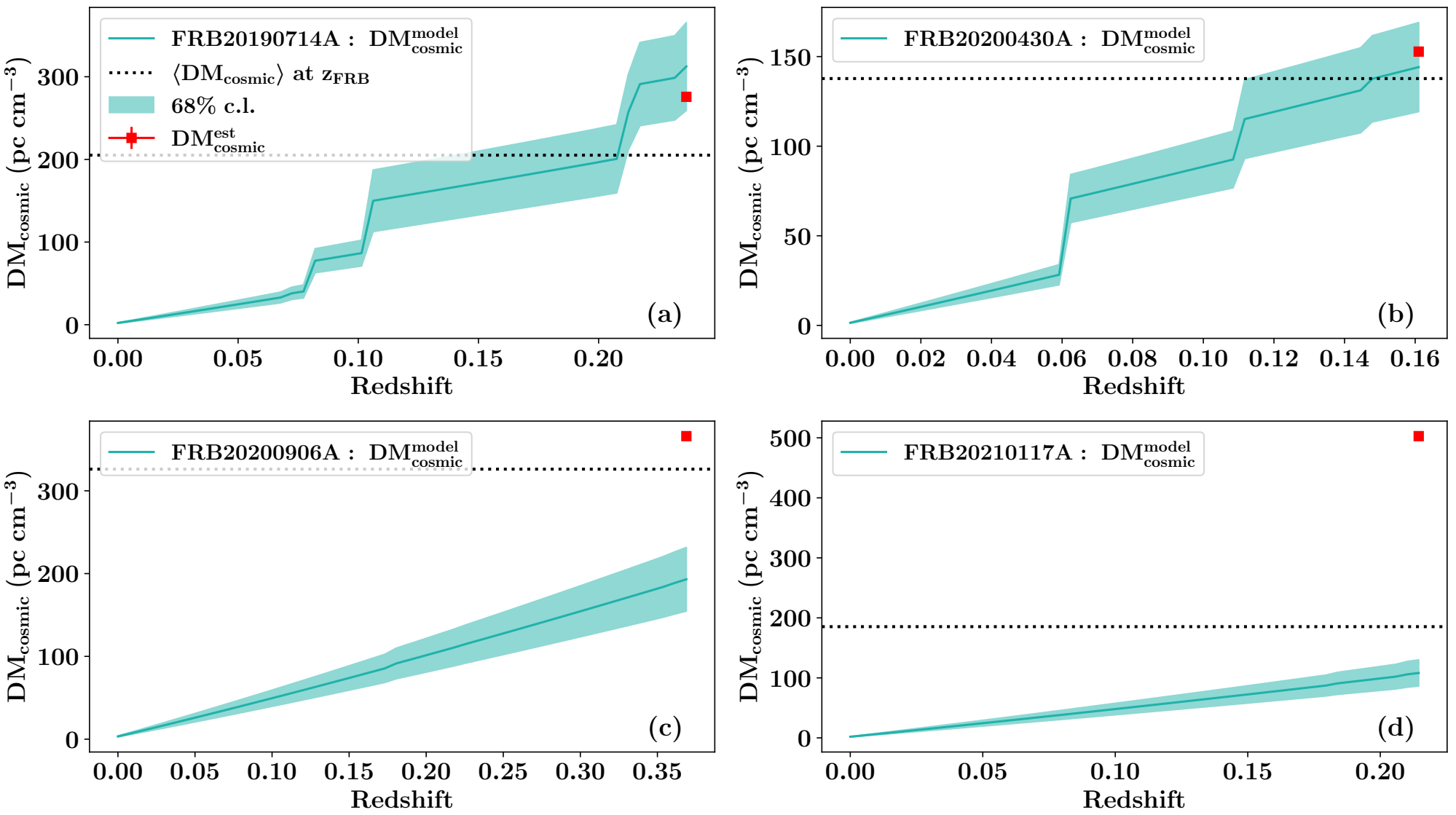

FIGURE 1.10: Foreground-galaxy accounting for FRBs with excess DM relative to the mean DM–$z$ relation. Figure adapted from Simha et al. (2023).

In parallel, recent simulation work has shown that baryonic physics modifies halo structure itself, not just the gas fraction. Feedback can reduce total halo masses and alter concentration–mass trends, implying that FRB halo-crossing statistics and CGM DM calibrations should be interpreted within baryon-modified halo models rather than dark-matter-only expectations (Sorini et al., 2025). This provides additional motivation to combine FRB measurements with simulation-based halo population modeling when inferring $f_{\rm IGM}$.

### 1.3.3 Host galaxy contribution

Beyond foreground structure, each FRB carries a host-galaxy contribution, $\mathrm{DM_{host}}$, that depends on host type, inclination, local ISM phase structure, and source location within the host potential. As localized samples expanded from the first few ASKAP events to large multi-survey catalogs, host demographics have become increasingly diverse, spanning MW-like spirals, dwarfs, and a smaller fraction of quiescent systems (Macquart et al., 2020, Heintz et al., 2020, Bhandari et al., 2022, Gordon et al., 2023, Bhardwaj et al., 2024, Law et al., 2024, Sharma et al., 2024, Connor et al., 2024, Shannon et al., 2024, Amiri et al., 2025).

Empirically, host-frame DM contributions are often of order $\sim 10^2\,\mathrm{pc\,cm^{-3}}$, though with substantial object-to-object scatter (Li et al., 2019b, Zhang, 2023). This scatter is expected from realistic ISM/CGM multiphase structure and from different source offsets relative to dense star-forming regions. Consequently, host modeling is a major uncertainty floor in precision FRB cosmology, particularly at low and intermediate redshift where host terms can rival or exceed diffuse-IGM variance.

To visualize this diversity, Figure 1.11 presents eight representative CHIME/PATH gold-sample systems using host-galaxy color images in a uniform layout (Chime/Frb Collaboration et al., 2025). This set includes the spiral-host case FRB 20231230A, the merger/starburst repeater FRB 20231204A, and the galaxy-cluster environment FRB 20231206A, together with several additional systems showing distinct morphologies. In addition, FRB 20121102 is a robust benchmark for host-galaxy diversity even though it is not shown in Figure 1.11. These examples highlight that host-galaxy DM contributions can differ substantially with both host type and the burst location within the host galaxy.

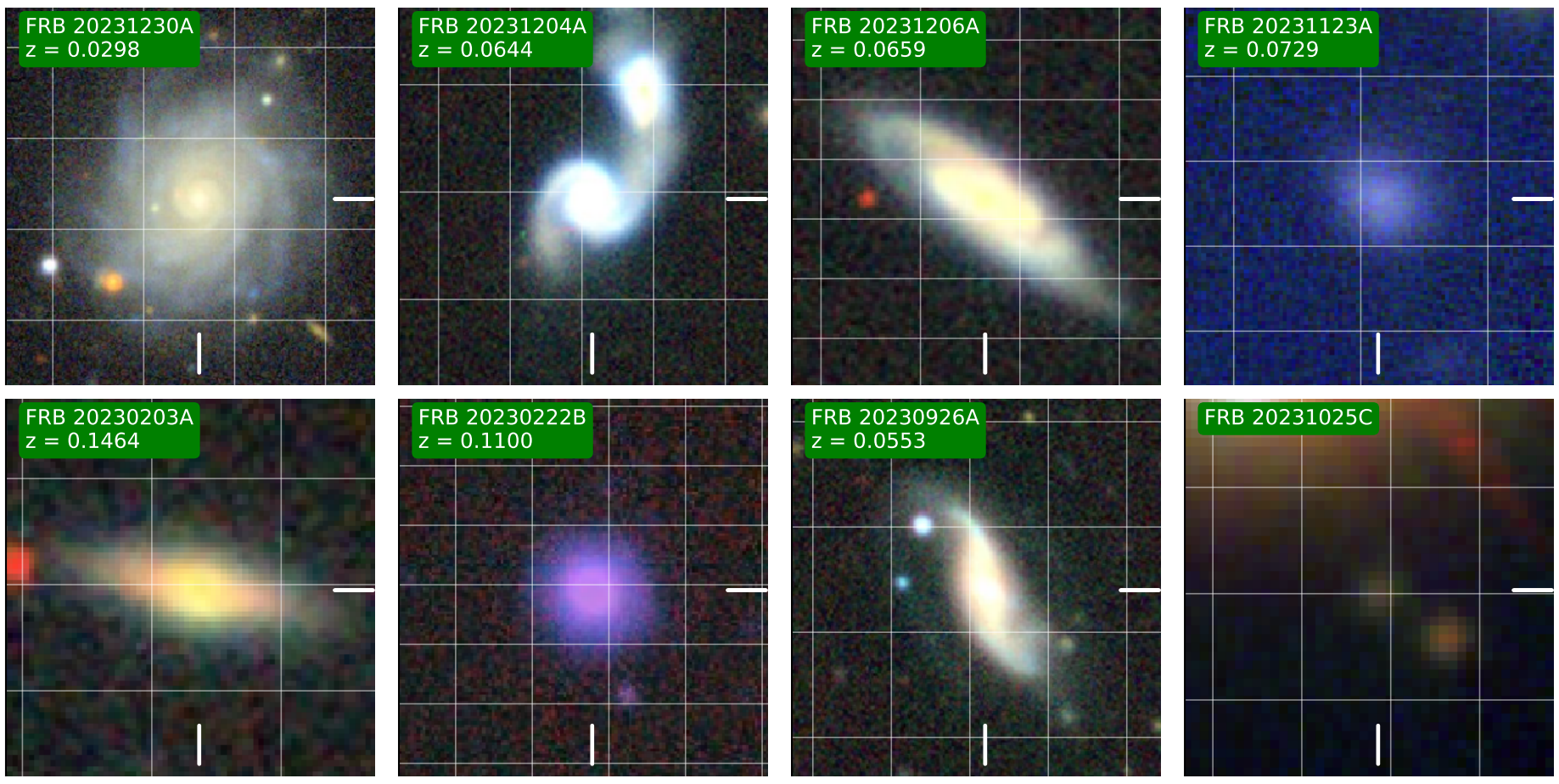

FIGURE 1.11: Representative host-galaxy images from the CHIME/PATH gold sample. The panels are shown in a uniform layout for visual comparison. Figure adapted from Chime/Frb Collaboration et al. (2025)

Simulation studies have therefore focused on statistically characterizing host DM distributions across galaxy populations and progenitor channels (Zhang et al., 2020b, Mo et al., 2023, Theis et al., 2024, Reischke et al., 2024). Benchmark efforts such as AGORA provide useful cross-code validation for galaxy-scale gas structure and strengthen confidence in host-contribution inferences (Kim et al., 2014, Roca-Fàbrega et al., 2021, 2024). In this thesis, we adopt this perspective and treat host DM as a structured, model-dependent distribution rather than a single nuisance constant.

### 1.3.4 Milky Way foreground contribution

Isolating extragalactic terms requires robust subtraction of the Milky Way foreground, including both disk and halo components. Widely used Galactic electron-density models such as NE2001 (Cordes & Lazio, 2002) and YMW16 (Yao et al., 2017) primarily describe the disk contribution and remain standard tools in FRB analyses.

Recent studies have emphasized that the Galactic halo can also contribute non-negligibly to the total foreground DM. In particular, Yamasaki & Totani (2020) presented a framework combining a disk component with an extended spherical hot-halo component motivated by X-ray constraints on the Milky Way CGM. In such models, the halo is treated as diffuse gas with a radial density profile, which can raise the expected Galactic DM, especially toward high Galactic latitudes.

Compared with disk-dominated estimates (typically of order $\sim 20$–$50\,\mathrm{pc\,cm^{-3}}$), models including a hot halo can yield larger foreground values (up to order $\sim 30$–$200\,\mathrm{pc\,cm^{-3}}$ depending on line of sight and parameter choices), highlighting the potential importance of the CGM in the Milky Way DM budget (Yamasaki & Totani, 2020, Prochaska &

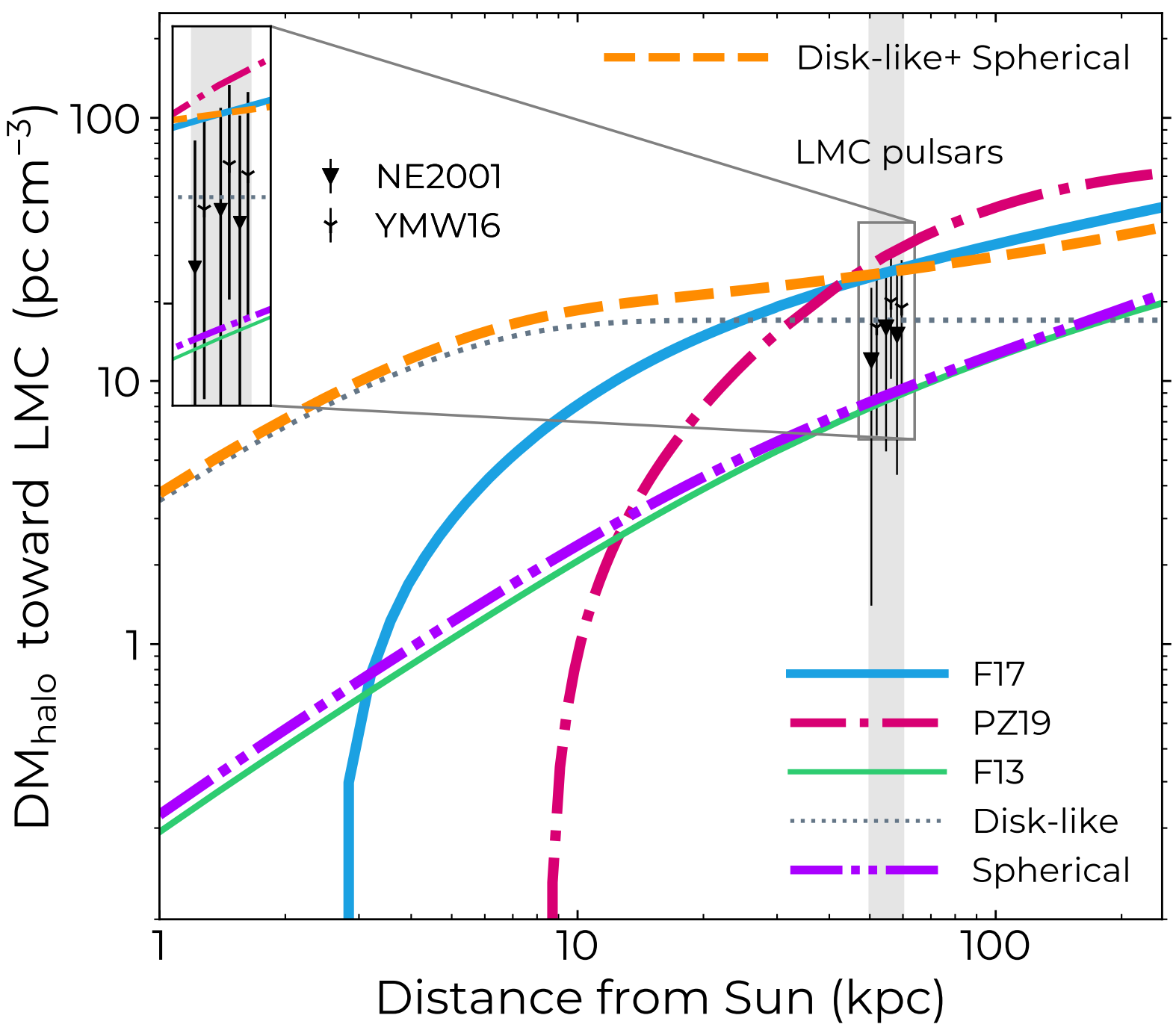

FIGURE 1.12: Illustrative Galactic DM profile for disk+halo foreground modeling. The extended halo component can provide a substantial additional DM contribution beyond disk-only models, especially for high-latitude sightlines. The curves labeled F17, F13, and PZ19 correspond to three commonly adopted halo models: an isothermal gas model with multiple gas phases (Faerman et al., 2017) (F17), an adiabatic gas model with a polytropic index of 5/3 (Fang et al., 2013) (F13), and a modified NFW profile with $\alpha = y_0 = 2$ (Prochaska & Zheng, 2019a) (PZ19). These models span a wide range of halo gas distributions and therefore predict significantly different DM contributions from the Galactic halo. Figure adapted from Yamasaki & Totani (2020).

Zheng, 2019b). At the same time, halo-based estimates remain model dependent because the spatial extent, density profile, and thermodynamic state of the halo gas are still uncertain. In addition, an increasing number of simulations of Milky Way-like galaxies have significantly advanced constraints on both the MW gas profile and the sightline-to-sightline DM distribution, including AGORA simulations (Kim et al., 2014, Roca-Fàbrega et al., 2021, 2024), AGORA-calibrated CROCODILE models (Granizo et al., 2026), NIHAO-UHD simulations (Buck et al., 2019), and FIRE simulations (Hopkins et al., 2018), among others.

Beyond modeling the Milky Way itself, increasing attention has been paid to the contribution of structures in the Local Universe, which play a key role in shaping the all-sky distribution of FRB DMs. In particular, cosmic variance in DM is expected to reflect the inhomogeneous matter distribution in the nearby Universe. Recent studies generally follow two complementary approaches: data-driven reconstructions, which combine galaxy catalogs with empirical scaling relations to estimate foreground DM along individual sightlines, and simulation-based analyses, which provide a statistical framework for quantifying how large-scale structures—including halos, filaments, and clusters—imprint on the DM field. For example, Huang et al. (2025) construct a data-driven model of the Local Universe tailored to specific sightlines, whereas Konietzka et al. (2025) use cosmological simulations to characterize the statistical impact of large-scale structures on the DM distribution.Together, these studies indicate that FRB DMs can exhibit significant anisotropy if the contribution from the Local Group is not properly accounted for, reflecting the strong influence of foreground structures in the local cosmic environment, with this effect being particularly pronounced at low redshift.

### 1.3.5 Source local environment contribution

On sub-galactic scales, the immediate source environment contributes an additional term, $\mathrm{DM}_{\mathrm{source}}$, which is both physically rich and potentially time variable. Candidate contributors include expanding supernova remnants (SNRs), pulsar/magnetar wind nebulae (PWN/MWN), compact H II regions, and small-scale plasma-lensing structures (Yang et al., 2017, Murase et al., 2016, Kashiyama & Murase, 2017, Yang & Zhang, 2017, Cordes et al., 2017, Main et al., 2018). Unlike the cosmological and most host-integrated components, these local terms can evolve measurably over human observing baselines.

Modern numerical modeling is now essential to move beyond idealized analytic scalings. Stellar-evolution calculations and CR-hydro-NEI SNR simulations enable self-consistent treatment of progenitor structure, ionization state, and shock evolution, which can significantly modify predicted $\mathrm{DM}_{\mathrm{source}}(t)$ relative to simple power-law forms (Paxton et al., 2011, 2013, 2015, 2018, 2019, Jermyn et al., 2023, Patnaude et al., 2010, Lee et al., 2012, 2013, 2014, Diesing, 2019, Diesing et al., 2024, Diesing & Gupta, 2025).

As shown in Figure 1.14, the modeled cases share a common late-time secular decline in local environmental contribution, but differ in short-timescale behavior. Motivated by the observed secular decreases in FRB 20190520B ($\mathrm{dDM}/\mathrm{d}t \simeq -12.4\,\mathrm{pc\,cm^{-3}\,yr^{-1}}$; Niu et al. 2025) and FRB 20121102 ($\mathrm{dDM}/\mathrm{d}t = -3.93\,\mathrm{pc\,cm^{-3}\,yr^{-1}}$; Wang et al. 2025a), we emphasize that FRB 20121102 is characterized by an initially flatter phase followed by sustained DM decrease, whereas FRB 20190520B shows a more steadily decreasing DM trend. A similar steadily decreasing DM has also been reported for FRB 20220529A, but with a shallower slope ($\sim -0.881\,\mathrm{pc\,cm^{-3}\,yr^{-1}}$; Pandhi et al. (2026)) than FRB 20190520B. In this comparison, we focus on DM evolution; the RM behavior of FRB 20190520B is not used here as the primary discriminator.

We also note recent attempts to place active repeaters within a unified framework, including FRB 20190417A, an active repeater with periodic activity that shows oscillatory RM with sign reversals and is generally interpreted in a magnetar–massive-star binary

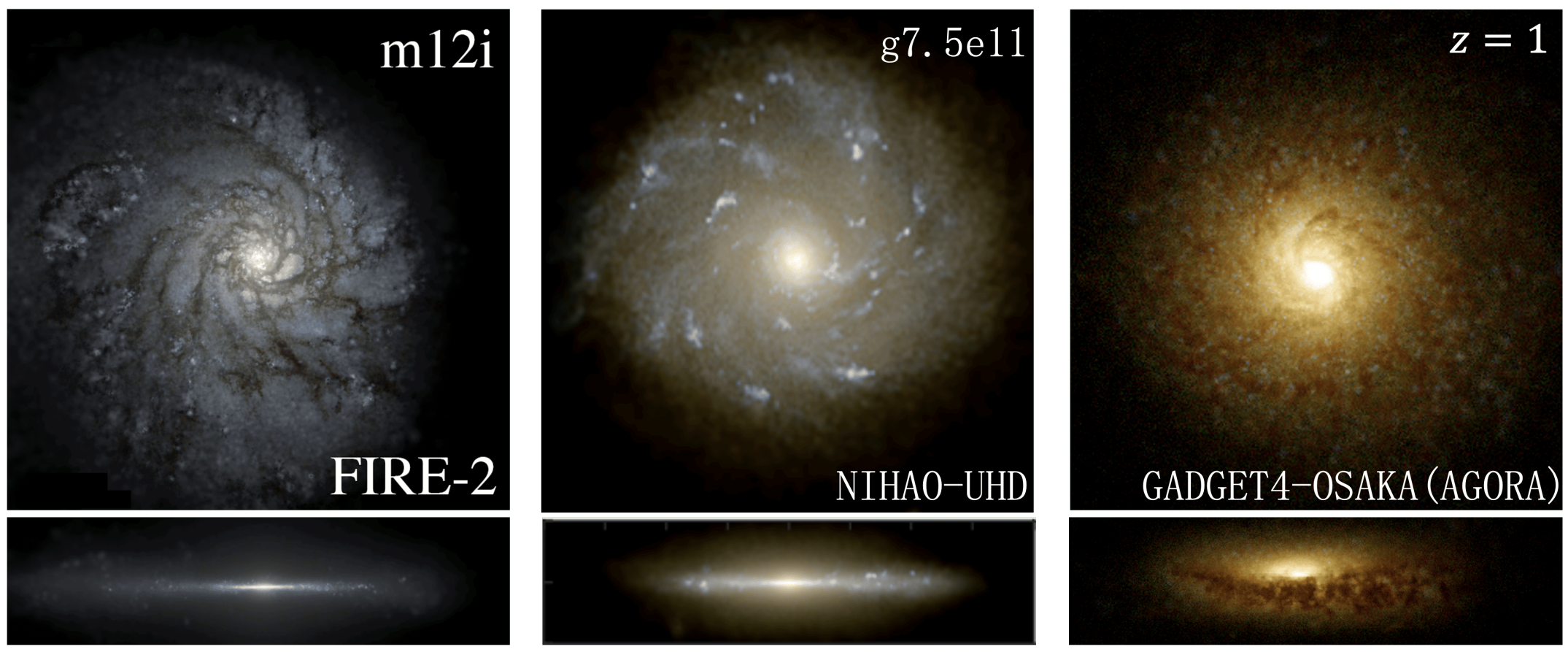

FIGURE 1.13: Comparison of high-resolution zoom-in simulations of Milky Way-like systems used to assess the Galactic foreground DM structure. From left to right: FIRE-2 (at $z = 0$, adapted from Hopkins et al. (2018)), NIHAO-UHD (at $z = 0$, adapted from Buck et al. (2019)), and AGORA GADGET4-OSAKA (at $z = 1$, adapted from Jung et al. (2025)).

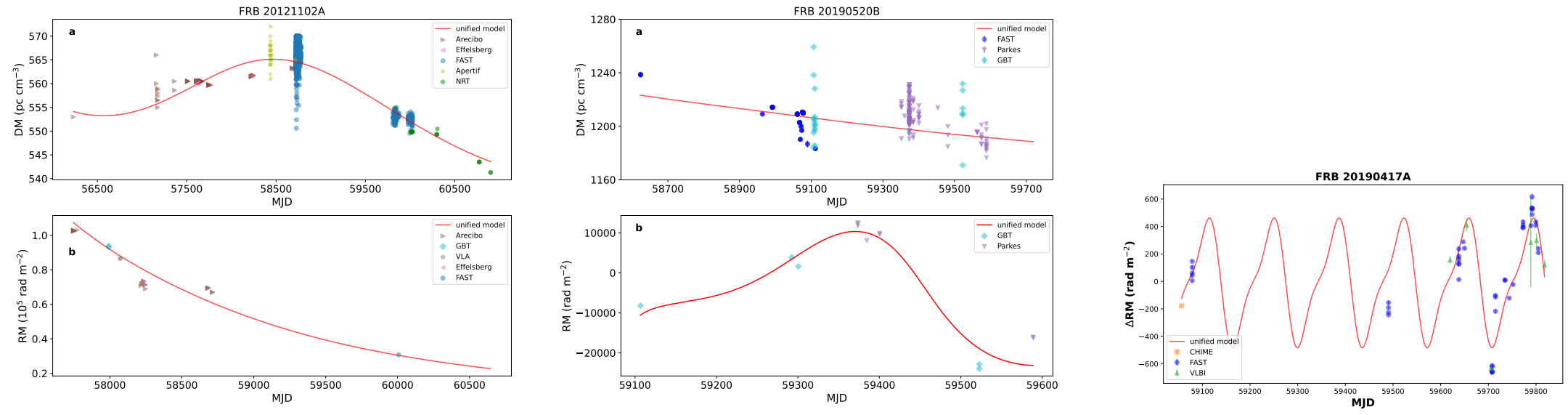

FIGURE 1.14: Illustrative examples of time-variable local-source contributions, showing coupled DM/RM evolution for FRB source-environment scenarios. Left: FRB 20121102-like case; middle: FRB 20190520-like case; right: generic binary-source setup. These examples highlight why $\mathrm{DM}_{\mathrm{source}}$ should be treated as dynamically evolving in repeater analyses. Figures adapted from Wang et al. (2025b).

environment, as well as persistent radio source (PRS)-associated systems (Wang et al., 2025b).

Therefore, the multi-scale FRB DM problem is inherently a coupled decomposition task: cosmological diffuse gas, intervening halos/CGM, host-galaxy structure, and local source physics all contribute with different redshift scalings, stochasticities, and temporal behaviors. In the following chapters, we use LSS and zoom-in simulations to constrain the non-local terms and dedicated SNR modeling to quantify the physically time-dependent local term, aiming for a unified interpretation of FRB dispersion across scales.

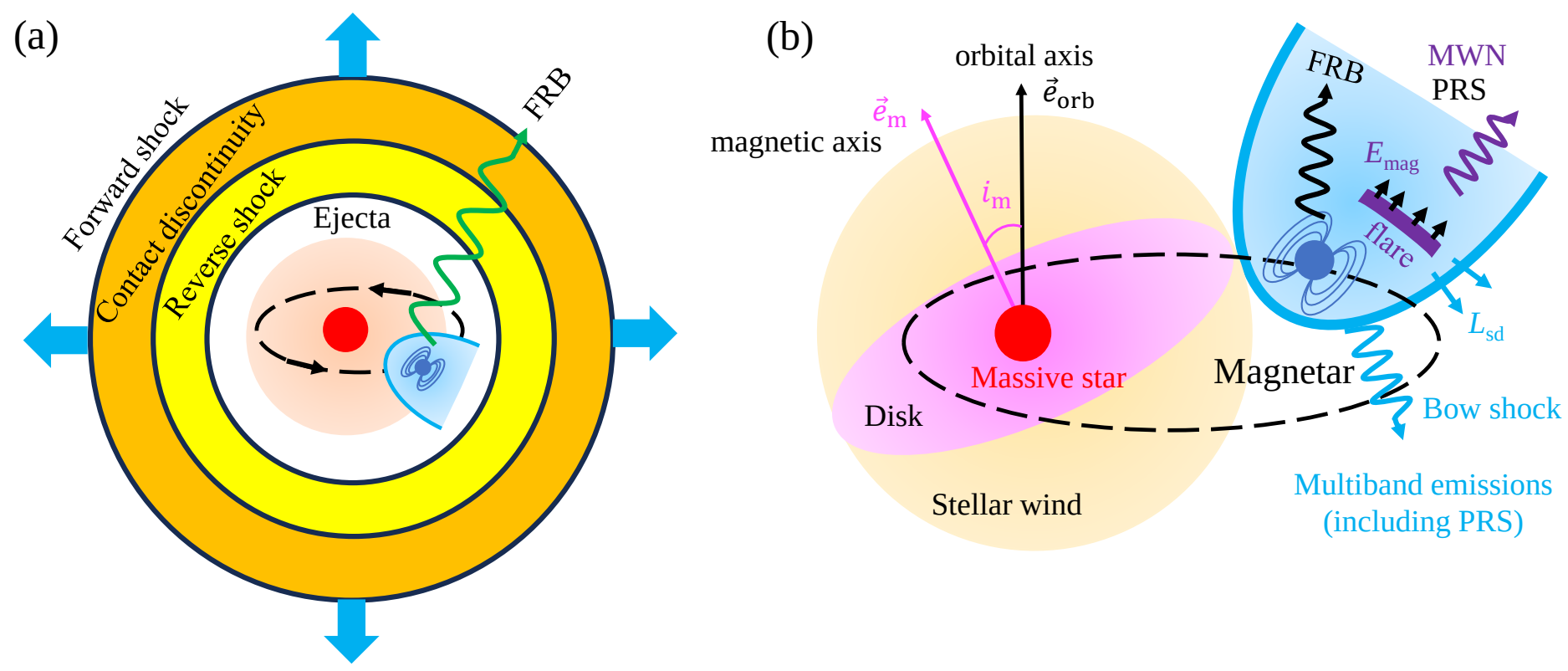

Figure 1.15: Schematic diagram of a unified source model for active repeaters. Panel (a): a young magnetar–massive-star binary embedded in a magnetar wind nebula and supernova remnant, where FRB signals propagate through stellar wind and supernova ejecta and can produce DM/RM variations, while magnetar energy injection powers persistent radio emission. Panel (b): geometry of the collision between the companion stellar wind and the magnetar wind; the bow shock can power multi-band emission, and an equatorial decretion disk (for Oe/Be-like companions) can add extra DM/RM contributions. Figure adapted from Wang et al. (2025b).

## 1.4 Thesis Objectives and Structure

In this thesis, the primary goal is to investigate how FRBs can be used as cosmological probes of baryons across multiple scales, from the large-scale cosmic web to local source environments. In particular, this work aims to quantify the contributions of different baryonic components to the observed DM and to assess the associated systematic uncertainties in FRB-based cosmological analyses.

The main objectives are summarized as follows:

(i) Characterize the distribution of baryons in cosmic web using cosmological large-scale simulations, establish the statistical properties of the DM–$z$ relation, and compare them with observational constraints.

(ii) Quantify the impact of foreground galaxies, especially the CGM contribution, on the observed DM and the anisotropy-induced cosmic variance.

(iii) Use zoom-in simulations to estimate host-galaxy DM contributions and their distributions over the full $4\pi$ solid angle, and test which probability density function (PDF) best describes these distributions.

(iv) Investigate contributions from the local source environment, especially SNRs, to assess whether SNR-only models can explain the observed temporal evolution of DM and RM, and evaluate how these trends constrain FRB escape and activity timescales.

To achieve these goals, this thesis combines cosmological simulations with detailed *forward-modeling* approch of local environments. Large-scale simulations from the CROCODILE project are used to construct the light-cone realizations and to study the statistical properties of FRB DMs, and using DM–$z$ relation to constrain the $f_{\rm diff}$ and searching mesures to estimate the intrinsic $f_{\rm IGM}$, meanwhile, different scales's zoom-in simulations are used to consider differnt host halo (HH) cases, from dwarf to cluster-size galaxies. On the other hand, the high-resolution one-dimensional (1D) hydrodynamical (HD) simulations of SNR are employed to investigated the evolution of the source-local DM contribution.

The structure of this thesis is organized as follows. Chapter 1 introduces the cosmological background, the missing baryon problem, and the use of FRBs as cosmological probes, as well as the multi-component nature and multi-scales contribution of DM. Chapter 2 describes the numerical methods and simulation framework,including the cosmological HD simulations, the GADGET-based code, and the construction of FRB light cones and DM calculations. Chapter 3 presents the main results on FRB cosmology from the CROCODILE simulations, including the DM–$z$ relation, the impact of AGN feedback, and the contributions from halos and host galaxies. Chapter 4 focuses on the local source environment, modeling the evolution of $\rm DM_{source}$ in supernova remnants and its implications for FRB observations. Finally, Chapter 5 summarizes the main results and presents a unified framework for understanding FRB DMs across different scales, along with a discussion of limitations and future prospects.

# Chapter 2

# Cosmological Numerical Simulation Framework and Data-analysis Methodology

## 2.1 Cosmological Hydrodynamical Simulations

The simulations are built on the GADGET code framework. The hydrodynamical evolution is solved using the standard smoothed particle hydrodynamics (SPH) formalism. The baseline runs use GADGET3-OSAKA, an evolved branch of GADGET-2 (Springel, 2005), with updated prescriptions for star formation and supernova feedback (Shimizu et al., 2019, Oku et al., 2022). For the CROCODILE simulations[1], GADGET4-OSAKA, the GADGET4-Osaka SPH code (Romano et al., 2022a,b), is used; this code is based on a modified version of GADGET-4 (Springel et al., 2021). In this setup, the updated SN feedback model of Oku et al. (2022) is adopted, and AGN feedback is included following Rosas-Guevara et al. (2015), Schaye et al. (2014), and Crain et al. (2015), providing a comprehensive set of runs with diverse feedback configurations. A detailed description of the CROCODILE simulation suite is presented in Section 2.2.

### 2.1.1 Smoothing Kernels and the Standard SPH Formalism

In standard SPH, fluid quantities are estimated through kernel interpolation. The smoothing kernel is written as

[1] https://sites.google.com/view/crocodilesimulation/home

$$W(\mathbf{r}, h) = \frac{1}{h^{\nu}} \, w\left(\frac{|\mathbf{r}|}{h}\right), \tag{2.1}$$

where $h$ is the smoothing length and $w(q)$ is a dimensionless kernel function with compact support. The three commonly used Wendland kernels in 3D are (Dehnen & Aly, 2012)

$$w_{\mathrm{C2}}(q) = \begin{cases} \frac{21}{\pi}(1-q)^4(4q+1), & 0 \le q \le 1, \\ 0, & q > 1, \end{cases}$$

$$w_{\mathrm{C4}}(q) = \begin{cases} \frac{495}{32\pi}(1-q)^6\left(\frac{35}{3}q^2 + 6q + 1\right), & 0 \le q \le 1, \\ 0, & q > 1, \end{cases} \tag{2.2}$$

$$w_{\mathrm{C6}}(q) = \begin{cases} \frac{1365}{64\pi}(1-q)^8\left(32q^3 + 25q^2 + 8q + 1\right), & 0 \le q \le 1, \\ 0, & q > 1. \end{cases}$$

In this work, the Wendland $C^4$ kernel is adopted. These kernels are compactly supported, continuously differentiable, and generally provide improved numerical convergence compared to traditional spline kernels. After defining the kernel function, any field value $F(\mathbf{r})$ at position $\mathbf{r}$ can be approximated by smoothed kernel interpolation as

$$F_{\mathrm{interp}}(\mathbf{r}) \simeq \sum_j \frac{m_j}{\rho_j} F_j \, W(\mathbf{r} - \mathbf{r}_j, h), \tag{2.3}$$

where $F_j = F(\mathbf{r}_j)$, and $j$ indexes the neighboring particles at $\mathbf{r}_j$ that contribute to the interpolation at position $\mathbf{r}$.

To resolve both clustered and diffuse regions with approximately constant local resolution quality, the smoothing length is adjusted so that each particle contains an approximately fixed number of neighbors:

$$\frac{4}{3}\pi h_i^3 n_i = N_{\mathrm{ngb}}, \tag{2.4}$$

where

$$n_i = \sum_{j=1}^{N} W_{ij}(h_i). \tag{2.5}$$

is the SPH particle number density, $h_i$ is the smoothing length of particle $i$, and $W_{ij}(h_i) \equiv W(|\mathbf{r}_i - \mathbf{r}_j|; h_i)$ is the kernel function. The parameter $N_{\mathrm{ngb}}$ is the target neighbor number that controls the effective spatial resolution and noise level: smaller $N_{\mathrm{ngb}}$ gives higher

spatial resolution but noisier estimates, while larger $N_{\rm ngb}$ gives smoother estimates at lower resolution. Following the CROCODILE setup, $N_{\rm ngb} = 120 \pm 2$ is adopted.

### 2.1.2 Gravity, hydrodynamics, cooling and star formation processes

#### 2.1.2.1 Gravity

In the CROCODILE runs, gravity is computed for both dark matter and baryons by solving the Poisson equation,

$$\nabla^2 \Phi = 4\pi G \rho, \tag{2.6}$$

with the GADGET4-OSAKA implementation based on the GADGET TreePM framework (Springel, 2005, Springel et al., 2021). Following the CROCODILE setup, the same gravitational softening is adopted for all particle species, with a comoving value $\epsilon_{\rm grav} = 3.38\,h^{-1}$ ckpc and a physical floor of $0.5\,h^{-1}$ kpc.

The gravitational force is split into a long-range PM part and a short-range tree part (Springel, 2005). The PM solver computes large-scale forces with CIC mass assignment and FFT-based Poisson inversion, which also handles periodic boundaries efficiently. The short-range correction is evaluated with an oct-tree walk using a relative opening criterion to maintain force accuracy in both dense and diffuse regions.

The split scale $r_s$ is introduced through the standard TreePM decomposition (Xu, 1995, Springel, 2005). In Fourier space, the long-range potential is written as

$$\phi_k^{\rm long} = \phi_k \exp\left(-k^2 r_s^2\right), \tag{2.7}$$

while the short-range real-space contribution takes the form

$$\phi^{\rm short}(\mathbf{x}) = -G \sum_i \frac{m_i}{r_i} \,\mathrm{erfc}\left(\frac{r_i}{2r_s}\right). \tag{2.8}$$

For the PM force calculation, the mesh acceleration is obtained by finite differencing the potential (four-point stencil), e.g.

$$\left.\frac{\partial \phi}{\partial x}\right|_{i,j,k} = \frac{1}{\Delta x}\left[\frac{2}{3}\left(\phi_{i+1,j,k} - \phi_{i-1,j,k}\right) - \frac{1}{12}\left(\phi_{i+2,j,k} - \phi_{i-2,j,k}\right)\right]. \tag{2.9}$$

Correspondingly, the short-range force is modulated by

$$f(r) = 1 - \text{erfc}\left(\frac{r}{2r_s}\right) - \frac{r}{\sqrt{\pi} r_s} \exp\left(-\frac{r^2}{4r_s^2}\right). \tag{2.10}$$

Here, $\text{erfc}(x)$ denotes the complementary error function,

$$\text{erfc}(x) = 1 - \text{erf}(x) = \frac{2}{\sqrt{\pi}} \int_x^\infty e^{-t^2}\, dt. \tag{2.11}$$

In the TreePM decomposition, this function suppresses the short-range contribution at large $r$, ensuring a smooth transition between the PM and tree forces around $r_s$. This hybrid strategy controls force errors over a wide dynamic range while significantly reducing the tree-walk cost.

#### 2.1.2.2 Hydrodynamics

Gas dynamics is evolved with the SPH formulation implemented in GADGET4-OSAKA. The hydrodynamic evolution is governed by

$$\frac{d\rho}{dt} = -\rho \nabla \cdot \mathbf{v}, \tag{2.12}$$

$$\frac{d\mathbf{v}}{dt} = -\frac{\nabla P}{\rho} + \mathcal{K}, \tag{2.13}$$

$$\frac{du}{dt} = -\frac{P}{\rho} \nabla \cdot \mathbf{v} + \Gamma - \Lambda + \mathcal{T}. \tag{2.14}$$

where $P$ and $\mathbf{v}$ are pressure and velocity, respectively, and $d/dt = \partial/\partial t + \mathbf{v} \cdot \nabla$ represents the Lagrangian (material) derivative. All quantities are evolved in comoving coordinates. Additional source terms include kinetic feedback ($\mathcal{K}$) and thermal feedback ($\mathcal{T}$) associated with supernovae (SNe) and AGNs, which are described in detail in the Section 2.1.3.

The pressure–energy formulation of SPH is adopted. The equation of motion is

$$\frac{d\mathbf{v}_i}{dt} = -\sum_{j=1}^{N} m_j \left[ f_{ij} \frac{u_j}{u_i} \frac{\bar{P}_i}{\bar{\rho}_i^2} \nabla_i W_{ij}(h_i) + f_{ji} \frac{u_i}{u_j} \frac{\bar{P}_j}{\bar{\rho}_j^2} \nabla_i W_{ij}(h_j) \right], \tag{2.15}$$

and the thermal-energy equation is

$$\frac{du_i}{dt} = \sum_{j=1}^{N} m_j f_{ij} \frac{u_j}{u_i} \frac{\bar{P}_i}{\bar{\rho}_i^2} (\mathbf{v}_i - \mathbf{v}_j) \cdot \nabla_i W_{ij}(h_i), \tag{2.16}$$

with

$$\bar{P}_i = \sum_{j=1}^{N} (\gamma - 1) m_j u_j W_{ij}(h_i), \tag{2.17}$$

and

$$f_{ij} = \left[ 1 + \frac{h_i}{3n_i} \frac{\partial n_i}{\partial h_i} \right]^{-1}. \tag{2.18}$$

Here, $\bar{P}_i$ is the smoothed pressure, and $f_{ij}$ is the correction factor that accounts for smoothing-length variation.

The viscous force added to the equation of motion as

$$\left. \frac{d\mathbf{v}_i}{dt} \right|_{\mathrm{visc}} = - \sum_{j=1}^{N} m_j \, \Pi_{ij} \, \nabla_i \bar{W}_{ij}, \tag{2.19}$$

where $\Pi_{ij}$ denotes the viscosity factor symmetric in $i$ and $j$ and

$$\bar{W}_{ij} = \frac{1}{2} \left[ W_{ij}(h_i) + W_{ij}(h_j) \right]. \tag{2.20}$$

is a symmetized kernel, which some research prefer to adopt $\bar{W}_{ij} = W_{ij}([h_i + h_j]/2)$. The thermal energy per unit mass is defined as

$$\left. \frac{du_i}{dt} \right|_{\mathrm{visc}} = \frac{1}{2} \sum_{j=1}^{N} m_j \, \Pi_{ij} \, \mathbf{v}_{ij} \cdot \nabla_i \bar{W}_{ij} \tag{2.21}$$

which conpensate the work done against the viscous force to conserve the energy, where $\mathbf{v}_{ij} = \mathbf{v}_i - \mathbf{v}_j$. To evaluate the viscosity contribution at particle $i$, the artificial-viscosity form of Frontiere et al. (2017) is adopted:

$$\Pi_i = \frac{-C_l \, c_i \, \mu_i^{\mathrm{AV}} + C_q \, (\mu_i^{\mathrm{AV}})^2}{\rho_i}, \tag{2.22}$$

$$\mu_i^{\mathrm{AV}} = \begin{cases} \dfrac{h_i \, \mathbf{v}_{ij} \cdot \mathbf{r}_{ij}}{|\mathbf{r}_{ij}|^2 + \varepsilon h_i^2}, & \mathbf{v}_{ij} \cdot \mathbf{r}_{ij} < 0, \\ 0, & \text{otherwise.} \end{cases} \tag{2.23}$$

Here, $C_l$ and $C_q$ are the linear and quadratic viscosity coefficients, $c_i$ is the local sound speed, and $\varepsilon = 0.1$ is introduced to avoid division by zero. The quantity $\mu_i^{\mathrm{AV}}$ is the artificial-viscosity kinematic compression term, which is activated only for approaching particle pairs ($\mathbf{v}_{ij} \cdot \mathbf{r}_{ij} < 0$) and vanishes otherwise; it is therefore not a molecular weight.

The superscript “AV” explicitly distinguishes it from the ionized-gas mean molecular weight $\mu_i$ used in subsequent section.

The velocity field is reconstructed using the high-order velocity-gradient estimator of Hu et al. (2014). Artificial conduction is implemented following Price (2008), with a velocity-based conductivity signal speed adopted from Wadsley et al. (2008) and the conduction limiter of Borrow et al. (2022). GADGET4-OSAKAfurther tracks dust production and destruction with the full grain-size distribution model of Romano et al. (2022a); detailed dust physics is beyond the scope of this thesis.

GADGET4-OSAKA further tracks dust production and destruction with the full grain-size distribution model of Romano et al. (2022a); however, detailed dust physics is beyond the scope of this thesis.

#### 2.1.2.3 Cooling

The heating ($\Gamma$) and cooling ($\Lambda$) terms are computed with the GRACKLE library (Smith et al., 2017). Its non-equilibrium primordial chemistry network (12 species: H, D, He, $H_2$, HD, and their ions) is employed, together with metal-line cooling precomputed with the photoionization code CLOUDY (Ferland et al., 2013) and photo-heating/photo-ionization under a spatially uniform UV background (UVB). For the UVB, the model of Haardt & Madau (2012) is adopted. In this setup, the UVB is switched on at $z = 8$ and reaches full strength by $z = 6$, implying that the main phase of reionization occurs around $z \sim 7$.

Additional heating from cosmic rays (CRs) or stellar radiation is not included, and neutral-hydrogen self-shielding of the UVB is neglected to avoid artificial overcooling. Chemical enrichment is tracked with 12 elements (H, He, C, N, O, Ne, Mg, Si, S, Ca, Fe, Ni) from core-collapse SNe, Type Ia SNe, and AGB stars. The metallicity floor is set to $Z_{\rm floor} = 10^{-6} Z_\odot$, with $Z_\odot = 0.0134$ (Asplund et al., 2009). Radiative cooling and photo-heating are included with tabulated rates under a redshift-dependent UV background. The cooling/heating source terms are coupled to the gas energy evolution, while feedback-related thermal regulation is discussed in Section 2.1.3.

#### 2.1.2.4 Star formation model

Star formation is allowed when the hydrogen number density exceeds the threshold density $n_{\rm SF} = 0.1\,{\rm cm}^{-3}$ and the temperature is below the threshold temperature, $T_{\rm SF} = 10^4\,{\rm K}$. Each gas particle is allowed to spawn up to $n_{\rm spawn}$ star particles, and the mass of each star particle is given by

$$m_* = \frac{m_{\rm gas}}{n_{\rm spawn}}, \tag{2.24}$$

where $m_{\rm gas}$ is the gas-particle mass; $n_{\rm spawn} = 2$ is adopted.

The simple stellar population (SSP) approximation is adopted, in which each star particle represents a coeval population with uniform metallicity. The initial mass function (IMF) is modeled as a function of metallicity and redshift following Chon et al. (2022). For $10^{-4} \leq Z \leq 10^{-1}$ and $0 \leq z \leq 20$, the excess high-mass fraction relative to a Salpeter-like IMF is

$$f_{\rm massive} = 1.07(1 - 2^x) + 0.04 \times 2.67^x\, z, \tag{2.25}$$

with

$$x = \min\{1 + \log_{10}(Z/Z_\odot),\, 0\}\,. \tag{2.26}$$

The metallicity is capped at $Z \leq 0.1\, Z_\odot$, above which the IMF approaches the present-day Salpeter-like form. The IMF is constructed by combining a Chabrier IMF (Chabrier, 2003) over $0.1\, M_\odot < M < 100\, M_\odot$ with an additional log-flat component over $5\, M_\odot < M < 100\, M_\odot$, weighted by $f_{\rm massive}$.

At star-particle formation, $f_{\rm massive}$ is set by the progenitor-gas metallicity and formation redshift, and is used to compute yields from supernovae (SNe) and asymptotic giant branch (AGB) stars. A metallicity floor for yields is adopted:

$$Z_{\rm yield,floor} = 0.03\, Z_\odot, \tag{2.27}$$

which accounts for unresolved early enrichment and reproduces the observed metallicity of galaxies with $M_* \sim 10^7\, M_\odot$.

The star formation rate is described by the local Schmidt law (Schmidt, 1959),

$$\dot{\rho}_* = \epsilon_* \frac{\rho}{t_{\rm ff}}, \tag{2.28}$$

where

$$t_{\rm ff} = \sqrt{\frac{3\pi}{32 G \rho}}, \tag{2.29}$$

with $\epsilon_* = 0.01$. Star particles are spawned stochastically according to this rate (Katz, 1992, Springel & Hernquist, 2003, Shimizu et al., 2019).

### 2.1.3 Baryonic feedback subphysics

#### 2.1.3.1 Stellar feedback

Core-collapse supernova (CCSN) feedback is modeled using the stellar-yield tables of Nomoto et al. (2013), which provide nucleosynthetic yields and explosion energies for progenitors in the mass range 13–40 $M_\odot$; massive stars ($\gtrsim 8\, M_\odot$) are assumed to explode as CCSNe with a typical energy of $\sim 10^{51}$ erg, while a fraction of progenitors with masses 20–40 $M_\odot$ are assumed to explode as hypernovae (HNe) with higher energies of $\sim 10^{52}$ erg. The HN fraction is treated as metallicity dependent and increases toward low metallicity, leading to a higher specific supernova energy through both enhanced HN contribution and a top-heavy initial mass function. The time evolution of energy and metal release is computed using stellar population synthesis via `CELib` (Saitoh, 2017), so feedback is distributed over multiple discrete events rather than injected instantaneously; feedback energy and metals are deposited into surrounding gas particles following

Oku et al. (2022), which includes both local mechanical feedback and galactic-wind feedback and captures gradual supernova energy injection, baryon redistribution through outflows, and metal enrichment of the interstellar and circumgalactic media.

The supernova (SN) feedback is implemented through a combination of a mechanical feedback model and an SN-driven galactic wind model (Oku et al., 2022). The mechanical feedback model accounts for the momentum injection from SN remnants during unresolved Sedov–Taylor and snowplow phases. The terminal momentum deposited by a single SN is given by

$$p_{\rm SN} = 1.75 \times 10^5 \, M_\odot \, {\rm km\,s^{-1}} \, n_{\rm H}^{-0.05} \left( \frac{Z}{Z_\odot} \right)^{-0.072}, \tag{2.30}$$

where $n_{\rm H}$ is the ambient hydrogen number density and $Z$ is the metallicity. This scaling reflects the weak dependence of the final momentum on radiative cooling efficiency, with higher density and metallicity leading to slightly reduced momentum injection. The SN energy and metals are distributed to neighboring gas particles using Voronoi-based weights, ensuring local conservation of momentum and capturing the cumulative effects of clustered SN explosions.

On larger scales, unresolved feedback is modeled through an SN-driven galactic wind prescription that describes hot ($T \gtrsim 10^6\,{\rm K}$) outflows. The wind properties are tied to local galaxy scaling relations, where the mass loading factor $\eta = \dot{M}_{\rm wind}/\dot{M}_\star$ depends on stellar mass and star formation rate, and the wind velocity is set proportional to the escape velocity of the host system. The wind phase is decoupled hydrodynamically to allow particles to escape dense regions, mimicking the breakout of SN-driven superbubbles.

The total SN energy injection is divided between local mechanical feedback and large-scale winds, capturing both the momentum-driven regulation of the interstellar medium (ISM) and the redistribution of baryons into the circumgalactic medium (CGM). This combined framework enables a self-consistent treatment of star formation suppression, metal enrichment, and baryon transport across multiple scales.

Type-Ia supernova (SN) feedback is modeled using a delay-time distribution $\Psi(t) \propto t^{-1}$ with a minimum delay of 40 Myr and a fixed explosion energy of $10^{51}\,{\rm erg}$ per event (Seitenzahl et al., 2013, Maoz et al., 2012), where the injected energy is converted into momentum following Eq. (2.30) without invoking a galactic wind component. Asymptotic giant branch (AGB) stars are included through stellar mass loss and metal enrichment using yield tables over 1–8 $M_\odot$ (Karakas, 2010, Doherty et al., 2014), with no mechanical feedback due to their dynamically weak winds. Overall, both Type-Ia SNe and AGB stars primarily contribute to chemical enrichment rather than driving significant gas redistribution, and their impact on the large-scale baryon distribution and FRB DM cosmology is subdominant compared to AGN feedback; therefore, they are not discussed further in this thesis.

#### 2.1.3.2 Black hole seeding and accretion

Following the EAGLE-style black-hole (BH) subgrid framework (Schaye et al., 2014, Crain et al., 2015), BH particles are seeded in sufficiently massive and resolved halos identified by a Friends-of-Friends (FoF) finder. The FoF catalog is updated at intervals of $\Delta \ln a = 0.005$, where $a$ is the cosmic scale factor. A gas particle at the minimum gravitational potential is converted into a BH seed when the host halo satisfies the mass criteria $M_{\rm FoF} > M_{\rm seed,FoF} = 10^{10}\,h^{-1}M_\odot$ and $M_\star > M_{\rm seed,\star} = 10^8\,h^{-1}M_\odot$, and when no BH is already present. To maintain numerical stability after halo mergers or dynamical perturbations, BH particles are repositioned to the potential minimum (if displaced by more than three gravitational softening lengths) and assigned the FoF-halo bulk velocity.

Gas accretion onto BHs is computed from a Bondi-like rate capped by the Eddington limit. The Bondi-like rate is written as

$$\dot{M}_{\rm Bondi} = \frac{4\pi G^2 M_{\rm BH}^2 \rho_l}{(c_{\rm s}^2 + v_{\rm BH}^2)^{3/2}}, \tag{2.31}$$

where $M_{\rm BH}$ is the BH mass, $\rho_l$ is the local gas density, $c_{\rm s}$ is the sound speed, and $v_{\rm BH}$ is the BH–gas relative velocity. The sound speed is evaluated from the effective equation of state,

$$c_{\rm s} = \sqrt{\frac{\gamma k_{\rm B} T_{\rm eos}}{\mu m_{\rm p}}}, \tag{2.32}$$

with $\gamma = 5/3$. We adopt

$$T_{\rm eos} = 8 \times 10^3\,{\rm K}\left(\frac{n_{\rm H}}{0.1\,{\rm cm}^{-3}}\right)^{1/3}, \tag{2.33}$$

and the primordial neutral-gas mean molecular weight $\mu = 1.2285$. The Eddington rate is

$$\dot{M}_{\rm Edd} = \frac{4\pi G M_{\rm BH} m_{\rm p}}{\epsilon_{\rm r} c \sigma_{\rm T}}, \tag{2.34}$$

where $\epsilon_{\rm r}$ is the radiative efficiency and $\sigma_{\rm T}$ is the Thomson cross section. The adopted accretion rate is

$$\dot{M}_{\rm acc} = \min\left(\dot{M}_{\rm Edd}, \mathcal{C}\dot{M}_{\rm Bondi}\right), \tag{2.35}$$

where $\mathcal{C} = \min\left[C_{\rm visc}^{-1}(c_{\rm s}/v_\phi)^3, 1\right]$ accounts for unresolved angular-momentum support in the accretion flow (Rosas-Guevara et al., 2015). In the fiducial run, we set $\epsilon_{\rm r} = 0.1$ and $C_{\rm visc} = 2\pi \times 10^2$. The physical BH growth rate is then

$$\dot{M}_{\rm BH} = (1 - \epsilon_{\rm r})\dot{M}_{\rm acc}. \tag{2.36}$$

As in standard particle-based implementations, the dynamical BH particle mass is synchronized with the subgrid BH mass by stochastic swallowing of neighboring gas particles, with probabilities weighted by the SPH kernel.

#### 2.1.3.3 AGN feedback

AGN feedback is implemented as stochastic thermal heating (Springel, 2005, Schaye et al., 2014). A feedback-energy reservoir is accumulated at a rate

$$\dot{E}_{\mathrm{res}} = \epsilon_{\mathrm{f}}\epsilon_{\mathrm{r}}\dot{M}_{\mathrm{acc}}c^2, \tag{2.37}$$

where $\epsilon_{\mathrm{f}}$ is the coupling efficiency between AGN radiative output and the surrounding gas. Once the reservoir exceeds the threshold required for a discrete heating event with temperature jump $\Delta T_{\mathrm{AGN}}$, energy is injected stochastically into neighboring gas particles. For a gas particle $j$, the heating probability is

$$P_{j,\mathrm{FB}} = \frac{(\gamma-1)\mu_i m_{\mathrm{p}}\, w_j\, E_{\mathrm{res}}}{k_{\mathrm{B}}\Delta T_{\mathrm{AGN}}\rho_l}, \tag{2.38}$$

where $w_j$ is the kernel weight and $\mu_i = 0.588$ is the primordial ionized-gas mean molecular weight. If selected, the thermal energy increment is

$$E_{j,\mathrm{FB}} = \frac{m_{\mathrm{gas}}k_{\mathrm{B}}\Delta T_{\mathrm{AGN}}}{(\gamma-1)\mu_i m_{\mathrm{p}}}. \tag{2.39}$$

This stochastic-heating strategy prevents immediate numerical overcooling and enables AGN to drive hot outflows that suppress central gas condensation, regulate star formation in massive systems, and redistribute baryons from halo centers into the CGM/IGM—all of which are essential for modeling FRB DM statistics at fixed halo mass.

For clarity, the fiducial black-hole and AGN-subgrid parameters used in this thesis are summarized in Table 2.1. More detailed physical implementations and discussions are provided by Oku & Nagamine (2024).

TABLE 2.1: Fiducial parameters of black-hole subgrid physics adopted in CROCODILE.

| Quantity | Symbol | Fiducial value |
|---|---|---|
| FoF mass threshold for BH seeding | $M_{\mathrm{seeding,FoF}}$ | $10^{10}\,h^{-1}M_\odot$ |
| FoF stellar-mass threshold for BH seeding | $M_{\mathrm{seeding,star}}$ | $10^{8}\,h^{-1}M_\odot$ |
| Number of gas neighbors heated per event | $N_{\mathrm{ngb,fb}}$ | $8 \pm 2$ |
| Radiative efficiency | $\epsilon_{\mathrm{r}}$ | 0.1 |
| AGN feedback efficiency | $\epsilon_{\mathrm{FB}}$ | 0.1 |
| AGN feedback temperature | $\Delta T_{\mathrm{AGN}}$ | $10^{8.5}$ K |
| Viscosity parameter | $C_{\mathrm{visc}}$ | $2\pi \times 10^2$ |
| Mean molecular weight (neutral gas) | $\mu$ | 1.2285 |
| Mean molecular weight (ionized gas) | $\mu_i$ | 0.588 |

## 2.2 The CROCODILE Simulation Suite

### 2.2.1 Initial Condition

Initial conditions are generated using MUSIC2 (Hahn & Abel, 2013), adopting a Planck 2018 cosmology (Planck Collaboration et al., 2020) with $H_0 = 67.74\,\mathrm{km\,s^{-1}\,Mpc^{-1}}$, $\Omega_m = 0.3099$, $\Omega_\Lambda = 0.6901$, and $\Omega_b = 0.04889$. We initialize the simulations at $z_{\rm ini} = 39$ using second-order Lagrangian perturbation theory. To suppress sample variance, we apply phase fixing for each model, while paired realizations are not used in this thesis. The fiducial volume is $L_{\rm box} = 50\,h^{-1}\,\mathrm{cMpc}$ with $N_{\rm p} = 2 \times 512^3$ particles (dark matter + gas), corresponding to mass resolutions of $m_{\rm dm} = 6.74 \times 10^7\,h^{-1}\,M_\odot$ and $m_{\rm gas} = 1.26 \times 10^7\,h^{-1}\,M_\odot$.

### 2.2.2 Simulation setup

TABLE 2.2: Simulation Parameters for Different Simulations

| Simulation | Box Size ($h^{-1}$ cMpc) | Particle Numbers | Dark Matter Mass ($h^{-1}\,M_\odot$) | Gas Mass ($h^{-1}\,M_\odot$) | $\epsilon_G^\dagger$ ($h^{-1}$ ckpc / $h^{-1}$ pkpc) | AGN Feedback |
|---|---|---|---|---|---|---|
| L25N512$_{\rm fiducial}$ | 25 | $2 \times 512^3$ | $8.43 \times 10^6$ | $1.58 \times 10^6$ | 1.63 / 0.25 | ✓ |
| L50N512$_{\rm fiducial}$ | 50 | $2 \times 512^3$ | $6.75 \times 10^7$ | $1.26 \times 10^7$ | 3.38 / 0.50 | ✓ |
| L50N512$_{\rm NoBH}$ | 50 | $2 \times 512^3$ | $6.75 \times 10^7$ | $1.26 \times 10^7$ | 3.38 / 0.50 | × |
| L100N1024$_{\rm fiducial}$ | 100 | $2 \times 1024^3$ | $6.75 \times 10^7$ | $1.26 \times 10^7$ | 3.38 / 0.50 | ✓ |
| L100N1024$_{\rm NoBH}$ | 100 | $2 \times 1024^3$ | $6.75 \times 10^7$ | $1.26 \times 10^7$ | 3.38 / 0.50 | × |
| L500N1024$_{\rm NoBH}$ | 500 | $2 \times 1024^3$ | $8.43 \times 10^7$ | $1.58 \times 10^7$ | 16 / 2.50 | × |

$^\dagger$ $\epsilon_G$ represents the gravitational softening length. The two values listed correspond to the softening comoving length (left, in $h^{-1}$ ckpc) and the softening max physical length (right, in $h^{-1}$ pkpc), which is the resolution limit of the simulation.

We analyze multiple cosmological simulations across a range of box sizes and feedback implementations: $25\,h^{-1}$Mpc with AGN feedback (L25N512$_{\rm fiducial}$), $50\,h^{-1}$Mpc with (L50N512$_{\rm fiducial}$) and without (L50N512$_{\rm NoBH}$) AGN feedback, $100\,h^{-1}$Mpc with (L100N1024$_{\rm fiducial}$) and without (L100N1024$_{\rm NoBH}$) AGN feedback, and $500\,h^{-1}$Mpc with stellar feedback but no AGN feedback (L500N1024$_{\rm NoBH}$)[2]. AGN feedback is not included in the current 500 Mpc run due to data availability.

Table 2.2 summarizes the parameters, including box size, total particle count $N_p$, dark matter and gas-particle masses ($m_{\rm DM}$, $m_{\rm gas}$), and gravitational softening length $\epsilon_{\rm grav}$. Initial conditions were generated using MUSIC2 (Hahn & Abel, 2013), with Planck 2018 cosmological parameters (Planck Collaboration et al., 2020): $H_0 = 67.74\,\mathrm{km\,s^{-1}\,Mpc^{-1}}$, $\Omega_m = 0.3099$, $\Omega_\Lambda = 0.6901$, and $\Omega_b = 0.04889$. SMBH seeding, accretion, and AGN thermal feedback follow the EAGLE-style implementation described in Section 2.1.3 (Oku & Nagamine, 2024).

Figure 2.1 shows the projected gas mass density distributions for different simulations. The L50N512 and L100N512 simulations include runs both with and without AGN feedback. In the present paper, we only consider the thermal AGN feedback without the jet (radio) mode. We plot the radial profiles of mass density and electron number density of the most massive halos in Figure 2.2. The simulation without AGN feedback (L50/L100 NoBH) shows a significantly higher central density, indicating that in the absence of AGN feedback, there is no mechanism to drive material outward, resulting in a more concentrated mass distribution at the center. In contrast, the simulation with AGN feedback (L50/L100 fiducial) has a lower central density but higher densities in the outer regions. This suggests that AGN feedback is effective in expelling material outwards into the CGM and even into the IGM, leading to a more extended distribution of mass and electron density.

[2]The current 500 Mpc simulation is based on GADGET3-OSAKA, not GADGET4-OSAKA.

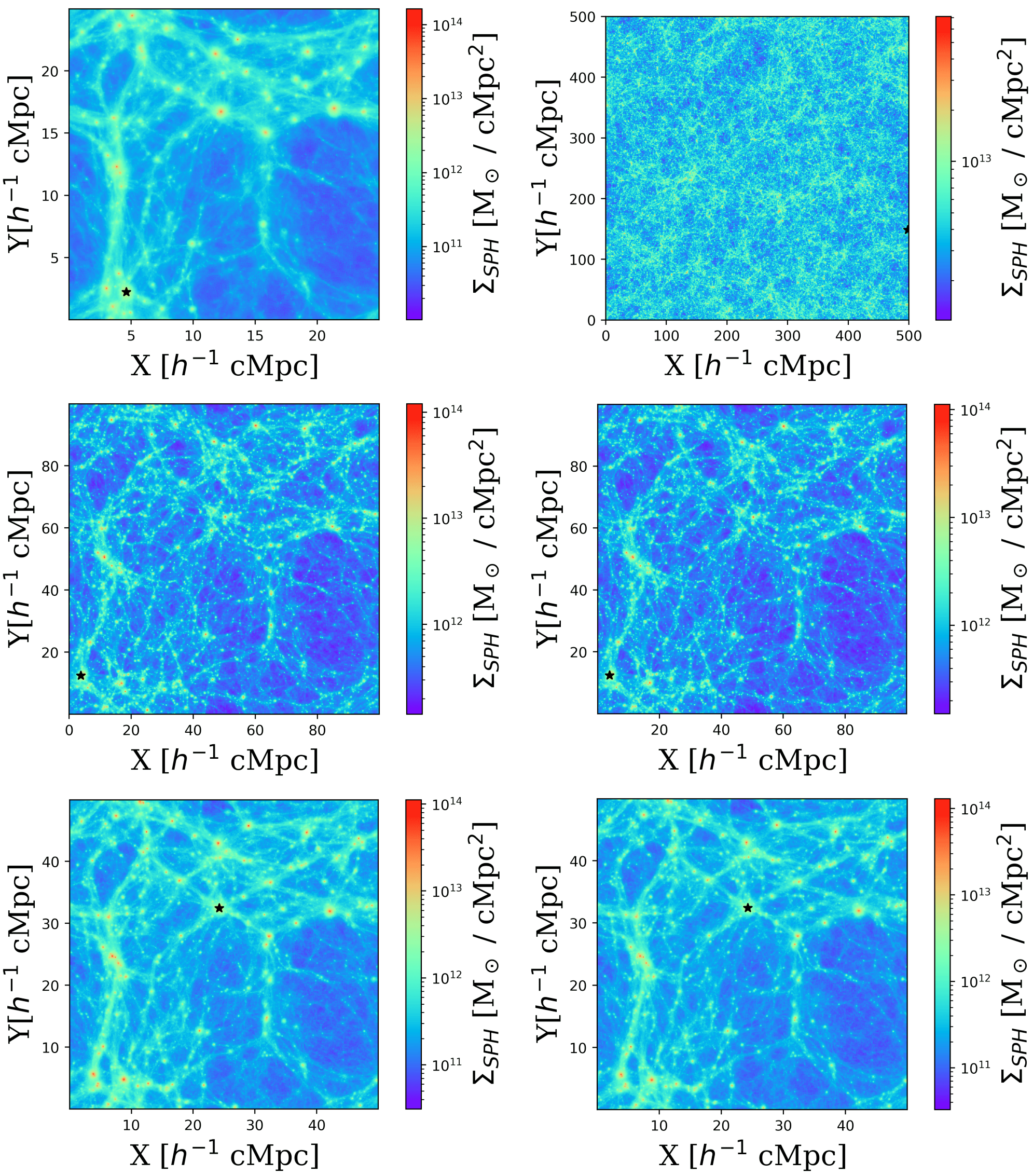

FIGURE 2.1: Projected gas mass density distributions for different simulation box sizes: L25N512 (top left), L500N1024 (top right), L100N1024$_{\text{fiducial}}$ (middle left), L100N1024$_{\text{NoBH}}$ (middle right), L50N512$_{\text{fiducial}}$ (bottom left) and L50N512$_{\text{NoBH}}$ (bottom right). The black stars indicate the locations of the most massive halos.

We also perform statistical analysis on all the halos in each simulation data to examine the HMF. Figure 2.3 shows the HMF for simulations with different box sizes, comparing the results with theoretical models from Press & Schechter (1974), Tinker et al. (2008).

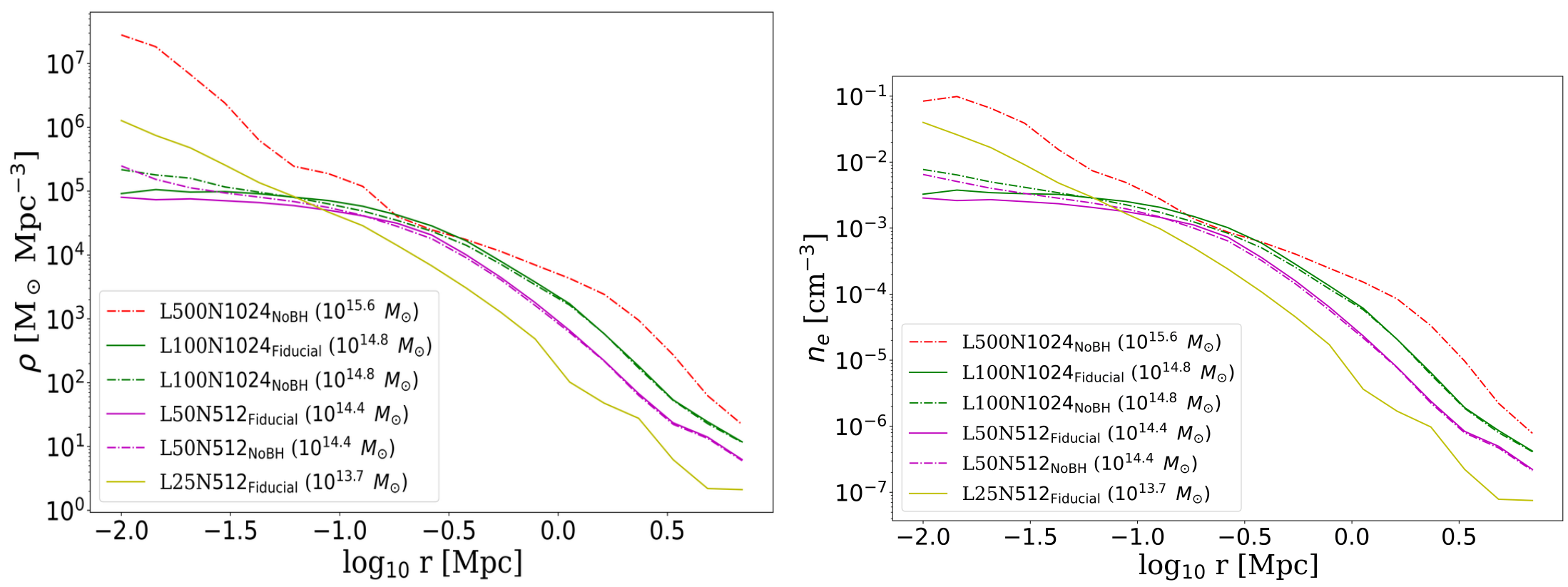

FIGURE 2.2: Radial profiles of gas mass density (left) and electron number density (right) for the most massive halos in simulations with different box sizes, as summarized in Table 2.2.

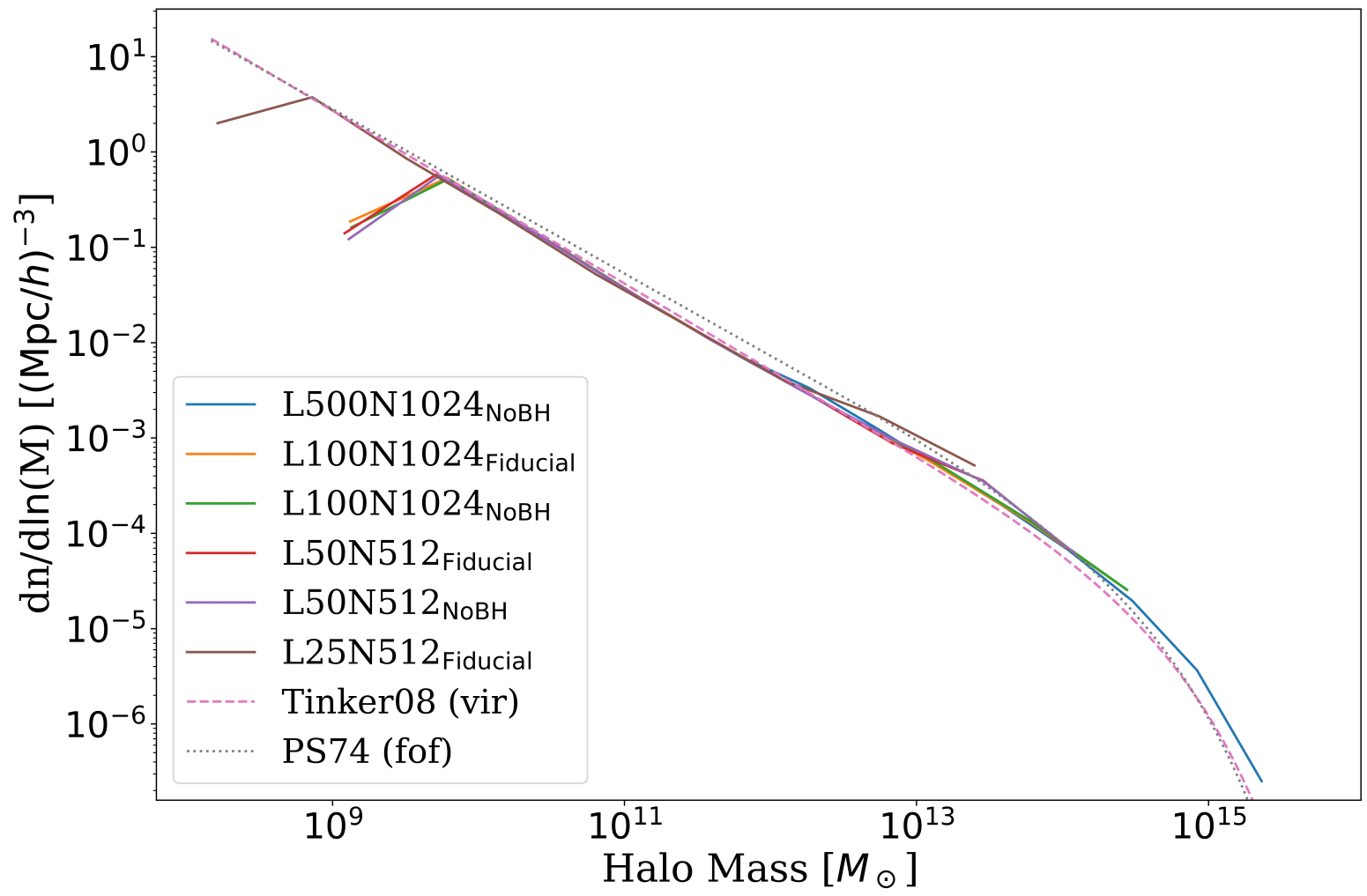

FIGURE 2.3: HMF for different simulations. The solid lines represent the simulation data, while the dashed and dotted lines indicate the theoretical models from Tinker et al. (2008) and Press & Schechter (1974).

The results agree well with the theoretical HMFs, indicating accurate halo formation processes. Notably, the L25 simulation shows a larger number of low-mass halos at $M < 10^{9.5}\,M_\odot$ due to higher resolution than in other runs. It also results in the absence of halos above $10^{14}\,M_\odot$ in the L25 run due to its small box size. In contrast, in larger simulation volumes, the effects of AGN feedback are more likely to be averaged out across a larger number of halos, leading to a more balanced distribution of halo masses. For the L50/L100 runs, differences in the HMF between simulations with and without AGN feedback are relatively small.

## 2.3 FRB Light-cone Construction and DM Integration

The Light-cone generation process involves: DM Calculation within a single snapshot: The electron number density $n_e$ along the light cone is computed using a gridding approach. The simulation box is divided into cells based on hierarchical levels: lv7 ($128^3$), lv8 ($256^3$), lv9 ($512^3$), and lv10 ($1024^3$). Each gas particle's mass and electron number density are distributed among the overlapping cells according to its smoothing length. The LoS is subdivided into bins (e.g., 400 bins for lv8). The electron density of a given LoS bin $i$ is taken from the corresponding cell $j$, and the DM contribution for that bin is computed as:

$$\mathrm{DM}_i = n_{e,j} \frac{l_i}{1+z}. \tag{2.40}$$

The total DM along the LoS is obtained by summing over all bins. Tests show that for cell levels lv8 and higher (cell size $\sim 400$ kpc), DM results are numerically converged. Thus, we adopt

### 2.3.1 Snapshot stitching and ray tracing

The ray-tracing path is extended across multiple snapshots and smoothed at snapshot interfaces to avoid discontinuities.

- Extending and shifting across snapshots: The light cone is extended beyond a single snapshot by assigning a new random angle for the next segment. This ensures diverse structure sampling while preserving continuity in the cosmic web. The process is repeated until the total comoving path length matches the required distance to the next snapshot (e.g., $z = 0.1$).
- Handling Snapshot Transitions and Smoothing: When transitioning between snapshots at different redshifts, an overlapping region of $5\,h^{-1}$ cMpc is introduced to prevent artificial discontinuities. The electron density in this overlap region is interpolated using a weighted linear approach:
$$n_{e,\mathrm{overlap}}(i) = \frac{w_{\mathrm{low}} n_{e,\mathrm{low}}(i) + w_{\mathrm{up}} n_{e,\mathrm{up}}(i)}{w_{\mathrm{low}} + w_{\mathrm{up}}}, \tag{2.41}$$
where the weights are defined as:
$$w_{\mathrm{low}} = \frac{N_{\mathrm{overlap}} - i}{N_{\mathrm{overlap}}}, \quad w_{\mathrm{up}} = \frac{i}{N_{\mathrm{overlap}}}. \tag{2.42}$$
Here, $N_{\mathrm{overlap}}$ represents the number of bins within the overlapping region (typically 13 bins for lv8). This interpolated region is then inserted at the interface between snapshots, and the DM calculation is repeated following the method described above.

### 2.3.2 Line-of-sight sampling and resolution control

We develop two complementary LoS sampling strategies (Figure 2.4).

**Method I (left panel: segment-by-segment directional sampling).** For each snapshot, the simulation box at a given redshift (e.g., $z = 0$) is placed on a reference plane (e.g., the $xy$-plane), and an initial LoS is defined by a randomly drawn direction vector $\mathbf{d}$. This construction provides an approximately uniform directional sampling in three-dimensional space. The

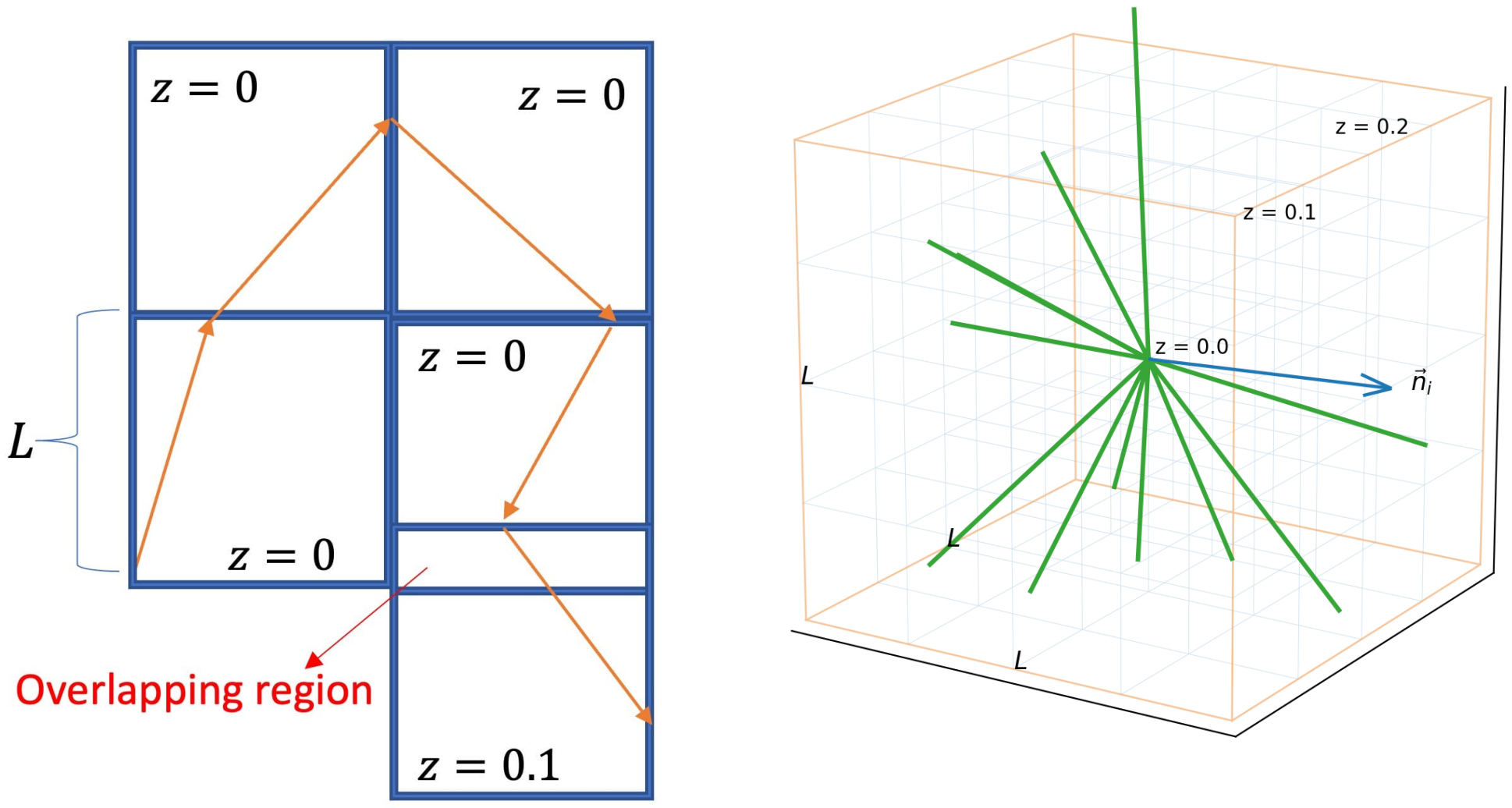

FIGURE 2.4: Two ray-tracing strategies developed in this work. Left: the segment-by-segment directional sampling method with per-segment direction updates across periodic boxes. Right: the full-sky ($4\pi$) fixed-direction sampling method, where rays are launched from one observer position and snapshots are stacked along each direction without rotating the grid. $L$ is the box size, and $\vec{n}_i$ is a random LoS vector.

components of $\mathbf{d}$ are given by:

$$d_i = \begin{cases} \text{random}(0,1) & \text{with probability 0.5,} \\ \text{random}(-1,0) & \text{otherwise.} \end{cases} \tag{2.43}$$

One component is set to $\pm 1$ to define the major axis. The corresponding LoS vector is

$$\mathbf{LoS} = \left(\frac{L}{m}, \frac{L}{n}, \frac{L}{o}\right), \tag{2.44}$$

where $L$ is the simulation box size, with $m = 1/d_x$, $n = 1/d_y$, and $o = 1/d_z$. As described in Section 2.3.1, the light cone is then extended snapshot by snapshot, and interface smoothing is applied in overlap regions.

**Method II (right panel: full-sky $4\pi$ fixed-direction sampling).** An observer is placed at a fixed position inside a full simulation box, selected in a region with local mass density in the range $10^{-4}$–$10^{-3}\,\mathrm{g\,cm^{-3}}$, (to minimize MW-like ISM contamination in cosmological DM). From this point, a set of unit vectors $\hat{\mathbf{n}}_i$ is generated to uniformly sample the full sky. For each direction, the ray propagates radially outward through periodic boxes without rotating the simulation grid; coordinates are wrapped periodically so each ray continuously traces large-scale structure while preserving the intrinsic density field of each snapshot. Snapshots are stacked in redshift order along the same direction (near segments from lower-$z$ outputs and distant segments from higher-$z$ outputs), and the same overlap-and-blending treatment in Section 2.3.1 is used to maintain continuity of $n_e$ and cumulative DM at snapshot boundaries.

In this work, we adopt Method I to maximize the use of large-scale-structure information in each simulation box while extending the light-cone to higher redshift, and to delay the onset of repeated structures as much as possible.

For resolution control, convergence tests show that lv8 and higher (cell size $\sim$ 400 kpc) give stable DM results; we therefore adopt lv8 in this work.

### 2.3.3 DM–$z$ relationship

To keep notation consistent with the Introduction, we adopt the same multi-component decomposition in Eqs. (1.14)–(1.18); see also the schematic summary in Figure 1.6. In particular, $\mathrm{DM}_{\mathrm{source}}$ is treated as a component separate from $\mathrm{DM}_{\mathrm{host}}$, consistent with the definitions used in Chapter 1.

Following Eq. (1.16), we define the diffuse component as $\mathrm{DM}_{\mathrm{diff}} \equiv \mathrm{DM}_{\mathrm{IGM}} + \mathrm{DM}_{\mathrm{halos}}$[3].

Using our cosmological simulations, we can compute the total FRB DM contribution from the diffuse baryonic medium as follows:

$$\mathrm{DM}_{\mathrm{diff}} = f_{\mathrm{diff}} \frac{\rho_{c,0}\Omega_b}{m_p} \frac{c}{H_0} \int_0^z \frac{\chi_e(z')(1+z')}{E(z')}\, dz' \tag{2.45}$$

where $\rho_{c,0}$ is the critical mass density and $\Omega_b$ is the baryon density parameter at $z = 0$. $f_{\mathrm{diff}} = f_{\mathrm{IGM}} + f_{\mathrm{Halos}}$ represents the baryon fraction in the diffuse medium (including both IGM and halos), $m_p$ is the proton mass, and $\chi_e(z)$ is the electron abundance from ionized hydrogen and helium, which is approximated by $\chi_e(z) \approx Y_H\, \chi_{e,\mathrm{H}}(z) + Y_{\mathrm{He}}\, \chi_{e,\mathrm{He}}(z)/2$, where $Y_{\mathrm{H}}$ and $Y_{\mathrm{He}}$ are the mass fractions of hydrogen and helium, respectively, assuming complete ionization and ignoring heavier elements.

This formulation aligns with previous works on the DM–$z$ relationship and allows us to constrain $f_{\mathrm{diff}}$ using our simulation results, providing a robust framework to understand the baryonic distribution in the Universe.

## 2.4 Zoom-in Simulations for Host Galaxies

To examine the contribution of the HHs of FRBs ($\mathrm{DM}_{\mathrm{host}}$), it is crucial to analyze FRB host environments with high-resolution simulations. This allows for a detailed separation of $\mathrm{DM}_{\mathrm{host}}$ contributions and helps in understanding how different types of host galaxies and the specific locations of FRBs within these galaxies (e.g., central regions versus outskirts) impact the total DM. To this end, we employ zoom-in cosmological hydrodynamic simulations performed by both GADGET3-OSAKA and GADGET4-OSAKA codes (simply due to the historical timeline). We employ three different zoom simulations to probe host galaxies of different mass scales and different environments: dwarf galaxy, Milky Way (MW)-like galaxy, and galaxy cluster. The initial conditions and parameters are summarized in Table 2.3.

[3]Note that Macquart et al. (2020) referred to this quantity as $\mathrm{DM}_{\mathrm{cosmic}}$, which is equivalent to our $\mathrm{DM}_{\mathrm{diff}}$. However, our $f_{\mathrm{diff}}$ explicitly excludes stars and compact stellar remnants (e.g. black holes), whereas $f_{\mathrm{cosmic}}$ may include them, since DM only traces ionized components along the LoS, while $f_{\mathrm{cosmic}}$ may account for all baryons regardless of their ionization state. See Appendix B.4 for a detailed discussion of $f_{\mathrm{diff}}$

Figure 2.5 presents the gas mass (left panels) and stellar mass (right panels) distributions for three different types of galaxies (dark matter halos) simulated using zoom-in techniques.

- Dwarf Galaxy: In the top row of Figure 2.5, the results for a dwarf galaxy (Tomaru et al., 2025) from GADGET4-Osaka are displayed. The gas mass distribution shows a concentrated core, with some surrounding substructures. Given the small overall mass of this galaxy ($M_{200} = 9.3 \times 10^9\,\mathrm{M}_\odot$), the gas distribution is primarily centered around the core, and the star formation is also concentrated in this region, reflecting the compact nature of this system. The stellar mass is $2.2 \times 10^7\,\mathrm{M}_\odot$, further emphasizing the compactness of the galaxy.
- MW-like Galaxy: The bottom row shows the results from a MW-like galaxy in the AGORA GADGET3 simulation (Roca-Fàbrega et al., 2021, 2024). The gas mass distribution reveals a well-defined spiral structure, typical of a disk galaxy. The stellar mass is concentrated in the disk and bulge regions, with significant star formation occurring throughout the spiral arms. With $M_{200} = 1.24 \times 10^{12} M_\odot$, this structure highlights the characteristic organization of stars and gas in spiral galaxies.
- Galaxy Cluster: The middle row displays the gas and stellar mass distributions for a galaxy cluster, also from GADGET3-OSAKA (Fukushima et al., 2023). The gas mass is widely distributed, showing the large scale of this system ($M_{200} = 9.23 \times 10^{14}\,\mathrm{M}_\odot$). The gas distribution covers a vast region, with the stellar mass being more concentrated in the central region, reflecting the dense core of the cluster. This extensive distribution and the filamentary LSS are characteristic of galaxy clusters, with star formation centered in the core but more dispersed compared to the dwarf galaxy.

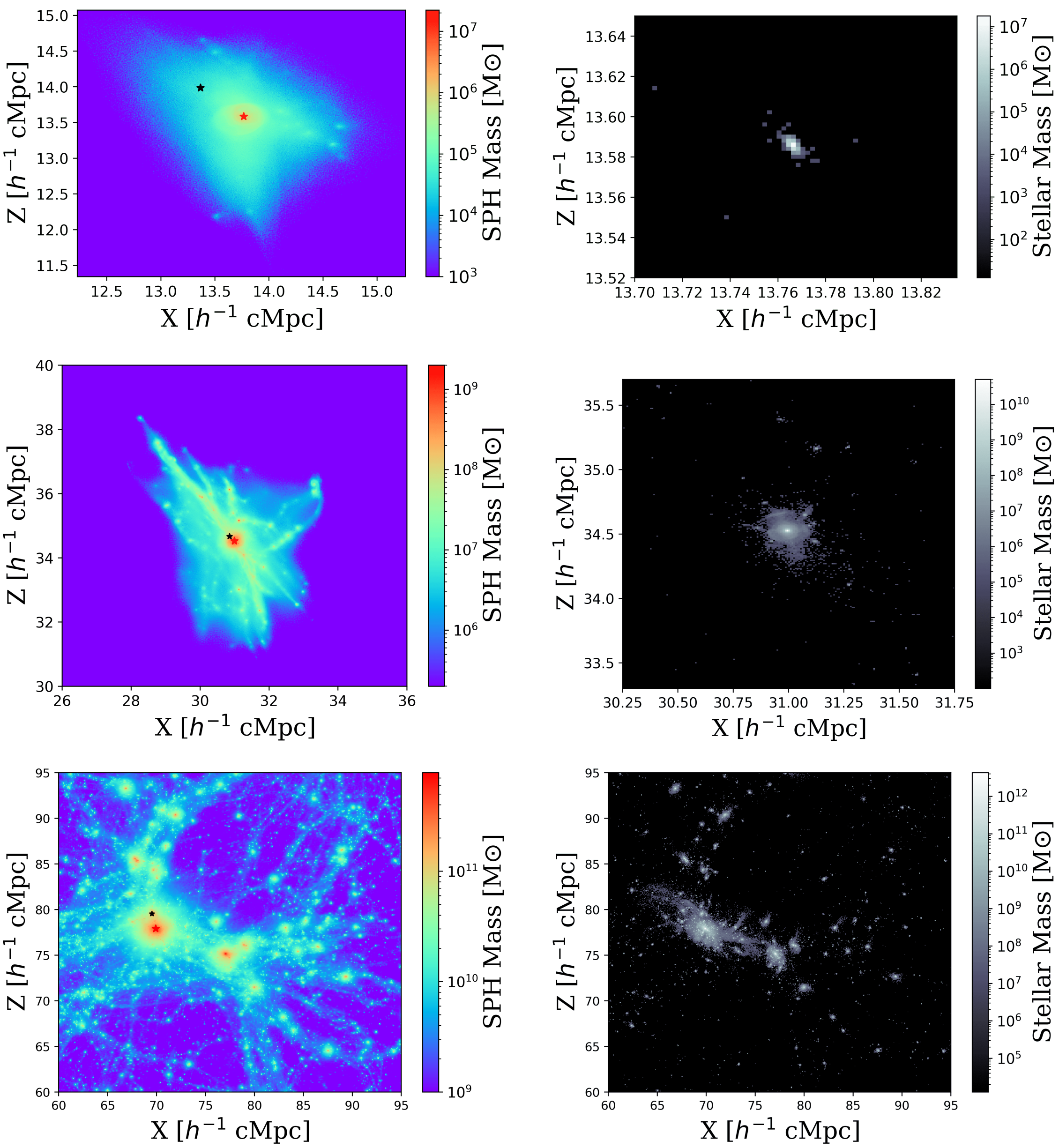

FIGURE 2.5: Distribution of gas mass (left panels) and stellar mass (right panels) for a dwarf galaxy (top row; Tomaru et al. in prep.), a MW-like galaxy from the AGORA simulation (middle row Roca-Fàbrega et al., 2021, 2024), and a galaxy cluster (bottom row) from Fukushima et al. (2023). The red stars indicate the center-of-mass of each target halo.

Table 2.3: Zoom-in Simulation Parameters

| Parameter | Dwarf Galaxy[†] | MW-like Galaxy[♯] | Galaxy Cluster[‡] |
|---|---|---|---|
| Simulation Code | GADGET4-Osaka | AGORA-GADGET3 | GADGET3-Osaka |
| $H_0$ [km s$^{-1}$ Mpc$^{-1}$] | 70.2 | 70.2 | 67.74 |
| $\Omega_m$ | 0.272 | 0.272 | 0.3099 |
| $\Omega_\Lambda$ | 0.728 | 0.728 | 0.6901 |
| $\Omega_b$ | 0.0455 | 0.0455 | 0.04889 |
| Box Size [$h^{-1}$ cMpc] | 20 | 60 | 100 |
| $m_{\rm DM}$ [$h^{-1}\,M_\odot$] | $7.32 \times 10^3$ | $2.8 \times 10^5$ | $6.73 \times 10^7$ |
| $m_{\rm gas}$ [$h^{-1}\,M_\odot$] | $1.47 \times 10^3$ | $5.65 \times 10^4$ | $1.26 \times 10^7$ |
| $\epsilon_{\rm grav}$ [$h^{-1}$ kpc] | 0.195 | 0.080 | 3.26 |
| $z_{\rm snapshot}^{\S}$ | 0 | 0.3 | 0 |
| $R_{200}$ [kpc] | 43.35 | 141.3 | 2052.94 |
| $M_{\rm halo,200}$ [$M_\odot$] | $9.32 \times 10^9$ | $1.24 \times 10^{12}$ | $9.23 \times 10^{14}$ |
| $M_{\rm FoF}$ [$M_\odot$] | $1.13 \times 10^{10}$ | $1.33 \times 10^{12}$ | $1.20 \times 10^{15}$ |
| Dark Matter Mass [$M_\odot$] | $1.11 \times 10^{10}$ | $1.17 \times 10^{12}$ | $1.05 \times 10^{15}$ |
| Gas Mass [$M_\odot$] | $2.10 \times 10^8$ | $1.02 \times 10^{11}$ | $1.22 \times 10^{14}$ |
| Stellar Mass [$M_\odot$] | $2.2 \times 10^7$ | $6.58 \times 10^{10}$ | $3.70 \times 10^{13}$ |
| Stellar Feedback | ✓ | ✓ | ✓ |
| AGN Feedback | ✗ | ✗ | ✗ |

[†] Tomaru et al. (2025) [♯] Roca-Fàbrega et al. (2021, 2024) [‡] Fukushima et al. (2023)
[§] $z_{\rm snapshot}$ is the redshift at which host galaxy properties are analyzed. All halo-related quantities
($R_{200}$, $M_{\rm halo,200}$, $M_{\rm FoF}$, and total masses of dark matter, gas, and stars) are computed at this redshift.

# Chapter 3

# Part I: FRB Cosmology from the Cosmic Web in CROCODILE

## 3.1 Simulated DM–$z$ Relation

### 3.1.1 Mock FRB population and statistics

To construct mock FRB sight lines, we rotate large-scale-structure (LSS) simulation boxes randomly and connect LoS segments continuously across redshift snapshots. This setup improves sampling of cosmic structures compared with fixed parallel directions (see Section 2.3.2).

In our light-cone realization, LoS origins are initialized from a $100 \times 100$ uniformly spaced grid on the $z = 0$ plane, yielding 10,000 sight lines in total. This choice ensures broad spatial coverage, while we do not require each LoS origin or endpoint to lie inside a halo. The implications of this modeling choice for DM estimates are discussed in Appendix B.2.

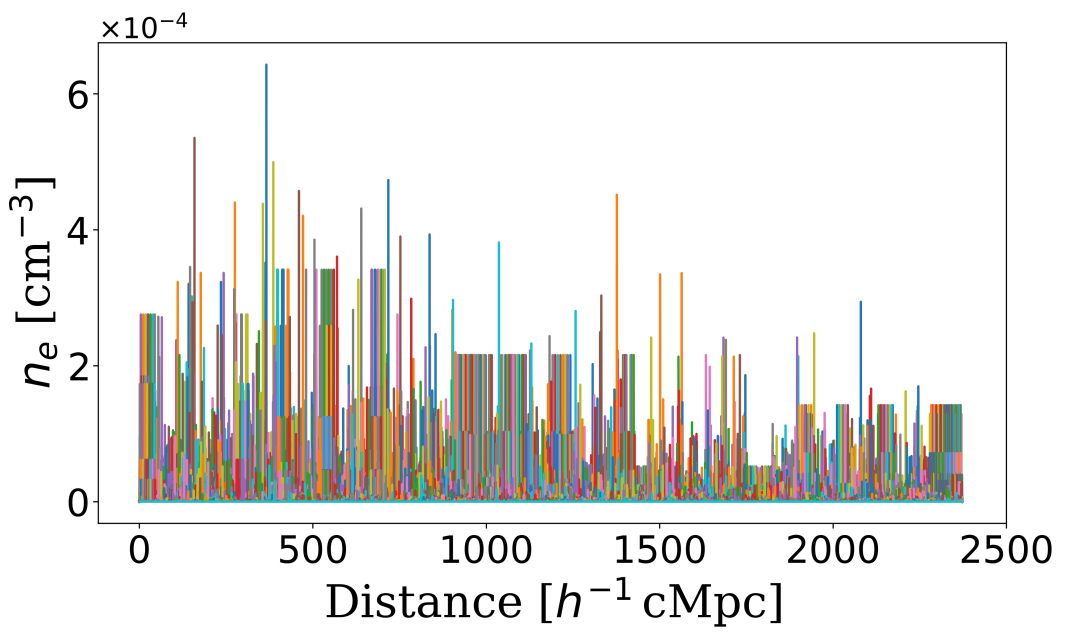

FIGURE 3.1: Electron density ($n_e$) distribution along 10,000 light cones in the fiducial model with AGN feedback. The integration spans 2500 Mpc in comoving space from $z = 0$ to $z \approx 1$. The NoBH model exhibits a similar distribution, with no visually significant differences, and is therefore omitted for clarity. Different colors represent different lines of sight.

Figure 3.1 shows the electron-density profiles along all 10,000 sight lines for the L100 fiducial run (with AGN feedback). Each colored curve corresponds to one individual sight line as a function of comoving distance. We have also analyzed the NoBH run, but its $n_e$ distribution is visually indistinguishable from the fiducial case and is therefore omitted for clarity.

### 3.1.2 Mean relation and scatter

Integrating $n_e$ along each LoS from $z = 0$ gives the cumulative DM as a function of redshift (Figure 3.2), spanning $z = 0$ to $z \approx 1$ (corresponding to a comoving distance of $\sim 2500$ Mpc). The fiducial model shows a slightly higher mean DM than the NoBH model, consistent with AGN-driven redistribution of baryons into diffuse phases. This difference is modest in CROCODILE because the current setup does not include a strong jet-feedback mode, so outflows are weaker than in simulations such as SIMBA or IllustrisTNG.

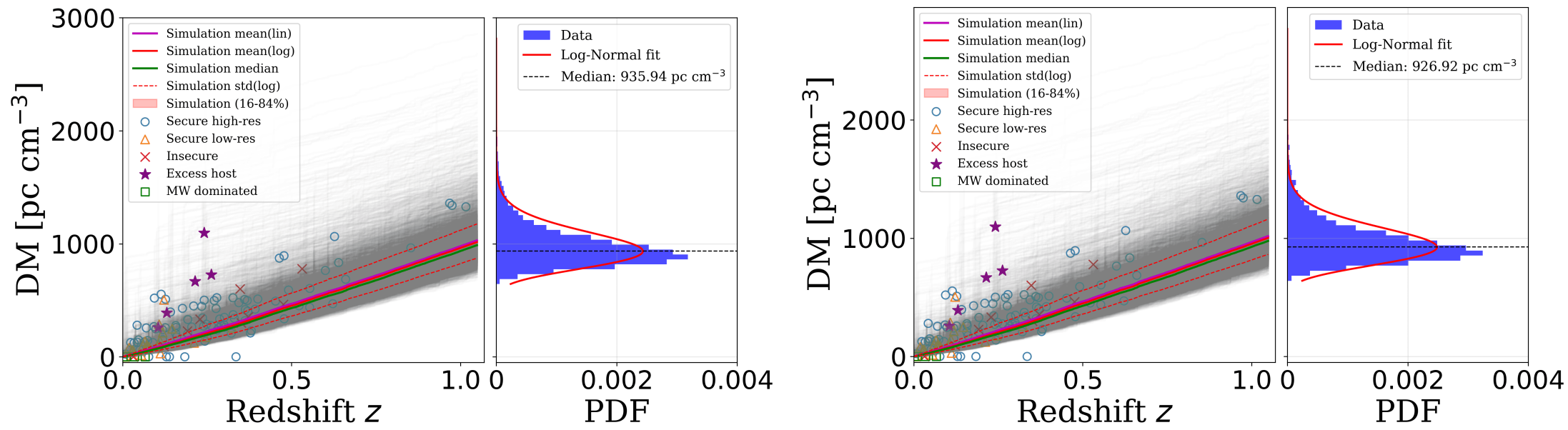

FIGURE 3.2: DM of all LoS are shown for the L100 fiducial model with AGN feedback (left panel), and the NoBH model (right panel). In each panel, the left subpanel shows the DM-$z$ relationship, with linear and logarithmic averages indicated, alongside the median and standard deviation of the DM values. The right sub-panel presents the histogram of DM values at $z = 1$, with the best-fit log-normal distributions shown in red curves, and the dashed lines mark the median DM values of 935.94 pc cm$^{-3}$ for the fiducial model and 926.92 pc cm$^{-3}$ for the NoBH model.

Because foreground halos are still included in the integration, the quantity measured here is $\mathrm{DM_{diff}}$ rather than purely $\mathrm{DM_{IGM}}$. Consequently, the inferred parameter is effectively $f_{\mathrm{diff}}$ (not a halo-subtracted $f_{\mathrm{IGM}}$).

The inset histograms in Figure 3.2 are well described by a log-normal distribution,

$$P(x) = \frac{1}{\sqrt{2\pi}\sigma x} \exp\left[-\frac{(\ln x - \ln \mu)^2}{2\sigma^2}\right], \tag{3.1}$$

where $\mu$ is the median DM and $\sigma$ is the standard deviation of ln(DM). In this parameterization,

$$E(\mathrm{DM}) = \mu e^{\sigma^2/2}, \quad \mathrm{Std}(\mathrm{DM}) = \mu e^{\sigma^2/2}\sqrt{e^{\sigma^2} - 1}. \tag{3.2}$$

The high-DM tails in both models are primarily produced by intersections with dense foreground halos.

For the fiducial model (integrated to $z = 1$), we obtain median DM $= 935.94\,\mathrm{pc\,cm^{-3}}$, log-normal mean $E(\mathrm{DM}) = 977.81\,\mathrm{pc\,cm^{-3}}$, and scatter $\mathrm{Std}(\mathrm{DM}) = 170.10\,\mathrm{pc\,cm^{-3}}$. For the NoBH model, the corresponding values are median DM $= 926.92\,\mathrm{pc\,cm^{-3}}$, $E(\mathrm{DM}) = 966.52\,\mathrm{pc\,cm^{-3}}$, and $\mathrm{Std(DM)} = 166.65\,\mathrm{pc\,cm^{-3}}$.

### 3.1.3 Comparison with analytic models

Figure 3.3 compares the simulated DM–$z$ trends with the Macquart-type model (Equation 2.45) for $f_{\mathrm{diff}} = 0.5$–$1.0$. Both fiducial and NoBH runs are broadly consistent with the analytic expectation. At $z = 1$, we infer $f_{\mathrm{diff}} = 0.865^{+0.101}_{-0.165}$ for fiducial and $0.856^{+0.101}_{-0.162}$ for NoBH.

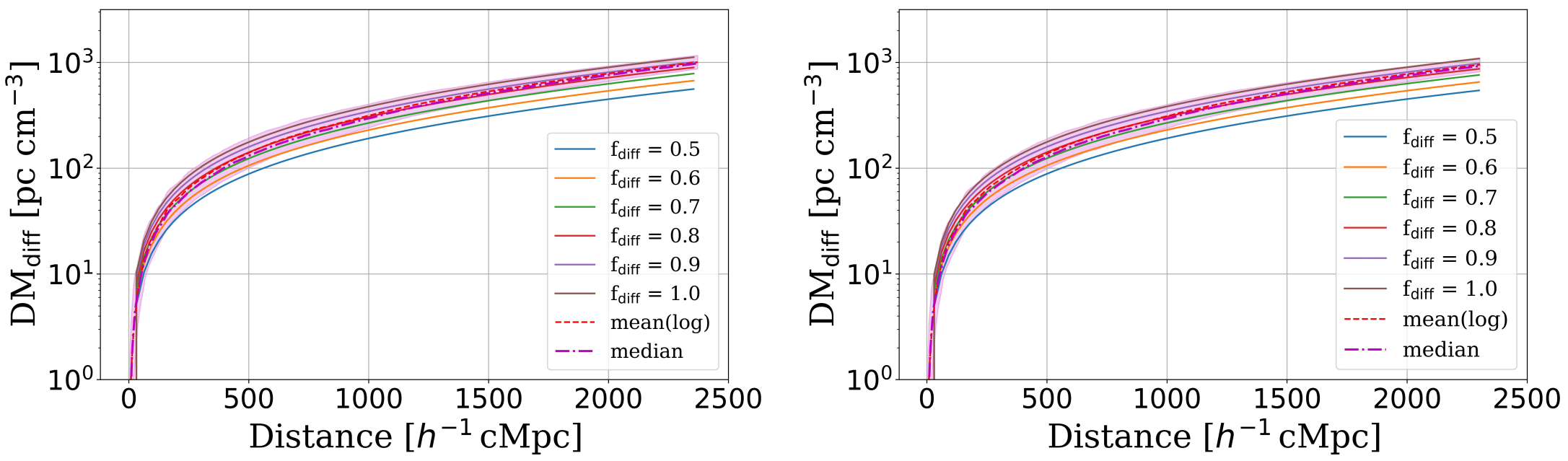

FIGURE 3.3: The dispersion measure up to certain distances, obtained from the light-cone data of CROCODILE simulation (log mean and median) and the model results calculated from Eq. 1.13 for different values of $f_{\mathrm{diff}}$ (for both IGM and intervening halo gas). The left panel shows results for the fiducial simulation, while the right panel corresponds to the NoBH simulation. The magenta-shaded region represents the 16th–84th percentile DM uncertainty range obtained from our simulations. The red dashed line and the purple dash-dotted line indicate the logarithmic mean and the median DM value from the simulation, respectively. The solid lines correspond to theoretical DM-$z$ relations derived from the Macquart relation for different values of $f_{\mathrm{diff}}$ ranging from 0.5 to 1.0.

Here the quoted uncertainty range corresponds to the 16th–84th percentiles across sight lines in log space, i.e., intrinsic cosmic variance from LoS-to-LoS structure (dominated by foreground-halo intersections), rather than the standard error on the mean. Because $\mathrm{DM_{halos}}$ is not subtracted in Figure 3.3, some sight lines yield overestimated $f_{\mathrm{diff}}$ (occasionally exceeding unity). This bias is reduced in later sections where we separate $\mathrm{DM_{IGM}}$ from $\mathrm{DM_{diff}}$.

## 3.2 Halo and CGM Contributions

As described in Section 3.1, contamination from foreground-galaxy density fluctuations along mock FRB sight lines can significantly bias the recovery of the intrinsic $f_{\mathrm{IGM}}$. This motivates a dedicated analysis of foreground-halo contributions to observed FRB DMs, a comparison between simulation results and observations, and an assessment of AGN-feedback effects.

### 3.2.1 Radial profiles and mass dependence

#### 3.2.1.1 Theoretical Model of Halo Density Profile

Because an FRB sight line can intersect a foreground halo at essentially any position, the corresponding halo DM contribution depends on the impact parameter, $b$, of that intersection. To compute the DM contribution at a given $b$, we must specify the underlying halo gas (and total-mass) density structure. We therefore adopt two widely used halo-profile models, the Navarro-Frenk-White (NFW) (Navarro et al., 1997) and modified NFW (mNFW) (Duffy et al., 2010), written as

$$\rho(r) = \begin{cases} \frac{\rho_0}{y(1+y)^2}, & \text{(NFW)} \\ \\ \frac{\rho_0}{y^{1-\alpha}(y_0+y)^{2+\alpha}}, & \text{(mNFW)} \end{cases} \tag{3.3}$$

where

$$\begin{aligned} y &= c_{200}\frac{r}{R_{200}}, \quad c_{200} = C\left(\frac{M_{\mathrm{halo},200}}{10^{14}M_\odot}\right)^{-\beta}, \\ \rho_0 &= \frac{(200\rho_c)}{3}c_{200}^3 f(c_{200}), \end{aligned} \tag{3.4}$$

and

$$f(c) = \ln(1+c) - \frac{c}{1+c}. \tag{3.5}$$

Here, $R_{200}$ is the halo radius where density is 200 times the critical density, $\rho_c = 8.6 \times 10^{-30}\,\mathrm{g\,cm^{-3}}$. $M_{\mathrm{halo},200}$ is defined as the total mass enclosed within $R_{200}$. The concentration parameter $c_{200}$ depends on halo mass $M_{\mathrm{halo},200}$, with $C$ as normalization. $\rho_0$ is set by $c_{200}$ and $f(c_{200})$. The mNFW profile modifies the inner slope via $\alpha$ to account for baryonic effects, with $y_0$ controlling the transition scale and $\beta$ describing the mass dependence of $c$.

While the NFW profile provides a standard description of dark matter halos, the mNFW model accounts for baryonic feedback, including AGN-driven outflows. We use these profiles to analyze the DM contributions from foreground halos along FRB sight lines.

#### 3.2.1.2 Density profiles derived from LSS simulations

We present density profiles derived from simulations, including gas, dark matter, stellar mass density, and $n_e$. To compute these profiles, we first categorize all halos in the LSS simulation into different mass bins, covering the range $10^{10.5} - 10^{15.5} M_\odot$. Within each mass bin, we randomly select 500 halos and compute their individual density profiles for various components. The median density profile is then extracted for each mass range, with the 16th-84th percentile range determined to represent the upper and lower bounds of the distribution. Additionally, we calculate the median $R_{200}$ for halos in each mass bin. Since the $R_{200}$ values for the fiducial and NoBH runs show negligible differences, we only retain the fiducial results. The geometric setup used for halo-centered LoS integration (halo center, $R_{200}$, impact parameter $b$, and the integration interval $[-2R_{200}, +2R_{200}]$) is illustrated in Figure 3.6. These profiles provide insights into the distribution of baryons and dark matter in halos of different masses and serve as an essential component for modeling the DM along FRB sight lines. The resulting density profiles are shown in Figures 3.4 and 3.5.

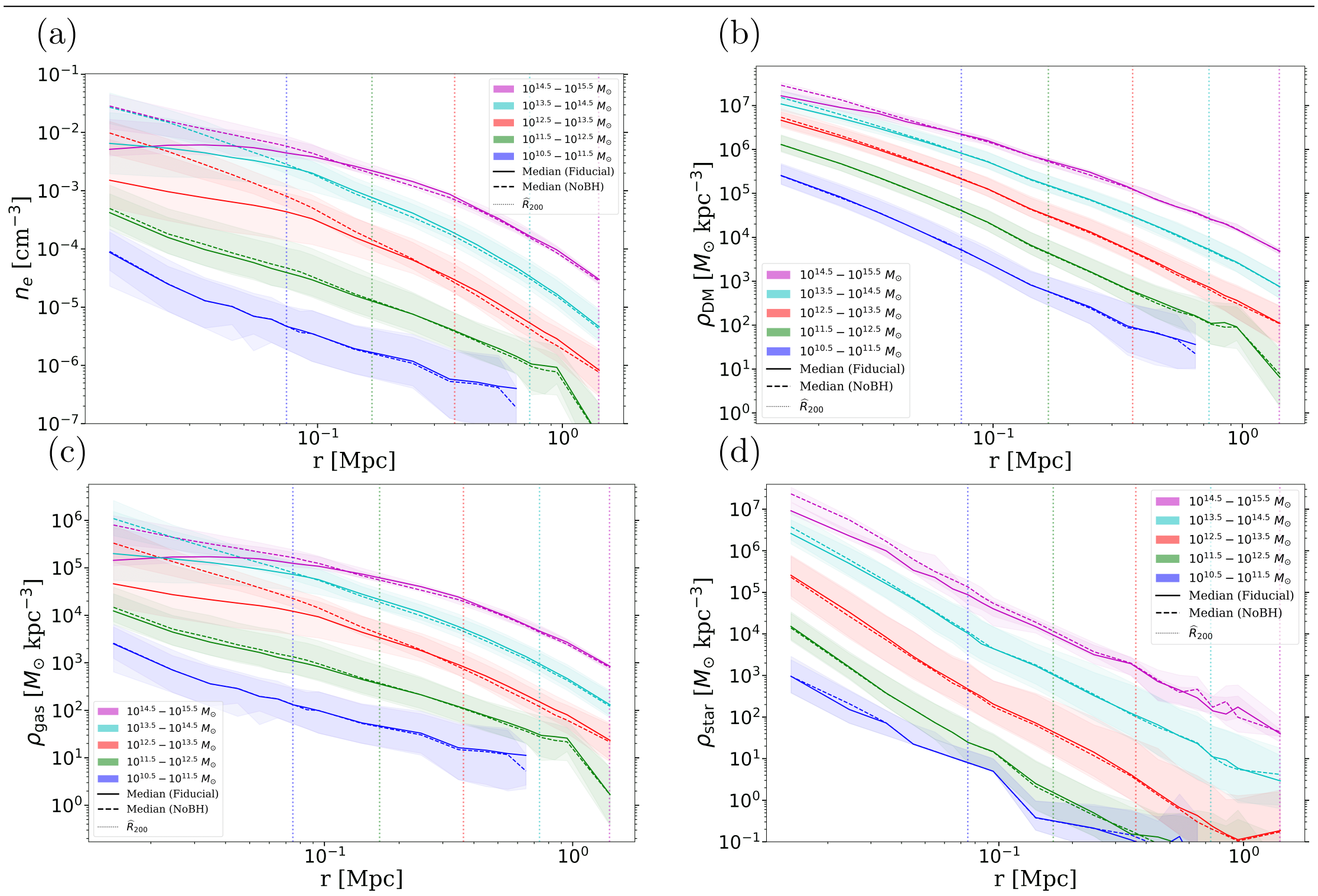

FIGURE 3.4: Comparison of density profiles for fiducial and NoBH simulations in the 100 Mpc box. The panels show the overlay of median density profiles for different components. The shaded regions represent the $1\sigma$ range of variations in each mass bin. Solid lines represent fiducial simulations, while dashed lines represent NoBH simulations. The dotted lines indicate the median $R_{200}$ for halos in each mass range in the fiducial results, with NoBH results showing nearly no difference.

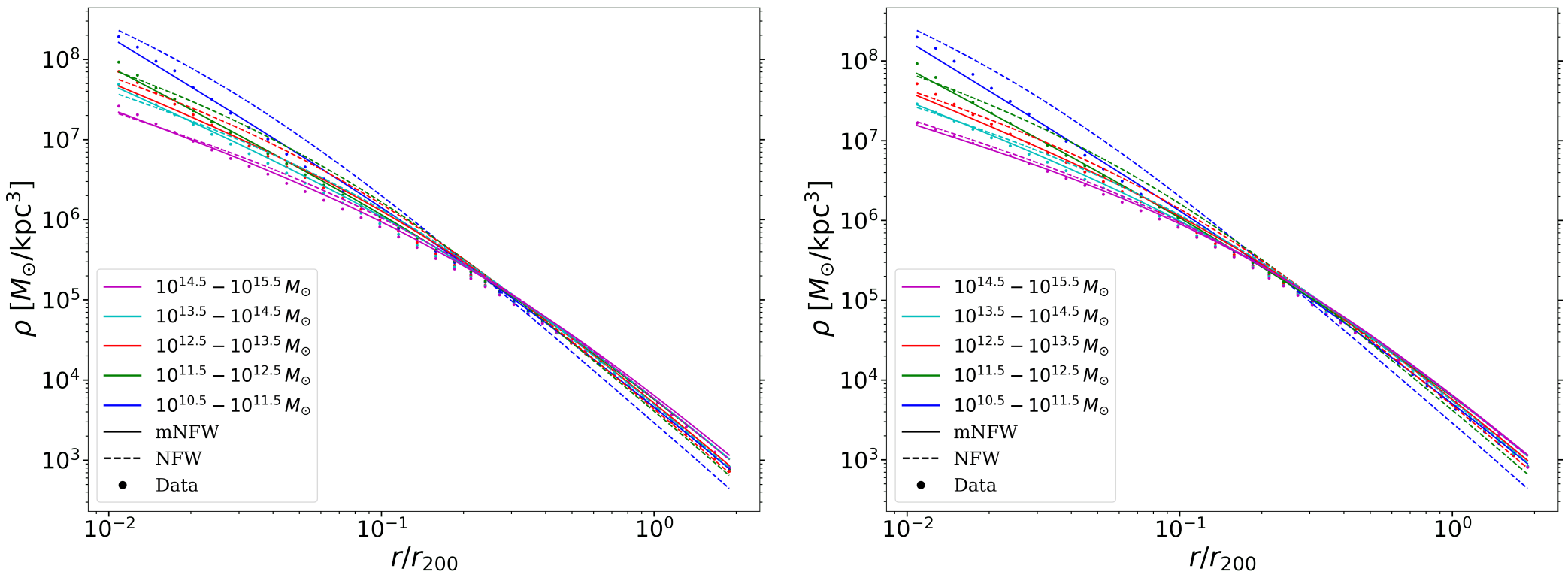

FIGURE 3.5: Comparison of the mNFW and NFW mass profiles for fiducial (top) and NoBH (bottom) for 100 Mpc box

Figure 3.4 show the median density profiles from the L100 simulation, respectively, for various mass ranges. (The L500 results are shown in the Appendix B.1). Panel (a) displays gas density, panel (b) $n_e$, panel (c) dark matter density ($\rho_{\rm DM}$), and panel (d) stellar mass density. The shaded regions indicate the 16th–84th percentile range of variations among the selected halos. Notably, the trends in gas and $n_e$ profiles are consistent, reflecting the high ionization fraction of the gas.

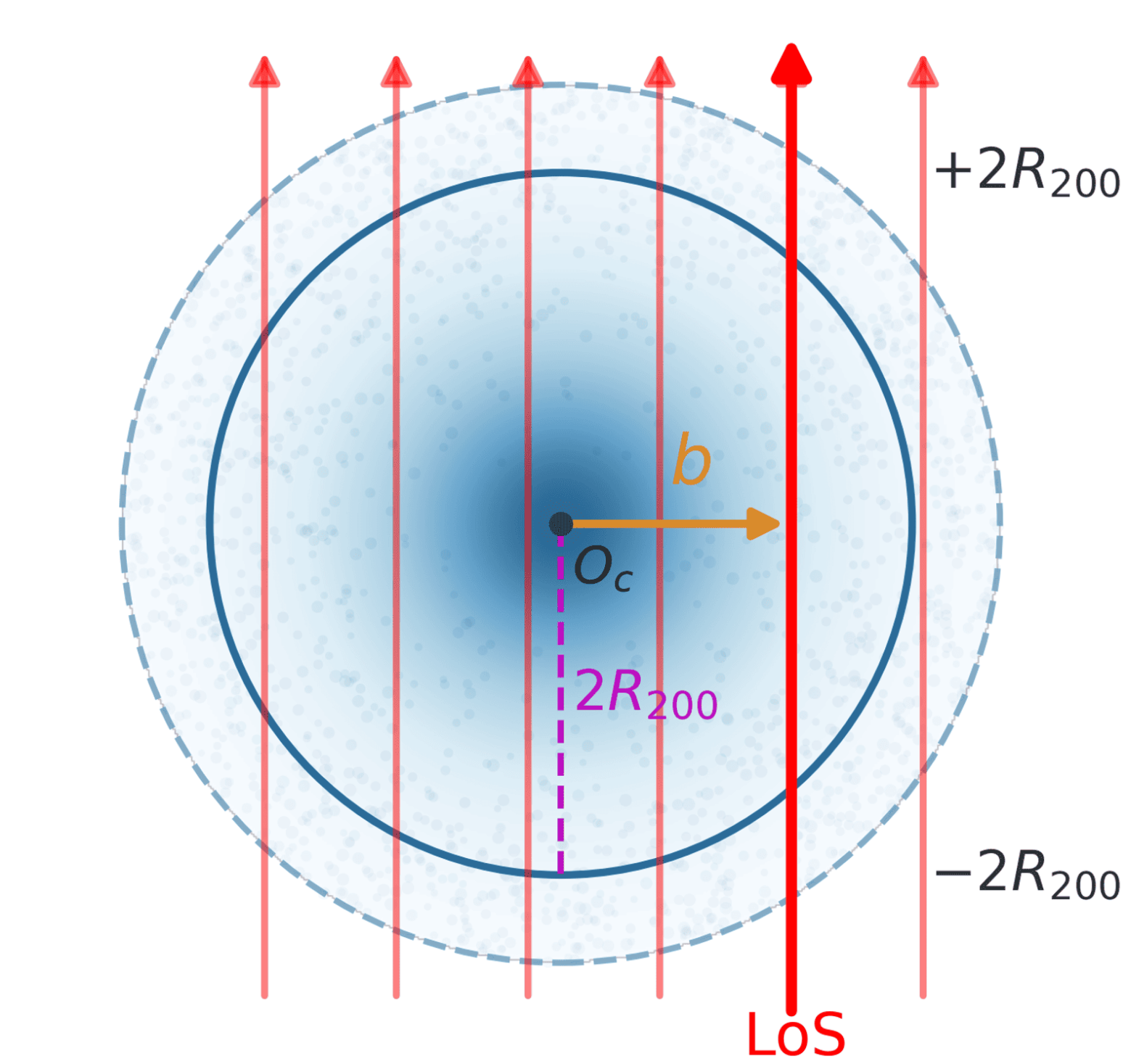

FIGURE 3.6: Schematic of the halo-centered line-of-sight (LoS) geometry adopted in this chapter. The halo center is marked by $O_c$, the virial radius by $R_{200}$, the impact parameter by $b$, and the integration path by $[-2R_{200}, +2R_{200}]$. The outermost dashed boundary indicates the actual particle-selection region used in the calculation, which is set to extend beyond $2R_{200}$. This enlarged boundary is adopted because, in SPH, particles outside $2R_{200}$ can still contribute to the physical properties of particles inside the integration region through kernel-weighted interactions.

To assess the effects of AGN feedback, we compare the density profiles of fiducial and NoBH simulations in the L100 simulation, as shown in Figure 3.4. Each panel compares the fiducial (solid lines) and NoBH (dashed lines) simulations across different halo mass ranges. AGN feedback significantly reduces the central densities of gas and $n_e$ in halos, redistributing baryonic matter toward the outskirts. This effect is particularly pronounced in halos with masses between $10^{12.5}\,M_\odot$ and $10^{13.5}\,M_\odot$. For lower-mass halos ($< 10^{12.5}\,M_\odot$), AGN feedback has minimal impact, while for more massive halos, gravitational forces limit its effectiveness.

Based on panels (c) and (d), despite the relatively weak influence of AGN feedback on dark matter and stellar mass, it slightly reduces the central densities of these components in high-mass halos. This subtle effect is likely due to the gravitational drag exerted by gas particles that are expelled by AGN feedback. As the gas is redistributed outward, its gravitational interaction modulates the distribution of non-baryonic components, including dark matter and stars. This effect has been thoroughly analyzed and discussed in the recent work by Sorini et al. (2025), which provides a detailed characterization of how baryonic processes shape dark matter density profiles. While this effect is not as prominent as the redistribution of gas and $n_e$, it nonetheless underscores the interconnected dynamics of baryonic and non-baryonic matter in high-mass halos influenced by AGN feedback.

We compare the dark matter density profiles for fiducial and NoBH simulations using NFW and mNFW models, as shown in Figure 3.5. Both models fit the dark matter density profiles well; however, the mNFW model provides a slightly better fit, capturing the subtle features in our simulation results more effectively.

### 3.2.2 DM profile as a function of impact parameter

To quantify the contribution of foreground halos to the DM, we determined the center of each halo and extracted all gas particles within a sphere of radius $2R_{200}$. sight lines were then generated perpendicular to a reference plane passing through the halo center, with the impact parameter $b$ representing the perpendicular distance from the sightline to the halo center.

The DM for each sightline was calculated using equation 1.12, which integrates the electron number density $n_{e,\mathrm{halo}}(l)$ along the sightline over a path length of $-2R_{200}$ to $2R_{200}$. This method was applied to 500 halos in each mass range to obtain the DM distribution as a function of $b$. Separate analyses were conducted for simulations with and without AGN feedback to investigate its influence. Figure 3.7 illustrates the DM distributions for halos of varying masses. For sight lines passing close to the halo center, DM values range from a few $\mathrm{pc\,cm^{-3}}$ for low-mass halos to several thousand $\mathrm{pc\,cm^{-3}}$ for massive halos. AGN feedback notably reduces the central DM while increasing the outer DM profiles, particularly in halos with masses $10^{12.5}\,M_\odot < M_H < 10^{13.5}\,M_\odot$.

Observational constraints from Wu & McQuinn (2023), Connor et al. (2023), and Lee et al. (2023), along with simulation-based fits from Medlock et al. (2024), are included for comparison. Although some deviations exist, the simulation results align well with Medlock's fits within similar mass ranges. The estimates from Lee et al. (2023) are derived using a semi-analytical gas profile model for two foreground clusters intersecting the sightline of FRB 20190520B, assuming different cutoff radii ($R_{200}$ and $2R_{200}$). The results from Connor et al. (2023) are based on excess DM values of two FRBs (20220914A and 20220509G) hosted by massive clusters A2310 and A2311, after subtracting Galactic and IGM contributions. These observational points fall within or slightly above the range predicted by our halo DM profiles across the corresponding mass scales. Furthermore, our analysis extends to higher halo masses ($M_H \gtrsim 10^{14} M_\odot$), surpassing those considered in CAMELS simulations, allowing us to explore the DM contribution from the most massive halos. This provides new insights into the CGM properties in cluster-scale environments.

### 3.2.3 2D DM Map and 1D Profiles of Individual Halos

The two-dimensional DM distribution (i.e. 2D map) of individual halos is often more complex than 1D profiles, especially for high-mass halos, such as galaxy clusters and massive galaxies.

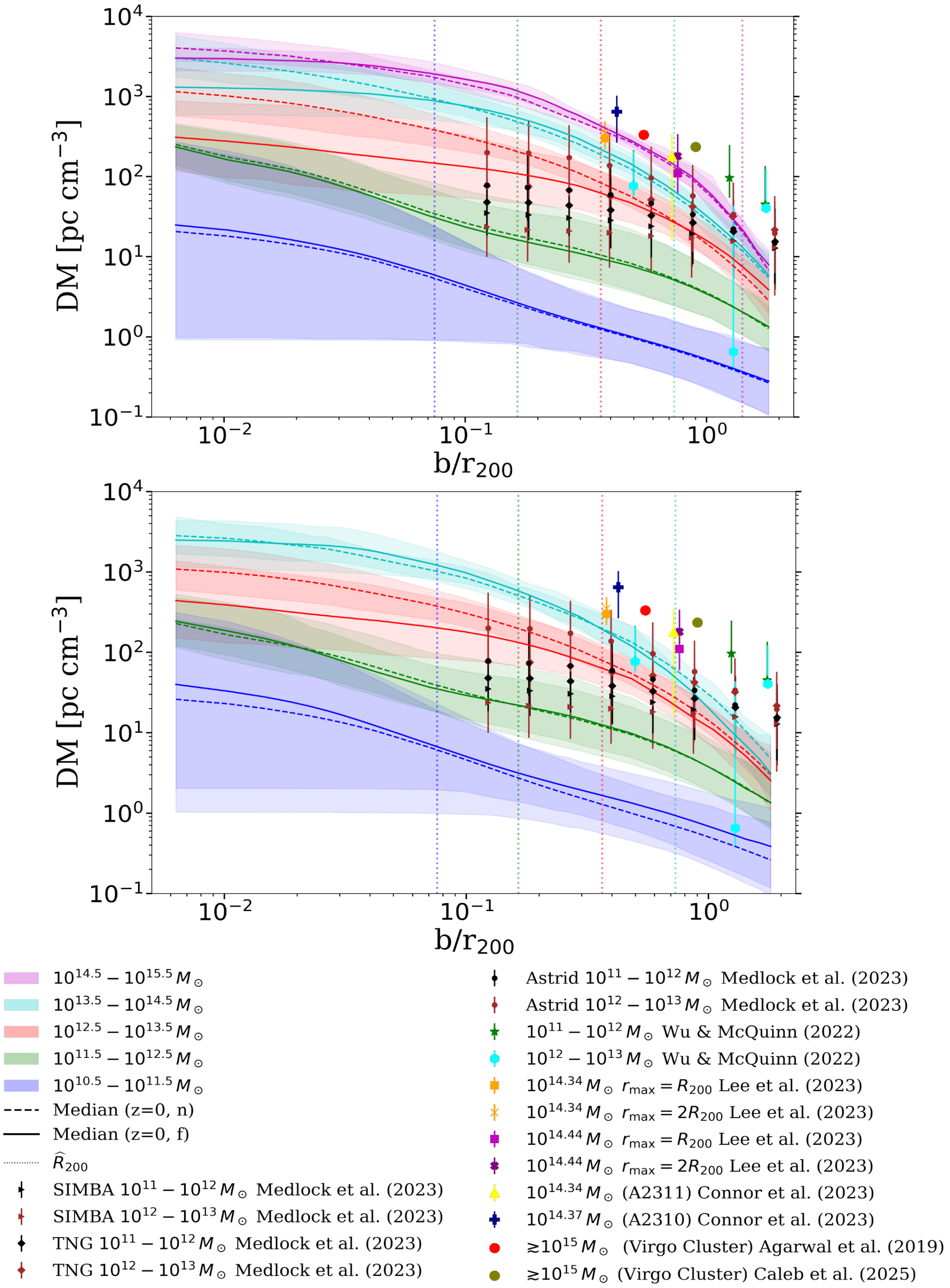

FIGURE 3.7: Dispersion measure (DM) as a function of impact parameter $b$ for halos in the L100 (top) and L50 (bottom) simulations, categorized by mass ranges. Solid and dashed lines represent our simulations with and without AGN feedback, respectively, while shaded regions denote the $1\sigma$ range of DM values for 500 halos in each mass range. The central curves correspond to the median DM values at each impact parameter. The DM values are obtained by integrating electron densities along the LoS over a path length of $-2R_{200}$ to $2R_{200}$. Observational constraints from Wu & McQuinn (2023), Connor et al. (2024), Lee et al. (2023), Agarwal et al. (2019) and Caleb et al. (2025), as well as simulation-based fits from Medlock et al. (2024), are overlaid for comparison. The Medlock data points include results from TNG, Astrid, and SIMBA simulations across different mass bins. The points from Connor et al. (2023) correspond to FRBs intersecting galaxy clusters A2310 and A2311. The Lee et al. (2023) estimates are derived from semi-analytical ICM models for two foreground clusters intersected by FRB 20190520B, assuming cutoff radii ($r_{\max}$) of $R_{200}$ and $2R_{200}$. The Agarwal et al. (2019) and Caleb et al. (2025) points are based on FRB 20180417 and FRB 20240304B, respectively; both sight lines pass through the Virgo Cluster and thus constrain DM at the corresponding impact parameters.

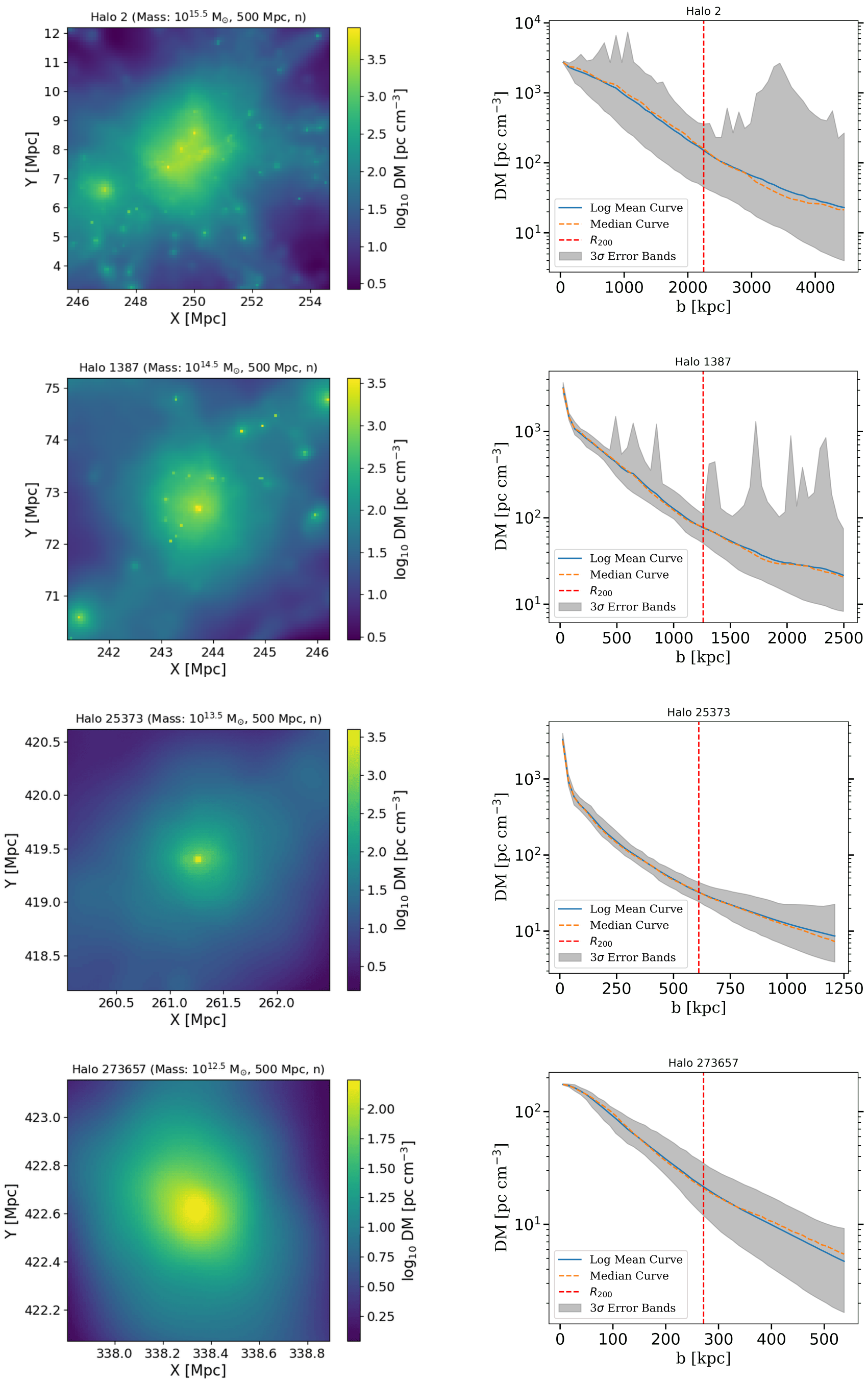

FIGURE 3.8: 2D DM maps (left column) and 1D DM profiles as a function of impact parameter $b$ (right column) for the most massive halos in each representative mass range, as shown in Figure B.1, from the L500 simulation. Each row corresponds to a halo, arranged in descending order of mass from top to bottom, with the halo ID and mass indicated in the title of each plot. The vertical red dashed lines in the right panels indicate the virial radius of the halo ($R_{200}$). Shaded regions in the 1D profiles represent the scatter of different sight lines at each $b$. All halos are from simulations without AGN feedback (labeled "n").

These halos can exhibit intricate structures, including satellite galaxies, varying degrees of gas dispersal (potentially influenced by AGN feedback), or even multiple halos in close proximity undergoing merger processes. In these cases, 1D spherically symmetric models may fail to accurately capture the true characteristics of the halos. Using such models requires caution when applied to such complex systems.

To investigate the contributions of dense substructures near halo centers and the surrounding gas environment, as well as to assess their deviations from 1D models, we analyzed the 2D DM map of halos. This analysis also enables us to examine the impact of AGN feedback on these distributions.

Figure 3.8 presents the 2D DM maps and 1D DM profiles of representative halos from different mass ranges in the L500 simulation. For high-mass halos like Halo 2 and Halo 1387, more complex structural distributions are observed. Numerous satellite galaxies surround these halos, leading to noticeable "peaks" in their 1D DM profiles at various locations. These fluctuations result in significantly elevated DM values and contribute to variations in the mean profile, although the overall impact on the average remains moderate. Halo 2, compared to Halo 1387, exhibits a larger number of substructures, and its surrounding gas environment shows greater overlap, resulting in a 1D profile with a broader scatter depending on the LoS location.

In contrast, low-mass halos, such as Halo 25373 and Halo 273657 (two lower sets), tend to be more isolated, lacking complex substructures. As a result, they are better described by 1D spherically symmetric models. Between these two halos, Halo 25373, having a higher mass, shows a sharp density increase near its center, appearing as a distinct peak in the 2D distribution. Conversely, the lower-mass Halo 273657 has a smoother central region, and its surrounding gas environment is more dispersed due to weaker gravitational influence.

To investigate the influence of AGN feedback on the contribution of foreground halos to the DM of FRBs, we compare results from the L100 simulations. Figure 3.9 presents the 2D DM maps and 1D DM–$b$ profiles of two representative massive halos ($\sim 10^{14.8} M_{\odot}$) in both the L100N1024$_{\mathrm{fiducial}}$ and L100N1024$_{\mathrm{NoBH}}$ runs.

An analysis of ten halos with masses exceeding $10^{14.5} M_{\odot}$ in the L100 simulations reveals that AGN feedback generally has a limited impact on the gas distribution within such massive halos, which exhibit strong self-gravitating systems. This is exemplified by Halo 1, where AGN feedback induces only minimal redistribution effects, with slight diffusion observed near the core in both the 2D map and 1D profile. However, among these massive halos, we identify an exceptional case—Halo 0. As seen in its 2D map and 1D profile, the NoBH simulation exhibits a highly dense, compact region approximately 1 Mpc from the halo center, with a gas density exceeding that of the central region, leading to an anomalous DM distribution. Such occurrences are possible in the absence of AGN feedback, where gas can over-concentrate locally. Similar phenomena are frequently observed in massive halos within the L500 simulation (Figure 3.8). In contrast, the fiducial run of Halo 0 shows that this dense clump is absent, as AGN feedback efficiently disperses the gas, resulting in a more diffuse central gas distribution.

Figure 3.10 presents a comparison of two halos in the $10^{12.5} - 10^{13.5} M_\odot$ mass range, where AGN feedback is expected to be more efficient. Our statistical analysis of over a hundred halos in this range confirms that AGN feedback is particularly effective within this mass interval. We highlight two representative cases: Halo 223 (fiducial) and Halo 222 (NoBH), both with masses of $\sim 10^{13.5} M_\odot$, collectively referred to as Halo $\alpha$; and Halo 1000 (fiducial) and Halo 886 (NoBH), with masses in the range of $10^{12.8} - 10^{12.9} M_\odot$, collectively referred to as Halo $\beta$.

Halo $\alpha$ exhibits three distinct structures: a primary halo, a massive subhalo in close proximity, and a significantly smaller subhalo. AGN feedback has a pronounced effect on both main components, leading to a substantial reduction in central DM in the fiducial run compared to the NoBH counterpart. Additionally, the highest AGN feedback efficiency is observed near $10^{13} M_\odot$, as demonstrated by Halo $\beta$. The NoBH simulation yields a central DM approaching $10^3$ pc cm$^{-3}$, whereas the fiducial counterpart shows a significantly lower central DM, barely exceeding $10^2$ pc cm$^{-3}$.

The qualitative AGN-driven differences seen in these representative halos motivate a dedicated quantitative analysis. We therefore defer the mass-dependent expulsion/suppression metrics and their physical interpretation to Section 3.3.1.

## 3.3 Impact of AGN Feedback on DM Statistics

### 3.3.1 Mass-dependent gas expulsion and DM suppression

To quantify the redistributing impact of AGN feedback on halo gas densities and DM, we compute the AGN-driven central gas expulsion fraction and central DM suppression fraction, defined as:

$$f_{\rm exp}^{15} = \frac{\widehat{\rho}_{\rm NoBH}(r = 15\,{\rm kpc}) - \widehat{\rho}_{\rm fiducial}({\rm r} = 15\,{\rm kpc})}{\widehat{\rho}_{\rm NoBH}(r = 15\,{\rm kpc})}, \tag{3.6}$$

$$f_{\rm suppr}^{c} = \frac{\widehat{\rm DM}_{\rm NoBH}(b = 0) - \widehat{\rm DM}_{\rm fiducial}(b = 0)}{\widehat{\rm DM}_{\rm NoBH}(b = 0)}, \tag{3.7}$$

where $\widehat{\rho}(r = 15\,{\rm kpc})$ represents the median gas density at 15 kpc for different halo mass bins, and $\widehat{\rm DM}(b = 0)$ represents the median DM at impact parameter $b = 0$. The computed expulsion and suppression fractions for different halo mass ranges are:

- $10^{10.5} - 10^{11.5} M_\odot$: $f_{\rm exp}^{15} = 2.3\%$, $f_{\rm suppr}^{c} = -20.6\%$,
- $10^{11.5} - 10^{12.5} M_\odot$: $f_{\rm exp}^{15} = 16.2\%$, $f_{\rm suppr}^{c} = 6.8\%$,
- $10^{12.5} - 10^{13.5} M_\odot$: $f_{\rm exp}^{15} = 86.1\%$, $f_{\rm suppr}^{c} = 73.0\%$,
- $10^{13.5} - 10^{14.5} M_\odot$: $f_{\rm exp}^{15} = 81.7\%$, $f_{\rm suppr}^{c} = 57.1\%$,
- $10^{14.5} - 10^{15.5} M_\odot$: $f_{\rm exp}^{15} = 81.9\%$, $f_{\rm suppr}^{c} = 25.3\%$.

This analysis demonstrates that AGN feedback is most efficient at expelling gas from halos in the mass range $10^{12.5} - 10^{13.5} M_\odot$, significantly reducing central gas densities. The strong correlation between $f^c_{\rm exp}$ and $f^c_{\rm suppr}$ indicates that AGN-driven gas removal directly impacts DM suppression along the LoS.

It is worth noting that although the AGN feedback in the $10^{13.5}-10^{14.5}M_\odot$ and $10^{14.5}-10^{15.5}M_\odot$ mass ranges also exhibits strong gas expulsion effects in the central region (at 15 kpc), it does not completely expel the gas from the halo. As shown in Figures 3.4 and 3.7, most of the redistributed gas remains within $R_{200}$, meaning that the expelled material largely stays within the so-called "CGM". Beyond $R_{200}$, the difference between the fiducial and NoBH simulations is relatively small. However, for halos in the $10^{12.5}-10^{13.5}M_\odot$ range, we observe that beyond $R_{200}$, the gas density profile, electron number density profile, and DM profile in the fiducial run are all significantly higher than those in the NoBH run. This suggests that halos in this mass range are the primary drivers of baryonic redistribution. In addition to exhibiting the highest AGN feedback efficiency, these halos also play a dominant role in facilitating material exchange between the IGM and CGM.

In the lowest-mass bin ($10^{10.5}-10^{11.5}M_\odot$), we observe a very small $f^c_{\rm exp}$ but a significant negative $f^c_{\rm suppr}$. This suggests that rather than AGN-driven expulsion, ionization effects are the primary factor responsible for the observed changes. The increased electron density and corresponding enhancement in DM could be attributed to the following possible mechanisms: (1) collisionally ionized gas due to heating within the halo itself, (2) shock heating effects from nearby massive halos, or (3) thermal energy injected by the SN-driven outflows. A more detailed investigation of these effects is beyond the scope of this paper.

The efficiency of AGN feedback in the $10^{12.5}-10^{13.5}M_\odot$ range is further validated in the appendix, where we analyze the relationship between AGN feedback energy and halo mass. As illustrated in Figures 3.13 and 3.15, the correlation between AGN feedback efficiency and halo mass provides additional evidence supporting our findings.

### 3.3.2 Impact of AGN Feedback on 2D DM Distributions in Massive Halos

Here we present additional visualizations of the DM distributions in halos from the L100 simulations. Figures 3.9 and 3.10 compare the effects of AGN feedback in two different halo mass ranges.

Figure 3.9 shows the 2D DM maps and 1D DM–$b$ profiles of two representative massive halos ($\sim 10^{14.8}M_\odot$) in both the fiducial and NoBH simulations. The analysis of these halos suggests that AGN feedback has a limited impact on the central gas distribution in the most massive halos but may still affect their outskirts.

Figure 3.10 illustrates the effect of AGN feedback in the $10^{12.5}-10^{13.5}M_\odot$ mass range, where feedback mechanisms are expected to be more effective. The comparison highlights the redistribution of gas and its impact on DM profiles due to AGN feedback.

These results support the findings discussed in Section 3.2.3 of the main text, emphasizing the varying efficiency of AGN feedback across different halo mass ranges.

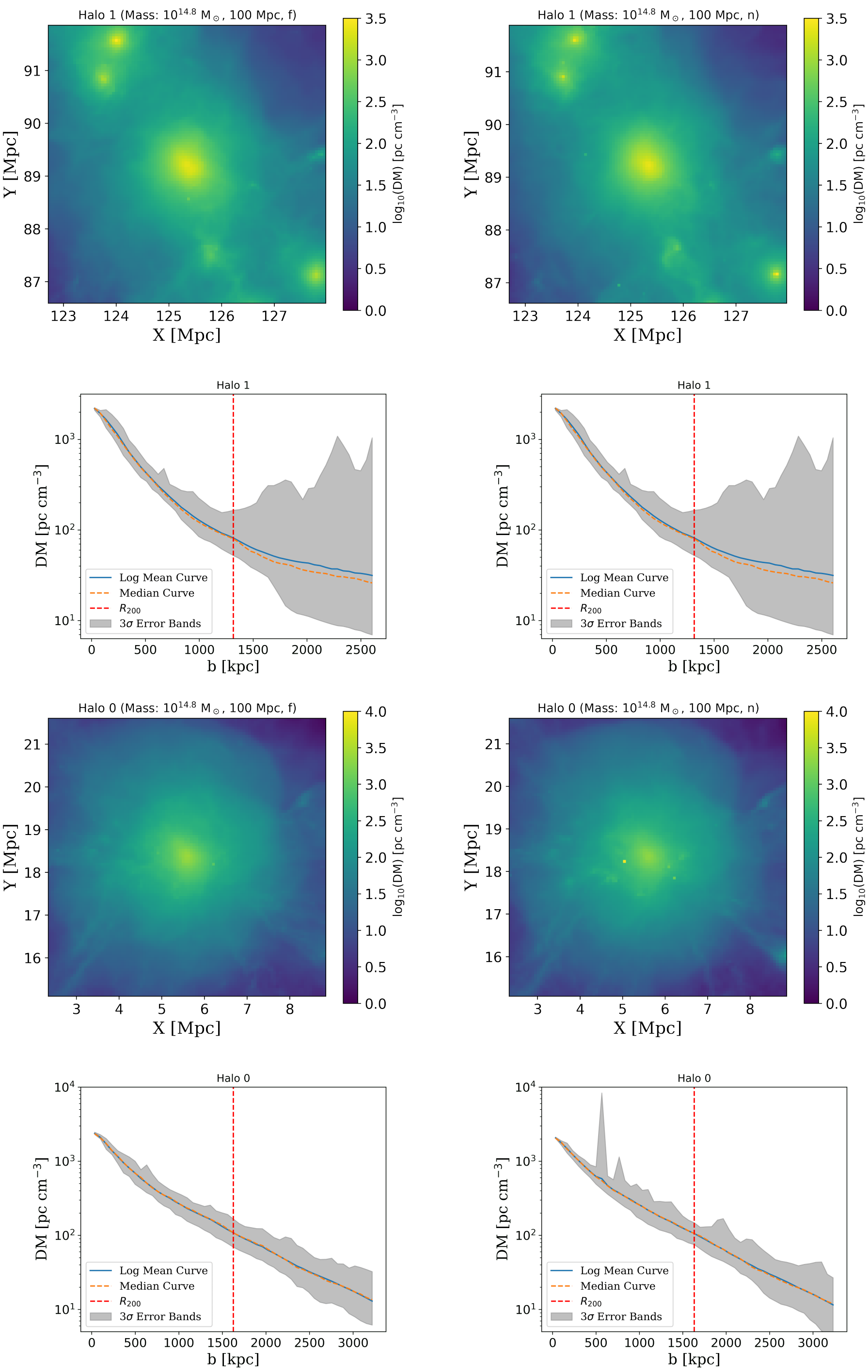

FIGURE 3.9: Similar to Figure 3.8, this figure compares 2D DM maps (top row) and 1D DM–$b$ profiles (bottom row) for halos in the highest mass range, focusing on Halo 0 and Halo 1. Results with AGN feedback (L100N1024$_{\text{fiducial}}$) are shown in the left column, and results without AGN feedback (L100N1024$_{\text{NoBH}}$) are shown in the right column.

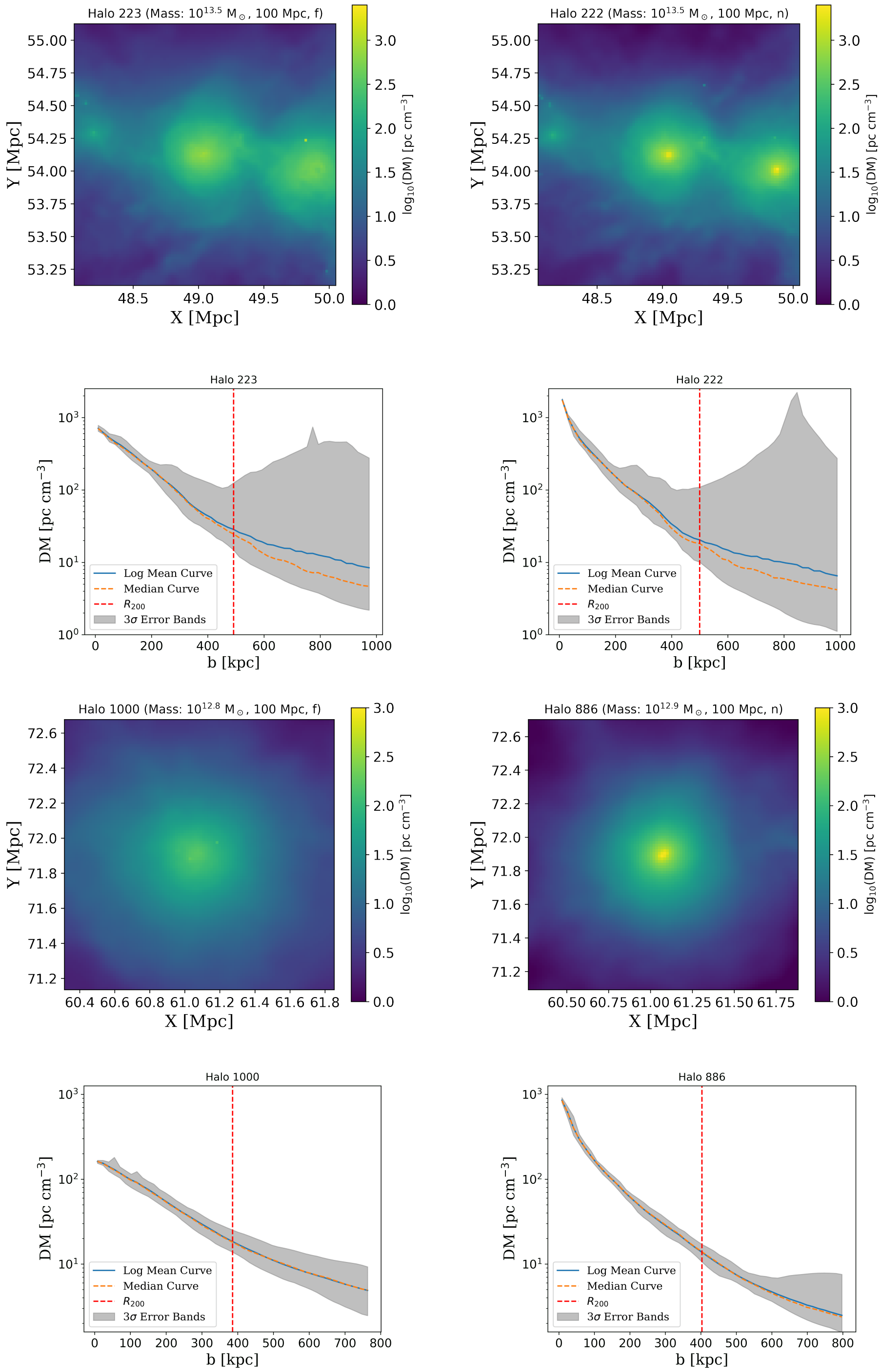

FIGURE 3.10: Similar to Figure 3.8, this figure is for halos in the mass range of $10^{12.5}$-$10^{13.5}$ $M_\odot$, where AGN feedback is expected to be more efficient. The left column shows results from the fiducial simulation (L100N1024$_{\text{fiducial}}$), while the right column corresponds to the NoBH simulation (L100N1024$_{\text{NoBH}}$). The impact of AGN feedback is evident, particularly in reducing central DM contributions.

### 3.3.3 Phase Diagram and Ionization Fraction

To investigate the impact of AGN feedback, Figure 3.11 presents the results from the L100 simulations, highlighting the relationship between gas temperature and density, as well as other key properties such as electron number density and hydrogen ionization state.

In the presence of AGN feedback, the phase diagram (left) exhibits a distinct difference from the NoBH scenario. In the temperature–density diagram (top row), AGN feedback leads to more hot gas with $T \gtrsim 10^8$ K in the low-density, high-temperature regions, while the NoBH case exhibits lower temperatures in these areas. In the high-density regions, the NoBH simulation shows a higher degree of condensation, with particles distributed more uniformly and extensively. This indicates that AGN feedback is actively expelling gas from the dense, collapsed regions, heating it to higher temperatures and disrupting the condensation process observed in the NoBH scenario.

In the hydrogen ionization fraction diagram (bottom row), while low-density hydrogen remains almost fully ionized in both scenarios, the NoBH simulation exhibits a broader distribution of highly ionized regions at high densities. This suggests that in the absence of AGN feedback, ionized gas remains more concentrated in dense environments, leading to a more extended ionization structure.

These differences highlight the role of AGN feedback in regulating the thermal and ionization states of gas, as it redistributes baryons from dense regions into a more diffuse, hotter intergalactic medium. This underscores the significant impact of AGN activity in shaping the balance between the circumgalactic medium (CGM) and the intergalactic medium (IGM).

### 3.3.4 Halo, Black Hole, and Stellar Mass Relations

To understand the broader impact of AGN feedback, we next examine how it influences the relationships between halo mass and both black hole mass and stellar mass. By exploring this connection, we can further elucidate the role of AGN feedback in the broader context of galaxy evolution, particularly in regulating black hole growth and quenching star formation.

Figure 3.12 compares the black hole mass vs. halo mass relations in the L50 (upper) and L100 (lower) simulations. In this context, $M_{\rm halo,200}$ refers to the total halo mass enclosed within a radius $R_{200}$. The dashed red line represents the relation from Booth & Schaye (2010), while the solid blue line indicates the best-fit log-linear relation derived from this work:

$$\log_{10} M_{\rm BH} = \alpha \log_{10} M_{\rm halo,200} + b, \tag{3.8}$$

where $\alpha$ and $b$ represent the slope and intercept of the log-linear fit.The observed differences might stem from the smaller sample size in the L50 box, which could enhance the impact of stochastic effects and deviations from the mean relation.

Additionally, the presence of two distinct black hole branches in both simulations might be attributed to the different merger histories of the black holes and growth histories. To characterize

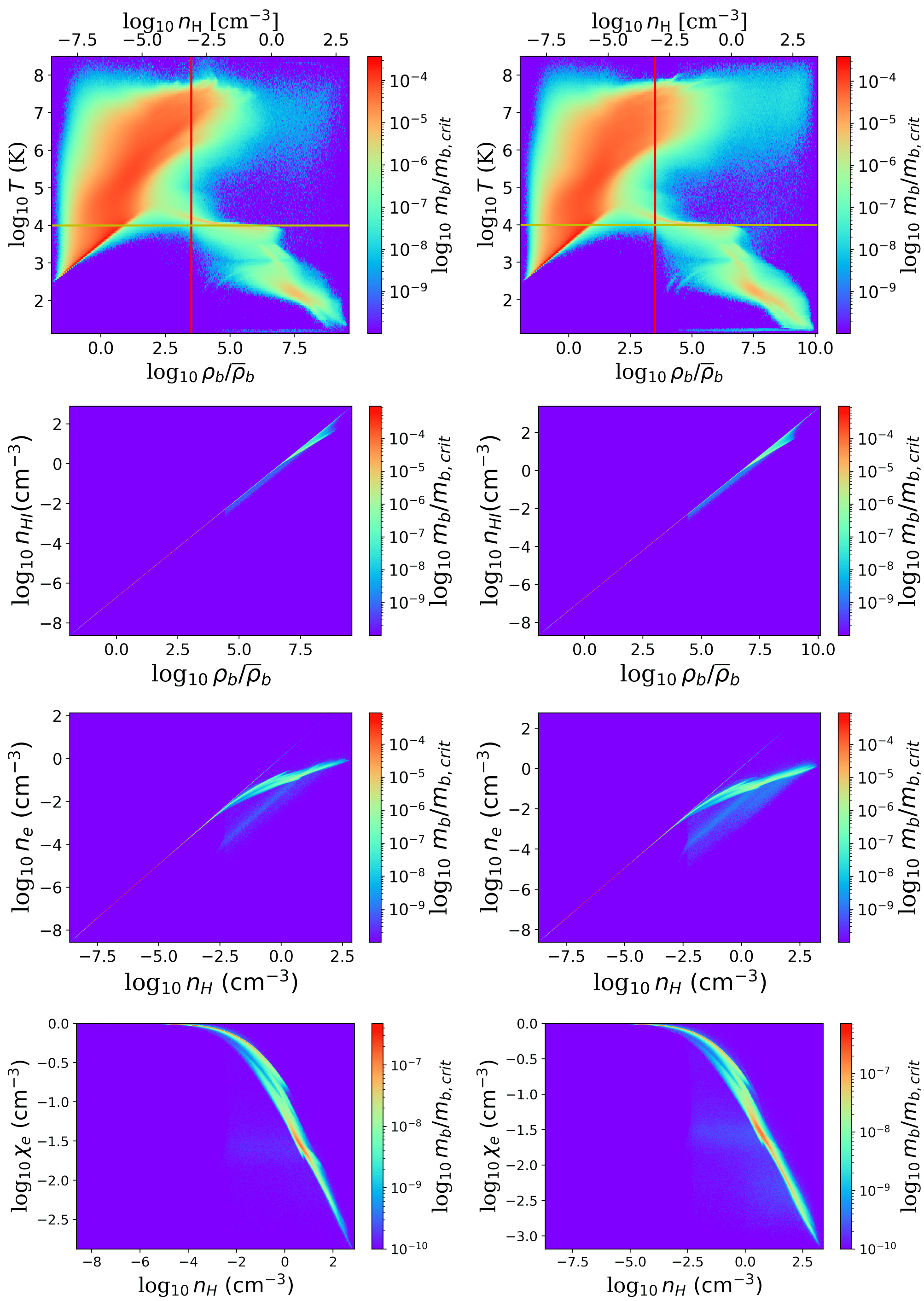

FIGURE 3.11: Comparison of various properties in fiducial and NoBH simulations for the L100 run. Each row represents a different set of properties: temperature vs. density (top row), baryon overdensity vs. hydrogen number density (second row), hydrogen number density vs. electron number density (third row), and hydrogen number density vs. ionization fraction (bottom row). The results from the fiducial simulation (left panels) are compared with those from the NoBH simulation (right panels).

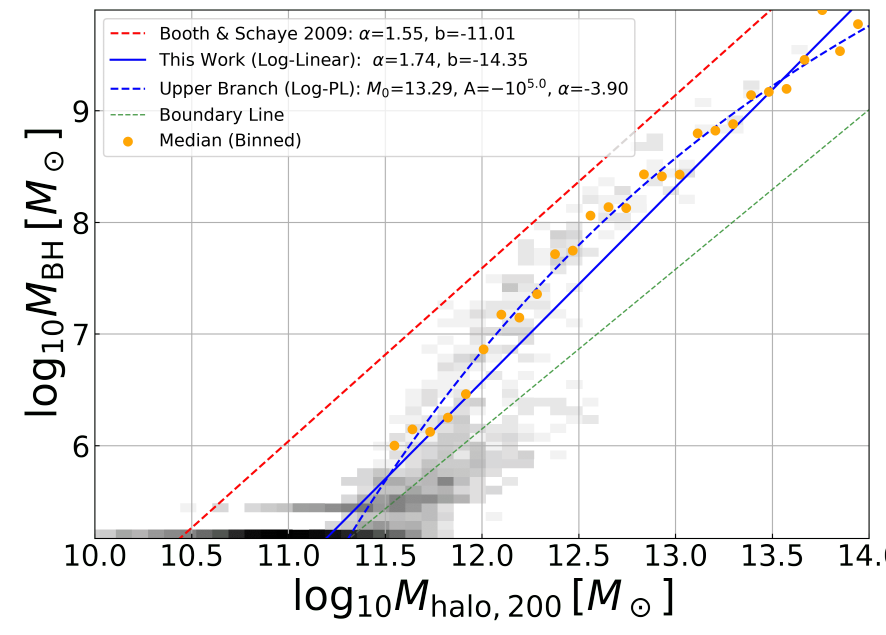

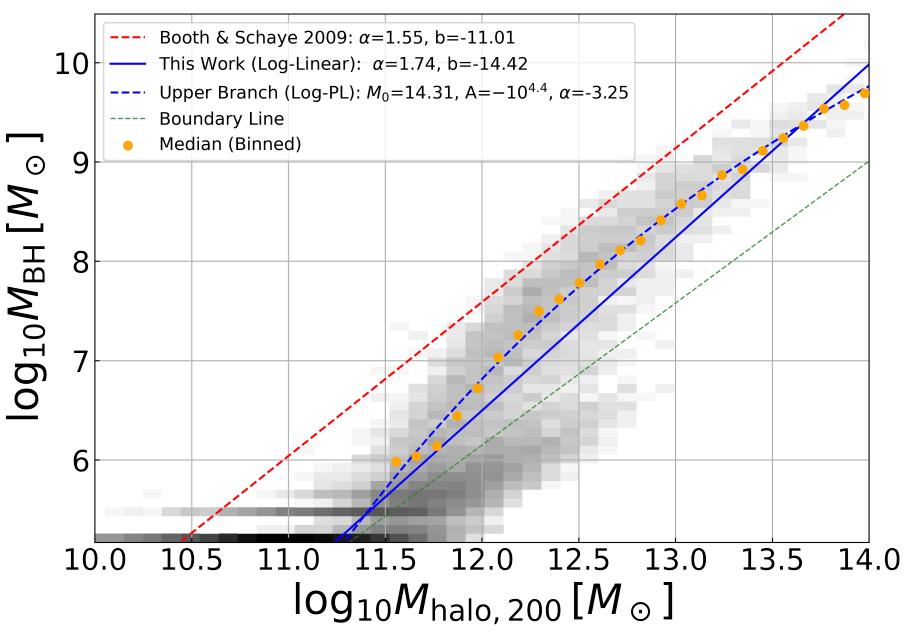

FIGURE 3.12: Black hole mass vs. dark matter halo mass relations for the L50N512$_{\text{fiducial}}$ (upper) and L100N1024$_{\text{fiducial}}$ (lower) fiducial simulations. The dashed red line represents the relation from Booth & Schaye (2010), while the solid blue line represents the best-fit relation from this work using a log-linear fit. The green dashed line represents the manual boundary line separating the upper and lower branches. The upper branch is fitted separately using a log-power-law (Log-PL) function. The best-fit values for each fitting method are annotated in each panel. The orange points indicate the binned median values used for fitting the upper branch.

the upper branch, we apply a log-power-law (Log-PL) fit to its median values:

$$M_{\mathrm{BH}} = 10^{M_0} + A M_{\mathrm{halo},200}^{\alpha}, \tag{3.9}$$

where $M_0$ accounts for the shift in normalization, $A$ is the amplitude, and $\alpha$ is the power-law exponent. The upper branch potentially results from major mergers, while the lower branch could be due to more isolated black hole systems that have experienced fewer mergers, which requires further analysis in our follow-up paper (D. Nishihama et al. 2025, in preparation).

### 3.3.5 Relation between gas mass and dark matter halo mass

Figure 3.13 presents the comparison between the halo mass ($M_{\mathrm{halo}}$) and gas mass ($M_{\mathrm{gas}}$) relations across different simulations with and without AGN feedback. The gas mass generally increases with halo mass across all simulations, as expected. However, AGN feedback tends to reduce the gas mass at a given halo mass, especially in more massive halos.

In all the upper panels of Figure 3.13, the relationship between $M_{\mathrm{gas}}$ and $M_{\mathrm{halo},200}$ shows a linear fit in log–log space, with minor differences in the slope $\alpha$. The fiducial run has a slightly shallower slope, particularly in the high-mass region ($M_{\mathrm{halo}} > 10^{12} M_\odot$), likely due to stronger AGN feedback expelling gas more efficiently in massive halos.

In the bottom panels of Figure 3.13, gas fraction $f_{\mathrm{gas}}$ is shown as a function of $M_{\mathrm{halo},200}$. As halo mass increases, the overall trend of $f_{\mathrm{gas}}$ shows a gradual increase. In the low-mass region ($M_{\mathrm{halo}} < 10^{11} M_\odot$), the $f_{\mathrm{gas}}$ distribution is relatively dispersed, with most $f_{\mathrm{gas}}$ below 0.1, while some retain higher gas fractions (exceeding 0.2). This suggests that lower-mass halos tend to lose more gas due to stellar feedback and supernova-driven winds, but there are still some halos capable of retaining a significant fraction of their baryonic content.

In the intermediate mass range ($M_{\rm halo} \in 10^{12} - 10^{13.5} M_\odot$), AGN feedback becomes the dominant mechanism in redistributing baryons between CGM and IGM. The efficiency of AGN-driven outflows in this range is high enough to expel gas from the CGM, effectively lowering the gas fraction of halos and causing a more dispersed distribution in $f_{\rm gas}$. This indicates that AGN feedback is most efficient at removing gas entirely from the halo potential well in this mass regime.

However, for massive halos beyond the galaxy group scale ($M_{\rm halo} > 10^{13.5} M_\odot$), the scatter in $f_{\rm gas}$ becomes significantly smaller. This suggests that despite stronger AGN feedback at higher masses, it is no longer strong enough to expel gas beyond the deep gravitational potential well of these halos. Instead, AGN feedback redistributes gas within the halo rather than fully removing it, likely pushing gas from the central regions to the outskirts of the CGM. This expelled gas remains gravitationally bound and will eventually reaccrete back into the halo center over time, contributing to the stable gas fraction observed in massive halos.

To further evaluate our results, we compare the baryon mass fraction ($f_{\rm baryon}$) in our simulations to previous theoretical models and observational constraints. Figure 3.14 illustrates the $f_{\rm baryon}$ as a function of halo mass at $z = 0$, including results from our fiducial and various feedback models described in Oku & Nagamine (2024), along with comparisons to EAGLE-Ref and TNG-100 simulations. Observational constraints from Das et al. (2023) are also shown. We can see a better agreement to the observed gas fraction in the intermediate halo mass range for the CROCODILE simulation compared to EAGLE-Ref and TNG-100, which could be pushing the gas too much by AGN feedback.

### 3.3.6 Feedback Energy, Potential Energy, vs. Halo Mass Relations

To identify the halo mass range where AGN feedback is most effective, we compare the energy from AGN feedback ($E_{\rm AGN}$) with the gravitational potential energy ($E_{\rm grav}$), the binding energy ($E_{\rm bind}$), and the quench energy ($E_{\rm quench}$) across different halo masses, as shown in Figure 3.15. We can compute each energy component by the following equations:

$$\dot{E}_{\rm AGN} = \epsilon_f \epsilon_r \dot{m}_{\rm acc} c^2 = \frac{\epsilon_f \epsilon_r}{1 - \epsilon_r} \dot{m}_{\rm BH} c^2, \tag{3.10}$$

$$E_{\rm grav} = \frac{3 G M_{200}^2}{5 R_{200}}, \tag{3.11}$$

$$E_{\rm bind} = \frac{1}{2} f_{\rm gas} \frac{G M_{200}^2}{R_{200}}, \tag{3.12}$$

$$E_{\rm quench} = f_B \times E_{\rm bind}. \tag{3.13}$$

Here, the parameter $\epsilon_f$ is the AGN feedback efficiency, where we adopt $\epsilon_f = 0.15$, as calibrated by Booth & Schaye (2010). $\epsilon_r$ is the radiative efficiency, typically derived from the standard model by Shakura & Sunyaev (1973), and in our work, we adopt $\epsilon_r = 0.1$. $\dot{m}_{\rm acc}$ represents the accretion rate of gas onto the black hole. $\dot{m}_{\rm BH}$ is the rate of mass accretion rate onto the black hole. $f_{\text{gas}}$ is the gas mass fraction within the halo. $f_B$ is a coupling efficiency factor defined as

$$f_B = \frac{\eta E_{\text{AGN}}}{E_{\text{bind}}}, \tag{3.14}$$

where $\eta$ encapsulates the inefficiencies in energy transfer, including radiative losses and the interaction between AGN feedback and halo gas. $f_B$ serves as a calibration factor to align theoretical models with observational constraints (Chen et al., 2020), reflecting the effective energy transfer required to overcome the binding energy and quench star formation or drive gas outflows.

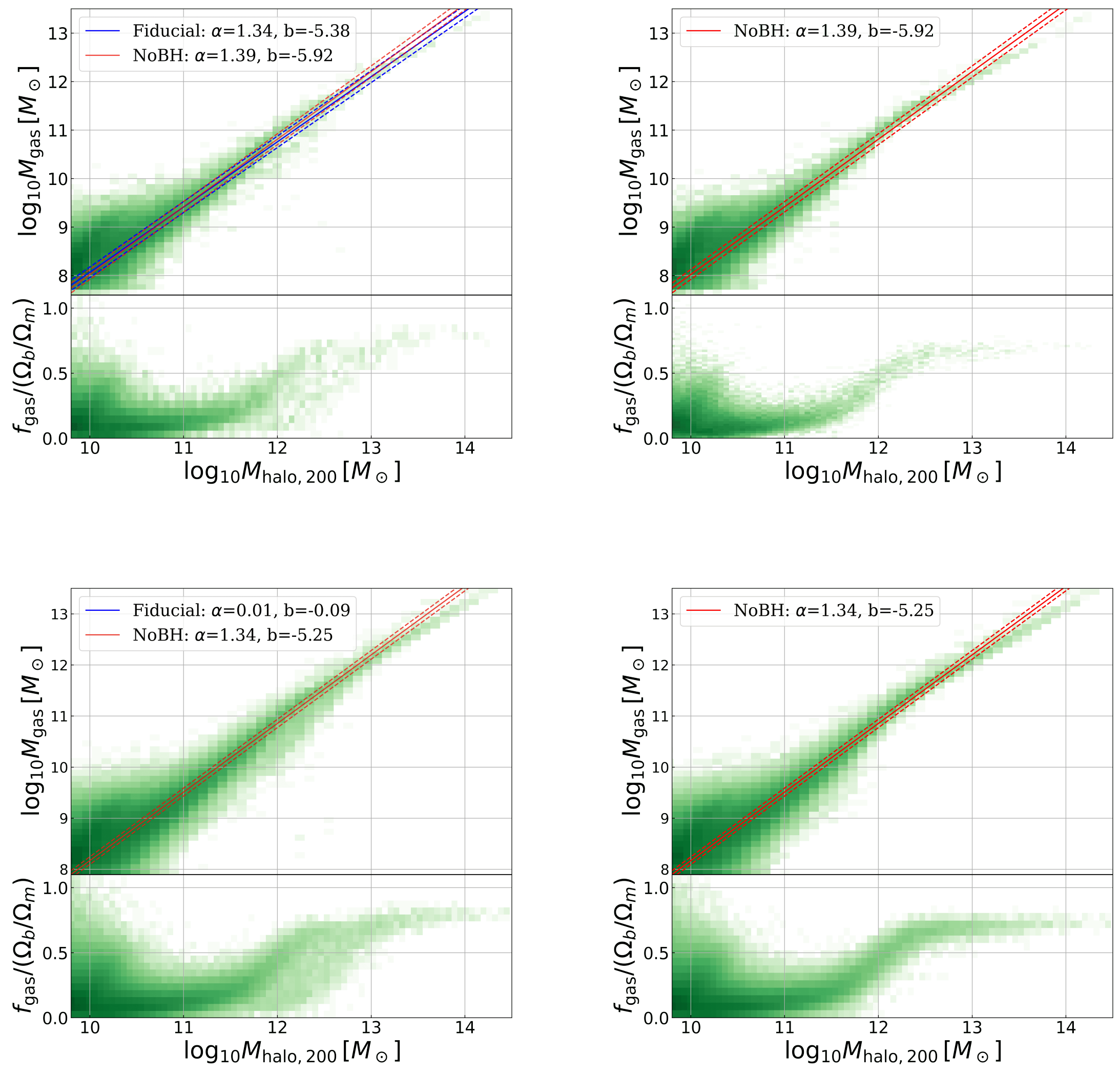

FIGURE 3.13: Comparison of dark matter halo mass vs. gas mass relations. The top row corresponds to the L50N512$_{\text{fiducial}}$ (left) and L50N512$_{\text{NoBH}}$ (right), while the bottom row corresponds to the L100N1024$_{\text{fiducial}}$ (left) and L100N1024$_{\text{NoBH}}$ (right). The color gradient represents the number of halos, and the solid lines indicate the best-fit relations for each case.

The comparison between $E_{\text{AGN}}$ and $E_{\text{quench}}$ highlights the halo mass range where AGN feedback is most effective. As shown in Figure 3.15, the largest difference between $E_{\rm AGN}$ and $E_{\rm quench}$ occurs in halos with masses around $10^{12.5} - 10^{13.5}\, M_\odot$, where AGN feedback efficiently overcomes the gravitational binding energy, facilitating the quenching of star formation and the removal of gas. In contrast, in halos with masses exceeding $10^{14}\, M_\odot$, the gravitational binding energy becomes dominant, significantly reducing the effectiveness of AGN feedback. Although AGN energy may still contribute to heating and driving outflows, gravity emerges as the primary factor regulating the gas and limiting the feedback's capacity to quench star formation.

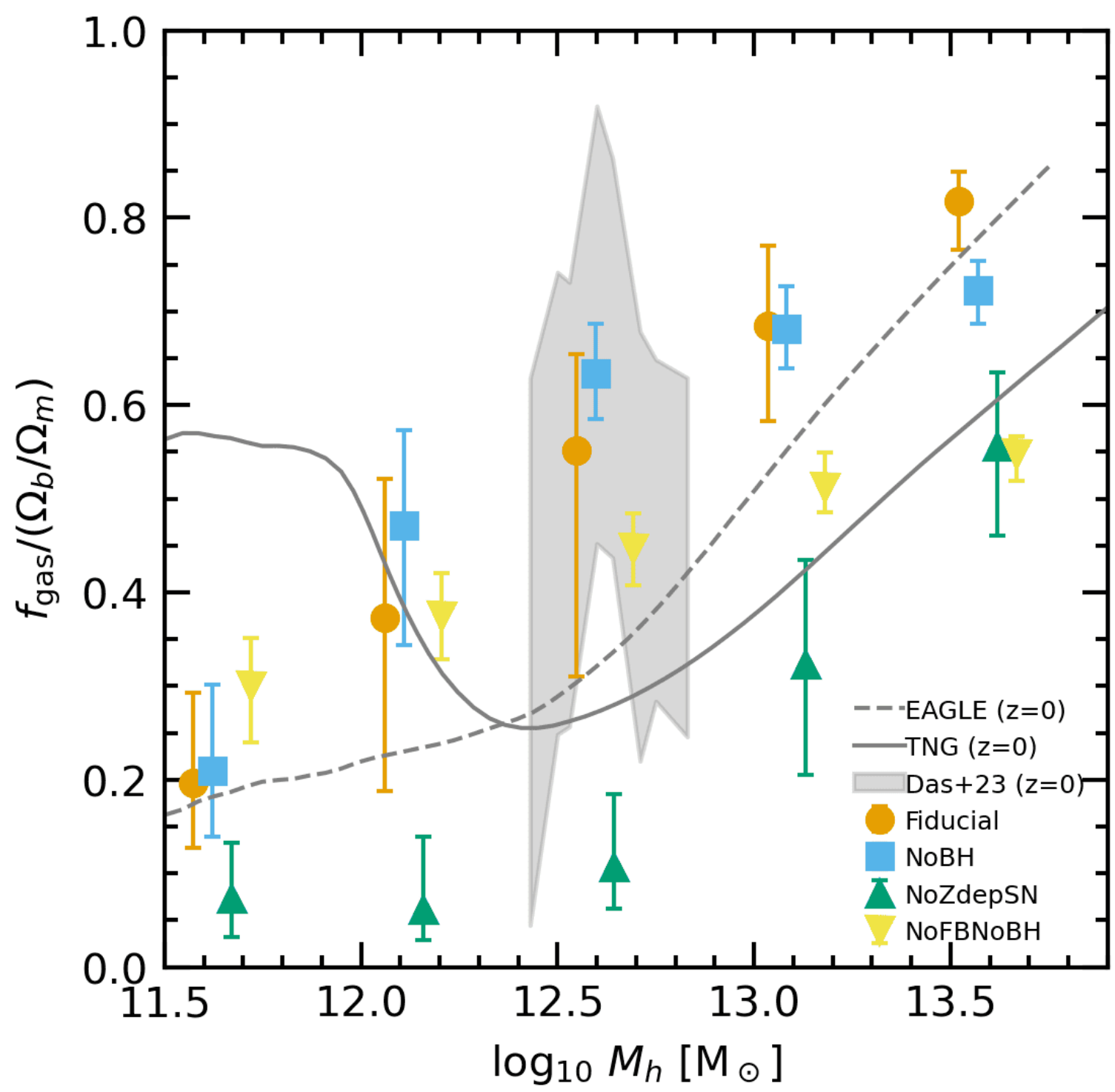

FIGURE 3.14: Comparison of CGM gas mass fraction ($f_{\rm gas}$) as a function of halo mass at $z = 0$. The gray dashed and solid lines indicate the median relations from the EAGLE-Ref and TNG-100 simulations, obtained from Davies et al. (2020). The shaded gray region represents observational constraints from Das et al. (2023). Data points represent the median values, with error bars indicating the 16th and 84th percentile. Different colors and symbols correspond to various feedback implementations in Oku & Nagamine (2024): fiducial (orange circles), no BHs (blue squares), fixed Chabrier IMF across different metallicities (green triangles), and no BHs and no SN feedback cases (yellow triangles). Our fiducial model better reproduces the baryon fraction in intermediate-mass halos compared to EAGLE-Ref and TNG-100.

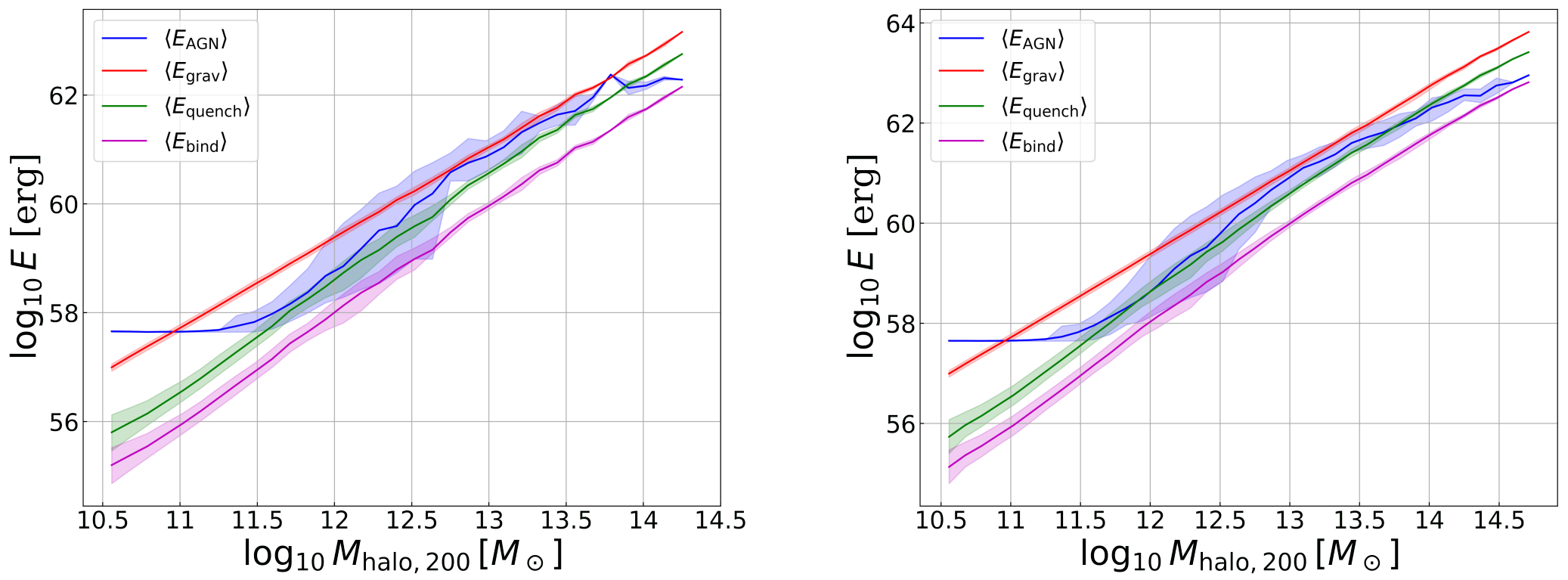

FIGURE 3.15: Energy vs. dark matter halo mass relations for the L100N1024$_{\rm fiducial}$ runs. The top panel shows results from the 50 $h^{-1}$ cMpc simulation, and the bottom panel shows results from the 100 $h^{-1}$ cMpc simulation. The different curves represent the average AGN energy $\langle E_{\rm AGN}\rangle$, gravitational energy $\langle E_{\rm grav}\rangle$, binding energy $\langle E_{\rm quench}\rangle$, and total binding energy $\langle E_{\rm bind}\rangle$.

## 3.4 IGM Baryon Fraction Analysis

In Section 3.1, we constrained an important parameter, $f_{\rm diff}$, using the DM–$z$ relation for FRBs. This parameter directly reflects the distribution of the "missing baryons" that we are investigating. As the dark matter halos grow, $f_{\rm diff}$ evolves over time. Additionally, based on numerical simulations, aside from the DM–$z$ relation, we can also estimate the gas mass fraction belonging to the IGM by excluding the gas in the halos, which is often called CGM. Here, we treat the $f_{\rm diff}$ inferred from the DM–$z$ relation as the observational estimate, and the simulation-derived $f_{\rm IGM}$ — which excludes the halo (CGM) gas (whose mass dependence is shown in figure 3.16) — as the theoretical reference or "true value." In the actual observations, it is usually difficult to subtract the intervening halo contribution, hence $f_{\rm diff} = f_{\rm IGM} + f_{\rm halo}$, as we defined in Sec 2.3.3. In some observational papers, the observed $f_{\rm diff}$ is written as $f_{\rm IGM}$ assuming that $f_{\rm Halos}$ is a part of the smooth IGM component. Such an estimate might lead to an overestimate of true $f_{\rm IGM}$, and a care is needed in its interpretation. In the present work, we attempt to clarify the distinction between $f_{\rm diff}, f_{\rm IGM}$, and $f_{\rm Halo}$ in the following sections.

### 3.4.1 "Observed" $f_{\rm diff}$

The calculation of the "observed value" ($f_{\rm diff,obs}$) follows the same method as in Section **??**, where we record the constrained $f_{\rm diff,obs}$ at each redshift when constructing the light cone and connecting the simulation box. We emphasize that the observationally constrained $f_{\rm diff,obs}$—derived through the integration of DM along light cones—is not strictly equivalent to the instantaneous simulation-based definition of $f_{\rm diff}(z)$ at a given redshift. The subtle differences are elaborated in Appendix B.4.

Since most FRB sources are observed at $z < 1$, we primarily focus on calculating the evolution of $f_{\rm diff}$ up to $z = 1$, as shown in Figure 3.17b, with the redshifts of our simulation snapshots including $z = [0.106, 0.287, 0.508, 0.754, 1.041]$. The black and orange circles represent data points constrained by L100N1024$_{\rm fiducial}$ and L100N1024$_{\rm NoBH}$, respectively, with values given as $f_{\rm diff,obs} = [0.706, 0.758, 0.810, 0.835, 0.865]$ (fiducial) and [0.685, 0.749, 0.804, 0.829, 0.856] (NoBH). The fiducial results are marginally higher than those of the NoBH case, reflecting the baryon redistribution driven by AGN feedback. A more detailed analysis is presented in Section 3.4.3.

At $z = 1$, our fiducial simulation yields $f_{\rm diff,fiducial} = 0.865$ and the NoBH simulation gives $f_{\rm diff,NoBH} = 0.856$, which are slightly lower than the latest FRB observational constraints from Connor et al. (2024), where the full sample of all localized FRBs (including detections from DSA-110, ASKAP, CHIME, MeerKAT, and VLA realfast) gives $f_{\rm diff} = 0.93^{+0.04}_{-0.02}$. However, our fiducial result is comparable to the constraint from the DSA-110 only sample, which gives $f_{\rm diff} = 0.89^{+0.06}_{-0.06}$. This agreement highlights the robustness of our simulation in capturing the baryonic distribution within large-scale structures and its impact on FRB DM statistics.

### 3.4.2 "True" $f_{\rm IGM}$

The estimation of $f_{\rm IGM,true}$ (the "true value" of $f_{\rm IGM}$ which does not include the halo component) depends critically on how we subtract the contributions from intervening halos. Unless we have a sufficiently detailed understanding of the foreground galaxies along the sight lines, which would allow us to identify all intervening halos and their density profiles, we cannot effectively remove their contributions to the IGM fraction. Thus, the calculation of $f_{\rm IGM,true}$ relies on the subtraction of CGM gas, which involves defining what gas belongs to the CGM. We employed multiple methods to identify the most appropriate approach for subtracting CGM gas fractions. We also find that the constraint on $f_{\rm IGM} = 0.8^{+0.08}_{-0.09}$ from Connor et al. (2024) falls within this range as well.

#### 3.4.2.1 Using the HMF to Calculate $f_{\rm IGM,true}$

We begin with the simplest approach: utilizing the Press & Schechter (1974) theoretical model to compute the halo mass function (HMF) at different redshifts (the result from our simulations at $z = 0$ is shown in Figure 2.3). We assume that the gas mass fraction in the CGM of each halo is a constant $f_{\rm gas} \sim \Omega_b/\Omega_m \approx 0.16$. Then, based on the HMF, we estimate the CGM gas fraction by summing over halos in different mass bins. This gives us an estimate for $f_{\rm IGM}$:

$$f_{\rm IGM}(z) = 1 - \frac{f_{\rm gas} \int M\Phi(M,z)dM}{\Omega_b \rho_{\rm crit}} \tag{3.15}$$

where $\Phi(M,z)$ represents the HMF at redshift $z$. $M$ denotes the halo mass, and the denominator, $\Omega_b\rho_{\rm crit}$, represents the cosmic mean baryon density, which serves as the total baryonic mass budget of the universe. $f_{\rm gas}$ accounts for the gas fraction within halos, which we define as the CGM gas fraction:

$$f_{\rm gas} = \frac{M_{{\rm gas},R_{\rm cut}}}{M_{{\rm halo},R_{\rm cut}}}. \tag{3.16}$$

Here, $M_{{\rm gas},R_{\rm cut}}$ is the total gas mass enclosed within a halo up to a cutoff radius $R_{\rm cut}$, and $M_{{\rm halo},R_{\rm cut}}$ is the total halo mass within the same radius. This definition explicitly incorporates the dependence on the chosen truncation radius, which we explore in later sections. The Press-Schechter (PS74) prediction is shown by the orange dot-dashed line in Figure 3.17a (diamond marker). It is clear that this result is much lower than $f_{\rm IGM}$ constrained by the Macquart relation, with $f_{\rm IGM,true} \sim 0.7$ at $z = 1$. This discrepancy likely stems from three factors: (1) we cannot approximate the CGM gas fraction by the cosmic average baryon-to-matter ratio, and

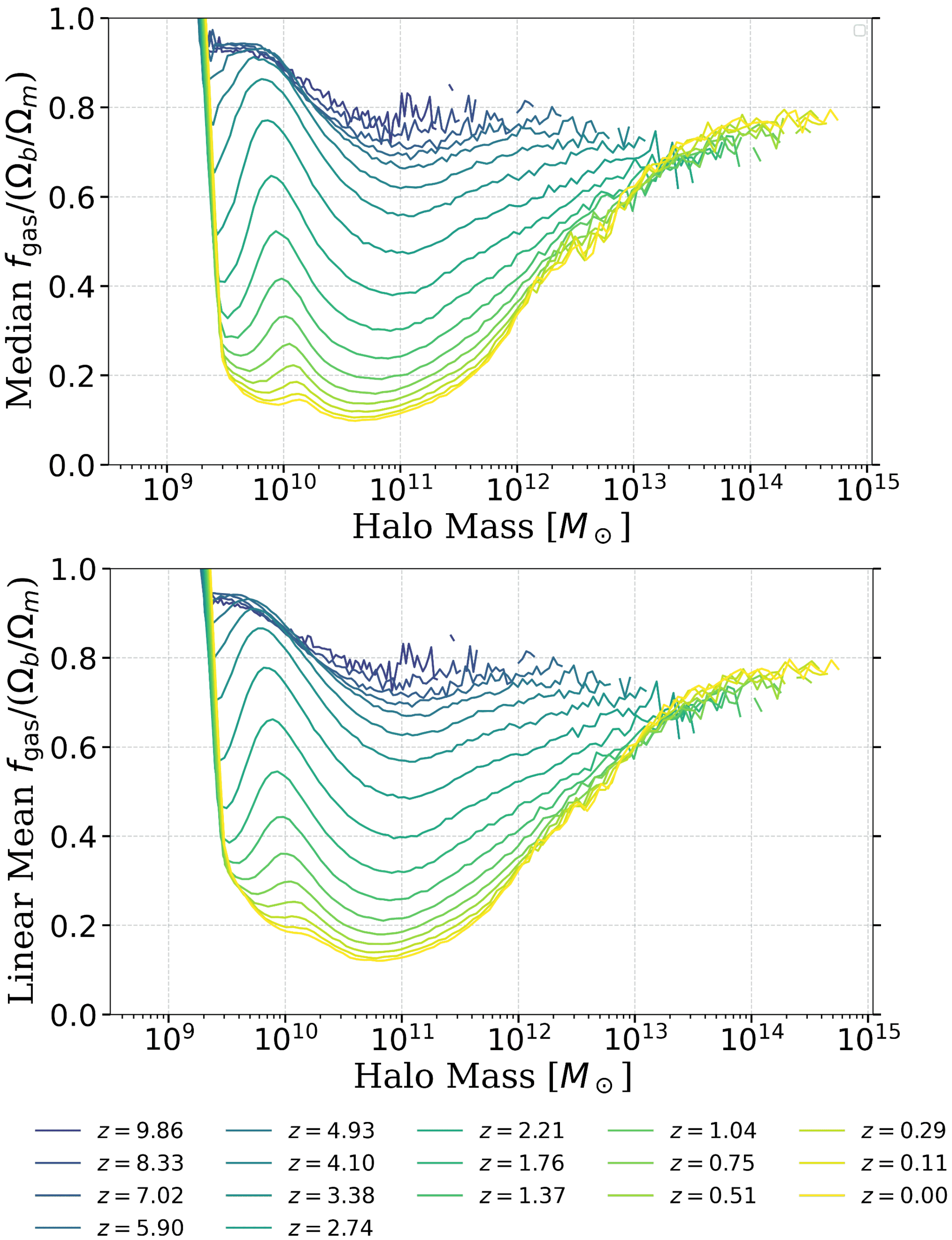

FIGURE 3.16: Evolution of CGM gas fraction ($f_{\text{gas}}$) as a function of halo mass for the L100N1024$_{\text{fiducial}}$ run. The top panel shows the median gas fractions at different redshifts for various mass ranges. The bottom panel presents the linear mean of $f_{\text{gas}}$ under the same conditions. Both plots reflect the redshift evolution and its dependence on halo mass, highlighting the role of AGN feedback in modifying the gas fraction.

(2) $f_{\rm gas}$ evolves with halo mass, and (3) random LoSs towards FRBs do not go through all halos at each redshift. The redistribution of baryons within halos reduces the CGM gas fraction below the cosmic average, and this effect is mass-dependent, as shown in Fig. 3.16 and the lower panels of Figure 3.13. To improve the accuracy of our estimate, We extract the median and mean $f_{\rm gas}$ evolution across redshifts and mass bins, as illustrated in Figure 3.16. We then employ them in a modified calculation:

$$f_{\rm IGM}(z) = 1 - \frac{\int f_{\rm gas}(M,z) M\Phi(M) dM}{\Omega_b \rho_{\rm crit}} \tag{3.17}$$

where $f_{\rm gas}(M,z)$ now explicitly accounts for mass- and redshift-dependent evolution. The resulting modified $f_{\rm IGM}$ is shown by the green and red dotted lines and diamond markers indicated with 'Modified PS' in Figure 3.17a. This result is closer to the constraints from the Macquart relation, though the evolution in the $z < 1$ range is somewhat smoother.

#### 3.4.2.2 Directly Subtracting CGM Gas to Calculate $f_{\rm IGM,true}$

Both the DM–$z$ relation and the Press-Schechter HMF-based calculations of $f_{\rm IGM}$ inherently rely on theoretical assumptions and uniform matter distribution. A more direct approach is to identify and precisely remove all CGM-associated gas particles within each halo in the simulation snapshot at each redshift. However, this method introduces a key challenge: defining the spatial extent of the CGM. The choice of CGM boundary significantly impacts the computed $f_{\rm IGM}$.

To address this, we adopt four different halo gas removal criteria. The first method involves directly excluding gas bound within FoF halos. The FoF algorithm identifies halos by linking particles within a fixed linking length, set to 0.2 times the mean interparticle separation. This method effectively captures the total mass and extent of a halo, including diffuse CGM. The results using this approach are shown as the blue solid line (square markers; fiducial) and the blue dashed line (circle markers; NoBH) in Figures 3.17 and 3.18 .

In addition to the FoF-based method, we further employ three alternative CGM subtraction approaches using different cutoff radii, listed in Table 3.1.

As shown in Figure 3.17a, all these methods, along with the $f_{\rm IGM}$ computed via the Press-Schechter HMF approach (Section 3.4.2.1), converge towards $f_{\rm IGM} = 1$ (gray dashed line) at high redshifts ($z \sim 10$), indicating their consistency in the early Universe when there are few dark matter halos. At lower redshifts, significant differences emerge as halos develop. When removing gas only within $R_{200}$, $f_{\rm IGM}$ remains the highest and exceeds the Macquart relation constraints, but at $z = 1$, it reaches $f_{\rm IGM} = 0.874$ (fiducial) and 0.865 (NoBH), aligning closely with the Macquart estimate (i.e., using Eq. (2.45) and regarding $f_{\rm diff}$ as $f_{\rm IGM}$). For $R_{\rm cut} = 2R_{200}$, the resulting $f_{\rm IGM}$ is lower than the Macquart estimate, yielding $f_{\rm IGM} = 0.738$ (fiducial) and 0.733 (NoBH) at $z = 1$. When extending $R_{\rm cut}$ to $3R_{200}$, the lowest $f_{\rm IGM}$ values are obtained, with $f_{\rm IGM} = 0.560$ (fiducial) and 0.558 (NoBH) at $z = 1$. These values are obtained only using a single snapshot of CROCODILE simulation at each redshift. The results derived from the FoF halo approach lie between the $R_{200}$ and $2R_{200}$ cases, suggesting that the effective radius defined by the FoF method is within this range. Although the $R_{200}$-based calculation yields $f_{\rm IGM}$ values closest to the Macquart relation at $z = 1$, the overall redshift evolution of $f_{\rm IGM}$ derived from the FoF halo method exhibits the best agreement with the Macquart relation at low redshifts. This suggests that if CGM is defined within the range of $R_{200}$ to $2R_{200}$, the $f_{\rm diff}$ constrained by the Macquart relation closely matches the true $f_{\rm IGM}$ at $z \lesssim 1$.

TABLE 3.1: CGM subtraction methods and their markers

| $R_{\rm cut}$ | Type | Line Style | Marker |
|---|---|---|---|
| $R_{200}$ | Fiducial | Light purple solid | $\times$ |
| | NoBH | Light purple dashed | $\triangle$ |
| $2R_{200}$ | Fiducial | Magenta solid | $\triangleright$ |
| | NoBH | Magenta dashed | $\triangledown$ |
| $3R_{200}$ | Fiducial | Olive-green solid | $\triangleleft$ |
| | NoBH | Olive-green dashed | $\star$ |

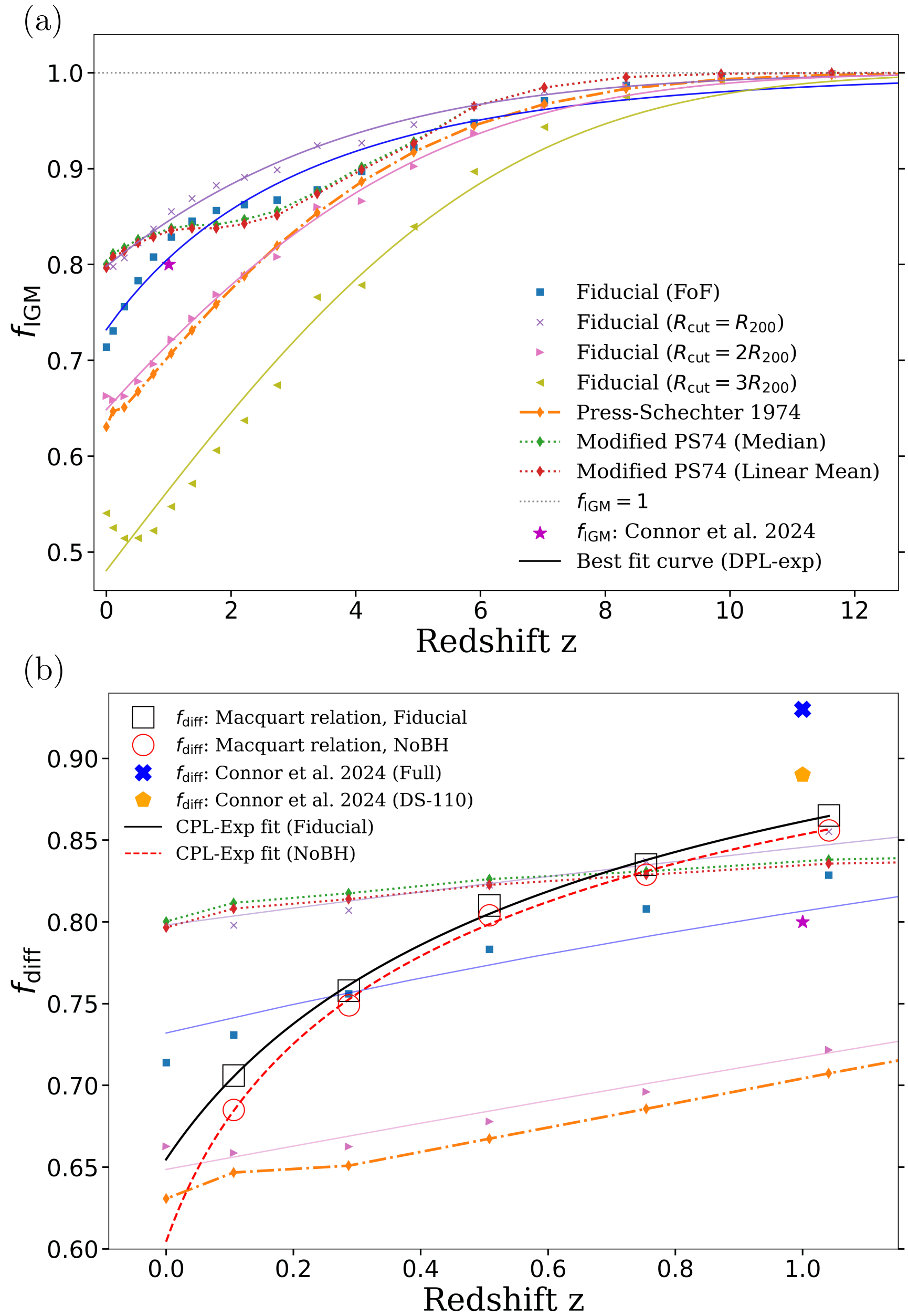

FIGURE 3.17: Redshift evolution of the baryon mass fractions in the intergalactic and diffuse media. (a) The baryon mass fraction in the intergalactic medium ($f_{\rm IGM}$), computed from the fiducial simulation under various CGM gas removal criteria, including radius cutoffs at $R_{\rm cut} = R_{200}, 2R_{200}, 3R_{200}$ and FoF-based halo boundaries. Solid and dashed lines represent results from the fiducial and NoBH models, respectively. The orange dot-dashed line shows the analytical prediction from the Press-Schechter HMF (Press & Schechter, 1974) with $\Omega_b/\Omega_m = 0.16$. Green and red dashed lines indicate "modified PS" estimates derived from the median and linear-mean gas fractions of halos in different mass bins. The orange pentagram shows the constraint on $f_{\rm IGM}$ from Connor et al. (2024) based on the full sample of all previously localized FRBs. The grey dashed line at $f_{\rm IGM}$ indicates the total cosmic baryon budget. (b) The baryon mass fraction in the diffuse medium ($f_{\rm diff}$) for $z \leq 1$, constrained using the Macquart relation. Black open squares and red open circles show simulation results from the fiducial and NoBH runs, respectively. The blue "X" and orange pentagon correspond to observational constraints from Connor et al. (2024) using the full sample and the DSA-110-only sample. For comparison, the $f_{\rm IGM}$ results from panel (a) are overplotted using thinner lines and smaller markers

However, the observed "$f_{\rm IGM}$" inferred from the FRB DM–$z$ relation is systematically higher than the true $f_{\rm IGM}$ due to the inclusion of foreground halos, which are not explicitly separated in most previous FRB-based studies (e.g., Yang et al., 2022, Wang & Wei, 2023). Scientifically, our primary interest lies in accurately determining $f_{\rm IGM}$ rather than $f_{\rm diff}$, as $f_{\rm IGM}$ more directly represents the baryon fraction in the IGM. This further underscores the necessity of constructing a foreground galaxy map to refine the Macquart relation's application in constraining $f_{\rm IGM}$, as emphasized by Li et al. (2019b), Lee et al. (2022, 2023) and Khrykin et al. (2024b).

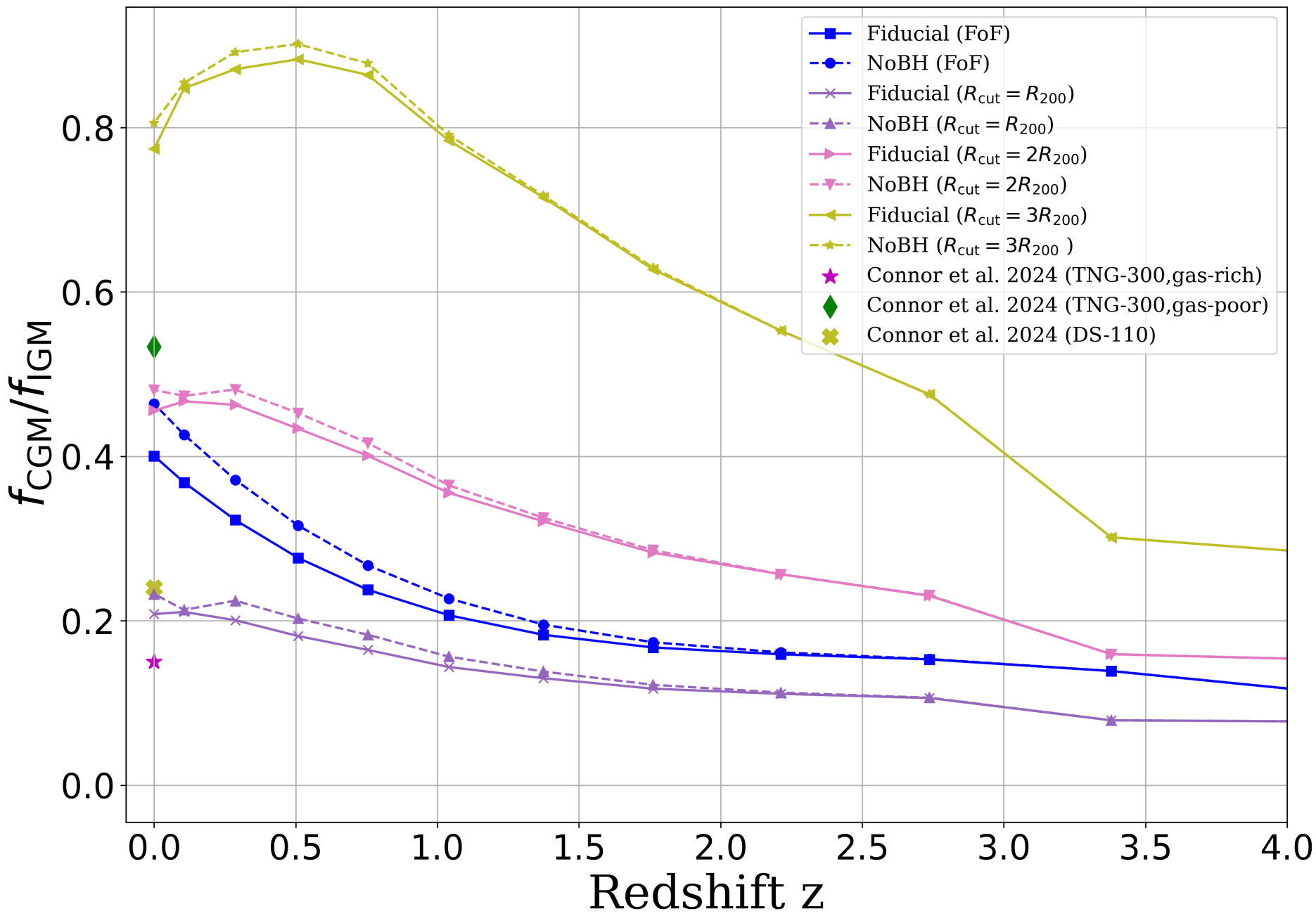

FIGURE 3.18: Redshift evolution of the baryon fraction ratio between the circumgalactic medium (CGM) and the intergalactic medium (IGM), $f_{\rm CGM}/f_{\rm IGM}$, under various CGM gas exclusion methods. Solid and dashed lines represent the fiducial and NoBH simulations, respectively. Observational estimates from Connor et al. (2024) (olive "X" symbols) are included for reference. The purple pentagram and dark green diamond indicate the expected CGM–IGM ratios from the TNG300 gas-rich and gas-poor simulations used in Connor's study for comparison with observations.

### 3.4.3 Physical interpretation of baryon redistribution

To more precisely analyze the impact of AGN feedback on baryon redistribution, we examine the evolution of halo gas relative to the IGM. Figure 3.18 shows the redshift evolution of the ratio [1] $f_{\rm CGM}/f_{\rm IGM}$ for both the fiducial and NoBH models under different $R_{\rm cut}$ definitions. In general, we see that this ratio is increasing with decreasing redshift, as the halos grow due to gravitational instability. Notably, the observational constraints from Connor et al. (2024) are broadly consistent with the predictions adopting $R_{\rm cut} = R_{200}$, suggesting that this structural definition may best capture the CGM scale probed by current FRB observations. At low redshifts, the ratio is systematically lower in the fiducial case, indicating that AGN feedback expels a portion of halo gas into the IGM. This trend confirms that AGN feedback plays a moderate but non-negligible role in redistributing baryons. This ratio therefore serves as a complementary tracer of AGN-induced baryon redistribution.

However, the relatively small gap between fiducial and NoBH simulations suggests that AGN feedback in our CROCODILE simulations has a modest impact on $f_{\rm diff,obs}$ and $f_{\rm CGM}/f_{\rm IGM}$. This result can be attributed to several factors:

- Absence of kinetic AGN feedback: Our current implementation of AGN feedback in the CROCODILE simulations primarily relies on thermal feedback and lacks AGN jet (kinetic) feedback. Without this additional kinetic mode, the efficiency of AGN-driven redistribution is likely underestimated.
- Effects of AGN feedback vary with halo mass: While high-mass halos ($M_{\rm halo} > 10^{12.5} M_\odot$) are expected to exhibit stronger AGN-driven outflows, lower-mass halos ($M_{\rm halo} < 10^{12} M_\odot$) retain a significant fraction of their CGM, leading to only a minor difference in $f_{\rm diff}$ between the two runs.
- Impact of cooling and recycling. Even if AGN feedback expels gas into the IGM, a fraction of this gas can return to the halo through radiative cooling and the gravity of the halo, partially counteracting the redistribution effect. This fallback might reduce the difference in $f_{\rm diff}$ between the fiducial and NoBH runs.
- Coupling between $f_{\rm IGM}$ and $f_{\rm CGM}$: While $f_{\rm diff}$ is defined as the sum of $f_{\rm IGM}$ and $f_{\rm CGM}$, these two components are intrinsically coupled, especially in the presence of AGN feedback. AGN-driven outflows can increase $f_{\rm IGM}$ by expelling gas from halos while simultaneously reducing $f_{\rm CGM}$ by depleting CGM gas. Observationally, constraining $f_{\rm IGM}$ using FRBs requires careful subtraction of foreground halo contributions along each LoS — a challenging but essential task for FRB-based cosmology. From the simulation perspective, although we have full knowledge of halo properties and provide a detailed comparison of CGM/IGM separation in Figure B.6, the precise definition of the CGM–IGM boundary remains model-dependent and nontrivial. This underscores the importance of joint observational and simulation-based efforts.

[1] After we define the CGM in Section 3.4.2.2, the concept of $f_{\rm Halos}$ can be naturally replaced by $f_{\rm CGM}$.

### 3.4.4 Fitting the Redshift Evolution of $f_{\mathrm{IGM}}$ and $f_{\mathrm{diff}}$

We provide fitting models for the redshift evolution of $f_{\mathrm{IGM}}$ and $f_{\mathrm{diff}}$, which could be useful for semi-analytic models of galaxy formation and IGM. A natural first approach is to use a power-law function, as many cosmic-scale evolutionary processes can be well-described by such relations. However, $f_{\mathrm{IGM}}$ asymptotically approaches 1 at high redshifts, suggesting that a simple power-law function may not be sufficient to capture this behavior. Consequently, we adopt an "exponential convergence model", ensuring that as $z$ increases, $f_{\mathrm{IGM}}$ naturally approaches 1.

To this end, we introduce three models of increasing complexity, each progressively improving the accuracy of the fit:

1. Convergence Exponential Model (C-Exp):

$$f(z) = 1 - \exp(-\kappa(z + \tau)) \tag{3.18}$$

   This is the simplest model, assuming that the growth of $f_{\mathrm{IGM}}$ or $f_{\mathrm{diff}}$ follows a straightforward exponential process governed by $\kappa$ and $\tau$. While it effectively describes the large-scale growth trend, it lacks flexibility in capturing the detailed evolution at both higher and lower redshifts ($z \lesssim 1$).

2. Convergence Power-Law Exponential Model (CPL-Exp):

$$f(z) = 1 - \exp(-\kappa(z + \tau)^{\zeta}) \tag{3.19}$$

   By introducing the exponent $\zeta$, this model gains additional flexibility, allowing the growth rate to vary with redshift instead of following a fixed exponential form. As a result, it better captures the evolution at lower redshifts while preserving the high-redshift convergence to 1.

3. Double Power-Law Exponential Model (DPL-Exp):

$$f(z) = 1 - \exp(-\kappa(z^{\gamma} + \tau)^{\zeta}) \tag{3.20}$$

   Further refining the model, we introduce an additional exponent $\gamma$ to allow the redshift dependence to be more flexible. This enables the model to better capture the growth behavior across different redshift ranges. In particular, it significantly improves the fit at lower redshifts, ensuring that the model accounts for early-stage evolution more effectively.

From C-Exp to CPL-Exp to DPL-Exp, each successive model builds upon the previous one by adding degrees of freedom, leading to a progressively better fit for $f_{\mathrm{IGM}}$ to the data. However, for $f_{\mathrm{diff}}$ the CPL-Exp model provides the best fit, as indicated by the $\chi^2/\mathrm{DOF}$ values (table B.3 in Appendix B.7).

As seen from the best-fit parameters listed in Table B.3, the evolution of $f_{\mathrm{IGM}}$ and $f_{\mathrm{diff}}$ follows a well-defined trend across different redshifts. Notably, we observe that the observationally inferred $f_{\rm diff}$, derived from FRB DM–$z$ constraints, exhibits a significantly higher growth rate compared to the intrinsic $f_{\rm IGM}$. This trend aligns with expectations:

- At low redshifts, $f_{\rm diff}$ remains low, closely matching the results obtained from applying a halo mass cut of $R_{\rm cut} = 2R_{200}$. This suggests that at lower redshifts, the probability of an FRB intersecting a massive foreground halo is low. As a result, the inferred $f_{\rm diff}$ is closer to the intrinsic $f_{\rm IGM}$, as it receives minimal contributions from foreground halos.

- As redshift increases, the probability of FRBs passing through massive foreground halos increases significantly. Consequently, the inferred $f_{\rm diff}$ includes a larger contribution from foreground halos, leading to a steeper increase. By $z \sim 1$, the inferred $f_{\rm diff}$ surpasses the FoF-halo-excluded results and closely matches the results obtained from a $1R_{200}$ mass cut at $z \sim 1$.

Furthermore, although direct DM–$z$ constraints cannot determine the exact value of $f_{\rm diff}$ at $z = 0$, we can use our best-fit models to extrapolate their asymptotic values as $z \to 0$. These extrapolated values are:

- C-Exp: $f_{\rm diff}(z \to 0) = 0.685$ (fiducial), 0.667 (NoBH)
- CPL-Exp: $f_{\rm diff}(z \to 0) = 0.655$ (fiducial), 0.604 (NoBH)
- DPL-Exp: $f_{\rm diff}(z \to 0) = 0.706$ (fiducial), 0.684 (NoBH)

For comparison, using the Press-Schechter HMF and the cosmic mean baryon density, the expected value of $f_{\rm IGM}$ at $z = 0$ is approximately $f_{\rm IGM}(z = 0) = 0.631$. This value closely matches the extrapolated values from the CPL-Exp model, supporting the validity of its extrapolations.

In summary, our results confirm that the inferred $f_{\rm diff}$ from FRB DM–$z$ constraints systematically increases with redshift due to the increasing contribution of foreground halos. Our progressively refined models effectively capture this trend and provide robust extrapolations for $f_{\rm diff}$ at $z = 0$. These results offer new insights into the distribution of ionized gas in the Universe and its redshift evolution.

### 3.4.5 From FRB Observables to Intrinsic $f_{\rm IGM}(z)$: A Semi-Analytic Reconstruction Framework

We now discuss how to observationally constrain the "true" IGM baryon content using FRB sight lines. The diffuse component of the DM was given by Eq. (2.45). From an observational perspective, assuming that FRBs are randomly distributed across the sky, and considering that $f_{\rm diff}$ evolves with redshift, what we constrain is actually its cumulative, redshift-averaged form as we discussed in Appendix B.4:

$$\langle {\rm DM}_{\rm diff}(< z) \rangle \propto \langle f_{\rm diff,\, obs}(< z) \rangle \int_0^z \frac{1+z'}{E(z')}\, dz', \tag{3.21}$$

where $E(z) = \sqrt{\Omega_m(1+z)^3 + \Omega_\Lambda}$. Here we assume a constant ionization fraction $\chi_e \approx 0.875$ for fully ionized diffuse medium. Explicitly considering the contribution from intersecting halos, the average DM contribution from both IGM and halos can be expressed as:

$$\langle \mathrm{DM}_{\mathrm{diff}}(<z)\rangle \propto \int_0^z \left[f_{\mathrm{IGM}}(z') + \langle f_{\mathrm{Halos}}^{\mathrm{intersect}}(z')\rangle\right] \frac{1+z'}{E(z')}\, dz', \tag{3.22}$$

where the term $\langle f_{\mathrm{Halos}}^{\mathrm{intersect}}(z)\rangle$ represents the average contribution from halos intersected by FRB sight lines at redshift $z$. To incorporate the extended gas density profiles of halos in a physically motivated manner, we express this halo term as an integral over a halo mass range $[M_{\mathrm{low}}, M_{\mathrm{up}}]$, with the projected baryonic column density:

$$\begin{aligned}\left\langle f_{\mathrm{Halos}}^{\mathrm{intersect}}(z)\right\rangle = \frac{1}{\rho_b(z)} \int_{M_{\mathrm{low}}}^{M_{\mathrm{up}}} n(M_h, z)\times \\ \left[\int_0^{R_{\mathrm{cut}}(M_h)} 2\pi b\, \Sigma_g(b; M_h, z)\, db\right] dM_h,\end{aligned} \tag{3.23}$$

where

$$\Sigma_g(b) = 2\int_b^{R_{\mathrm{cut}}(M_h)} \frac{r f_{\mathrm{gas}} \rho_{\mathrm{model}}(r)}{\sqrt{r^2 - b^2}}\, dr. \tag{3.24}$$

Here $\Sigma_g(b; M_h, z)$ is the projected gas mass column density of a halo with mass $M_h$ at impact parameter $b$, $n(M_h, z)$ is the HMF, $\rho_{\mathrm{model}}(r)$ is the modeled density profile of the dark matter halo (e.g., Equation 3.3), and $\rho_b(z)$ is the mean cosmic baryon density at redshift $z$. The upper limit $R_{\mathrm{cut}}(M_h)$ is the assumed maximum cutoff radius of halo gas (e.g., $R_{200}$ or $2R_{200}$). This formulation properly accounts for both the extended gas structure and the geometric probability of intersecting a halo, making it a physically motivated expression for the average halo contribution to the FRB dispersion measure.

By differentiating both sides of Equations 3.21 and 3.22, we derive the following expression for the IGM baryon fraction:

$$\begin{aligned}f_{\mathrm{IGM}}(z) = \langle f_{\mathrm{diff,obs}}(<z)\rangle - \left\langle f_{\mathrm{Halos}}^{\mathrm{intersect}}(z)\right\rangle \\ + \left(\frac{d}{dz}\langle f_{\mathrm{diff,obs}}(<z)\rangle\right) \cdot \left[\frac{1}{X(z)}\int_0^z X(z')\, dz'\right]\end{aligned} \tag{3.25}$$

where

$$X(z) = \frac{(1+z)}{E(z)}. \tag{3.26}$$

Here, $\langle f_{\mathrm{diff,obs}}(<z)\rangle$ can be inferred from the observation of the Macquart relation up to $z$, which statistically links FRB DMs and redshift. Its analytic form can be approximated using the empirical fitting functions (e.g., C-exp) described in Section 3.4.4, allowing us to compute the derivative term $\frac{d}{dz}\langle f_{\mathrm{diff,obs}}(<z)\rangle$.

The sightline-weighted halo baryon fraction $\langle f_{\mathrm{Halos}}^{\mathrm{intersect}}(z)\rangle$ can also be constrained observationally, by modeling the density field of foreground galaxies along FRB sight lines (e.g., Lee et al., 2022).

Figure 3.19 compares the reconstructed IGM baryon fraction $f_{\mathrm{IGM,th}}(z)$—recoverd from the cumulative observational constraint $\langle f_{\mathrm{diff,obs}}(<z)\rangle$—with the baryon fraction directly computed from simulations, $f_{\rm IGM,sim}(z)$. The black curve in the figure corresponds to the CPL-Exp fit to the fiducial $\langle f_{\rm diff,obs}(<z)\rangle$ shown in Table B.3.

To calculate $\langle f_{\rm Halos}^{\rm intersect}(z)\rangle$ (Equation 3.23), we adopt two standard HMFs—PS74 and Tinker08. For the gas density profiles, we use the fitted parameters of the mNFW profile at $z=0$, evaluated in different halo mass bins (listed in Table B.4), assuming a gas distribution of the form $f_{\rm gas}\rho_{\rm model}(r)$. The resulting values of $\langle f_{\rm Halos}^{\rm intersect}(z)\rangle$ are shown in Table 3.2.

TABLE 3.2: Comparison between theoretical $f_{\rm Halos}$ and simulated $f_{\rm CGM}$

| Redshift $z$ | $f_{\rm Halos}^{\rm Press}$ | $f_{\rm Halos}^{\rm Tinker}$ | $f_{\rm CGM}^{\rm sim}$ |
|---|---|---|---|
| 0.000 | 0.1899 | 0.1407 | 0.1662 |
| 0.106 | 0.1758 | 0.1318 | 0.1683 |
| 0.287 | 0.1527 | 0.1176 | 0.1619 |
| 0.508 | 0.1268 | 0.0991 | 0.1493 |
| 0.754 | 0.1021 | 0.0805 | 0.1378 |
| 1.041 | 0.0790 | 0.0624 | 0.1230 |

As shown in Figure 3.19, $f_{\rm IGM,th}(z)$ lies closer to the simulated $f_{\rm IGM,sim}(z)$ obtained with a CGM exclusion radius of $2R_{200}$. Compared to the $1R_{200}$ case, it shows a steeper evolution with redshift and lies inbetween the $R_{200}$ and $2R_{200}$ definitions at $z\sim 1$.

It is important to note several sources of systematic uncertainties in this reconstruction:

1. The analytic form of $\langle f_{\rm diff,obs}(<z)\rangle$ significantly affects the inferred IGM fraction and depends on the validity of the chosen fitting function.
2. The HMF used in our model is based on analytical (PS74) and empirical (Tinker08) relations. However, as shown in Figure 2.3, these HMFs can deviate from the simulation results, especially at the high-mass end. In principle, one could interpolate the HMF directly from simulation outputs at each redshift for higher accuracy, but this would undermine the purpose of using FRBs to test theoretical models — as in reality, we do not know the true HMF of the Universe. Therefore, we opt for established analytic HMFs instead of interpolated simulation values.
3. The mNFW fitting parameters are derived from median density profiles at $z=0$ in each mass bin. These do not represent individual halos and are used as characteristic profiles for population-level estimation.
4. The concentration parameter $c_{200}$ evolves with redshift and alters the relationship between $M_{200}$ and $R_{200}$, thereby modifying the density profile shape of the halo. This is particularly important for gas, whose distribution is further affected by baryon–DM interactions (see Ludlow et al. (2016), Sorini et al. (2025)).
5. The gas density profiles are more complex than simply scaling dark matter profiles by a constant $f_{\rm gas}$ — baryonic processes such as feedback and thermal pressure modulate the gas structure differently from dark matter (Sorini et al., 2025).

Despite the limitations from these sources of uncertainty, this semi-analytic framework provides a direct bridge between FRB cosmological observables, such as $f_{\rm diff,\,obs}$, and physically motivated quantities predicted by theoretical models (e.g., in the standard ΛCDM cosmology), enabling improved interpretation of FRB-derived constraints on the cosmic baryon distribution.

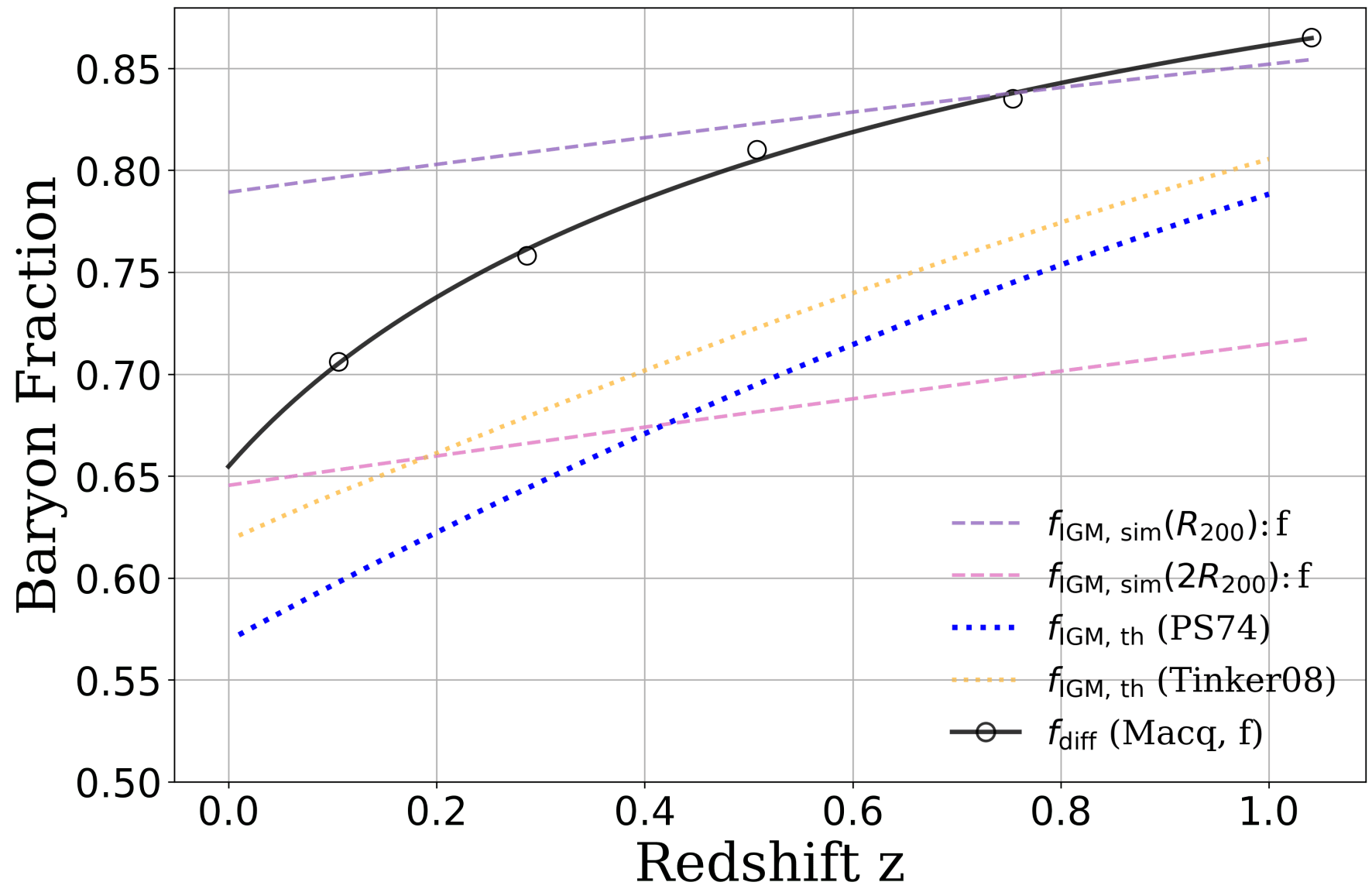

FIGURE 3.19: Comparison of the redshift evolution of the average diffuse baryon fraction $\langle f_{\mathrm{diff}}^{\mathrm{obs}} \rangle$ (black) obtained from the DM–$z$ relation against our theoretical modeling. The dashed lines represent simulation results of $f_{\mathrm{IGM}}$ under two different CGM boundary definitions: $R_{200}$ (light purple) and $2R_{200}$ (pink). The dotted lines show the reconstructed IGM baryon fraction recovered from $f_{\mathrm{diff}}$ using PS74 (blue) and Tinker08 (yellow) HMFs.

## 3.5 Host Galaxy Contribution from Zoom-in Simulations

Building on the zoom-in simulation framework described in Section 2.4, this section examines the contribution of host-galaxy to FRB dispersion measures. We focus on how gas structure (ISM/CGM), source position (central versus off-center), and host type shape the host-DM distribution. Figure 2.5 presents the gas-mass (left) and stellar-mass (right) distributions of the three representative hosts analyzed in this chapter: a dwarf galaxy, a Milky Way-like galaxy, and a galaxy cluster. Using these systems, we derive line-of-sight DM statistics and compare central and offset source scenarios.

### 3.5.1 FRBs in Dwarf Galaxies

Many repeating FRBs, such as FRB 121102 and FRB 20190520B, are located in low-metallicity, star-forming dwarf galaxies (Tendulkar et al., 2017, Niu et al., 2022). These FRBs are believed to originate from highly magnetized environments, possibly driven by young neutron stars or magnetars as discussed in Appendix B.6. The stellar masses of these dwarf galaxies range approximately from $10^7 M_\odot$ to $10^9 M_\odot$, with star formation rates not exceeding $1 M_\odot \mathrm{yr}^{-1}$. Here, we select a relatively low-mass dwarf galaxy to reflect its characteristic features, with a stellar mass of only $2.2 \times 10^7 M_\odot$ and a star formation rate of $0.03 M_\odot \mathrm{yr}^{-1}$. On the other hand, although BNS mergers are no longer considered a dominant FRB formation mechanism, they may still contribute a minor subset of the FRB population, as discussed in Appendix B.6. We compute the DM distribution along 10000 different sight lines originating from both the central and off-center locations of the host galaxies (Figure 3.20). For the central case, the DM integration length endpoint for each galaxy type is set to $2R_{200}$, and the LoS resolution length is chosen as $\epsilon_{\rm grav}$, both listed in Table 2.3. For the off-center case, the DM integration length is set to 2 Mpc, and the LoS resolution length is chosen as 1 kpc. Unlike the central case, where the dominant contribution comes from the dense interstellar medium (ISM), the off-center DM primarily originates from the CGM and the extended halo. A shorter integration length might lead to an artificially narrow or underestimated DM distribution, failing to capture the full contribution from the CGM and halo gas.

To account for off-center FRB sources caused by binary neutron star (BNS) mergers, we consider that these sources can be offset due to the kick velocity imparted to the system during supernova explosions. Here we simply assume that the source is located at an off-center location of 580 kpc for this dwarf galaxy.

In Figure 3.20, the blue component represents the dwarf galaxy throughout the figure, while the orange and pink components correspond to the MW-like galaxy and galaxy cluster, respectively, which will be discussed in later sections. Panel (a) shows that the electron number density in the central region of the dwarf galaxy can exceed $10^{-3}\,{\rm cm}^{-3}$ but rapidly declines within tens of kpc. The corresponding DM profile in panel (c) shows that the median DM rapidly converges in the high-density region within a few tens of kpc and is primarily distributed (16th-84th percentile) in the range of 2.3-9.6 pc cm$^{-3}$. The DM distribution at $2R_{200}$ ($\mathrm{DM}_{2R_{200}}$) in the dwarf galaxy is best described by a Gaussian Mixture Model (GMM), indicating that multiple underlying components contribute to the observed distribution rather than a single Gaussian or log-normal function. The GMM is a probabilistic model that represents the data as a weighted sum of multiple Gaussian distributions:

$$P(x) = \sum_{i=1}^{n} w_i \cdot \mathcal{N}(x|\mu_i, \sigma_i^2), \tag{3.27}$$

where $n$ is the number of Gaussian components, $w_i$ is the weight of each Gaussian component, and $\mathcal{N}(x|\mu_i, \sigma_i^2)$ is a Gaussian distribution defined as:

$$\mathcal{N}(x|\mu, \sigma^2) = \frac{1}{\sqrt{2\pi\sigma^2}} \exp\left(-\frac{(x-\mu)^2}{2\sigma^2}\right). \tag{3.28}$$

We apply a three-component GMM ($n = 3$) to fit the DM distribution of the dwarf galaxy, effectively capturing the observed structure, as shown in panel (e). The best-fit parameters, including the means ($\mu_i$), covariances ($\sigma_i^2$), and weights ($w_i$), are summarized in Table 3.3.

Among the three Gaussian components, the dominant one is centered at $\mu_1 = 3.65^{+8.42}_{-0.02}$ pc cm$^{-3}$, with the highest weight of $w_1 = 0.80^{+0.01}_{-0.66}$, indicating that this component contributes the most to the overall distribution. This suggests that the majority of sight lines in the dwarf galaxy have DM values close to this peak, likely corresponding to the bulk of the ISM within the galaxy.

TABLE 3.3: Gaussian Mixture Model Fit for $DM_{2R_{200}}$ in Dwarf and MW-like Galaxies.

| | Dwarf Galaxy | | | MW-like Galaxy (AGORA) | | |
|---|---|---|---|---|---|---|
| Component | Mean $\mu_i$ (pc cm$^{-3}$) | Variance $\sigma_i^2$ (pc$^2$ cm$^{-6}$) | Weight $w_i$ | Mean $\mu_i$ (pc cm$^{-3}$) | Variance $\sigma_i^2$ (pc$^2$ cm$^{-6}$) | Weight $w_i$ |
| 1 | $3.65^{+8.42}_{-0.02}$ | $2.32^{+16.56}_{-0.05}$ | $0.80^{+0.01}_{-0.66}$ | $120.03^{+323.91}_{-1.48}$ | $2571.98^{+52246.88}_{-140.31}$ | $0.60^{+0.01}_{-0.33}$ |
| 2 | $25.50^{+0.31}_{-13.67}$ | $9.84^{+1.76}_{-3.10}$ | $0.070^{+0.07}_{-0.01}$ | $448.16^{+1423.77}_{-37.11}$ | $56537.28^{+5.38\times10^4}_{-5.38\times10^4}$ | $0.27^{+0.32}_{-0.11}$ |
| 3 | $12.11^{+3.25}_{-0.52}$ | $19.13^{+1.57}_{-11.58}$ | $0.1340^{+0.01}_{-0.06}$ | $1847.64^{+1420.32}_{-326.08}$ | $917426.72^{+8.67\times10^5}_{-1.11\times10^5}$ | $0.15^{+0.11}_{-0.01}$ |

The second and third components (see Table 3.3), which correspond to higher DM values with relatively small weights, likely represent more extreme sight lines. These may trace regions with higher gas densities or outer halo structures

The median DM value for the dwarf galaxy is found to be $DM_{median} = 4.13\,\text{pc cm}^{-3}$, aligning closely with the primary Gaussian component, further supporting the interpretation that this component dominates the overall distribution.

For the off-center case, the electron number density profile is more dispersed, hovering around the cosmic mean electron density of $2.2 \times 10^{-7}\,\text{cm}^{-3}$. Consequently, its contribution to the FRB's $DM_{Host}$ is negligible and does not fit a log-normal distribution. Its symmetry is weaker, with most values concentrated near the lower boundary. As a result, the $DM_{2Mpc}$ distribution exhibits a shape characterized by a rapid rise followed by a slow decline. Therefore, we applied the Generalized Pareto Distribution (GPD) to fit the distribution.

$$f(x; k, \sigma, \theta) = \begin{cases} \frac{1}{\sigma}\left(1 + k\frac{x-\theta}{\sigma}\right)^{-\frac{1}{k}-1}, & k \neq 0 \\ \frac{1}{\sigma}e^{-\frac{x-\theta}{\sigma}}, & k = 0 \end{cases} \tag{3.29}$$

Where the shape parameter is $k = 0.65^{+0.01}_{-0.02}$, and the scale parameter is $\sigma = 0.02^{+0.0004}_{-0.0001}$. The location parameter is tightly constrained at $\theta = 0.01$ with negligible uncertainty. The median DM for the off-center case is only $0.02\,\text{pc cm}^{-3}$.

This result indicates that, if the FRB source is located in the star-forming region of this dwarf galaxy, its DM contribution would generally remain within $4\,\text{pc cm}^{-3}$. In contrast, for the off-center case, the DM contribution can be considered negligible.

Although our analysis indicates that the DM contribution from this dwarf galaxy is quite limited, it is important to note that the stellar mass of our selected FRB host galaxy is only $2.2 \times 10^7\,M_\odot$, which is relatively low even among dwarf galaxies. This mass is also close to the lower end of the dwarf galaxy sample studied in Zhang et al. (2020b). While even lower-mass dwarf galaxies exist, FRB hosts below this mass range are rarely identified, likely due to both observational limitations and intrinsically lower FRB occurrence rates. Their faintness makes precise localization challenging, and their lower star formation rates may result in fewer FRB events. Considering that some more massive dwarf galaxies can reach up to $10^9\,M_\odot$, the DM contribution from this galaxy can be regarded as a lower bound for identifiable FRB host dwarf galaxies.

### 3.5.2 FRBs in Spiral Galaxies

Repeating FRBs such as FRB 180916 and FRB 20201124A are also believed to be magnetar-driven, but their host galaxies differ from FRB 121102. These FRBs are located in the star-forming regions of spiral galaxies, whose stellar masses range from $10^9 - 10^{10}\,M_\odot$ and star formation rates exceeding $0.01\,M_\odot\,\mathrm{yr}^{-1}$.

For this study, we select a MW-like galaxy with a stellar mass of $1.21 \times 10^{11}\,M_\odot$ and a star formation rate of $6.6\,M_\odot\,\mathrm{yr}^{-1}$, as shown in Table 2.3, to represent its key characteristics. The analysis method follows the same procedure as in Section 3.5.1. Results are presented in Figure 3.20. However, due to the higher mass of the spiral galaxy compared to the dwarf galaxy, the BNS merger timescale is expected to be shorter. Therefore, we adopt a smaller off-center distance  194 kpc.

In Figure 3.20, panel (a) shows that the differences in electron density profiles between central and off-center sight lines are within two orders of magnitude. In the central region, the electron density exceeds $10^{-2}\,\mathrm{cm}^{-3}$, while in the off-center case, it is around $10^{-4}\,\mathrm{cm}^{-3}$. Despite this, the electron density curves exhibit noticeable differences: in the central case, $n_e$ decreases more steeply, whereas in the off-center case, the decline is more gradual. Additionally, the off-center profiles exhibit larger deviations near the source. Both cases display numerous peaks in their profiles, indicative of the complex structures within spiral galaxies.

Panel (c) presents the $\mathrm{DM}_{2\mathrm{Mpc}}$ distribution. The DM for the central case is primarily distributed in $83 - 813\,\mathrm{pc\,cm}^{-3}$, we also use the GMM (equation 3.27) with $n = 3$, to fit the distribution, as shown in panel (e). The best-fit parameters are listed in Table 3.3. The dominant component is centered at $\mu_1 = 120.03^{+323.91}_{-1.48}$ pc cm$^{-3}$, with the highest weight of $w_1 = 0.60^{+0.01}_{-0.33}$, indicating that this component contributes the most to the overall distribution.

For the off-center case (panel e), the DM values range between 6.38-28.42 $\mathrm{pc\,cm}^{-3}$. The distribution's shape is similar to the off-center case for dwarf galaxies, with the values more concentrated near the lower boundary. Consequently, we use the GPD (Equation 3.29) to fit the distribution. The best-fit parameters are: $k = 0.21^{+0.03}_{-0.01}$, $\sigma = 11.43^{+0.37}_{-0.87}$, and $\theta = 3.91^{+0.19}_{-0.20}$.

The median DM values are $163.53\,\mathrm{pc\,cm}^{-3}$ and $11.59\,\mathrm{pc\,cm}^{-3}$ for the central and off-center cases, respectively. The median value for the central case is comparable to the Galactic ISM contribution $\mathrm{DM}_{\mathrm{MW}} \approx 140\,\mathrm{pc\,cm}^{-3}$ or $\mathrm{DM}_{\mathrm{MW}} \approx 200\,\mathrm{pc\,cm}^{-3}$, according to the NE2001 model (Cordes & Lazio, 2002) and the YMW16 model (Yao et al., 2017).

### 3.5.3 FRBs in Galaxy Clusters

Galaxy clusters represent the most massive gravitationally bound structures in the Universe, and their vast gravitational potential significantly impacts the DM distribution of any hosted FRB. Unlike spiral or dwarf galaxies, the massive gravitational potential of a galaxy cluster confines its BNS mergers to regions very close to the cluster center. This results in a negligible difference between central and off-center sources. Therefore, instead of studying the off-center case, we analyze the DM contributions from a satellite galaxy located within the galaxy cluster.

For this analysis, we select the largest subhalo near the galaxy cluster halo center as the representative satellite galaxy. This subhalo is located approximately 1.86 Mpc from the cluster center. The analysis follows the same methodology as that used for spiral galaxies, as outlined in 3.5.2, but the off-center case is replaced by the satellite case. The results are presented using pink component in Figure 3.20.

In Figure 3.20, panel (a) shows the electron density profile along central sight lines, while panel (b) displays the profile for the satellite galaxy. The central case exhibits electron densities approaching $10^{-1}\,\mathrm{cm}^{-3}$ near the cluster center, gradually declining over a distance of several thousand kiloparsecs. The satellite case, while showing lower electron densities near the center, still reflects a complex density structure, including peaks indicative of substructures within the cluster. These profiles suggest that galaxy clusters possess extremely dense and complex baryonic distributions.

Panels (c) and (d) show the DM profiles for the central and satellite cases, respectively. The central case demonstrates a steep increase in DM near the center, and the 16th-84th percentile DM range is distributed between $1477 - 2003\,\mathrm{pc\,cm^{-3}}$, with a median DM value of $1670.16\,\mathrm{pc\,cm^{-3}}$. The log-normal distribution (Equation 3.1) provides a good fit to the DM distribution. The best-fit parameters for the central case are $\sigma = 0.1466^{+0.0008}_{-0.0010}$ and $\mu = 1698.05^{+2.31}_{-2.18}\,\mathrm{pc\,cm^{-3}}$. The corresponding mean and standard deviation (Equation 3.2) of DM are $E(\mathrm{DM}) = 1716.40\,\mathrm{pc\,cm^{-3}}$ and $\mathrm{Std(DM)} = 252.98\,\mathrm{pc\,cm^{-3}}$.

For the satellite case, the DM distribution exhibits a more gradual increase, with a median value of $1472.86\,\mathrm{pc\,cm^{-3}}$. The log-normal fit for this case gives $\sigma = 0.1063^{+0.0014}_{-0.0015}$ and $\mu = 1490.15^{+1.34}_{-1.50}\,\mathrm{pc\,cm^{-3}}$. The corresponding mean and standard deviation of DM are $E(\mathrm{DM}) = 1498.59\,\mathrm{pc\,cm^{-3}}$ and $\mathrm{Std(DM)} = 159.75\,\mathrm{pc\,cm^{-3}}$.

The results reveal that if a galaxy cluster serves as the host for an FRB, its DM contribution ($\mathrm{DM_{HG}}$) is expected to dominate the total $\mathrm{DM_{FRB}}$, potentially exceeding the contributions from all other components combined. Even for FRBs originating in satellite galaxies within the cluster, the $\mathrm{DM_{HG}}$ remains significant, highlighting the importance of considering the environment of galaxy clusters in FRB studies.

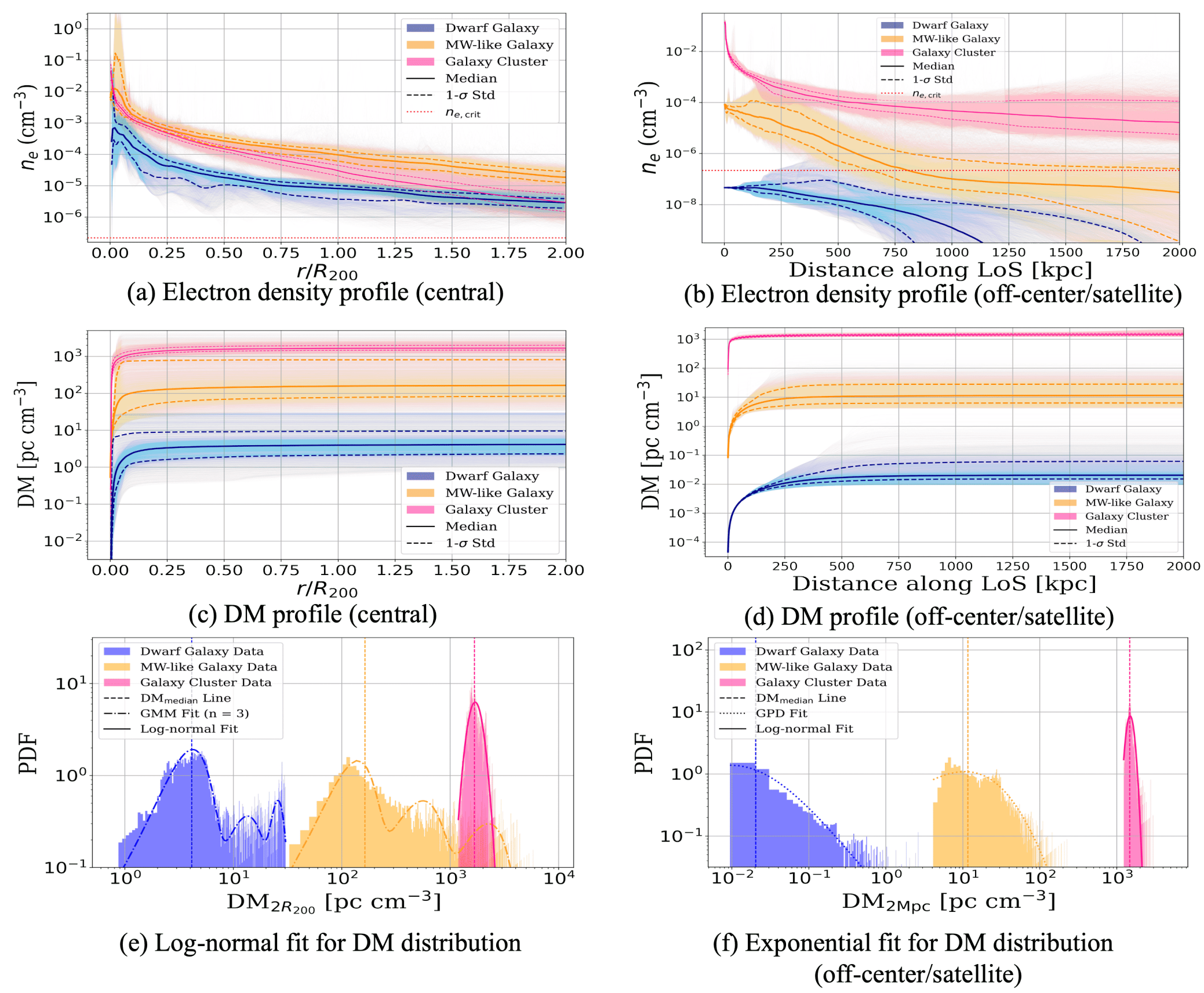

FIGURE 3.20: Consolidated analysis for different galaxy environments. The blue, orange, and pink colors represent the dwarf galaxy, MW-like galaxy, and galaxy cluster cases, respectively. We utilize 10000 sight lines, with different integration limits: for the central cases (panels a, c, e), the integration extends only up to $2R_{200}$ for higher resolution, while for the off-center/satellite cases (panels b, d, f), the integration extends to 2000 kpc. The solid lines in panels (a) and (b) indicate the median values, while the dashed lines represent the uncertainty range of 16th–84th percentile. The red dotted line in panels (a) and (b) denotes the critical electron density ($n_{e,\mathrm{critical}}$). Panels (c) and (d) display the DM profiles along the LoS. For the central cases (panel e), the DM distributions are fitted using a Gaussian Mixture Model (GMM) with 3 components $n = 3$ for the dwarf and MW-like Galaxies, and a log-normal distribution for the galaxy cluster. For the off-center/satellite cases (panel f), the dwarf and MW-like galaxies represent off-center DM values, with $\mathrm{DM}_{\mathrm{median}}$ values of $0.02\,\mathrm{pc\,cm^{-3}}$ and $11.59\,\mathrm{pc\,cm^{-3}}$, respectively, both fitted with a Generalized Pareto Distribution (GPD). The galaxy cluster instead represents a satellite halo near the cluster center, fitted with a skewed log-normal distribution, with a $\mathrm{DM}_{\mathrm{median}}$ of $1472.86\,\mathrm{pc\,cm^{-3}}$.

## 3.6 Implications for Precision FRB Cosmology

For precision FRB cosmology, our analysis indicates that inferred DM statistics can vary substantially when the same methodology is applied to simulations with different physical prescriptions. This section therefore evaluates how specific physical ingredients and modeling choices propagate into DM estimates, and outlines priorities for improving both simulation frameworks and observational constraints.

- *AGN Feedback and DM Estimation:* AGN feedback is a primary driver of baryon exchange between the CGM and IGM, and thus directly shapes the DM measured along FRB sight lines. Stronger feedback transfers a larger gas mass into the IGM and lowers the CGM DM contribution, whereas weaker feedback keeps more baryons in the CGM and enhances its imprint on FRB sight lines.

  Because AGN feedback prescriptions differ across simulation suites, the predicted DM statistics also differ. SIMBA, for instance, adopts strong kinetic jet feedback that efficiently evacuates halo gas, while IllustrisTNG and EAGLE (Schaye et al., 2014) apply different thermal–kinetic feedback mixes that produce distinct CGM structures. These structural differences map directly onto the free-electron column density and therefore the DM signal.

  In our CROCODILE setup, AGN jet feedback is currently absent, which likely contributes to the relatively small DM-statistics difference between the fiducial and NoBH runs. This interpretation aligns with Medlock et al. (2025b), who examined CAMELS feedback energetics and showed that different AGN implementations produce different baryon-redistribution efficiencies. Specifically, kinetic models generally transport gas farther into the IGM than thermal models, strongly modifying the CGM gas fraction. How and when kinetic feedback is activated (e.g., via a black-hole mass threshold) is a key control parameter for its low-redshift impact ($z = 0$). Together, these findings motivate adding kinetic AGN feedback in future CROCODILE developments to better capture gas redistribution.

  In addition, Medlock et al. (2025a) investigated how $f_{\rm IGM}$ affects suppression in the matter power spectrum, demonstrating that feedback-driven baryon redistribution modifies matter clustering across scales. This is directly relevant for FRB-based baryon-fraction constraints because it changes the statistics of DM fluctuations along sight lines. A key next step is to test whether this suppression is detectable in FRB datasets and to compare the signal with predictions from large-volume simulations such as CROCODILE.

- *$F$-parameter and Baryon Spread:* The $F$-parameter statistically encodes feedback strength through the scatter of the Macquart relation, following the framework introduced by McQuinn (2014) and further developed in Macquart et al. (2020). It is defined from the fractional DM standard deviation ($\sigma_{\rm DM}$) as:

$$\sigma_{\rm DM}(\Delta) \propto F z^{-1/2} \tag{3.30}$$

  where $\sigma_{\rm DM}$ is the DM standard deviation at redshift $z$, and $F$ measures the level of feedback-induced baryon redistribution. For fixed $\Omega_m$ and $\sigma_8$, smaller $F$ implies stronger feedback, with more baryons displaced into the IGM and consequently lower DM scatter. Larger $F$ indicates weaker feedback, with more baryons retained in the CGM and thus larger DM scatter across FRB sight lines.

Using CAMELS, Medlock et al. (2024, 2025b) showed that $F$ is sensitive to the adopted feedback model. Their analysis indicates that SIMBA's kinetic feedback yields lower $F$ values (stronger baryon redistribution), whereas IllustrisTNG and Astrid produce higher $F$ values (weaker feedback). They also stress that limited box sizes constrain interpretation, since small volumes may miss large-scale structure contributions to DM statistics; this echoes the box-size conclusions of Batten (2020) based on EAGLE.

Observational constraints on $F$ are still sparse. A recent study by Baptista et al. (2024) derived an empirical lower bound on $F$ from localized FRBs. Cross-comparing such measurements with simulation outputs can help identify which feedback channels dominate IGM shaping. The correlation between $F$ and baryon spread (Figure 1 of Medlock et al. (2025a)) further emphasizes the need for improved models that include AGN-driven gas expulsion.

- *Impact of Initial Conditions.* Changes in cosmological parameters (e.g., $\sigma_8$ and $\Omega_m$) alter large-scale-structure formation and therefore modify gas partitioning between the CGM and IGM (Khrykin et al., 2024a,b); the inferred $F$-parameter is likewise affected. Such shifts can induce systematic offsets in DM-based measurements.

  Stochastic differences in the initial perturbation field also generate sight-line-to-sight-line DM variance, increasing scatter in statistical inferences. In parallel, galaxy assembly histories—including merger frequency and cooling timescales—shape CGM structure and thus FRB DM contributions. More realistic treatments of these ingredients in future simulations should improve the accuracy of DM-based baryon-fraction estimates.

- *Numerical Methodology and Computational Effects.* Computational methodology itself introduces DM-statistics differences. Smoothed Particle Hydrodynamics (SPH), for example, tends to generate smoother CGM distributions—particularly in low-density regions—and may therefore bias DM low. Likewise, choices in radiative-transfer treatment, UV background intensity, and AGN radiative feedback modify CGM ionization and can shift DM predictions. Although these numerical factors are important, the present discussion emphasizes the underlying physical drivers of baryon redistribution.

- *Bridging Simulations and Observations.* Progress in both simulations and datasets is essential for tightening DM-based baryon-fraction constraints. Larger simulated volumes—including TNG-300 and CAMELS—enable broader statistical characterization of DM variability. Increased resolution and improved galaxy-formation modeling can also refine halo density profiles and therefore foreground-galaxy DM estimates.

  On the observational side, upcoming wide-field FRB programs—such as DSA-2000 (Hallinan et al., 2019) and BURSTT (Lin et al., 2022)—will deliver much larger samples and thus tighter constraints on the CGM DM contribution. Complementary deep spectroscopy, including Prime Focus Spectrograph campaigns on Subaru (Tamura et al., 2016), will better characterize foreground galaxies intersecting FRB sight lines. These data will constrain

CGM gas structure, ionization state, and metallicity, thereby improving DM modeling fidelity.

Ultimately, resolving the missing-baryon problem requires tight coupling between simulations and observations: simulations provide physically grounded baryon-distribution models, and observations provide the constraints needed to calibrate and discriminate among them. Continued progress on both fronts will further sharpen our understanding of how baryons are distributed across the Universe.

### 3.6.1 Part I Synthesis and Transition

This subsection summarizes the *first part* of the thesis, which focuses on cosmological-scale FRB DM modeling in the CROCODILE framework. Using light-cone realizations from the GADGET3/4-OSAKA SPH simulations (including star formation, SN feedback, and AGN feedback), we computed gas profiles and DM statistics along FRB sight lines to constrain baryon distributions across large-scale structures.

At the physical-process level, AGN feedback is identified as a key regulator of baryon redistribution between the CGM and IGM. It lowers central halo gas densities and reshapes the CGM–IGM transition, with the strongest impact in $10^{12.5} - 10^{13.5} M_{\odot}$ halos, where the central gas density at 15 kpc is reduced by 86.1%. Consistently, AGN feedback suppresses the median central halo DM contribution by $\sim 73\%$ (relative to NoBH) in the same mass range, while enhancing contributions at larger radii. The modified NFW model provides a better description of these simulated profiles than the standard NFW form, indicating the importance of AGN-driven gas expulsion in halo-scale DM modeling.

At the cosmological-observable level, the light-cone DM–$z$ relation yields $f_{\rm diff}(z = 1) = 0.865^{+0.101}_{-0.165}$ (fiducial) and $0.856^{+0.101}_{-0.162}$ (NoBH), in agreement with current FRB-based constraints and supporting the robustness of the Macquart relation when halo-associated gas is properly included. This also provides a useful cross-check of the ΛCDM framework. We further quantified the redshift evolution of $f_{\rm diff}$ and $f_{\rm IGM}$ under different CGM cutoff radii, and provided fitting functions that can be used in semianalytic galaxy-formation and IGM studies.

Finally, we extended the cosmological context to host-galaxy environments using zoom-in simulations of dwarf galaxies, MW-like galaxies, and galaxy clusters. The host contribution shows strong environmental dependence: central host DMs are typically below 100 pc cm$^{-3}$ in dwarfs, but can exceed 1300 pc cm$^{-3}$ in clusters, where host terms may dominate the total FRB DM budget. Overall, this first part establishes a simulation-informed framework for FRB baryon census work on cosmic scales and motivates the second part of the thesis, which isolates source-local contributions ($\rm DM_{source}$) in supernova-remnant environments.

# Chapter 4

# Part II: FRB Source Environment: Supernova Remnants

Chapter 3 established the cosmological utility of FRBs and quantified contamination from foreground large-scale structures (e.g., intervening galaxies and clusters) along the LoS. For precision FRB cosmology, however, these foreground terms are not the only source of bias: an additional and potentially significant contribution can arise locally at the source. FRB 20190520B provides a representative example, with a source-dominated interpretation (Zhao & Wang, 2021) versus a substantial foreground-halo/cluster contribution (Lee et al., 2023).

Section 1.3.5 established the observational motivation for source-local plasma around FRB progenitors. Building on that context, this chapter focuses on the SNR-physics side of the problem: how the structure and evolution of young remnants regulate $\mathrm{DM}_{\mathrm{source}}$ and its time variability. Unlike cosmic-web and foreground-halo terms, which evolve weakly over human observing baselines, source-local DM can vary measurably on year-scale timescales. We therefore adopt time-dependent SNR modeling to quantify this term and to evaluate how strongly progenitor and circumstellar assumptions affect interpretation.

## 4.1 SNR modeling landscape and chapter scope

Analytical frameworks based on the McKee–Truelove description of remnant evolution have been widely used to estimate SNR contributions to FRB dispersion across the ejecta-dominated (ED), Sedov–Taylor (ST), and radiative phases (McKee & Truelove, 1995, Truelove & McKee, 1999, Piro, 2016, Piro & Gaensler, 2018). These models provide transparent scaling relations and have been useful for placing order-of-magnitude constraints on source age and ambient density.

Numerical approaches now extend these scalings by resolving the coupled hydrodynamics, ionization state, and radiative processes of remnants. Modern stellar-evolution calculations with MESA (version 12115; Paxton et al., 2011, 2013, 2015, 2018, 2019, Jermyn et al., 2023) provide physically motivated progenitor structures from zero-age main sequence to core collapse, which can be mapped into CCSNR simulations. Complementary SNR calculations range from 1D CR-hydro-NEI (ChN) models with detailed shock microphysics and cooling (Patnaude et al., 2010, Lee et al., 2012, 2013, 2014, Diesing, 2019, Diesing et al., 2024, Diesing & Gupta, 2025) to multidimensional studies that capture PWN interaction, clumping, and asymmetric ejecta–CSM structure (Blondin et al., 2001, Ferrand et al., 2010, Orlando et al., 2012, Ferrand et al., 2012, 2019).

Despite this progress, a key uncertainty remains the diversity of progenitor channels and circumstellar media. Single-star and binary-stripped progenitors produce different mass-loss histories and wind-structured CSM profiles (Patnaude et al., 2017, Jacovich et al., 2021, Laplace et al., 2021, Farmer et al., 2023, Kawashima et al., 2026), which can generate substantially different density, ionization, and therefore DM evolution patterns compared with idealized homogeneous-media models.

Motivated by this gap, we use detailed 1D HD simulations with non-equilibrium ionization and radiative cooling to model the SNR contribution to $\mathrm{DM_{source}}$. The goal is to determine whether self-consistent remnant models can reproduce the observed ranges of $\mathrm{DM_{source}}$, $\mathrm{dDM}/\mathrm{d}t$, and $\mathrm{RM_{source}}$, and to identify which combinations of ZAMS mass, ejecta mass, and CSM structure (for single-star and binary-stripped channels) are consistent with current FRB constraints.

## 4.2 Motivation: Importance of $\mathrm{DM_{source}}$

### 4.2.1 Magnetar origin scenario

As described by Equation 1.18, the total $\mathrm{DM_{FRB}}$ can be decomposed into a non-local term, $\mathrm{DM_{nonloc}}$, and a source-local term, $\mathrm{DM_{source}}$. The latter arises from the immediate environment of the FRB engine, spanning scales from $\lesssim 0.1$ pc for PWN/MWN components to several parsecs for the SNR shell and the surrounding H II region. Under the young-magnetar scenario, a substantial fraction of $\mathrm{DM_{source}}$ may originate in a dense, recently formed CCSNR, whose plasma structure evolves dynamically on year-to-decade timescales. In this picture, $\mathrm{DM_{source}}$ provides a natural explanation for the secular DM evolution reported in some repeating FRBs.

Moreover, because $\mathrm{DM_{source}}$ depends sensitively on progenitor mass loss, explosion energetics, and circumstellar structure, it may retain information about the evolutionary channel of the source. In particular, differences between single-star (SS) and binary-stripped (BS) progenitors can, in principle, imprint distinct signatures on the magnitude and temporal evolution of the local DM contribution.

### 4.2.2 Candidate local environments: SNR, PWN, H II region, plasma lensing

Previous work has explored several possible local environments that could contribute significantly to the FRB DM and its time evolution:

- *SNR:* An expanding SNR ejecta and its forward shock shell provide a natural reservoir of dense ionized gas whose column density decreases over time. In previous FRB–SNR theoretical studies, the temporal evolution of the SNR has often been modeled using the self-similar solutions of McKee & Truelove (1995), which describes the ED, ST, and radiative snowplow phases. Within this analytic framework, the resulting DM evolution is commonly written as $\mathrm{DM}_{\mathrm{SNR}} \propto t^{-\alpha}$ with $0 \lesssim \alpha \lesssim 2$[1], and the corresponding $\mathrm{dDM}/\mathrm{d}t$ values can fall in the range discussed by Piro (2016), Piro & Gaensler (2018), Yang & Zhang (2017).

- *PWN / magnetar wind nebula (MWN):* A relativistic wind from the central neutron star can inflate a nebula whose internal plasma contributes to $\mathrm{DM}_{\mathrm{source}}$. Two types of PWN models have been discussed. In rotation-powered pulsar–wind bubbles (e.g., Murase et al., 2016, Kashiyama & Murase, 2017), the nebula is dominated by $e^{\pm}$ pairs, and the lepton density is set by the Goldreich–Julian density, the pair multiplicity, and the spin-down power, generally yielding modest DM. In contrast, flare-powered, baryon-loaded nebulae (e.g., Margalit & Metzger, 2018) contains an ion–electron plasma whose density depends on the baryon-loading factor and energy-injection history, and can produce substantially larger DM (and RM). The PWN/MWN can not only enhance the $e^{\pm}$ pair-plasma density near the magnetosphere, but also inject baryon-loaded plasma into the surrounding SNR ejecta/CSM, further modifying the local DM and RM.

- *H 2 region and star-forming complex:* If the FRB source is located in a compact H 2 region or a dense star-forming clump, the associated ionized gas can contribute a substantial, but relatively slowly varying, DM.

- *Plasma lensing structures:* Small-scale overdense plasma sheets or filaments can both refract and disperse FRB radiation, potentially producing burst-to-burst DM variations that are not strictly secular. Although dense plasma anywhere along the line of sight can, in principle, cause plasma lensing, "strong" lensing requires extremely compact, overdense plasma structures (Cordes et al., 2017, Main et al., 2018), which are found almost exclusively in the immediate local environment of the FRB (from sub-AU scales to a few–tens of parsecs); hence their DM contribution is naturally assigned to $\mathrm{DM}_{\mathrm{source}}$.

Yang & Zhang (2017) presented a unified framework for these different contributors and emphasized that the cosmological and host-galaxy terms are nearly time-independent, while the SNR and PWN components can produce measurable $\mathrm{dDM}/\mathrm{d}t$. In this study, we concentrate on the SNR[2] scenario and ask: given realistic SNR evolution, what range of $\mathrm{DM}_{\mathrm{source}}$ and $\mathrm{dDM}/\mathrm{d}t$ can be produced, and how does this depend on the progenitor channel? We further examine whether the time evolution of $\mathrm{DM}_{\mathrm{SNR}}$ alone can account for the observed secular decrease of FRB 20190520B.

[1] We emphasize, however, that these analytic trends need not hold for our numerical 1D SNR simulations, whose detailed ionization, shock structure, and density evolution may deviate from the idealized self-similar solutions.

[2] Central compact-object effects, such as PWN/MWN contributions, are discussed only as an ionization-enhancement upper-limit scenario in this work.

## 4.3 The SNR framework: hydrodynamics, progenitor imprint, and shock physics

### 4.3.1 Hydrodynamic evolution and governing equations

For the purposes of this thesis, the most relevant description of a young SNR is the standard hydrodynamic picture in which freely expanding ejecta drive a forward shock into the surrounding medium while a reverse shock propagates back into the ejecta, with a contact discontinuity separating the two shocked regions. As the remnant evolves, it passes from an ejecta-dominated stage toward the Sedov–Taylor regime and, at still later times, toward radiative evolution (McKee & Truelove, 1995, Truelove & McKee, 1999, Chevalier, 1982). For FRB applications, however, the main point is not simply the sequence of dynamical phases, but the fact that $\mathrm{DM_{source}}$ is set by the evolving density, ionization state, composition, and geometry of the unshocked ejecta, shocked ejecta, and shocked circumstellar/interstellar gas. In that sense, the forward shock and reverse shock matter primarily because they regulate where the electrons are, how ionized they are, and how long the source remains radio-transparent.

In the one-dimensional, spherically symmetric HD limit adopted for the production models in this thesis, the ejecta–CSM interaction is governed by the Euler equations with radiative losses,

$$\frac{\partial \rho}{\partial t} + \frac{1}{r^2}\frac{\partial}{\partial r}\left(r^2 \rho u\right) = 0, \tag{4.1}$$

$$\frac{\partial (\rho u)}{\partial t} + \frac{1}{r^2}\frac{\partial}{\partial r}\left[r^2\left(\rho u^2 + P\right)\right] = 0, \tag{4.2}$$

$$\frac{\partial E}{\partial t} + \frac{1}{r^2}\frac{\partial}{\partial r}\left[r^2 u(E+P)\right] = -\mathcal{L}_{\rm rad}, \tag{4.3}$$

where $\rho$ is the mass density, $u$ is the radial velocity, $P$ is the thermal pressure, and $E = P/(\gamma-1) + \rho u^2/2$ is the total fluid energy density for an ideal gas with adiabatic index $\gamma$. These equations provide the hydrodynamic backbone of the present SNR calculations and naturally generate the double-shock structure that later determines the DM contribution from the unshocked ejecta, shocked ejecta, and shocked CSM/ISM (Truelove & McKee, 1999, Patnaude et al., 2017, Kawashima et al., 2026). Here $\mathcal{L}_{\rm rad}$ denotes the radiative energy-loss rate per unit volume (e.g., in $\mathrm{erg\,s^{-1}\,cm^{-3}}$), which represents cooling by the thermal plasma; in the adiabatic limit one simply sets $\mathcal{L}_{\rm rad} = 0$.

The associated shock physics can be summarized through the Rankine–Hugoniot relations. For a shock of Mach number $\mathcal{M}_1$ propagating into an ideal gas,

$$\frac{\rho_2}{\rho_1} = \frac{(\gamma+1)\mathcal{M}_1^2}{(\gamma-1)\mathcal{M}_1^2 + 2}, \tag{4.4}$$

which approaches $\rho_2/\rho_1 \to 4$ in the strong-shock limit for $\gamma = 5/3$. The post-shock ion temperature then scales approximately as

$$kT_i \simeq \frac{3}{16}\mu_i m_p v_s^2, \tag{4.5}$$

where $v_s$ is the shock speed and $\mu_i m_p$ is the mean ion mass. These relations are especially relevant here because shock compression sets the local electron column while the post-shock temperature controls the subsequent ionization evolution and radiative cooling.

### 4.3.2 CCSN evolutionary branches relevant to FRB–SNR systems.

A useful schematic sequence for the present problem is: main-sequence H burning, post-main-sequence expansion, one of several pre-SN surface configurations (extended RSG, conditional BSG/blue-loop excursion, or stripped He-star/WR-like state), advanced burning up to iron-core collapse, successful shock revival and SN explosion, early SN–CSM interaction, ejecta-dominated SNR evolution, and finally the Sedov–Taylor stage. Two caveats are important. First, not every progenitor passes through a BSG or blue-loop stage; such excursions are conditional rather than universal and depend sensitively on initial mass, metallicity, mixing, mass loss, and the core–envelope structure. Second, only successful explosions produce a classical SNR, whereas failed explosions may terminate the sequence before an observable remnant develops. Likewise, strong ejecta–CSM interaction can shape some optical SNe, but it is not a universal ingredient of the SN light curve. For thesis purposes, this means that any flow-chart description should be understood as a physically motivated branch diagram rather than a single mandatory path.

Equally important is that the ambient medium into which the SNR expands is not generic. It is shaped by the progenitor's pre-supernova evolution, including wind mass loss, binary stripping, and the chemical structure produced by stellar evolution and nucleosynthesis. These factors strongly affect the ejecta mass, density stratification, elemental abundances, and circumstellar density profile, and therefore directly influence the predicted DM and, once magnetic fields are modeled, the RM as well (Nomoto et al., 2013, Jacovich et al., 2021, Farmer et al., 2023, Kawashima et al., 2026). This is precisely why the single-star (SS) and binary-stripped (BS) progenitors considered in this thesis are physically meaningful beyond being different initial conditions: they represent different pre-SN evolutionary channels, with correspondingly different ejecta compositions and surrounding media. More broadly, the SNR population is expected to span a wide range of environments—from remnants expanding into relatively dense nearby material to systems evolving inside wind-blown cavities—so a realistic FRB interpretation must allow for substantial diversity rather than a single canonical remnant model.

### 4.3.3 Characteristic pre-SN wind states.

The time-dependent wind histories used to define the progenitor environments in this work are summarized in Figure 4.1. In this thesis, we adopt progenitors with initial masses of $11\,M_\odot$ and $30\,M_\odot$, for which corresponding stellar-evolution models have been provided by Kawashima et al. (2026). This enables a direct and self-consistent mapping between the stellar-evolution calculations and the pre-supernova wind profiles used in our simulations.

A useful point of emphasis is that the differences among the models are not merely differences in normalization, but reflect qualitatively distinct evolutionary states. Lower-mass single progenitors tend to approach core collapse with slower, denser RSG-like outflows, which raise $\dot{M}$ while keeping $v_w$ low. By contrast, binary stripping can remove most of the H-rich envelope before collapse and expose a more compact, hotter star, producing a faster and typically more rarefied pre-SN wind.

At higher initial mass, both the single and binary channels can experience multiple wind-state transitions prior to collapse, so both $\dot{M}(t)$ and $v_w(t)$ exhibit sharp features associated with structural readjustment of the envelope, changes in surface temperature, and the onset or disappearance of line-driven hot-star winds (Farmer et al., 2023, Kawashima et al., 2026).

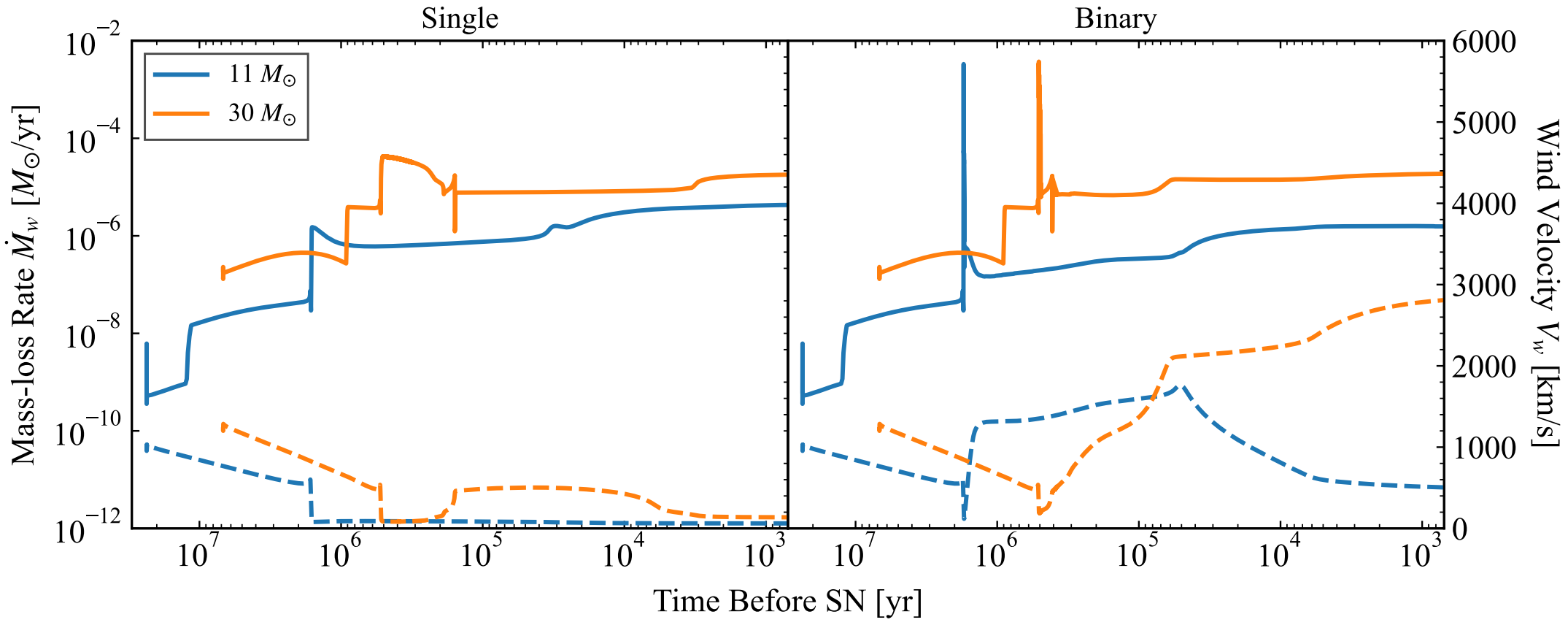

FIGURE 4.1: Time evolution of the pre-supernova mass-loss rate $\dot{M}_w(t)$ (solid) and wind velocity $v_w(t)$ (dashed) for the single-star (left) and binary-stripped (right) progenitors adopted in this thesis. The figure illustrates that the dominant difference between channels is not simply stronger versus weaker winds, but a systematic shift in wind state: low-mass single progenitors tend to end as slow, dense RSG-like wind sources, whereas binary stripping exposes hotter, more compact stars with faster pre-SN outflows. Higher-mass models exhibit multiple wind-state transitions prior to collapse, naturally generating the structured CSM later inherited by the SNR. The underlying stellar-evolution calculations are adapted from Kawashima et al. (2026); the 11 $M_\odot$ case is directly shared with the present model set, while the higher-mass cases are not in exact one-to-one correspondence.

The pre-SN wind taxonomy is also best regarded as a set of characteristic channels rather than a rigid sequence, and the commonly quoted velocities and mass-loss rates should be interpreted as order-of-magnitude ranges rather than exact boundaries. It is also important to keep in mind that these wind states are associated with different evolutionary times and progenitor regimes. OB-star winds are characteristic of earlier hot-star phases and are fast and relatively tenuous. RSG winds arise after substantial envelope expansion and are slow and dense, often dominating the immediate CSM of lower-mass or hydrogen-rich single-star progenitors; some such stars may undergo core collapse directly from the RSG stage. BSG winds are intermediate and usually appear only for a subset of progenitors during transitional or blue-loop excursions, rather than as inevitable endpoints for all stars. By contrast, WR/stripped-He-star winds are again fast but comparatively tenuous and are more naturally associated with strongly stripped binaries or more massive/hotter progenitors, some of which may pass only briefly through an RSG state—or bypass it almost entirely—before collapse. In interacting binaries, Roche-lobe overflow and non-conservative mass loss add further structure beyond any simple steady-wind prescription and can shift the star into a stripped, fast-wind state substantially earlier than in the single-star channel.

This qualitative picture is consistent with the progenitors adopted here: in the Kawashima et al. (2026) models, the $11\,M_\odot$ SS progenitor ends with a dense, slow RSG-like wind, whereas the $11\,M_\odot$ BS progenitor retains a much faster pre-SN outflow and therefore a more evacuated local medium; the higher-mass models show multiple wind-state transitions and are best interpreted as representative examples of how cavity-and-shell structures can be assembled prior to core collapse, rather than as a universal sequence followed by every progenitor.

### 4.3.4 Forward-looking note on CR acceleration and magnetic-field amplification

Although the production calculations used in this thesis do not include cosmic-ray (CR) back-reaction, nonlinear diffusive shock acceleration (NLDSA), or self-consistent magnetic-field amplification, it is still useful to briefly note why these processes matter for future work. In the broader CR-hydro/`ChN` framework, efficient particle acceleration can modify the precursor compression, post-shock temperature, and ionization history, while amplified magnetic fields can strongly affect synchrotron emission, polarization, and especially $\mathrm{RM}_{\mathrm{source}}$ (Caprioli et al., 2010, Lee et al., 2012, 2014). Therefore, the CR-related discussion in this chapter should be read not as an ingredient of the present DM calculations, but as motivation for a future extension in which DM and RM are modeled self-consistently in a magnetized shock environment.

In a generalized CR-enabled formulation, the momentum flux across the precursor may be written as

$$\rho_0 u_0^2 + P_{\mathrm{g},0} + P_{\mathrm{CR},0} + P_{\mathrm{w},0} = \rho(x)u(x)^2 + P_{\mathrm{g}}(x) + P_{\mathrm{CR}}(x) + P_{\mathrm{w}}(x), \tag{4.6}$$

where $P_{\mathrm{CR}}$ is the CR pressure and $P_{\mathrm{w}}$ is the magnetic-wave pressure. Correspondingly,

$$P_{\mathrm{w}}(x) = \frac{\delta B(x)^2}{8\pi}, \tag{4.7}$$

so that amplified turbulence contributes both dynamically and magnetically to the shock structure. In the present thesis, however, the production models effectively adopt the HD+NEI limit by switching off these CR-related terms, i.e., we set $P_{\mathrm{CR}} = 0$ and $P_{\mathrm{w}} = 0$ in practice. Keeping Equations (4.6) and (4.7) in view is nevertheless useful because they show exactly how the current framework could be extended in future RM-focused work.

This distinction is important: the results presented here isolate the effects of hydrodynamics, composition, and NEI, while CR acceleration and magnetic amplification remain prospective ingredients for the next stage of source-environment modeling.

### 4.3.5 Thermal Plasma and Non-Equilibrium Ionization (NEI)

The thermal and ionization states of the plasma are evolved self-consistently using a non-equilibrium ionization (NEI) framework.

First, the temperatures of electrons and ions are evolved via Coulomb interactions:

$$T_i(t + dt) = T_i(t) + \frac{T_j(t) - T_i(t)}{t_{\rm eq}(i, j)} dt, \tag{4.8}$$

where $t_{\rm eq}$ is the equilibration timescale given as

$$t_{\rm eq}(i, j) = \frac{3}{8\sqrt{2\pi}} \frac{m_i m_j}{n_j Z_i^2 Z_j^2 e^4 \ln \Lambda} \left( \frac{kT_i}{m_i} + \frac{kT_j}{m_j} \right)^{3/2}, \tag{4.9}$$

where $n$ is the number density, $Z$ is the charge number, and $\ln \Lambda$ is the Coulomb logarithm. Given the temperature, the ionization state of each element is then computed by solving a coupled set of rate equations:

$$\frac{df_{X,i}}{dt} = n_e \left[ C_{X,i-1} f_{X,i-1} - (C_{X,i} + \alpha_{X,i-1}) f_{X,i} + \alpha_{X,i} f_{X,i+1} \right], \tag{4.10}$$

where the index $X$ denotes the chemical element (e.g., H, He, O), and $i$ denotes the ionization stage, with $i = 0$ corresponding to the neutral state and $i = Z_X$ to the fully ionized state of element $X$. Here, $C_{X,i}$ and $\alpha_{X,i}$ are the collisional ionization and recombination coefficients between adjacent ionization states, respectively.

The resulting free-electron density is given by

$$n_e = \sum_X \sum_i i \, f_{X,i} \, n_X, \tag{4.11}$$

where $n_X$ is the total number density of element $X$. This expression shows that the free-electron density depends not only on the total gas density but also on the detailed ionization state of each element.

### 4.3.6 The ChN 1D Code

To model the time-dependent plasma structure around a young FRB source, we employ the 1D `ChN` code family developed for SNR studies (Patnaude et al., 2010, Lee et al., 2012, 2013, 2014, Jacovich et al., 2021, Court et al., 2024, Kawashima et al., 2026). In the broader code framework, `ChN` can couple spherically symmetric hydrodynamics to non-equilibrium ionization (NEI), radiative cooling, and, in some implementations, nonlinear diffusive shock acceleration (NLDSA). The specific models adopted in this thesis, following Kawashima et al. (2026), use the 1D HD+NEI configuration without CR back-reaction or magnetic-field evolution. This is the configuration that is directly relevant for our DM calculations, because $\mathrm{DM}_{\rm source}$ depends first and foremost on the coupled evolution of density, composition, temperature, and ionization in the ejecta–circumstellar-medium (CSM) interaction region.

Accordingly, the production runs used in the FRB analysis do *not* include self-consistent NLDSA, CR-driven precursor modification, or magnetic-field amplification; equivalently, the CR-related terms in Equations (4.6) and (4.7) are turned off for the models analyzed here. These omissions are intentional rather than accidental: they allow the present work to isolate the hydro+NEI contribution to $\mathrm{DM}_{\rm source}$ and to trace how it changes with progenitor structure, ejecta composition, and circumstellar environment. At the same time, these omitted ingredients remain highly relevant for future RM-oriented studies. Once magnetic-field evolution is included self-consistently, the same framework can be extended to address how shock-amplified fields and accelerated particles reshape the magneto-ionic environment probed by repeating FRBs.

## 4.4 Analytic benchmarks for DM evolution in SNR environments

To establish a reference baseline for the subsequent comparison with our numerical SNR simulations, we summarize four commonly used (semi-)analytic DM prescriptions drawn from the literature. These models capture different physical regimes, including (i) Entire ionized ejecta in free expansion, (ii) a shocked ionized ejecta contribution, and (iii–iv) wind contributions in the presence of a forward shock, either excluding or including the unshocked wind outside the shock. The corresponding expressions are summarized in Eqs. (4.12)–(4.20).

**(1) Entire Ionized ejecta.** In this scenario, the ejecta shell (including the unshocked component) is assumed to be ionized, and the DM is estimated as the electron column across the ejecta layer, $\mathrm{DM}_{\rm SNR} \sim n_e \Delta R$, where $\Delta R = R_b - R_r$, with $R_b$ and $R_r$ denoting the blast-wave and reverse-shock radii, respectively. Piro (2016) provided an analytic treatment of $\mathrm{DM}_{\rm SNR}$ and its temporal evolution in this free-expansion limit. Here we adopt the thin-shell approximation described by Yang & Zhang (2017), which provides an explicit time scaling, given by

$$\mathrm{DM}_{\rm SNR}(t) \simeq 2.6 \times 10^4 \, \chi_e \left(\frac{\mu_m}{1.2}\right)^{-1} M_1^2 \, E_{51}^{-1} \, t_{\rm yr}^{-2} \, \mathrm{pc\ cm^{-3}}. \tag{4.12}$$

where $\chi_e$ is the average ionization fraction of the medium in the SNR, $\mu_m$ is the mean molecular weight (for reference, $\mu_m \simeq 1.2$ corresponds to a solar-composition, approximately neutral gas), $M_1 \equiv M_{\rm ej}/(10\, M_\odot)$, $E_{51} \equiv E/(10^{51}\,\mathrm{erg})$ and $t_{\rm yr} \equiv t/(1\,\mathrm{yr})$.

**(2) Shocked ionized ejecta.** In this scenario, only the shocked ejecta layer between the reverse shock and the contact discontinuity is ionized, i.e., $\Delta R = R_c - R_r$. Assuming a uniform ambient ISM with constant number density $n_0$, this prescription leads to a more gradual temporal decay in the DM contribution from the ionized region (Piro & Gaensler, 2018) :

$$\mathrm{DM}_{\rm SNR}(t) = 52.6 \left(\frac{\mu_m}{\mu_e}\right) E_{51}^{-1/4} \, M_1^{3/4} \, n_0^{1/2} \, t_{\rm yr}^{-1/2} \ \mathrm{pc\ cm^{-3}}, \tag{4.13}$$

where $\mu_e$ is the mean molecular weight per electron, respectively.

**(3) Wind case: shocked-ejecta contribution in a circumstellar wind.** If a proto-stellar wind is present, the wind density profile follows $\rho_w = Kr^{-2}$, where $K = \dot{M}/(4\pi v_w)$, $\dot{M}$ is the mass loss rate and $v_w$ is the wind velocity. In this case, the dominant DM contribution is also the ionized ejecta layer from the reverse shock. Compared to the constant-density case, the resulting DM declines more steeply with time (Piro & Gaensler, 2018):

$$\mathrm{DM_{SNR}}(t) \simeq 1.3 \times 10^4\, \mu_e^{-1}\, E_{51}^{-3/4}\, M_1^{5/4}\, K_{13}^{1/2}\, t_{\mathrm{yr}}^{-3/2}\ \mathrm{pc\ cm^{-3}}, \tag{4.14}$$

where $K_{13} \equiv K/(10^{13}\,\mathrm{g\,cm^{-1}})$. The contribution from the shocked circumstellar medium, $\mathrm{DM_{sh,CSM}}$, arises from the forward-shock–heated wind material and exhibits a shallower temporal decay. In this case, $\mathrm{DM_{sh,CSM}}$ becomes important only at sufficiently late times, while it remains subdominant over the time range of interest considered in this work.

**(4) Wind case: shocked ejecta and shocked CSM with self-similar evolution.** In this Scenario, the SNR DM receives contributions from both the reverse-shock–heated ejecta and the forward-shock–heated CSM. Following the self-similar solution of ejecta–wind interaction (Zhao et al., 2021, Zhao & Wang, 2021, Tang & Chevalier, 2017), with the relevant characteristic scales and self-similar constants summarized in Appendix C.2, we write

$$\mathrm{DM_{SNR}}(t) = \mathrm{DM_{sh,ej}}(t) + \mathrm{DM_{sh,CSM}}(t) + \mathrm{DM_{un,ej}}(t)\,, \tag{4.15}$$

where $\mathrm{DM_{sh,ej}}$ and $\mathrm{DM_{sh,CSM}}$ denote the contributions from the shocked ejecta and shocked CSM, respectively, and $\mathrm{DM_{un,ej}}$ is the DM contribution from the unshocked ejecta (cf. Zhao et al. 2021).

The shocked-ejecta contribution is

$$\mathrm{DM_{sh,ej}}(t) = \frac{2K(n-3)(n-4)}{\mu_a m_p} \left(\frac{1}{r_2} - 1\right) \frac{1}{\zeta_c R_{\mathrm{ch}}} \left(\frac{t_{\mathrm{ch}}}{t}\right)^{\frac{n-3}{n-2}}\,, \tag{4.16}$$

where $r_2 \equiv R_{\mathrm{r}}/R_{\mathrm{c}}$, $\mu_a$ is the mean atomic weight, $m_p$ is the proton mass, $R_{\mathrm{ch}}$ is the SNR characteristic radius (i.e., the characteristic length scale in the SSDW normalization), and $t_{\mathrm{ch}}$ is the corresponding characteristic timescale, as defined by Eqs. (C.2) and (C.3), while the dimensionless constant $\zeta_c$ follows Eq. (C.11).

The shocked-CSM (i.e., forward-shocked wind) contribution is

$$\mathrm{DM_{sh,CSM}}(t) = \frac{4K}{\mu_a m_p} \left(1 - \frac{1}{r_1}\right) \frac{1}{\zeta_c R_{\mathrm{ch}}} \left(\frac{t_{\mathrm{ch}}}{t}\right)^{\frac{n-3}{n-2}}\,, \tag{4.17}$$

where $r_1 \equiv R_1/R_{\mathrm{c}}$ is the ratio of the forward-shock radius to the contact-discontinuity radius for the wind case ($s = 2$), and the numerical values of $r_1$ (and $r_2$) for a given $(n, s)$ can be taken from Chevalier (1982).

The unshocked-ejecta contribution can be decomposed into the DM of the unshocked core and the DM of the unshocked outer power-law envelope,

$$\mathrm{DM}_{\mathrm{un,ej}}(t) = \mathrm{DM}_{\mathrm{core}}(t) + \mathrm{DM}_{\mathrm{pl}}(t)\,. \tag{4.18}$$

For the broken power-law ejecta profile in Eq. (C.7), we may write

$$\begin{aligned}\mathrm{DM}_{\mathrm{un,ej}}(t) = &\int_0^{R_{\mathrm{core}}} \frac{\chi_e\, M_{\mathrm{ej}}}{\mu_a m_p\, R_{\mathrm{ej}}^3}\, f_0\, \mathrm{d}r \\ &+ \int_{R_{\mathrm{core}}}^{R_{\mathrm{ej}}} \frac{\chi_e\, M_{\mathrm{ej}}}{\mu_a m_p\, R_{\mathrm{ej}}^3}\, f_0 \left(\frac{r}{R_{\mathrm{core}}}\right)^{-n} \mathrm{d}r\,.\end{aligned} \tag{4.19}$$

where $R_{\mathrm{core}} = w_{\mathrm{core}} R_{\mathrm{ej}}$ and we take $\chi_e$ to be the average electron ionization fraction of the unshocked ejecta. In the free-expansion (FE) solution, $R_{\mathrm{ej}} = R_{\mathrm{c}}$, and the unshocked-core contribution becomes

$$\mathrm{DM}_{\mathrm{core}}(t) = \frac{M_{\mathrm{ej}}}{\mu_a m_p} \chi_e f_0 w_{\mathrm{core}}\, \lambda_c^{-2}\, v_{\mathrm{ch}}^{-2}\, t^{-2}\,. \tag{4.20}$$

The reverse shock radius in the FE solution can be written as $R_{\mathrm{r}} = q_r R_{\mathrm{c}}$, where $q_r \equiv q_b/\ell_{\mathrm{ED}}$. For an "apples-to-apples" comparison with the semi-analytic model of Zhao et al. (2021), hereafter Zhao+21, we evaluate the self-similar constants using their fiducial parameters $(n, s) = (10, 2)$, even though our numerical simulations adopt $n = 11$ (see Section 4.5.2).

Although an additional DM contribution from the unshocked CSM is present, it is subdominant in this regime. We therefore use Eqs. (4.16)–(4.20) to provide a more complete analytic estimate for the wind scenario.

## 4.5 SNR Model Setup

### 4.5.1 Progenitor channels: single vs binary

We adopt progenitor models computed by Kawashima et al. (2026), which are primarily based on the detailed stellar evolution calculations of Farmer et al. (2023) using MESA from the ZAMS to core collapse. These models self-consistently follow the mass-loss history, pre-supernova structure, and explosion properties of massive stars, and provide physically motivated initial conditions for the subsequent SNR evolution.

We consider two main evolutionary channels for producing CCSNe that host young FRB central engines. For both channels, we adopt progenitors with initial masses $M_{\mathrm{ZAMS}} = 11$ and $30\,M_\odot$, which serve as representative low- and high-mass cases:

1. *Single-star (SS) channel:* Massive stars that evolve in isolation, losing part of their envelopes via stellar winds and exploding as hydrogen-rich Type II SNe. Their wind velocity, mass-loss history, and stellar radius vary significantly over the stellar lifetime and are taken directly from the `MESA` outputs, in contrast to the steady-wind approximations adopted in some previous studies (e.g.,Jacovich et al. 2021). The final pre-SN structure and ejecta mass $M_{\rm ej}$ therefore reflect the integrated history of wind-driven mass loss.

2. *Binary-star (BS) channel:* Massive binary systems in which the primary star loses most of its hydrogen-rich envelope via Roche-lobe overflow (RLOF) and winds prior to core collapse, becoming a stripped helium star prior to core collapse (Type Ib/c). The binary mass ratio is set to $M_2/M_1 = 0.8$ followed by Farmer et al. (2023) and Kawashima et al. (2026). Binary evolution is treated in a simplified yet physically motivated manner: the companion is modeled as a point mass until the end of core helium burning, after which the primary continues its evolution without further interaction. Mass transfer during RLOF is parameterized by a mass-loss condition (e.g., Paczyński, 1967, van den Heuvel, 1969) and because the exact fraction of mass retained by the companion is uncertain, we adopt the assumption that half of the transferred mass is accreted while the remaining half is ejected from the system and contributes to the CSM. This treatment captures the key effect of binary interaction—a substantially reduced $M_{\rm ej}$ a more compact progenitor at collapse—without introducing additional free parameters.

Note that we include both single and binary progenitors because a substantial fraction of massive stars reside in interacting binaries (Sana et al., 2012), and for each $M_{\rm ZAMS}$, the SS and BS channels can yield markedly different ejecta masses and composition structures, which directly affect the density normalization of the SNR and the resulting electron density ($n_e$) profile relevant for $\mathrm{DM}_{\rm source}$. Table 4.1 summarizes the main properties of the progenitor models used in the present paper.

### 4.5.2 SNe to SNR transition

Modeling the transition from the SN explosion to the SNR phase requires specifying the initial ejecta structure at the onset of the remnant evolution. Following common practice (e.g., Truelove & McKee 1999), we do not attempt to model the explosion mechanism itself. Instead, we assume that the ejecta have already reached a homologous expansion (FE) phase. The ejecta density structure is characterized by a prescribed density profile defined, consisting of a flat inner core and a steep outer envelope, defined by Eq. (C.7) and detailed in Appendix C.2.2. The outer ejecta follow a power-law profile $\rho_{\rm ej} \propto r^{-n}$, and we adopt $n = 11$ for both evolutionary channels.

TABLE 4.1: Summary of SS and BS progenitor models.

| Channel | $M_{\rm ZAMS}$ ($M_\odot$) | $\Delta M_{\rm wind}$ ($M_\odot$) | $\Delta M_{\rm RLOF}$ ($M_\odot$) | $M_{\rm ej}$ ($M_\odot$) | $E_{\rm SN}$ ($10^{51}$ erg) | $M_{\rm rem}$ ($M_\odot$) | SN type (expected) | Fate |
|---|---|---|---|---|---|---|---|---|
| Single | 11 | 1.64 | – | 7.83 | 1.0 | 1.53 | IIP-like | NS |
| Binary | 11 | 0.70 | 7.12 | 1.81 | 1.0 | 1.62 | Ib/c-like | NS |
| Single | 30 | 15.74 | – | 12.27 | 1.0 | 1.99 | IIb-like | NS |
| Binary | 30 | 7.50 | 10.94 | 9.85 | 1.0 | 1.71 | Ib/c-like | NS |

SN-type labels are indicative and based on pre-SN envelope composition (especially residual H/He), not on radiative-transfer light-curve/spectral modeling.

This choice is motivated by the compact pre-supernova progenitors considered in this work, particularly for the binary-stripped channel.

The SNR simulations are initialized at $t_{\rm init} = 3$ yr after explosion, by which time the ejecta are well described by homologous expansion. The total ejecta mass $M_{\rm ej}$ and explosion energy $E_{\rm SN}$ are set by the progenitor models and are listed in Table 4.1.

### 4.5.3 CSM

The CSM is constructed by explicitly following the interaction between the progenitor outflows and the surrounding ISM, rather than prescribing an analytic density profile (e.g., $\rho \propto r^{-2}$ for a wind-shaped medium), as commonly adopted in both analytical models and some numerical treatments (see Section 4.4). As described in Kawashima et al. (2026), time-variable mass-loss rates and wind velocities during different evolutionary stages can generate complex CSM structures, including piled-up shells, which strongly influence the subsequent SNR evolution.

The CSM structure is computed using 1D HD simulations of stellar winds expanding into a uniform ISM, employing the 1D hydrodynamics code `VH-1` (Blondin et al., 2001, Blondin & Ellison, 2001). Radiative cooling is included, allowing dense, geometrically thin shells to form as wind material accumulates. The adopted ISM number density is $n_{\rm ISM} = 1.0\,{\rm cm}^{-3}$, and both the wind and ISM temperatures are set to $T = 10^4\,{\rm K}$. The resulting CSM profiles provide the ambient density structure into which the SN ejecta expand in our subsequent SNR simulations. Further details of the CSM construction and parameter choices are described in Kawashima et al. (2026).

### 4.5.4 SNR simulation settings

The time-dependent evolution of the SNR is computed using a one-dimensional, spherically symmetric hydrodynamic code coupled with non-equilibrium ionization (HD+NEI), based on the ChN framework with CR acceleration turned off.

The code follows the NEI of 30 elements and accounts for radioactive decay by converting 162 unstable isotopes from the MESA yields to their stable counterparts using the Python library `radioactivedecay` (Amaku et al., 2010), based on the ICRP decay tables (ICRP, 2008).

The code follows the coupled evolution of the forward and reverse shocks as they propagate through the CSM and SN ejecta, and self-consistently tracks the energy partition, electron–ion temperature equilibration, and the time-dependent ionization states of multiple elements in each radial zone. In this work, we disable the cosmic-ray (CR) evolution/acceleration module; consequently, there is no self-consistent magnetic-field ($B$) evolution. This simplified setup is nevertheless sufficient for our purpose, since we focus on the thermal plasma structure— $n_e$, $T$, and ionization fraction $\chi_e \equiv n_e/(\rho/m_p)$—that determines the FRB DM, rather than on detailed non-thermal emission or magnetic-field diagnostics.

The simulations output 1D radial profiles of the hydrodynamical and microphysical quantities in each zone, including the mass and radius coordinates, density, pressure, velocity, electron and proton temperatures, electron and ion number densities, and element-by-element ion fractions. These profiles are used in post-processing to compute the FRB DM and optical depths as functions of time (See section 4.6.1).

To adequately resolve the early-time evolution of the SNR contribution while maintaining computational efficiency, we adopt a non-uniform output cadence. Motivated by the fact that the most significant and rapid evolution of the SNR contribution to the FRB DM occurs at early times (e.g., Piro & Gaensler 2018) we use a finer time spacing during the first $\sim 300$ yr after explosion, followed by a coarser spacing at later times. Specifically, we start from the simulation initialization time $t_{\rm init} = 3$ yr, snapshots are recorded at $\Delta t = 1$ yr intervals from $t = 3$ to 300 yr, and at $\Delta t = 4$ yr intervals from 300 to 500 yr, resulting in a total of $\sim 350$ output time bins. This sampling captures the relevant DM evolution and its temporal variability while keeping the overall computational cost tractable.

## 4.6 Calculation Methods of FRB observables

### 4.6.1 DM and optical depth calculations

Given the radial profiles of electron density $n_e(r, t)$ produced by the numerical simulations, we compute the DM contributed by the SNR as

$$\mathrm{DM_{SNR}}(t) = \int_{R_{\rm min}(t)}^{R_{\rm max}(t)} n_e(r, t)\, \mathrm{d}\ell, \tag{4.21}$$

where $R_{\rm min}$ and $R_{\rm max}$ denote, respectively, the inner and outer boundaries of the ionized region in the simulation. For the shocked region, we take $R_{\rm min} = R_r(t)$ and $R_{\rm max} = R_b(t)$, i.e., the reverse- and forward-shock radii. For the entire ionized region, we take $R_{\rm min} = R_{\rm un,min}(t)$ (the inner boundary of the unshocked ejecta) and $R_{\rm max} = R_{\rm CSM,max}$ (the fixed outer boundary of the unshocked CSM). In our setup, $R_{\rm CSM,max} = 39.9\,\mathrm{Mpc}$ is set by the initial conditions. Here $\mathrm{d}\ell$ is the line element along the FRB line of sight. For now we assume a magnetar as the FRB's central engine located at the origin[3] of the SNR, such that the DM is independent of the line-of-sight direction and can be written as a radial integral of the $n_e$.

Given the extremely high energy budget of FRBs, viable progenitors are generally expected to possess ultra-strong magnetic fields and highly active crustal or magnetospheric environments, pointing to a young magnetar origin. As a result, FRB radio emission can escape only once the SNR has expanded to become sufficiently optically transparent, while remaining dynamically young. Therefore, by modeling the time evolution of the optical depth of the SNR, one can place constraints on the FRB escape timescale ($t_{\rm esc}$).

[3]In this work, we neglect the effect of magnetar natal kick, as our analysis is restricted to very early SNR evolution ($\lesssim 500\,\mathrm{yr}$), during which the kick-induced displacement remains much smaller than the radial extent of ionized shock region

We consider two sources of opacity in the SNR: electron scattering and free–free absorption, such that the total optical depth is given by

$$\tau_{\rm tot} = \tau_{\rm es} + \tau_{\rm ff}. \tag{4.22}$$

At early times, when the SNR is still young and the ionized region remains dense, electron scattering can contribute significantly to the optical depth. In this regime, FRB propagation is primarily affected by scattering-induced temporal broadening rather than true absorption. We approximate the electron scattering optical depth using Thomson scattering,

$$\tau_{\rm es} = \int_{R_{\rm min}(t)}^{R_{\rm max}(t)} n_e \sigma_T \, dl, \tag{4.23}$$

where $\sigma_T$ is the Thomson cross section.

As the SNR expands and the electron density decreases, the contribution from electron scattering rapidly diminishes. At sufficiently late times, free–free absorption becomes the dominant opacity source and ultimately determines whether FRB radio emission can escape. We therefore compute the free–free optical depth at an observing frequency $\nu$ as

$$\tau_{\rm ff}(\nu, t) = \int_{R_{\rm min}(t)}^{R_{\rm max}(t)} \alpha_{\rm ff}(\nu, T_e, n_e, n_{ij}, Z_{ij}) \, {\rm d}\ell, \tag{4.24}$$

where $\alpha_{\rm ff}$ is the thermal free–free absorption coefficient. For radio frequencies satisfying $h\nu \ll k_{\rm B}T_e$, we adopt the standard expression from Rybicki & Lightman (1979),

$$\alpha_{\rm ff}(\nu) = 1.9 \times 10^{-2} \, T_e^{-3/2} \, \nu^{-2} \, n_e \sum_{i,j} Z_{ij}^2 \, n_{ij} \, g_{\rm ff}(\nu, T_e, Z_{ij}), \tag{4.25}$$

where $Z_{ij}$ denotes the ionic charge of element $i$ in its $j$-th ionization state, and $n_{ij}$ is the corresponding number density. where $Z_{ij}$ is the ion charge, $n_i$ is the ion number density, and $g_{\rm ff}$ is the Gaunt factor, whose detailed behavior is described in Appendix C.1. Here, in practice, we evaluate the Gaunt factor using the mean ionic charge $\langle Z \rangle = \sum_{ij} Z_{ij} n_{ij} / \sum_{ij} n_{ij}$ at each radius without loss of accuracy. Equation (4.25) can then be rewritten as

$$\alpha_{\rm ff}(\nu) = 1.9 \times 10^{-2} \, T_e^{-3/2} \, \nu^{-2} \, n_e \, g_{\rm ff}(\nu, T_e, \langle Z \rangle) \sum_{i,j} Z_{ij}^2 \, n_{ij}, \tag{4.26}$$

The FRB emission is able to escape once $\tau_{\rm tot} \lesssim 1$ at the observing frequency. Therefore, the combination of ${\rm DM}_{\rm SNR}(t)$ and $\tau_{\rm tot}(t)$ provides both the instantaneous DM contribution and an estimate of the "$t_{\rm esc}$" for GHz emission (throughout this work, we adopt $\nu = 1\,{\rm GHz}$ as a representative frequency for the GHz observing band for FRBs).

Finally, we compute the first and second time derivatives of the SNR DM numerically:

$$\frac{\mathrm{dDM}}{\mathrm{d}t}(t_i) \approx \frac{\mathrm{DM}_{i+1} - \mathrm{DM}_{i-1}}{t_{i+1} - t_{i-1}},$$
$$\frac{\mathrm{d}^2\mathrm{DM}}{\mathrm{d}t^2}(t_i) \approx \frac{\mathrm{DM}_{i+1} - 2\mathrm{DM}_i + \mathrm{DM}_{i-1}}{(t_{i+1} - t_i)^2}, \tag{4.27}$$

which will be compared with observationally inferred dDM/d$t$ and constraints on higher-order variations.

### 4.6.2 RM Calculation and Observational Matching Method

The rotation measure (RM) is defined as

$$\mathrm{RM} = \frac{e^3}{2\pi m_e^2 c^4} \int n_e \, B_\parallel \, dl, \tag{4.28}$$

which is commonly approximated as

$$\mathrm{RM} \simeq 0.81 \int n_e \, B_\parallel \, dl \quad (\mathrm{rad\,m^{-2}}), \tag{4.29}$$

when $n_e$ is in cm$^{-3}$, $B_\parallel$ is in $\mu$G, and $l$ is in pc.

In this work, we compute the RM evolution using only shocked-region profiles. The magnetic field is parameterized as a fixed fraction of the ram pressure,

$$\frac{B^2}{8\pi} = \epsilon_B \frac{\rho v^2}{2}, \tag{4.30}$$

so that $B = \sqrt{4\pi \, \epsilon_B \, \rho v^2}$, evaluated cell by cell in the shocked layer. We also track the shocked-region mean magnetic field as an electron-density-weighted, cell-by-cell average,

$$\langle B_{\mathrm{sh}} \rangle = \frac{\sum_i n_{e,i} B_i}{\sum_i n_{e,i}}, \tag{4.31}$$

where $i$ labels the $i$th cell in the shocked region. The RM is then obtained by integrating $n_e B_\parallel$ along the line of sight over $r_r < r < r_b$. We exclude the unshocked ejecta from the RM integral because, in this prescription, it is approximately in free expansion and therefore does not carry a ram-pressure-supported magnetic field. For ionization, we directly use the NEI-simulated ionization state in the shocked cells. We explore $\epsilon_B = 0.01$–$0.3$, with a fiducial value of $\epsilon_B = 0.1$, consistent with Piro & Gaensler (2018). The adopted range $\epsilon_B = 0.01$–$0.3$ spans values commonly inferred in young supernova remnants and shock-powered synchrotron sources, where magnetic amplification by turbulence and cosmic-ray streaming can bring the post-shock magnetic energy density to a few percent up to nearly equipartition with the ram pressure.

To identify the simulated RM interval that matches the observed evolution of FRB 20121102, we adopt a slope-matching approach based on the logarithmic RM decay index. We define the

logarithmic slope

$$\beta(t) \equiv \frac{d \log \mathrm{RM}(t)}{d \log t}. \tag{4.32}$$

For the simulated RM evolution, the instantaneous slope is computed numerically in log–log space using finite differences,

$$\beta_{\mathrm{sim}}(t_i) \approx \frac{\log \mathrm{RM}_{i+1} - \log \mathrm{RM}_{i-1}}{\log t_{i+1} - \log t_{i-1}}, \tag{4.33}$$

and interpolation is applied to obtain a smooth representation of $\beta_{\mathrm{sim}}(t)$. Using the RM measurements reported in Hilmarsson et al. (2021) and Wang et al. (2025a), we fit the post-peak observational data in log–log space and derive an observed decay slope of

$$\beta_{\mathrm{obs}} \approx -0.36. \tag{4.34}$$

We then search for the epoch $t_0$ in each single-star (SS) model that satisfies

$$\beta_{\mathrm{sim}}(t_0) = \beta_{\mathrm{obs}}, \tag{4.35}$$

which yields

$$t_0 \approx 9.54 \text{ yr} \tag{4.36}$$

for the $11\,M_\odot$ model and

$$t_0 \approx 9.99 \text{ yr} \tag{4.37}$$

for the $30\,M_\odot$ model. To compare the temporal evolution directly, we shift the observational time axis such that the median observed RM is placed at $t_0$. The mapped time is

$$t_{\mathrm{mapped}} = t_0 + (t_{\mathrm{obs}} - t_{\mathrm{ref}}), \tag{4.38}$$

where $t_{\mathrm{ref}}$ corresponds to the epoch of the median observed RM, while the observed RM amplitudes are kept unchanged.

## 4.7 Results

Motivated by the analytical considerations in Section 4.4—which suggest that $\mathrm{DM}_{\mathrm{SNR}}$ may be dominated either by the shocked ejecta/shell or by the entire ionized region (including unshocked material)—we carry out the analysis for both cases.

### 4.7.1 Shocked region

#### 4.7.1.1 DM evolution in the shocked region

Fig. 4.2 shows the temporal evolution of the characteristic radii in our 1D SNR simulations, including the reverse shock ($R_r$), contact discontinuity ($R_c$), and forward shock ($R_b$), for both the SS and BS progenitors. These radii define the extent of the shocked layer and its effective thickness, $\Delta R_{\rm sh} \equiv R_b - R_r$ (see Fig. 4.3c), and therefore set the integration boundaries used in our DM and optical-depth calculations. The radii increase rapidly at early times and then evolve more smoothly as the remnant expands.

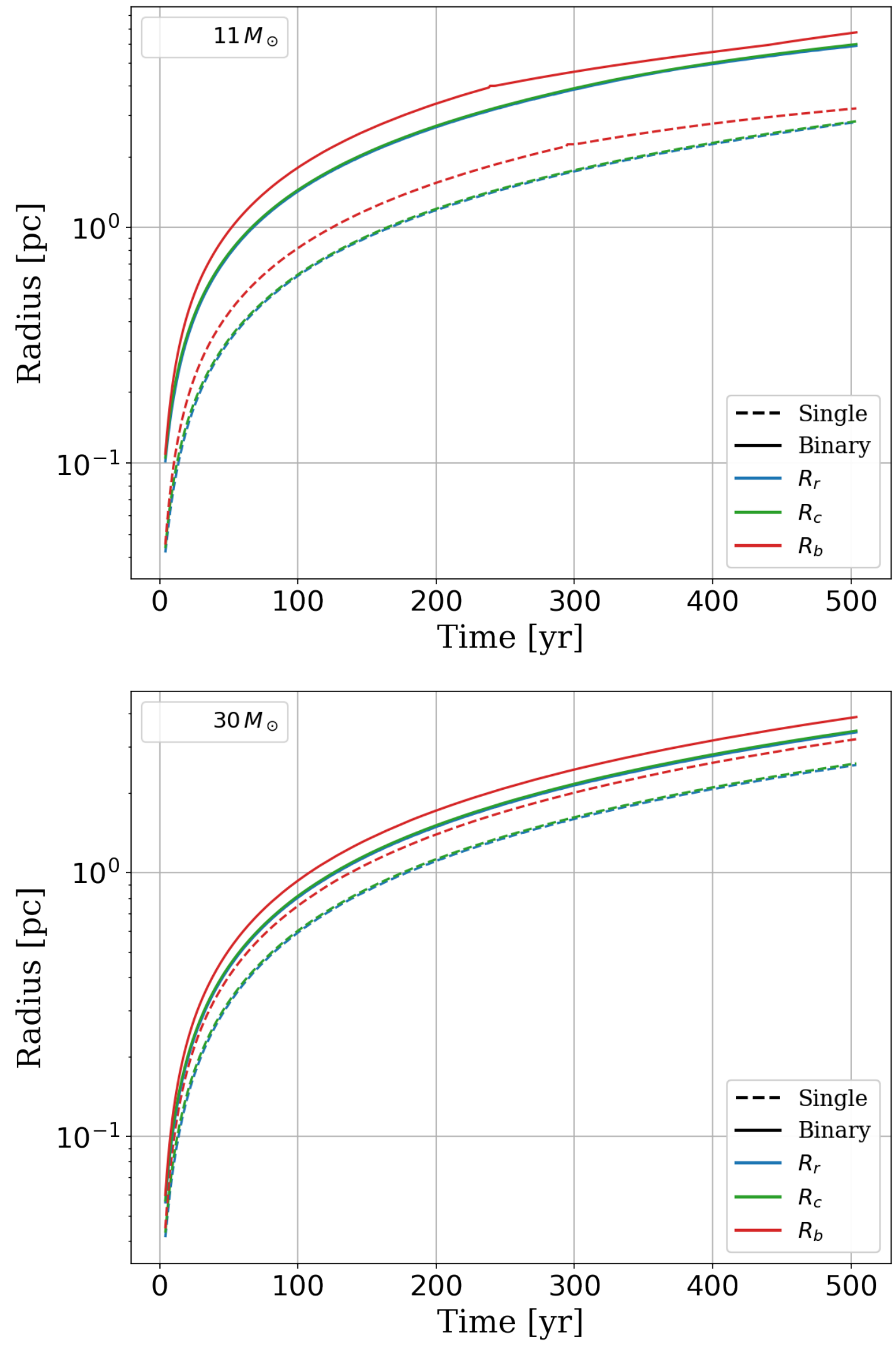

FIGURE 4.2: Radius evolution for the $11\,M_\odot$ (top) and $30\,M_\odot$ (bottom) models. Dashed: single-star; solid: binary-stripped. Blue, green, and red curves indicate the $R_r$, $R_c$, and $R_b$, respectively.

Fig. 4.3 compares the DM evolutions of each case. Aside from the initial peak due to the NEI during the earliest expansion stage, both models exhibit a monotonic decline in $\mathrm{DM_{SNR}}(t)$ as the ejecta expand and their density decreases. For both $M_{\rm ZAMS} = 11$ and $30\,M_\odot$, the SS channel produces a larger DM than the BS channel, consistent with its larger ejecta mass (Table 4.1) and hence a denser ionized ejecta layer.

Within the SS channel, the $11\,M_\odot$ model produces a larger DM than the $30\,M_\odot$ model because $\mathrm{DM_{SNR}} \propto \int n_e\,d\ell$ depends on both the electron density ($n_e \propto \chi_e\,\rho$) and the effective path length through the shocked/ionized layer (characterized here by $\Delta R_{\rm sh}$). However, in the thin-shell approximation, these two factors are not independent. Using $\langle\rho\rangle_{\rm sh} \sim M_{\rm sh}/V_{\rm sh}$ with $V_{\rm sh} \simeq (4\pi/3)(R_b^3 - R_r^3) \approx 4\pi R^2 \Delta R_{\rm sh}$, we obtain $\langle n_e\rangle \propto \chi_e\, M_{\rm sh}/(4\pi R^2 \Delta R_{\rm sh})$ and, therefore, $\mathrm{DM_{SNR}} \sim \langle n_e\rangle\, \Delta R_{\rm sh} \propto \chi_e\, M_{\rm sh}/(4\pi R^2)$, i.e., the explicit dependence on $\Delta R_{\rm sh}$ largely cancels. Consequently, the DM difference between the $11\,M_\odot$ and $30\,M_\odot$ SS cases is primarily controlled by (i) the ionization degree $\chi_e$ and (ii) the ratio $M_{\rm sh}/R^2$.

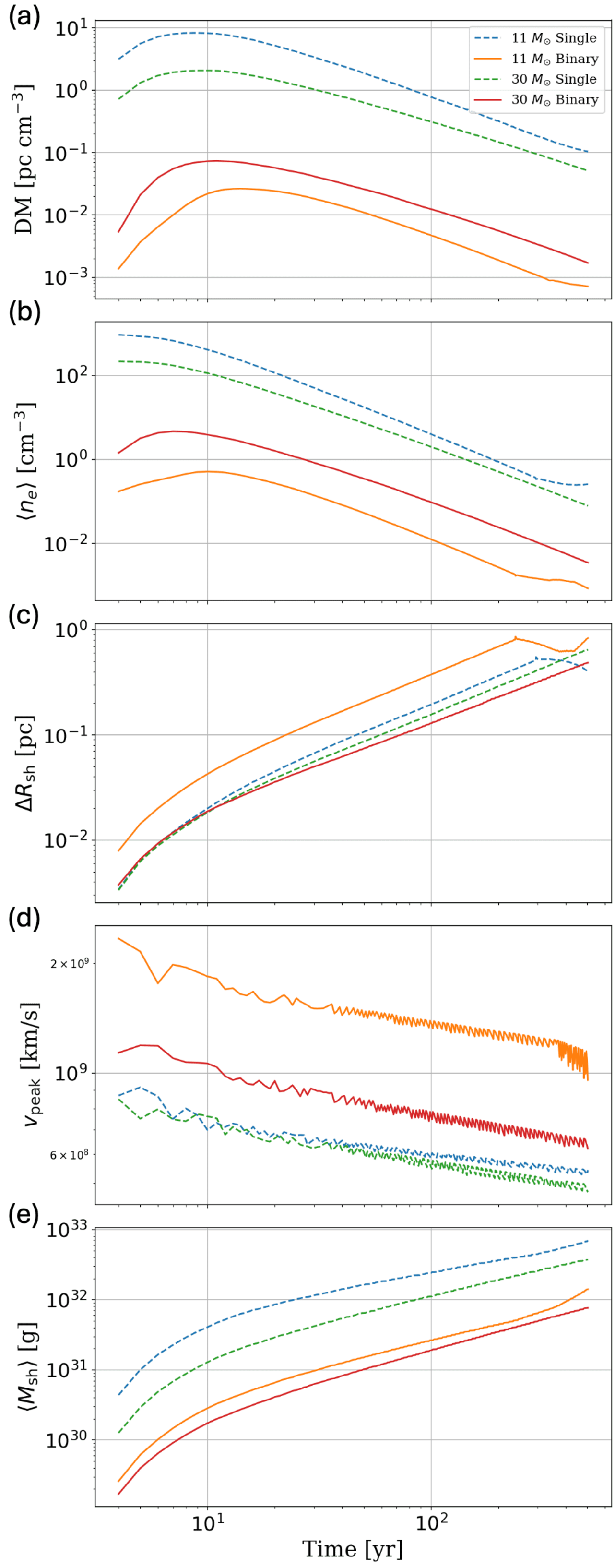

FIGURE 4.3: Comparison of the time evolution for 11 $M_\odot$ and 30 $M_\odot$ progenitor models in the SS and BS channels. (a) DM. (b) Volume-averaged electron density $\langle n_e \rangle$. (c) Effective thickness of the shocked ionized region, $\Delta R_{\rm sh}$. (d) Peak (maximum) SNR expansion velocity. (e) Cumulative swept-up (shocked) shell mass. Solid curves correspond to the fiducial unshocked-ionization case $\chi_{e,\rm unsh} = 1$ (HH in Table 4.3).

The shocked volume is dominated by the geometric factor $R^2$ rather than by the shell thickness. At the same age the $11\,M_\odot$ SS remnant has a noticeably larger characteristic radius than the $30\,M_\odot$ case (see Fig. 4.2); for a fixed shocked mass, this larger $R$ would reduce $\langle\rho\rangle_{\rm sh}$ and hence DM. In our models, however, the $11\,M_\odot$ remnant has a larger characteristic expansion speed (Fig. 4.3d), so at the same age it sweeps up more material and attains a higher shocked mass $M_{\rm sh}$ (Fig. 4.3e) and hence a higher $\rho_{\rm sh}$, despite the larger shocked volume.

In addition, the systematically larger $M_{\rm sh}$ in the SS channel than in the BS channel can be physically understood from binary stripping. In BS progenitors, Roche-lobe overflow removes much of the pre-SN H-rich envelope and suppresses dense pre-SN winds, yielding a more rarefied surrounding medium. The weaker external density profile leads to weaker SNR-shell deceleration and therefore a slower reverse-shock penetration into the ejecta, while the lower ambient density also reduces the shocked CSM component. Equivalently, writing $M_{\rm sh} = M_{\rm sh,ej} + M_{\rm sh,CSM}$, both terms are reduced in BS models, which naturally drives $M_{\rm sh,BS} < M_{\rm sh,SS}$ even when the total ejecta mass is not the sole controlling factor.

We also find that the $11\,M_\odot$ model reaches a higher ionization degree (e.g., $\chi_e \sim 0.9$) than the $30\,M_\odot$ model ($\chi_e \sim 0.7$; see Fig. 4.7). This is consistent with NEI physics, since the approach to ionization equilibrium is governed by the ionization timescale parameter $n_e t$: the higher post-shock electron density in the $11\,M_\odot$ shell leads to a larger $n_e t$ at a given age, and thus a higher $\chi_e$. Aside from the early-time NEI-driven rise, the long-term DM evolution resembles a power-law decay over hundreds of years, consistent with expectations from expanding SNR dynamics.

#### 4.7.1.2 Optical depth evolution in the shocked region

Within the binary-stripped channel, on practical FRB timescales (tens to hundreds of years) the DM contribution from the SNR is negligible and exhibits essentially no measurable secular evolution. The $30\,M_\odot$ stripped model nevertheless yields a higher DM than the $11\,M_\odot$ stripped model (Fig. 4.3a). This difference is mainly driven by the much smaller ejecta mass in the $11\,M_\odot$ stripped case, which results in a particularly low post-shock electron density. The larger $\Delta R_{\rm sh}$ in that model further increases the shocked volume; under the thin-shell scaling, $V_{\rm sh} \propto R^2 \Delta R_{\rm sh}$, this increase tends to dilute $n_e$.

Fig. 4.4 shows the time evolution of the optical depths and several mass-weighted averages in the shocked region. The total optical depth remains $\tau_{\rm tot} \ll 1$ at GHz frequencies throughout the simulation, implying essentially no free–free "escape time" when considering only the shocked region, i.e., GHz FRB emission could escape almost immediately after the explosion. While the shocked region makes only a limited contribution to the total DM in our current setup, it provides a useful diagnostic of the plasma state: the mass-weighted electron temperature $\langle T_e\rangle$ declines with time in all cases, reflecting ongoing radiative cooling. We also find that $\langle Z \rangle$ is slightly above unity and differs modestly among the progenitor channels, while the Gaunt factor (evaluated at $\nu = 10^9\,$Hz) stays in the range $\langle g_{\rm ff} \rangle \sim 7$–$10$. We defer a more detailed discussion of the composition/ionization origin of these trends to Section 4.7.5.

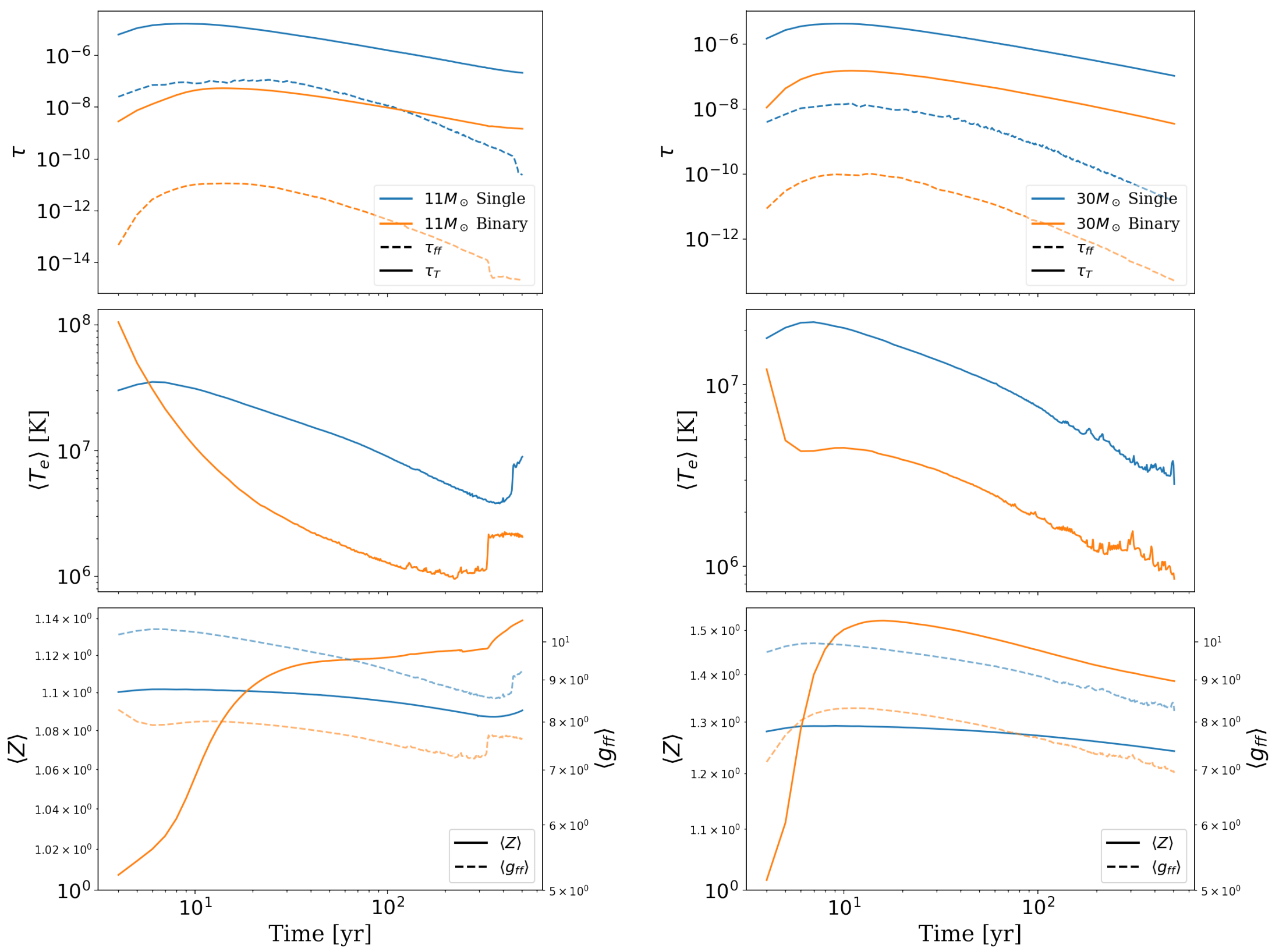

FIGURE 4.4: Time evolution of optical depths and mass-weighted averages in the shocked region for the $11\,M_\odot$ (left) and $30\,M_\odot$ (right) progenitor models. Each panel shows $\tau_{\rm es}$ and $\tau_{\rm ff}$, together with $\langle T_e \rangle$, $\langle Z \rangle$, and $\langle g_{\rm ff} \rangle$, comparing the SS and BS channels. Throughout this work, the Gaunt factor is evaluated at $\nu = 10^9\,$Hz, representative of the GHz band relevant for FRBs.

These results highlight the qualitative difference between the SS and BS channels: for the same $M_{\rm ZAMS}$, a stripped progenitor with a small ejecta mass produces both a much lower DM and a substantially more transparent shocked environment. Even at early times ($t \sim 10$ yr) when the shocked-region contribution reaches its maximum, the peak value remains modest: for the $11\,M_\odot$ SS model we find $\mathrm{DM_{peak}} \simeq 8.3\,\mathrm{pc\,cm^{-3}}$, while the binary-stripped models exhibit significantly smaller peaks ($< 0.1\,\mathrm{pc\,cm^{-3}}$) that occur during the earliest evolution.

### 4.7.2 Entire region

#### 4.7.2.1 DM evolution in the entire ionized region

As demonstrated above, the shocked region alone cannot account for the large inferred $\mathrm{DM_{source}}$ and its observed secular evolution in FRB 20190520B or FRB 20121102. This motivates us to extend the calculation to the entire ionized region of the SNR, incorporating contributions from the unshocked ejecta and the surrounding unshocked CSM/ISM.

Because our SNR simulations are primarily designed to resolve the detailed thermodynamic and ionization evolution in the shocked region, the unshocked ejecta and the surrounding unshocked CSM/ISM are not evolved with a physically self-consistent ionization history. In these regions, the simulation provides the density structure and elemental abundances, but does not self-consistently track the temperature evolution and ionization balance. We therefore treat the ionization fraction in the unshocked regions as a parameter and explore physically motivated bounds.

For the unshocked ejecta, Chevalier & Fransson (2017) estimated an ionization fraction of $\chi_{e,\rm unej} \sim 0.03$ for SN 1993J, in which the ionization is primarily driven by photoionization from radiation produced at the reverse shock.

More recently, Laming & Temim (2020) modelled the pre–reverse-shock (i.e., unshocked) inner ejecta of Cas A under a photoionization–recombination (PR) equilibrium, with the photoionizing radiation field provided by UV-to-X-ray emission from both the forward- and reverse-shocked plasma. In their PR models, the inferred charge-state distributions that produce the observed IR fine-structure lines imply that the unshocked ejecta can maintain a non-negligible electron fraction, potentially at the level of $\chi_{e,\rm unej} \gtrsim 0.1$ depending on density/temperature conditions and the strength of the ionizing radiation field.

Furthermore, the ionization state of the unshocked ejecta in young core-collapse supernova remnants can depend sensitively on several factors, including the progenitor mass, mass-loss rate, wind velocity, and the resulting shock luminosity. In addition, the presence of a central compact object (CCO), such as a neutron star or magnetar, may provide additional high-energy radiation through spin-down-powered emission or a wind nebula (MWN/PWN), potentially enhancing the ionization level beyond that inferred from reverse-shock irradiation alone.

Given these uncertainties, and in the absence of a fully self-consistent radiative transfer calculation in our hydrodynamic models, we treat the ionization fraction of the unshocked ejecta parametrically and adopt a conservative yet physically motivated range:

$$0.01 \leq \chi_{e,\rm unej} \leq 1, \tag{4.39}$$

where $\chi_{e,\rm unej} \sim 0.01$ is taken as the weak-ionization lower limit and $\chi_{e,\rm unej} \sim 1$ represents a fully ionized case, motivated by PR-equilibrium scenarios under strong photoionizing conditions (e.g., Laming & Temim 2020) and/or potential additional ionizing contributions from a CCO.

For the unshocked CSM/ISM, we adopt

$$10^{-4} \leq \chi_{e,\rm ISM} \leq 1, \tag{4.40}$$

spanning environments from cold neutral media to fully ionized hot plasma. which spans plausible astrophysical environments. Specifically, $\chi_{e,\rm ISM} \sim 10^{-4}$ corresponds to a cold neutral medium (CNM), $\chi_{e,\rm ISM} \sim 10^{-2}$ to a warm neutral medium (WNM), $\chi_{e,\rm ISM} \sim 0.1$–$0.5$ to a partially ionized warm ionized medium (WIM), and $\chi_{e,\mathrm{ISM}} \sim 1$ to a fully ionized hot ionized medium (HIM). These bounds allow us to bracket the resulting DM contribution from the unshocked components and to quantify the systematic uncertainty associated with the poorly constrained ionization state in these regions. Table 4.3 summarizes a subset of the ionization configurations explored in this work.

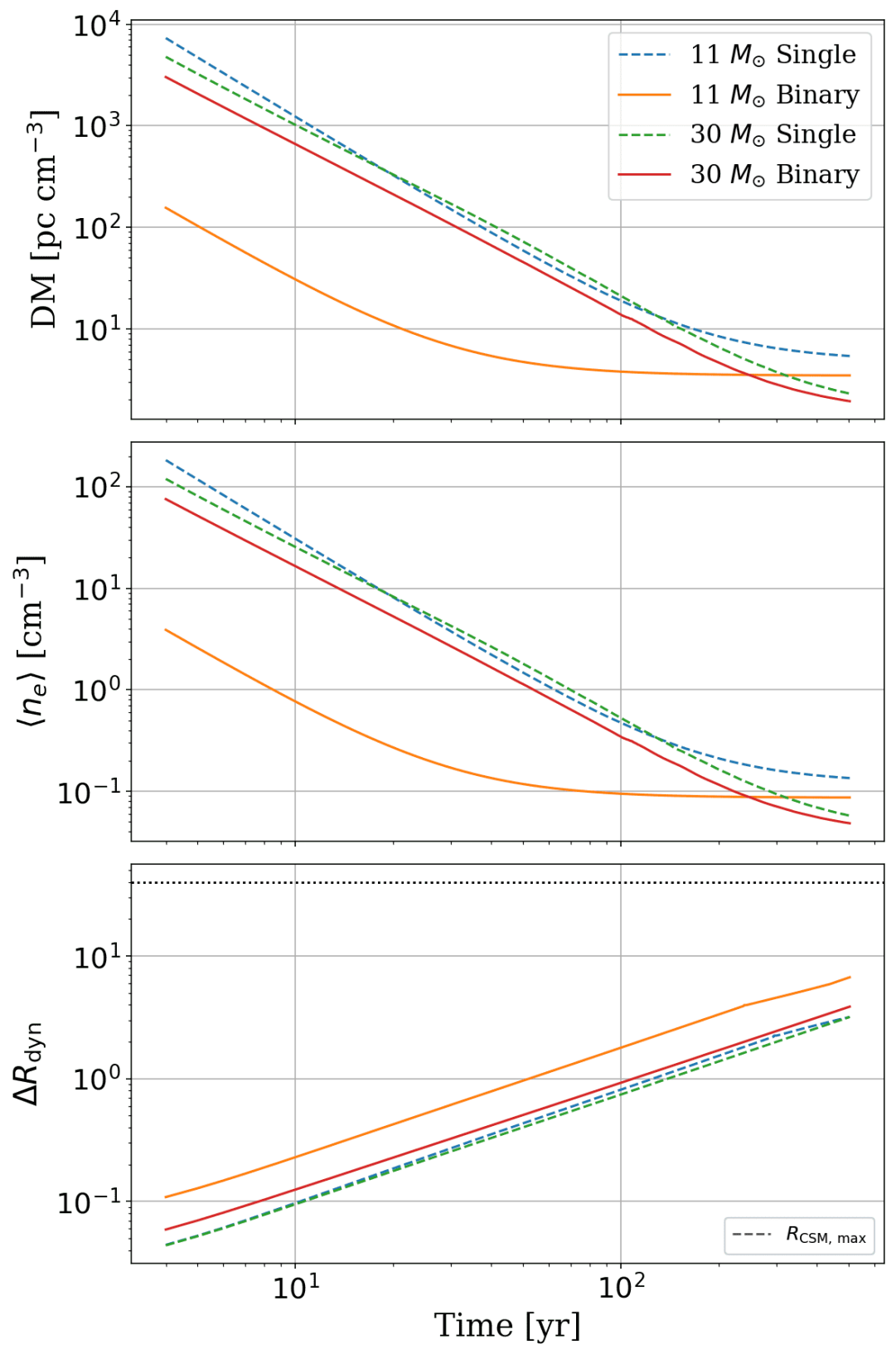

FIGURE 4.5: Time evolution of the characteristic quantities for the full-region DM calculation, analogous to Fig. 4.3. The results correspond to the fiducial HH case with $\chi_{e,\mathrm{unej}} = \chi_{e,\mathrm{ISM}} = 0.1$. Here we adopt a dynamical path length $\Delta R_{\mathrm{dyn}} = \Delta R_{\mathrm{ej}} + \Delta R_{\mathrm{sh,CSM}}$. In the third panel, $R_{\mathrm{CSM,max}}$ is a constant maximum integration radius for the ISM/CSM contribution.

Figure 4.5 shows the time evolution of the characteristic quantities entering our full-region DM estimate. In this setup, the ionization fraction in the unshocked components is treated as a fixed parameter, and we use $\mu_a$ for the unshocked regions. The fiducial case adopted in this work is HH, with $\chi_{e,\mathrm{unej}} = \chi_{e,\mathrm{ISM}} = 0.1$ (hereafter $\chi_{e,\mathrm{unsh}} = 0.1$, unless otherwise stated). We also adopt a low-ionization approximation for the unshocked ejecta in which each atom contributes at most one free electron (effectively singly ionized), so the composition dependence enters primarily through $\mu_a$ (i.e., $\mu_e \simeq \mu_a$).

During the ejecta-dominated phase, the SS models yield larger DMs than the BS models, while the relative ordering between the $11\,M_\odot$ and $30\,M_\odot$ SS cases changes with time. In the later CSM-dominated phase, the ordering is set by the CSM contribution and does not necessarily follow the same SS–BS ranking. At early times ($t \lesssim 20$ yr), the $11\,M_\odot$ SS model produces a larger DM than the $30\,M_\odot$ SS model even though the latter has a higher unshocked-ejecta mass density (see the middle panels of Fig. 4.7a,b). This arises because the electron density scales as $n_e = \chi_e \rho/(\mu_e m_p)$: the 30 $M_\odot$ ejecta has a larger $\mu_e$ (due to its thinner H envelope and higher He/metal fraction; see the bottom panels of Fig. 4.7a,b), which reduces $n_e$ and offsets its density advantage, leading to $n_{e,30} < n_{e,11}$ and hence $\mathrm{DM}_{30} < \mathrm{DM}_{11}$ at early times. As the remnant expands, dynamical dilution becomes dominant; the lower ejecta mass in the 11 $M_\odot$ model implies a higher characteristic expansion velocity and a faster density drop, so the DM curves cross around $t \simeq 20$ yr and $\mathrm{DM}_{11} < \mathrm{DM}_{30}$ thereafter. Throughout the evolution, the DM decline remains approximately power-law, with the unshocked ejecta dominating at early times (to $\lesssim 100$ yr) before the CSM contribution grows in importance; for the 11 $M_\odot$ BS case, this transition occurs much earlier, at $\sim 20$ yr.

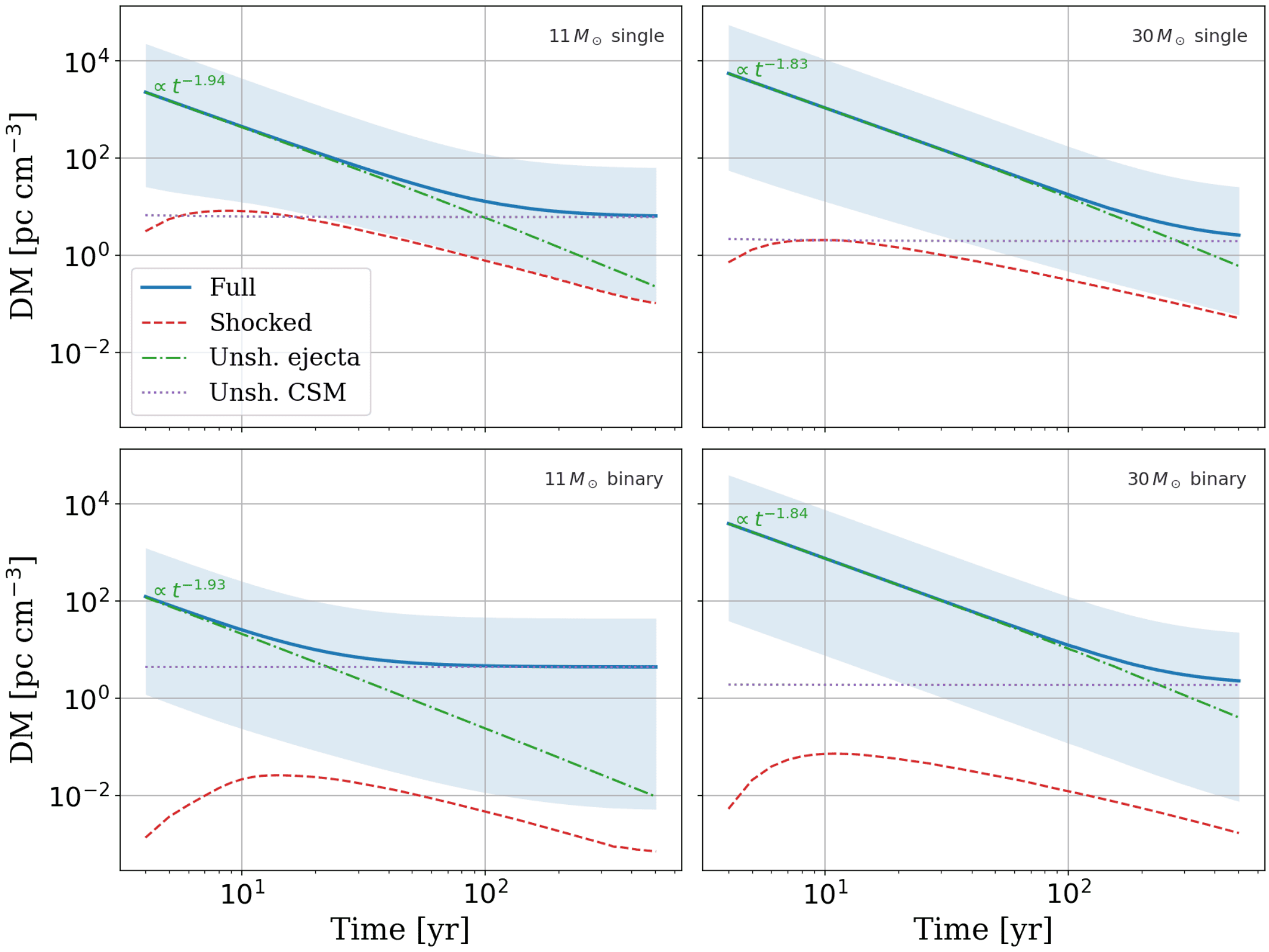

FIGURE 4.6: Decomposition of the total SNR DM into individual components in the full-ionized-region calculation. The time-scaling shown for the unshocked-ejecta component is obtained from our fit to the unshocked-ejecta evolution.

All models exhibit an initial power-law-like decline in DM and a progressively smoother evolution at late times. These two behaviors correspond, respectively, to the transition from the ED phase to the later stage in which the swept-up CSM increasingly governs the remnant dynamics.

To identify which physical region dominates $\mathrm{DM}_{\mathrm{SNR}}$, and when the dominance transitions in each model, we decompose $\mathrm{DM}_{\mathrm{SNR}}$ into contributions from the unshocked ejecta, the unshocked CSM, and the shocked region (Fig. 4.6). At early times, the unshocked ejecta provide the dominant contribution and exhibit an approximately power-law decline, $\mathrm{DM} \propto t^{-\alpha}$, with best-fit $-\alpha$ slopes of $-1.93$ and $-1.94$ for the 11 $M_\odot$ SS and BS models, respectively, and $-1.83$ and $-1.84$ for the $30\,M_\odot$ SS and BS models. The fact that these slopes are close to $-2$ is physically expected: in this epoch the dominant unshocked ejecta are still in free expansion, so if $\chi_e$ is approximately constant and $\mu_e$ evolves only weakly, then $\langle n_e \rangle \propto \rho \propto R^{-3} \propto t^{-3}$, while the characteristic integration length scales as $\Delta R_{\rm dyn} \propto t$. Therefore DM $\sim \langle n_e \rangle \Delta R_{\rm dyn} \propto t^{-2}$. At later times, the unshocked CSM becomes increasingly important and eventually dominates; the characteristic transition times differ across models (e.g., $\sim 30$ yr and $\sim 10$ yr for the $11\,M_\odot$ SS and BS models, respectively, and $\sim 100$ yr for the $30\,M_\odot$ model). By contrast, the shocked region remains a subdominant contribution throughout the simulated time span.

Figure 4.7 provides an independent cross-check of the above interpretation by showing the radial cumulative DM (top panels) together with the corresponding mass-density (middle panels) and ionization-fraction (bottom panels) profiles at several representative epochs. To facilitate comparison with analytic expectations, we interpret the radial structure in a set of characteristic regimes, starting from the density profiles.

(1) *Inner-core region.* As an illustrative example, we focus on $t = 3\,$yr (the initial condition of our SNR calculations). At this epoch, the ejecta are extremely dense and retain a core–envelope structure close to the adopted ejecta profile (Eq. C.7), i.e., an approximately flat (nearly uniform) inner core surrounded by a steeply declining power-law envelope. As the remnant evolves, the inner-core region gradually develops a modest radial gradient (i.e., a weak decrease with radius). In the inner-core region, the cumulative DM therefore rises rapidly at small radii and is dominated by the unshocked ejecta.

(2) *Rapidly declining unshocked-ejecta envelope.* Outside the nearly uniform core, the unshocked ejecta enter a steep power-law envelope (with the initial decline set by the ejecta index $n$). The density therefore decreases rapidly with radius, and the cumulative DM correspondingly flattens once the line of sight exits the core.

(3) *Shocked region.* After the unshocked-ejecta envelope, the profiles enter the shocked layer. Although the shocked plasma is highly ionized (see $\chi_e$), its DM contribution remains small because the shocked region has a limited path length and a comparatively low mass density, consistent with the decomposition in Fig. 4.6.

(4) *Wind-dominated region (pre-SN wind$\times$CSM).* Outside the shocked layer, the density profile transitions to the circumstellar environment shaped by pre-SN mass loss, exhibiting an approximately wind-like scaling $\rho \propto r^{-s}$ over a broad radial range (with $s = 2$ in most analytic wind models).

(5) *Pre-SN wind–CSM interaction morphologies: SS vs. BS.* "Beyond the wind-like region, the SS and BS channels develop distinct pre-SN wind–CSM interaction morphologies, including a hot cavity excavated by fast winds and density bumps produced by fast–slow wind interactions.

- *SS channel: hot bubble + "double" wall + WR-driven bump.* In the SS models (Fig. 4.7a,b), the hot bubble is primarily generated by the interaction between the fast OB-type wind and the ambient ISM: the OB wind sweeps up the ambient medium into an outer dense shell ("ISM wall"), while the reverse shock propagating back into the wind heats the shocked wind material and inflates a high-pressure hot bubble. At later stages, the dense and slower RSG wind is confined by the hot-bubble thermal pressure, producing an additional inner density enhancement (“bubble wall”). A WR wind, when present (in the 30 $M_\odot$ case), can interact with the RSG wind and imprint a modest bump in the cavity profile.

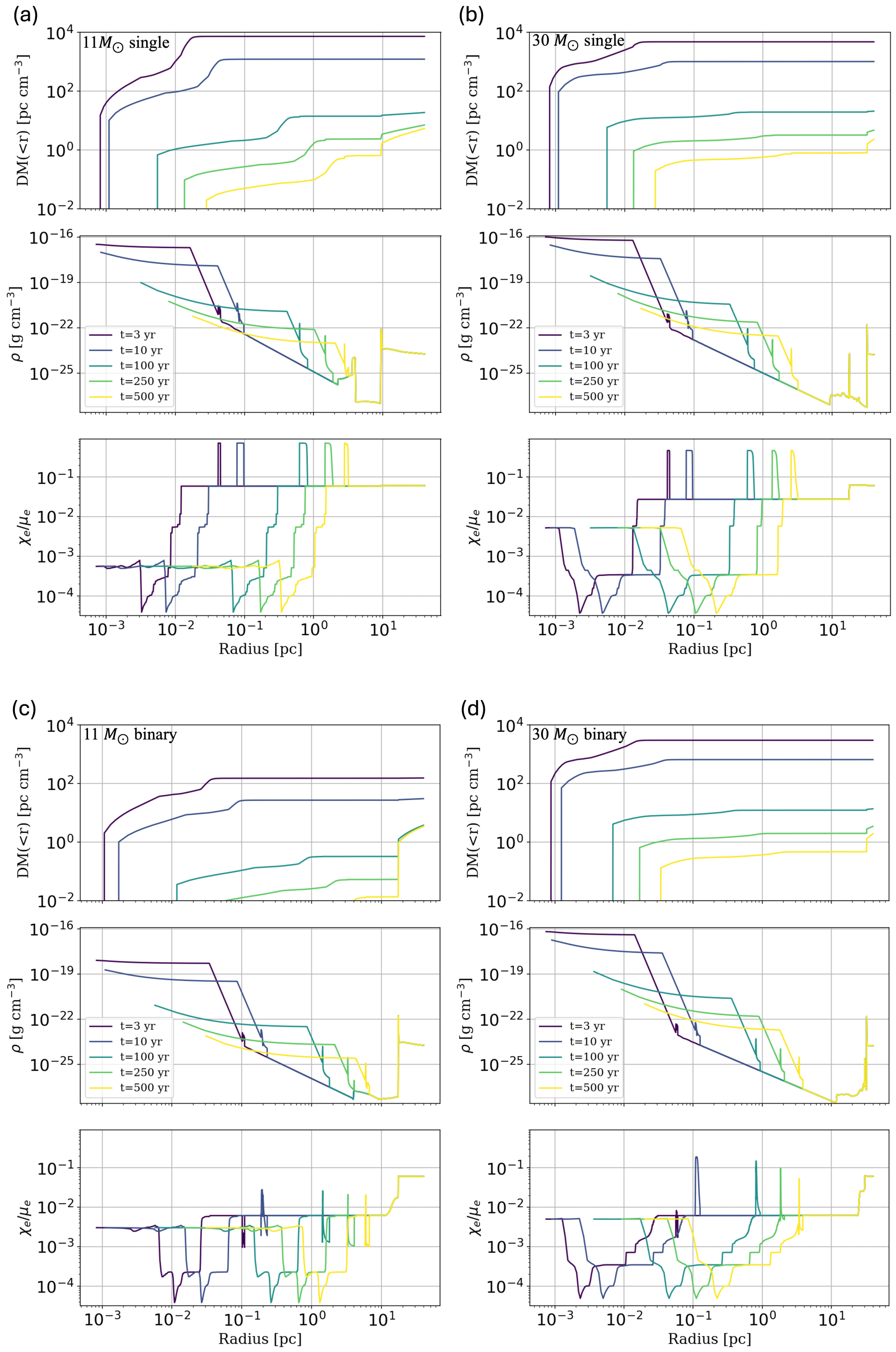

FIGURE 4.7: Radial cumulative profiles for the full-region calculation, including the cumulative DM as well as the mass density and ionization fraction as functions of radius, evaluated at several characteristic epochs. The top and bottom panels correspond to the single-star (SS) and binary-stripped (BS) models, respectively.

- *BS channel: ISM wall + WR-like bump(with an absent/weak bubble wall).* In the BS models (Fig. 4.7c,d), the H envelope is stripped quickly, so the progenitor largely skips the RSG-wind stage (especially in the $11\,M_\odot$ case) and the evolution is dominated by a WR-like He-star wind, broadly similar to the WR-wind phase in the $30\,M_\odot$ SS model. In the $30\,M_\odot$ BS case in particular, a brief yet physical RSG-like slow-wind phase can still occur and form an inner "bubble wall," but the subsequent WR-like fast wind rapidly disrupts this structure and drives it outward, leaving it nearly coincident with the "ISM wall" and thus hard to distinguish in the radial profiles. The remaining prominent feature is typically a WR-like bump.

(6) *Outer ambient ISM: late-time growth of the cumulative DM.* At sufficiently large radii—beyond the wind-bubble structure—the ambient medium approaches an approximately constant density. At early times ($t \sim 3$–$10\,\mathrm{yr}$) this outer material contributes little to the cumulative DM, whereas at late times ($t \gtrsim 100\,\mathrm{yr}$) the declining ejecta density becomes comparable to that of the ambient medium and the cumulative DM can increase again toward the outer boundary.

#### 4.7.2.2 The derivatives of DM

To characterize the secular evolution of $\mathrm{DM_{SNR}}(t)$ and identify the observational time window in which an FRB can exhibit a measurable DM drift, we compute the first and second time derivatives of the simulated DM using Eq. (4.27). Figure 4.8 shows that both $\mathrm{dDM}/\mathrm{d}t$ and $\mathrm{d}^2\mathrm{DM}/\mathrm{d}t^2$ approximately follow a power-law decline in time over the smooth part of the evolution. At $t \gtrsim 100\,\mathrm{yr}$, small-amplitude fluctuations become visible, and they are especially prominent in the second derivative. This behavior is expected because finite-difference estimates of higher-order derivatives amplify snapshot-to-snapshot irregularities in $\mathrm{DM}(t)$; physically, additional non-smooth features can also be introduced when the shock structure encounters density inhomogeneities (shells/bubbles) in the progenitor-shaped CSM.

Motivated by the observed secular decreases in FRB 20190520B ($\mathrm{dDM}/\mathrm{d}t \simeq -12.4\,\mathrm{pc\,cm^{-3}\,yr^{-1}}$; Niu et al. 2025) and FRB 20121102 ($\mathrm{dDM}/\mathrm{d}t = -3.93\,\mathrm{pc\,cm^{-3}\,yr^{-1}}$; Wang et al. 2025a), we identify the corresponding time window after the SN explosion at which an FRB would need to occur. We denote this FRB occurrence time as $t_{\mathrm{occur}}$, i.e., the elapsed time between the SN explosion and the observed FRB activity. Assuming the DM drift is dominated by the local SNR contribution and using the fiducial parameter values adopted in this section, we find $t_{\mathrm{occur}} \simeq 27\,\mathrm{yr}$ (SS) and $t_{\mathrm{occur}} \simeq 8\,\mathrm{yr}$ (BS) for the $11\,M_\odot$ models, and $t_{\mathrm{occur}} \simeq 27\,\mathrm{yr}$ (SS) and $t_{\mathrm{occur}} \simeq 23\,\mathrm{yr}$ (BS) for the $30\,M_\odot$ models for FRB 20190520B. For FRB 20121102, the same fiducial matching gives systematically later epochs, $t_{\mathrm{occur}} \simeq 40\,\mathrm{yr}$ (11 SS), $11\,\mathrm{yr}$ (11 BS), $42\,\mathrm{yr}$ (30 SS), and $35\,\mathrm{yr}$ (30 BS). The corresponding fiducial $\mathrm{DM_{SNR}}$ values for both sources are summarized in Table 4.2. Notably, the significantly lower DM in the $11\,M_\odot$ BS case compared with the other progenitors is primarily driven by its much smaller ejecta mass (see Table 4.1), which directly lowers the unshocked-ejecta electron column. These values correspond to the very early SNR phase, close to the transition from optically thick to optically thin conditions for GHz radio emission (see

Section 4.7.2.3 and Table 4.3). The short duration of this "ambiguous" phase may naturally contribute to the rarity of FRBs showing a clear SNR-like secular DM signature.

The simulated second-derivative DM curvature evaluated at $t_{\rm occur}$ is summarized in Table 4.3 for FRB 20190520B and in Appendix Table C.1 for FRB 20121102. At these early epochs, the curvature for FRB 20190520B is typically $\mathrm{d}^2\mathrm{DM}/\mathrm{d}t^2 \sim 1$–$4\,\mathrm{pc\,cm^{-3}\,yr^{-2}}$ across the four models, while for FRB 20121102 it is generally smaller, $\sim 0.1$–$1.4\,\mathrm{pc\,cm^{-3}\,yr^{-2}}$. Following the order-of-magnitude estimate used by Niu et al. (2025), we define $\left|\mathrm{d}^2\mathrm{DM}/\mathrm{d}t^2\right|_{\rm est} \equiv \alpha\,|\mathrm{dDM}/\mathrm{d}t|/t_{\rm SNR}$ and evaluate it using $t_{\rm SNR} \simeq t_{\rm occur}$. For both FRB 20190520B and FRB 20121102, the estimator remains consistent with the simulated trends at the factor-of-few level, with agreement within $\sim 1.1$–$1.6$.

#### 4.7.2.3 Optical depth evolution in the full region

To assess whether FRB 20190520B and FRB 20121102 can plausibly escape from an SNR environment during the inferred occurrence time $t_{\rm occur}$, we compute the time evolution of the optical depth in the full-region calculation and indicate the epoch at which the system becomes optically thin. Figure 4.9 shows that at early times the total optical depth is dominated by free–free absorption, so we define the GHz escape time $t_{\rm esc}$ by the condition $\tau_{\rm ff}(t_{\rm esc}) = 1$.

For the models shown in Figure 4.9, we find $t_{\rm esc} = 13\,\mathrm{yr}$ for the $11\,M_\odot$ SS model and $t_{\rm esc} = 17\,\mathrm{yr}$ for the $30\,M_\odot$ SS model. For the $30\,M_\odot$ BS model, we find $t_{\rm esc} = 13\,\mathrm{yr}$. In contrast, the $11\,M_\odot$ BS model is essentially transparent from the beginning ($t_{\rm esc} \simeq 0\,\mathrm{yr}$; $\tau_{\rm tot} < 1$ at all simulated times).

### 4.7.3 RM results

The RM and magnetic-field evolutions are shown in Fig. 4.10. In this shocked-region-only RM framework, only the $11\,M_\odot$ SS model reproduces the observed RM range and decay trend of FRB 20121102; the $30\,M_\odot$ SS model and both BS models remain below the data and do not satisfy the joint amplitude–slope constraints.

The strong RM contrast between SS and BS models is physically consistent with the shocked-shell dynamics discussed in Section 4.7.1. In particular, the dominant driver is the large difference in shocked mass $M_{\rm sh}$ (Fig. 4.3e), which translates into a large density contrast in the shocked region. Although BS models can have a higher shock velocity $v_{\rm sh}$ in a more rarefied environment (Fig. 4.3d), this does not compensate for their much lower shocked density. As a result, the ram pressure $\rho v_{\rm sh}^2$—and hence the amplified magnetic field in our prescription—remains significantly smaller than in SS models. In addition, the denser SS shocked layer produces a substantially higher post-shock electron density $n_e$ (see Fig. 4.3b), which further boosts $\int n_e B_\parallel\,\mathrm{d}l$ and amplifies the final RM separation between SS and BS channels.

Therefore, in a deliberately conservative sense, if we include only magnetic-field amplification in the SNR shock region and neglect other nonlinear effects or additional magnetic/ionization enhancement (e.g., from a MWN or photoionization), the current RM data for FRB 20121102 select the $11\,M_\odot$ SS case as the only successful model in our grid, despite studies arguing for a strong MWN environment for this source (Beloborodov, 2017, Yang & Dai, 2019).

TABLE 4.2: Fiducial (HH) local SNR DM values at $t_{\rm occur}$ for FRB 20190520B and FRB 20121102.

| Progenitor model | $\mathrm{DM_{SNR}}$ for FRB 20190520B ($\mathrm{pc\,cm^{-3}}$) | $\mathrm{DM_{SNR}}$ for FRB 20121102 ($\mathrm{pc\,cm^{-3}}$) |
|---|---|---|
| $11\,M_\odot$ SS | 183.6 | 88.5 |
| $11\,M_\odot$ BS | 44.9 | 26.2 |
| $30\,M_\odot$ SS | 202.8 | 97.7 |
| $30\,M_\odot$ BS | 167.5 | 83.0 |

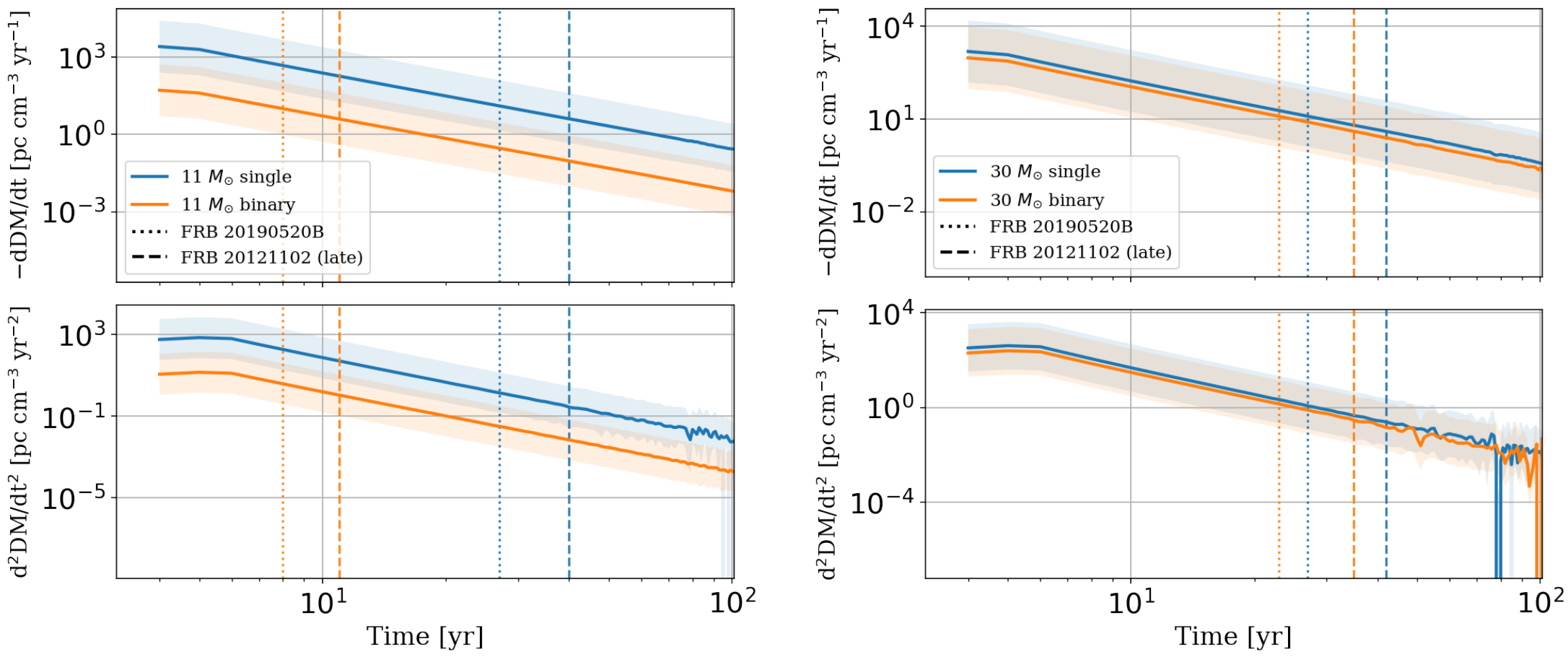

FIGURE 4.8: Time evolution of the DM time derivative in the full-region calculation for the $11\,M_\odot$ (left) and $30\,M_\odot$ (right) models. For the fiducial (HH) configuration, $t_{\rm occur}$ for FRB 20190520B (Table 4.3) is 27 (SS) and 8 (BS) yr for $11\,M_\odot$, and 27 (SS) and 23 (BS) yr for $30\,M_\odot$; for FRB 20121102 (Table C.1), it is 40 (SS) and 11 (BS) yr for $11\,M_\odot$, and 42 (SS) and 35 (BS) yr for $30\,M_\odot$.

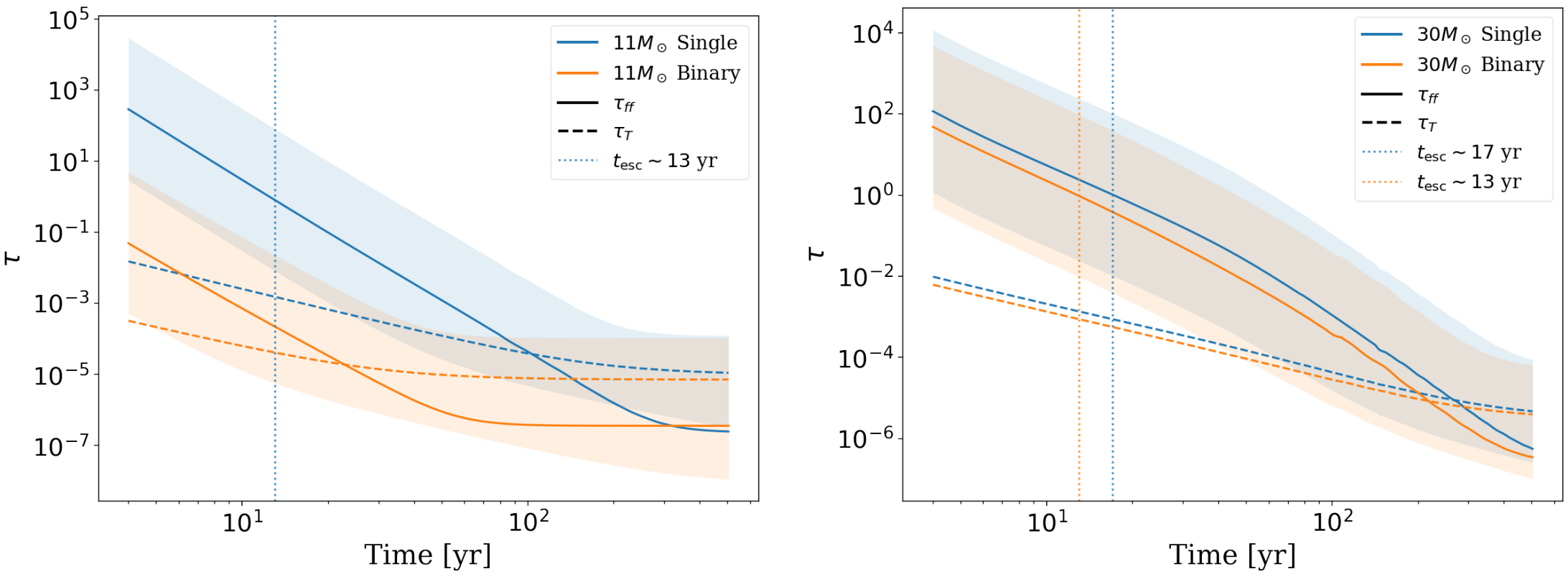

FIGURE 4.9: Time evolution of the optical depth in the full region for the $11\,M_\odot$ (left) and $30\,M_\odot$ (right) models, including contributions from free–free absorption and Thomson scattering.

TABLE 4.3: Summary of characteristic timescales and DM derivatives evaluated at $t_{\rm occur}$ for different assumed ionization configurations of the unshocked ejecta ($\chi_{e,\rm unej}$) and the unshocked ISM ($\chi_{e,\rm ISM}$).

| Ionization config. | Quantity | $11\,M_\odot$ SS | $11\,M_\odot$ BS | $30\,M_\odot$ SS | $30\,M_\odot$ BS |
|---|---|---|---|---|---|
| FF: $\chi_{e,\rm unej}=1$ $\chi_{e,\rm ISM}=1$ | $t_{\rm esc}$ (yr) | 32 | 6 | 62 | 46 |
| | $t_{\rm occur}$ (yr) | 60 | 16 | 65 | 55 |
| | $d^2{\rm DM}/dt^2$ (pc cm$^{-3}$ yr$^{-2}$) | 0.65 | 2.45 | 0.51 | 0.73 |
| | $\lvert d^2{\rm DM}/dt^2\rvert_{\rm est}$ (pc cm$^{-3}$ yr$^{-2}$) | 0.40 | 1.50 | 0.35 | 0.41 |
| HF: $\chi_{e,\rm unej}=0.1$ $\chi_{e,\rm ISM}=1$ | $t_{\rm esc}$ (yr) | 13 | 0 | 17 | 13 |
| | $t_{\rm occur}$ (yr) | 27 | 8 | 27 | 23 |
| | $d^2{\rm DM}/dt^2$ (pc cm$^{-3}$ yr$^{-2}$) | 1.40 | 3.88 | 1.17 | 1.40 |
| | $\lvert d^2{\rm DM}/dt^2\rvert_{\rm est}$ (pc cm$^{-3}$ yr$^{-2}$) | 0.89 | 3.01 | 0.84 | 0.99 |
| MF: $\chi_{e,\rm unej}=0.01$ $\chi_{e,\rm ISM}=1$ | $t_{\rm esc}$ (yr) | 5 | 0 | 4 | 0 |
| | $t_{\rm occur}$ (yr) | 13 | – | 11 | 10 |
| | $d^2{\rm DM}/dt^2$ (pc cm$^{-3}$ yr$^{-2}$) | 2.60 | – | 3.28 | 3.04 |
| | $\lvert d^2{\rm DM}/dt^2\rvert_{\rm est}$ (pc cm$^{-3}$ yr$^{-2}$) | 1.84 | – | 2.06 | 2.28 |
| FH: $\chi_{e,\rm unej}=1$ $\chi_{e,\rm ISM}=0.1$ | $t_{\rm esc}$ (yr) | 32 | 6 | 62 | 46 |
| | $t_{\rm occur}$ (yr) | 60 | 16 | 65 | 55 |
| | $d^2{\rm DM}/dt^2$ (pc cm$^{-3}$ yr$^{-2}$) | 0.065 | 2.45 | 0.51 | 0.73 |
| | $\lvert d^2{\rm DM}/dt^2\rvert_{\rm est}$ (pc cm$^{-3}$ yr$^{-2}$) | 0.40 | 1.50 | 0.35 | 0.41 |
| HH: $\chi_{e,\rm unej}=0.1$ $\chi_{e,\rm ISM}=0.1$ | $t_{\rm esc}$ (yr) | 13 | 0 | 17 | 13 |
| | $t_{\rm occur}$ (yr) | 27 | 8 | 27 | 23 |
| | $d^2{\rm DM}/dt^2$ (pc cm$^{-3}$ yr$^{-2}$) | 1.41 | 3.88 | 1.17 | 1.40 |
| | $\lvert d^2{\rm DM}/dt^2\rvert_{\rm est}$ (pc cm$^{-3}$ yr$^{-2}$) | 0.89 | 3.01 | 0.84 | 0.99 |
| MH: $\chi_{e,\rm unej}=0.01$ $\chi_{e,\rm ISM}=0.1$ | $t_{\rm esc}$ (yr) | 5 | 0 | 4 | 0 |
| | $t_{\rm occur}$ (yr) | 13 | – | 11 | 10 |
| | $d^2{\rm DM}/dt^2$ (pc cm$^{-3}$ yr$^{-2}$) | 2.59 | – | 3.28 | 3.03 |
| | $\lvert d^2{\rm DM}/dt^2\rvert_{\rm est}$ (pc cm$^{-3}$ yr$^{-2}$) | 1.84 | – | 2.06 | 2.28 |

In the ionization configuration labels, H/M/L denote High/Middle/Low ionization levels, and F denotes Fully ionized.
$t_{\rm occur}$ is determined by matching the observed DM derivative of FRB 20190520B, i.e., imposing ${\rm dDM}/{\rm d}t=-12.4\,{\rm pc\,cm^{-3}\,yr^{-1}}$. "–" indicates that no solution for $t_{\rm occur}$ exists under this condition for the stated ionization assumption.
$t_{\rm esc}$ is defined by $\tau_{\rm ff}(t_{\rm esc})=1$; $t_{\rm esc}=0$ indicates the model is transparent at all simulated times.

### 4.7.4 Simulation v.s. analytic models

Fig. 4.11 compares the DM evolution obtained from our numerical SNR simulations with several commonly adopted analytic prescriptions. Hereafter, we denote Yang & Zhang (2017) as YZ17 and Piro & Gaensler (2018) as PG18.

We evaluate the analytic prescriptions using the same explosion and wind parameters adopted in our progenitor setup (Table 4.1). For the remaining analytic-model parameters, we follow the values adopted in the original literature where they are explicitly specified (Table 4.4). Here we only restate the mathematical definitions: $n_e=\rho/(\mu_e m_p)$, $n_{\rm tot}=\rho/(\mu_m m_p)$, and $\mu_a=\sum_i n_i A_i/\sum_i n_i$, where $n_{\rm tot}$ denotes the total particle number density (ions + electrons).

For specific models, Yang & Zhang (2017) adopt $\mu_m=1.2$ (solar-composition, approximately neutral material), while Zhao & Wang (2021) state $\mu_a\simeq 1$ for their fiducial H-dominated ejecta. When required, we compute $\mu_a$ directly from the simulated elemental composition. For the PG18 analytic curves, which do not explicitly state $\mu_m$ or $\mu_e$, we adopt fiducial composition assumptions; for a fully ionized H/He mixture, where $X$ and $Y$ are the hydrogen and helium mass fractions ($X+Y=1$), the standard relations are $\mu_e=2/(1+X)$ and $\mu_m=1/(2X+3Y/4)$.

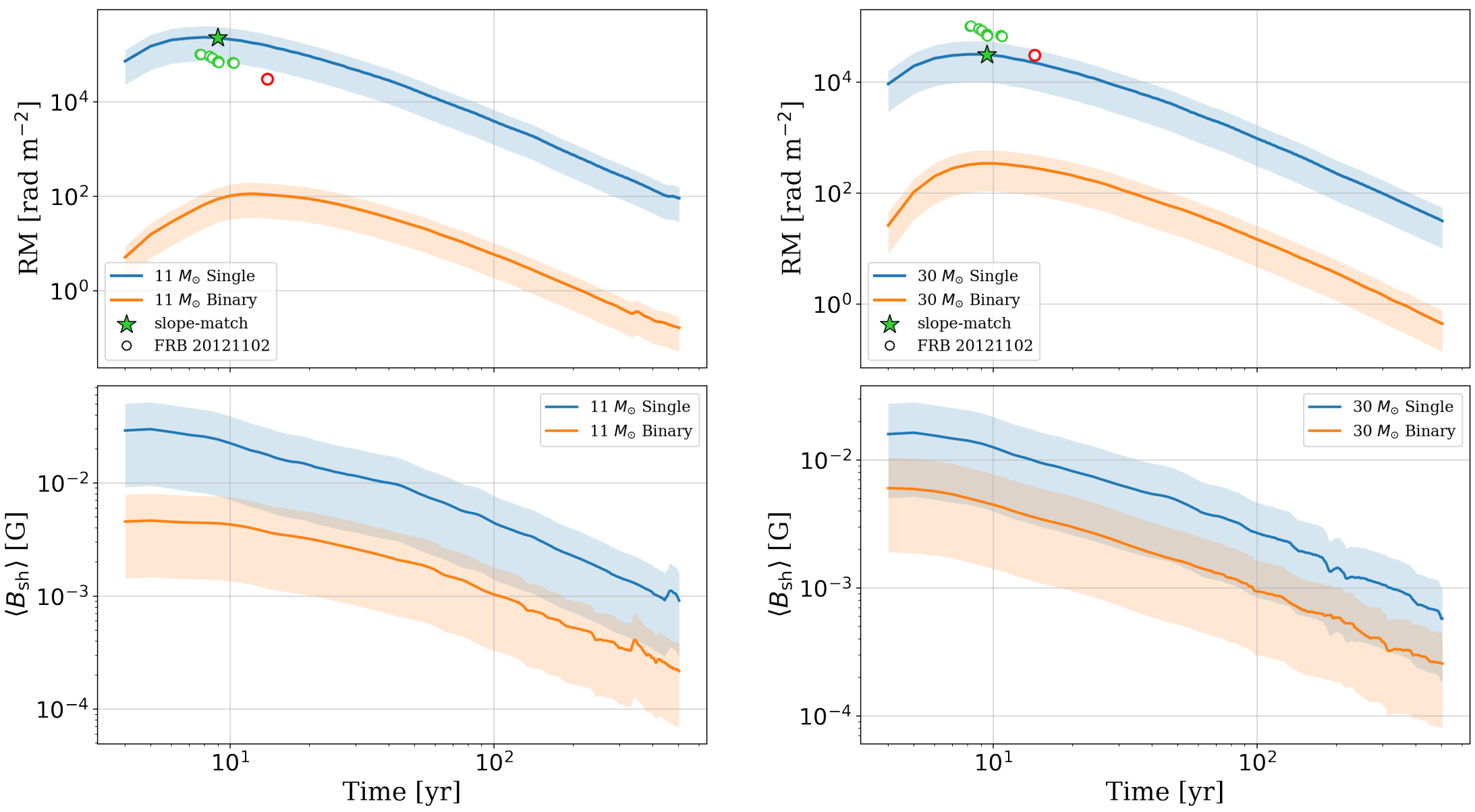

FIGURE 4.10: Time evolution of RM and magnetic field for the 11 $M_\odot$ (left) and 30 $M_\odot$ (right) models, with the FRB 20121102 RM data mapped onto the model time axis (see text). In the top panels, blue circles denote RM measurements from Hilmarsson et al. (2021), while red circles denote the newly added RM measurements from Wang et al. (2025a). The top panels show RM evolution, and the bottom panels show magnetic-field evolution, where $\langle B_{\rm sh} \rangle$ is the electron-density-weighted mean field in the shocked region. Shaded bands mark the observed RM range and the allowed $\langle B_{\rm sh} \rangle$ range.

TABLE 4.4: Physical parameters adopted in different analytic/semi-analytic DM prescriptions.

| Model | Parameters in formula | Adopted values |
|---|---|---|
| YZ17: ionized ejecta (FE; Eq. 4) | $M_1$, $E_{51}$, $\chi_e$, $\mu_m$ | $\mu_m = 1.2$ |
| PG18: shocked ionized (Eq. 5–6) | $M_1$, $E_{51}$ shared<br>ISM: $n_0$, $\mu_m$, $\mu_e$<br>wind: $\mu_e$, $K(\dot{M}, v_w)$ | ISM: $n_0 = 1$ cm$^{-3}$<br>SS: $\mu_m = 0.62$, $\mu_e = 1.18$<br>BS: $\mu_m = 1.34$, $\mu_e = 2.0$<br>wind: $\dot{M} = 10^{-5}\,M_\odot\,{\rm yr}^{-1}$<br>$v_w = 10$ km s$^{-1}$ |
| Zhao+2021: SSDW wind (Eq. 8–12) | $M_{\rm ej}$, $E_{51}$, $n$, $s$,<br>$\mu_a$, $\chi_e$, $\dot{M}$, $v_w$ | $n = 10$, $s = 2$<br>$w_{\rm core} = 0.1$, $\mu_a = 1$<br>$\dot{M} = 10^{-5}\,M_\odot\,{\rm yr}^{-1}$<br>$v_w = 10$ km s$^{-1}$ |

We then use: (i) SS case with H+He-dominated ejecta, $X \simeq 0.7$, $Y \simeq 0.3$, giving $\mu_m \simeq 0.62$ and $\mu_e \simeq 1.18$; (ii) BS case with He-dominated ejecta (H envelope lost), giving $\mu_m \simeq 1.34$ and $\mu_e \simeq 2.0$. These adopted values are the ones used in our PG18 comparisons.

We also evaluate the analytic prescriptions using the simulated, composition-dependent $\mu_a$, $\mu_e$, and $\mu_m$. These results are presented in Appendix C.3 and Fig. C.1.

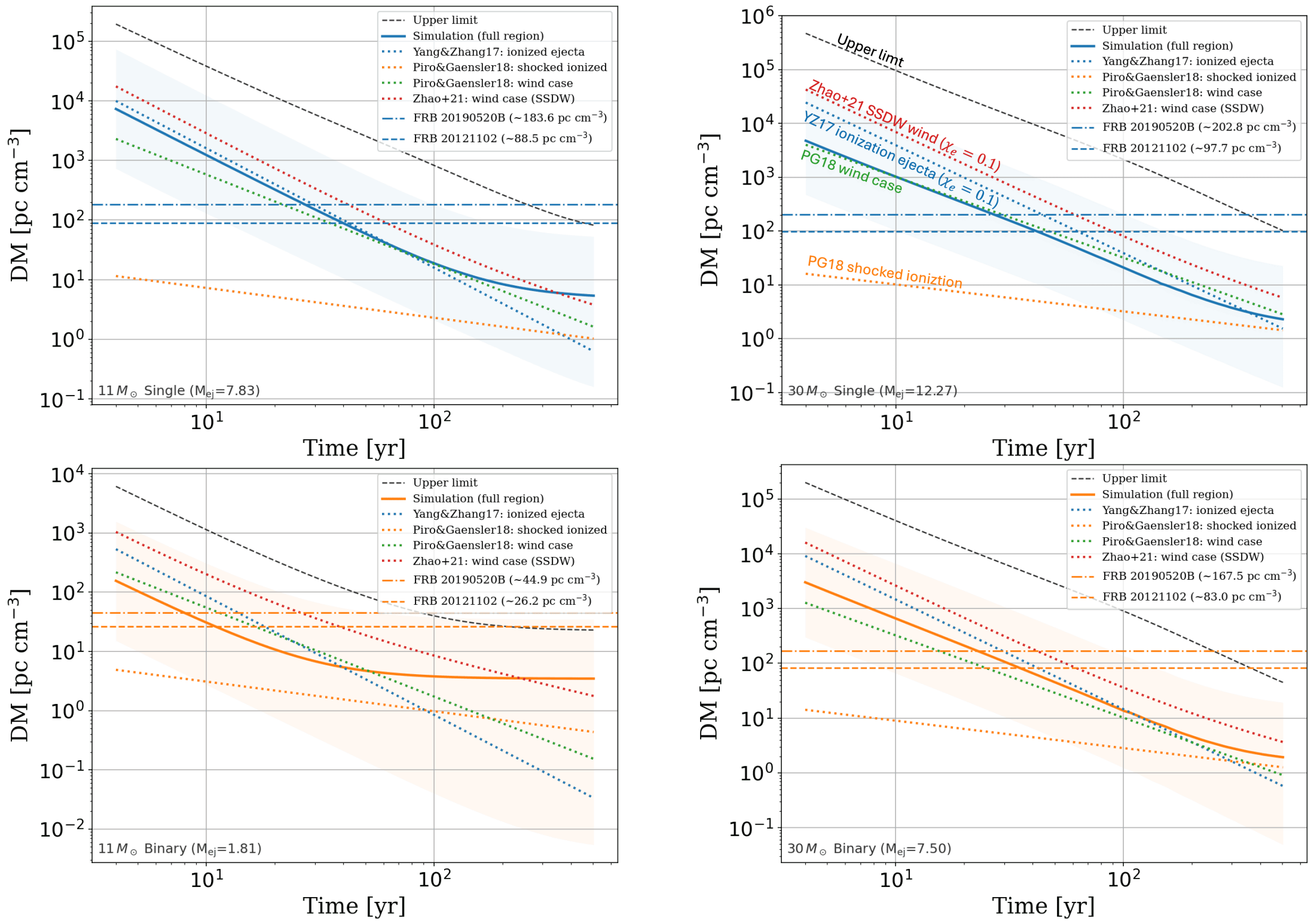

FIGURE 4.11: Time evolution of the SNR DM comparing numerical simulations with analytic expectations. Left panels show 11 $M_\odot$ models and right panels show 30 $M_\odot$ models; the top row corresponds to SS progenitors and the bottom row to BS progenitors. The "upper limit" curve assumes that the entire SNR material is fully ionized and that the ejecta are composed purely of hydrogen. For Zhao+21 and YZ17, we adopt the fiducial assumption $\chi_{e,\mathrm{unsh}} = 0.1$

In Fig. 4.11, the black dashed curve denotes the upper limit, which assumes H-dominated unshocked ejecta and CSM ($\mu_e = \mu_a = 1$) that are fully ionized ($\chi_e = 1$). Among the cases shown, the largest deviation from the numerical results occurs for the shocked-ionized ejecta scenario, whereas most analytic prescriptions fall within the upper and lower envelopes of our DM evolution. Under our fiducial assumptions in Fig. 4.11, YZ17 most closely matches the 11 $M_\odot$ SS case, while the PG18 wind case is closer for the 30 $M_\odot$ SS case. These "best-match" identifications are dependent on progenitor-model choices and parameter assumptions and are not intended to indicate a general preference among analytic prescriptions. Notably, when we adopt simulation-based, composition-dependent $\mu_a$, $\mu_e$, and $\mu_m$ (Fig. C.1), the 11 $M_\odot$ SS case becomes more consistent with the Zhao+2021 prediction.

Using the $t_{\mathrm{occur}}$ values constrained by Fig. 4.8, we obtain the corresponding $\mathrm{DM}_{\mathrm{SNR}}$ values of 183.6 pc cm$^{-3}$ (11 $M_\odot$ SS), 202.8 pc cm$^{-3}$ (30 $M_\odot$ SS), 44.9 pc cm$^{-3}$ (11 $M_\odot$ BS), and 167.5 pc cm$^{-3}$ (30 $M_\odot$ BS). This indicates that the DM constraint varies significantly not only with model parameters but also with progenitor channel. Nevertheless, all inferred contributions lie in the tens-to-hundreds pc cm$^{-3}$ range, representing a non-negligible extra component that can materially affect FRB cosmological analyses.

### 4.7.5 Shocked-region electron source and ionization states

In addition to the total $\mathrm{DM_{sh}}$, our simulations track the contribution to the electron density from different elements and ionization stages in the shocked region physically. Fig. 4.12 shows the radial number-density profiles in the shocked region at the $\mathrm{DM_{sh}}$ peak (see Fig. 4.3a). A clear channel dependence is visible: in the SS models, the shocked composition is generally H-rich with He as the secondary contributor, whereas in the BS models He dominates and metals remain subdominant. Notably, compared with the $11\,M_\odot$ SS model, the $30\,M_\odot$ SS model has a smaller hydrogen fraction and is therefore more Type IIb-like than a typical H-rich Type IIP case. This difference ultimately reflects the distinct progenitor-envelope structures: SS ejecta are largely composed of the progenitor's H-rich envelope, while in the BS channel binary interaction efficiently removes the H envelope prior to core collapse (e.g., via common-envelope evolution and Roche-lobe overflow), leaving a He-rich progenitor whose ejecta contain comparatively little H. As a result, in the BS models the post-shock electron budget is primarily supplied by ionized He, whereas in the SS models it is dominated by ionized H.

This composition difference also helps explain why the BS models reach a larger $\langle Z\rangle$ at the peak (bottom panels of Fig. 4.4) than the SS models: helium can contribute up to two electrons per ion once ionized, raising the typical ionic charge and hence $\langle Z \rangle$ relative to an H-dominated composition.

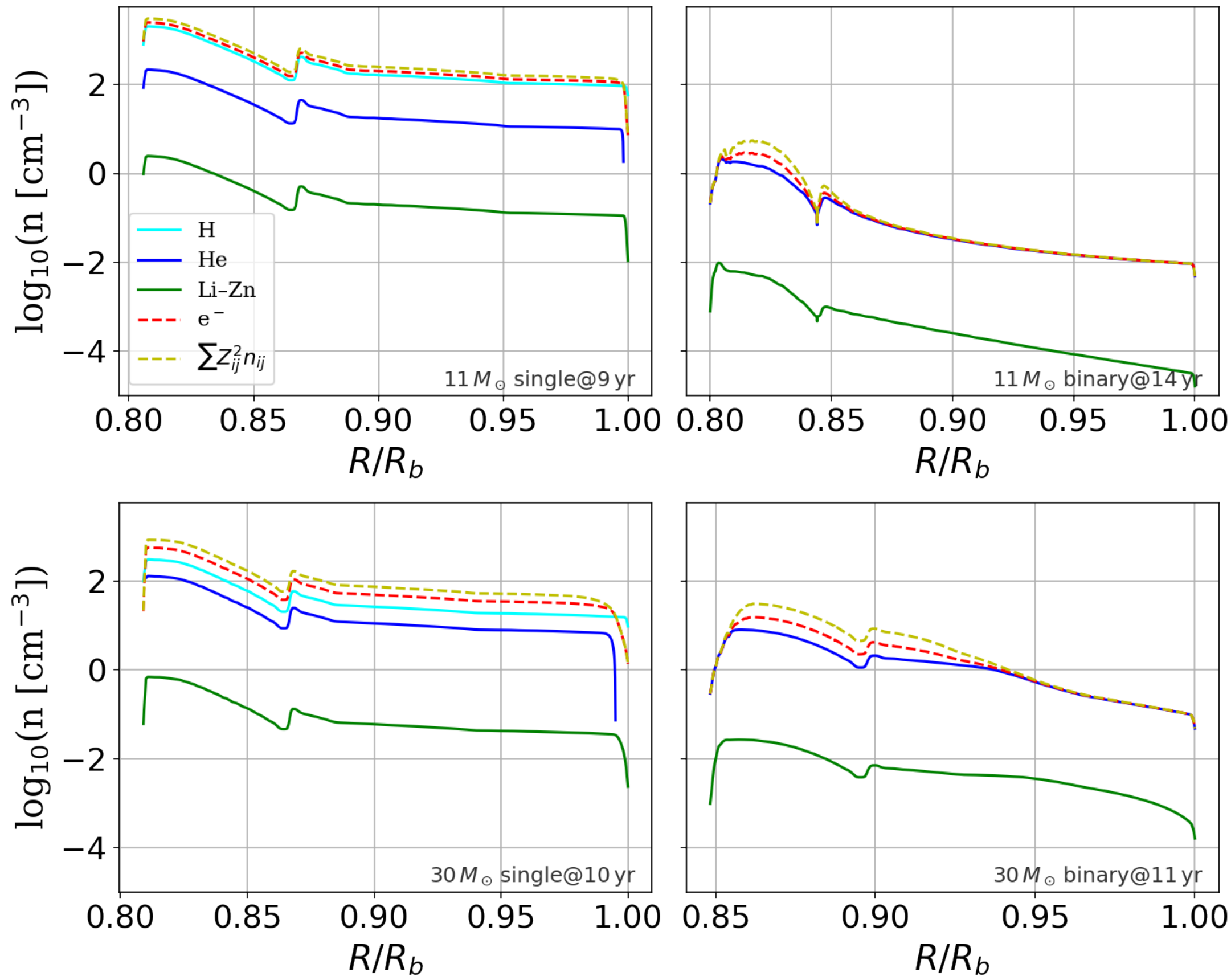

FIGURE 4.12: Radial profiles in the *shocked region* at the $\mathrm{DM_{sh}}$ peak, showing number densities of electrons (red) and selected elements for the four models considered in this work. The cyan and blue curves correspond to H and He, respectively.

This interpretation is further supported by the ionization-state structure of the shocked region in Fig. 4.15: in panel (a), corresponding to the $11\,M_{\odot}$ SS model, the peak number density of $H^{+}$ exceeds that of $He^{2+}$, whereas in panel (b), corresponding to the $11\,M_{\odot}$ BS model, the shocked region is instead dominated by $He^{+}$ and $He^{2+}$.

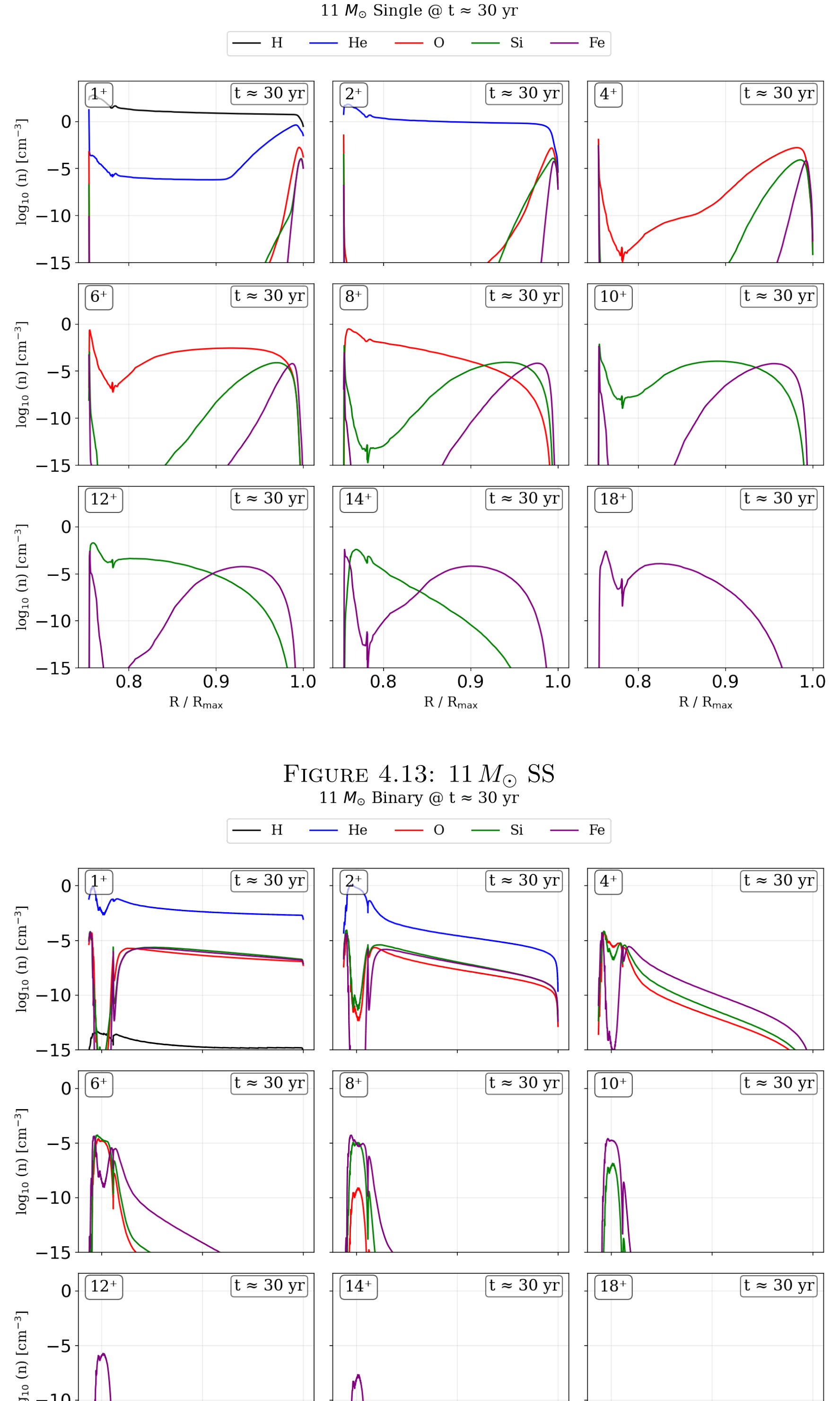

FIGURE 4.13: 11 $M_\odot$ SS

FIGURE 4.14: 11 $M_\odot$ BS

FIGURE 4.15: Ionization-state profiles in the shocked region at $t = 30\,\mathrm{yr}$ for the 11 $M_\odot$ SS and BS models.

## 4.8 Model Limitations and Future Directions

In this section, we outline some of the main implications of our preliminary results and discuss directions for future refinement.

- The ionization and thermal evolution history of the unshocked ejecta is treated approximately in the current setup. In particular, we do not yet include a fully self-consistent ionization–recombination balance, nor possible CCO-related heating/ionization channels, in the unshocked region. This limitation introduces a relatively broad uncertainty when constraining source-environment parameters from DM evolution, and should be improved in future work.
- In our current 1D SNR simulations we do not enable cosmic-ray (CR) acceleration and transport. As a result, potentially important CR-related processes—including escaping CR particles and CR-driven magnetic-field amplification (e.g., streaming instabilities and Bell-type instabilities)—are absent. Incorporating these effects in future calculations may provide new physical insight into the magnitude and time evolution of RM.
- The present 1D framework assumes spherical symmetry and a stationary central engine. Real SNR evolution is intrinsically 3D and often asymmetric: binary stripping through Roche-lobe overflow is non-spherical, the SN explosion and external CSM can be nonuniform and anisotropic, and a young magnetar may have a substantial natal kick velocity. In addition, the binary-stripped channel considered here represents only one pathway within a much broader binary-interaction parameter space (e.g., different mass ratios, orbital separations, mass-transfer histories, and common-envelope outcomes). These effects can produce significant sightline-to-sightline variance in shell structure, and therefore in DM and RM. Extending the model toward multidimensional calculations and a broader binary-model grid will be essential to quantify these geometric and progenitor-channel systematics.
- The apparent rarity of FRBs with a clear SNR-like secular DM signature can be naturally understood as a short observational window effect, consistent with the population-level transient-association and timescale arguments presented by Dong et al. (2025).

## 4.9 Part II Synthesis

In this part, we synthesize the main results on the local SNR contribution to FRB DM, based on 1D HD+NEI simulations for both $11\,M_\odot$ and $30\,M_\odot$ progenitors in single-star (SS) and binary-stripped (BS) channels. Our main conclusions are as follows.

- Considering only the shocked region, the DM contribution is limited ($\lesssim 10\,\mathrm{pc\,cm^{-3}}$) and is especially negligible for BS models. By contrast, even under this conservative setup—using the simulated shocked-plasma ionization state and ram-pressure-based magnetic amplification—the predicted RM can still reach the high values observed in FRB 20121102 for the $11\,M_\odot$ SS model. Within our model grid, this shocked-region RM comparison selects only the $11\,M_\odot$ SS case and disfavors the $30\,M_\odot$ SS and BS progenitor models as primary explanations of the FRB 20121102 RM data.

- For DM evolution, the dominant contribution comes from the unshocked ejecta. Its early-time decline is governed by free expansion and follows DM $\propto t^{-\alpha}$ with $\alpha \simeq 1.8$–$1.9$ (slightly shallower than 2), while the late-time behavior transitions toward a regime increasingly influenced by the approximately constant-density ambient ISM/CSM contribution.
- BS models generally yield smaller DM than SS models at the same $M_{\rm ZAMS}$ because H-envelope stripping reduces ejecta mass and therefore the available electron column in the unshocked ejecta, which dominates $\mathrm{DM_{SNR}}$ in our simulations. This SS–BS contrast is additionally modulated by reverse-shock development and CSM structure: BS models (and also the massive SS model) can evolve in lower-density wind-blown cavities, delaying strong reverse-shock formation at ages of a few hundred years and further reducing the shocked-ejecta DM contribution.
- The total optical depth is generally dominated by free–free absorption. Using the condition $\tau_{\rm ff} = 1$, we infer $t_{\rm esc} \lesssim 70\,\mathrm{yr}$ across the explored ionization configurations. For weakly ionized ejecta ($\chi_{e,\rm unej} \sim 0.01$), the SNR is nearly transparent to FRB emission from very early times.
- Using the observed DM decay rates of FRB 20190520B and FRB 20121102 to constrain $t_{\rm occur}$, we find $t_{\rm occur} \lesssim 100\,\mathrm{yr}$ (measured from the SN explosion) in all fiducial models, and FRB 20121102 consistently occurs later than FRB 20190520B within the same progenitor model; in the great majority of cases, both sources can be transparent to GHz FRB escape.
- The mass dependence is not always monotonic. In our models, DM is dominated by weakly ionized unshocked ejecta, so under the low-ionization assumption (approximately fixed $\chi_e$, effectively at most one free electron per atom), larger mean atomic/electron weights in metal-rich ejecta can reduce $n_e$ and produce non-monotonic trends across progenitors. This behavior can change or even reverse when the reverse shock penetrates more ejecta and/or reverse-shock-related radiation (photoionization/PR) further ionizes heavy elements, increasing the electron yield.
- A common conclusion from both FRB 20190520B and FRB 20121102 is that the inferred local SNR contribution falls in the tens-to-hundreds $\mathrm{pc\,cm^{-3}}$ range. At the fiducial-value $t_{\rm occur}$ constrained from the observed DM slopes, the inferred $\mathrm{DM_{SNR}}$ shows a clear model split for both sources: for FRB 20190520B, the $11\,M_\odot$ BS model gives only $\sim 44.9\,\mathrm{pc\,cm^{-3}}$, while the other models are $\sim 200\,\mathrm{pc\,cm^{-3}}$; for FRB 20121102, the $11\,M_\odot$ BS model gives $\sim 26.2\,\mathrm{pc\,cm^{-3}}$, while the others are $\sim 100\,\mathrm{pc\,cm^{-3}}$. The significantly smaller DM in the $11\,M_\odot$ BS case is mainly due to its much smaller ejecta mass. Therefore, the early CCSN/SNR local contribution to $\mathrm{DM_{source}}$ is non-negligible and can materially affect FRB cosmological inferences.
- For FRB 20121102, the coexistence of strong RM evolution and a two-stage DM-slope evolution (early weak rise and late decline) suggests that a pure single-component interpretation is incomplete. A hybrid SNR+MWN picture is favored, where the late-time DM decline is consistent with SNR expansion while additional local plasma from MWN-related processes can contribute to the early-stage slope turnover and magneto-ionic variability.

# Chapter 5

# Bridging Scales: Unified Framework and Conclusions and Prospects

## 5.1 Summary of Main Results

This thesis has addressed a single central question from two complementary directions: how much of an observed FRB dispersion measure is produced by the large-scale baryon distribution of the Universe, and how much is generated by the dense environment close to the source itself. The main message is that these two aspects cannot be cleanly separated by intuition alone. A reliable physical interpretation of FRB DM requires a multi-scale framework in which cosmological structure, intervening haloes, host-galaxy gas, and source-local plasma are all treated as measurable components of the same signal.

Part I showed that the "cosmic" DM term is not simply a smooth IGM background. In hydrodynamical simulations, the baryons contributing to FRB sightlines are redistributed by halo growth, environmental structure, and especially AGN feedback. As a result, the DM signal depends not only on redshift, but also on halo mass, impact parameter, and the partition of gas between the IGM and the CGM. In practical terms, this means that FRBs can do more than recover a mean DM–$z$ relation: with sufficient samples, they can constrain where the missing baryons reside and how feedback reshapes the cosmic web.

Part II showed that the source-local contribution can also be astrophysically important, especially for young systems. In the SNR models explored here, the dominant DM contribution generally comes from the unshocked ejecta rather than the shocked shell, and its evolution is controlled by expansion, ionization state, and free–free transparency. This naturally leads to observable secular DM evolution on year-to-decade timescales and implies that a non-negligible fraction of $\mathrm{DM}_{\mathrm{source}}$ can remain in the tens-to-hundreds of $\mathrm{pc\,cm^{-3}}$ range for physically plausible FRB ages. For FRB 20190520B and FRB 20121102, the comparison with observed DM/RM evolution further suggests that source-local plasma is not a minor correction, but a component that can materially affect physical and cosmological interpretation.

Taken together, the two parts of this thesis support a simple picture: FRB dispersion measures are most powerful when interpreted as layered signals. The cosmological value of FRBs does not come from assuming away foregrounds and local environments, but from modeling them well enough to isolate the scale of interest. Conversely, the astrophysical value of FRBs does not come from treating source environments in isolation, but from embedding them in a realistic line-of-sight baryon census. In this sense, FRBs are not merely distance indicators; they are multi-scale baryon probes linking the cosmic web, gaseous haloes, host galaxies, and compact-object environments within one observable.

## 5.2 Unified Multi-scale DM Decomposition

The central goal of resolving the multi-scale DM budget is to turn FRBs into more precise probes of both cosmology and baryon cycling across scales. In that sense, intervening galaxies/haloes and source-local environments should not be viewed as mutually exclusive explanations, but as physically distinct components that must be modeled simultaneously. Combined with the main results from Part I (Chapter 3) and Part II (Chapter 4), a useful bookkeeping scheme is

$$\mathrm{DM_{obs}} = \mathrm{DM_{MW,ISM}} + \mathrm{DM_{MW,halo}} + \mathrm{DM_{local\,volume}} + \mathrm{DM_{fg\,halos}} + \mathrm{DM_{IGM}} + \mathrm{DM_{host}} + \mathrm{DM_{source}}, \tag{5.1}$$

where the exact boundary between terms depends on data quality and the science goal. All terms in Equation (5.1) are understood in the observer frame; when host- or source-local contributions are modeled in the source rest frame, their observed values are reduced by a factor of $(1+z)^{-1}$. Below we summarize how each component can be constrained in a unified framework.

- **Host galaxy and host halo.** The DM–$z$ relation is best interpreted as a population-mean relation with substantial sightline-to-sightline scatter, rather than as a precision distance indicator for any single burst (Macquart et al., 2020). Even so, for poorly localized FRBs it provides a useful redshift prior that can eliminate implausible host candidates. For well-localized FRBs—whether repeaters or apparently one-off bursts—the host can be identified directly, and in favorable cases the burst position within the host can be constrained to sub-galactic scales (Tendulkar et al., 2017, Bassa et al., 2017, Heintz et al., 2020, Bhandari et al., 2022, Gordon et al., 2023). In that regime, host stellar mass can be estimated from multi-band photometry or SED fitting and related to halo mass through abundance matching or empirical stellar-to-halo mass relations, while the gas distribution can be constrained using morphology, inclination, star-formation tracers, nebular emission, IFU kinematics, and—for sufficiently massive hosts, groups, or stacked samples—X-ray/tSZ information together with analytic or simulation-motivated halo-gas profiles. Because the observer-frame host term obeys $\mathrm{DM_{host,obs}} = (1+z)^{-1}\mathrm{DM_{host,rest}}$ and depends on the path length through both the host ISM and host halo, the FRB position inside the galaxy should be modeled probabilistically rather than assumed to lie at a single characteristic radius. Future gains will come from better radio localization, deeper optical/IR imaging and spectroscopy, and multi-wavelength follow-up of well-localized FRBs.

- **Foreground haloes.** The logic for intervening haloes is similar, but the key geometric variable is the impact parameter $b$. A single FRB can intersect multiple foreground galaxies, groups, or clusters, and each intersection probes a different part of the baryon profile (Prochaska & Zheng, 2019a). With sufficiently complete imaging and redshift information, one can associate these systems with halo-mass proxies and then use the DM–$b$ trends in Figure 3.7 to estimate or marginalize over their contributions. In practice, a single sightline usually constrains the allowed gas profile at fixed halo-mass information more robustly than it constrains halo mass itself, whereas large samples are needed to break degeneracies between halo mass, gas fraction, and feedback history (Lee et al., 2022, Khrykin et al., 2024b). Recent localized sightlines already show that foreground structures can be important, including halo-gas detections and group/cluster intersections in individual FRB sightlines (Prochaska et al., 2019, Lee et al., 2023, Agarwal et al., 2019). At the same time, incompleteness in faint satellites or dwarf galaxies remains a systematic uncertainty, so ensemble analyses will ultimately be more powerful than any single burst for constraining baryon profiles and testing AGN-feedback signatures in the CGM/ICM.
- **Milky Way foreground and the nearby Universe.** Galactic subtraction is usually anchored by the NE2001 and YMW16 electron-density models (Cordes & Lazio, 2002, Yao et al., 2017), but the relevant foreground is not limited to the Galactic disk alone (Yamasaki & Totani, 2020). The Milky Way halo, the Magellanic Clouds, the Local Group medium, and nearby structures in the local volume can all introduce anisotropic DM contributions. On some sightlines, nearby massive systems such as the Virgo Cluster should be treated explicitly as extragalactic foreground structures rather than absorbed into a generic Milky Way term (Agarwal et al., 2019, Huang et al., 2025). Therefore, improving all-sky foreground subtraction requires combining Galactic ISM models with Milky Way halo models and constrained reconstructions of the nearby Universe (Faerman et al., 2017, Huang et al., 2025). This step is especially important because even a modest mean foreground term can generate large directional variance in precision DM analyses.
- **Diffuse IGM and the cosmic web.** After subtracting the best-estimated Milky Way, nearby-structure, foreground-halo, host, and source terms, the residual DM retains the cosmological signal. Its ensemble mean follows the DM–$z$ relation (Macquart et al., 2020), but real sightlines exhibit non-negligible variance because baryons are distributed inhomogeneously among halo outskirts, filaments, sheets, and voids rather than in a perfectly uniform medium (Simha et al., 2020, Zhang et al., 2025). In that sense, FRBs probe not only the mean baryon census, but also the scatter induced by the topology of the cosmic web. However, a single DM measurement rarely maps a unique filament or void; robust inference about the WHIM and filamentary baryon distribution will require large samples, cross-correlation with galaxy surveys, absorption-line tomography, X-ray/tSZ maps, and hydrodynamical simulations (Lee et al., 2022, Khrykin et al., 2024b, Zhang et al., 2025).
- **Source-local environment.** The source term is the component most naturally identified through time variability. In the framework developed in Part II, a young SNR contribution predicts a secular DM decline associated with expanding ejecta and evolving ionization/free–free opacity (Piro, 2016, Yang & Zhang, 2017, Piro & Gaensler, 2018, Niu et al., 2025, Pandhi et al., 2026). By contrast, an engine-powered nebula such as a PWN/MWN can produce flatter or more complex DM/RM evolution, especially when coupled to a persistent radio source or a strongly magnetized local environment (Dai et al., 2017, Margalit & Metzger, 2018, Yang & Dai, 2019, Michilli et al., 2018). The sign, amplitude, and duration of $\mathrm{DM}_{\mathrm{source}}$ evolution are therefore model-dependent, so multi-epoch measurements of DM, RM, scattering, and polarization are needed to break degeneracies. This is precisely where the two parts of this thesis are complementary: cosmic-web and foreground-halo modeling define the slowly varying large-scale baseline, while SNR/MWN modeling isolates the rapidly evolving local term. In that sense, the hybrid SNR+MWN interpretation advocated in Part II for FRB 20121102 is fully consistent with the multi-scale bookkeeping proposed here.

Overall, for studies targeting the missing baryons in the WHIM, these results further suggest that the most valuable "golden-sample" FRBs are likely those with well-constrained host-galaxy identifications, minimal foreground contamination, and relatively weak local source contributions. In particular, FRBs located in dwarf galaxies or at large galactocentric offsets are especially advantageous because they are expected to suffer less contamination from dense host ISM components. Sightlines with minimal intervening foreground galaxies and large angular separation from the Milky Way disk are likewise preferable for reducing Galactic foreground uncertainties. In addition, FRBs whose source environments exhibit relatively small and slowly evolving DM contributions — for example, systems observed after the surrounding MWN and SNR have sufficiently expanded and become dilute — are especially favorable for isolating the diffuse IGM component. Such systems allow the WHIM contribution to be constrained with fewer model-dependent assumptions and provide cleaner probes of the baryonic content of the cosmic web.

## 5.3 Future Prospects

The framework developed in this thesis naturally suggests that the next stage of FRB science will move in two directions at once: toward larger, statistically useful samples for cosmological baryon studies, and toward richer multi-epoch characterization of individual sources for local-environment physics. The most important future progress will come not from treating these as separate programs, but from combining them. In that spirit, several directions appear especially promising.

This observational frontier is already beginning to move outward in redshift. FRB 20240304B, reported by Caleb et al. (2025), currently provides an important demonstration case: it is presented as the highest-redshift FRB detected so far and is localized to a host galaxy in the first few billion years of cosmic history. As shown in Figure 5.1, such systems begin to probe a part of the DM–$z$ plane where the interplay among the IGM, host contribution, and foreground subtraction becomes especially informative for cosmological baryon studies.

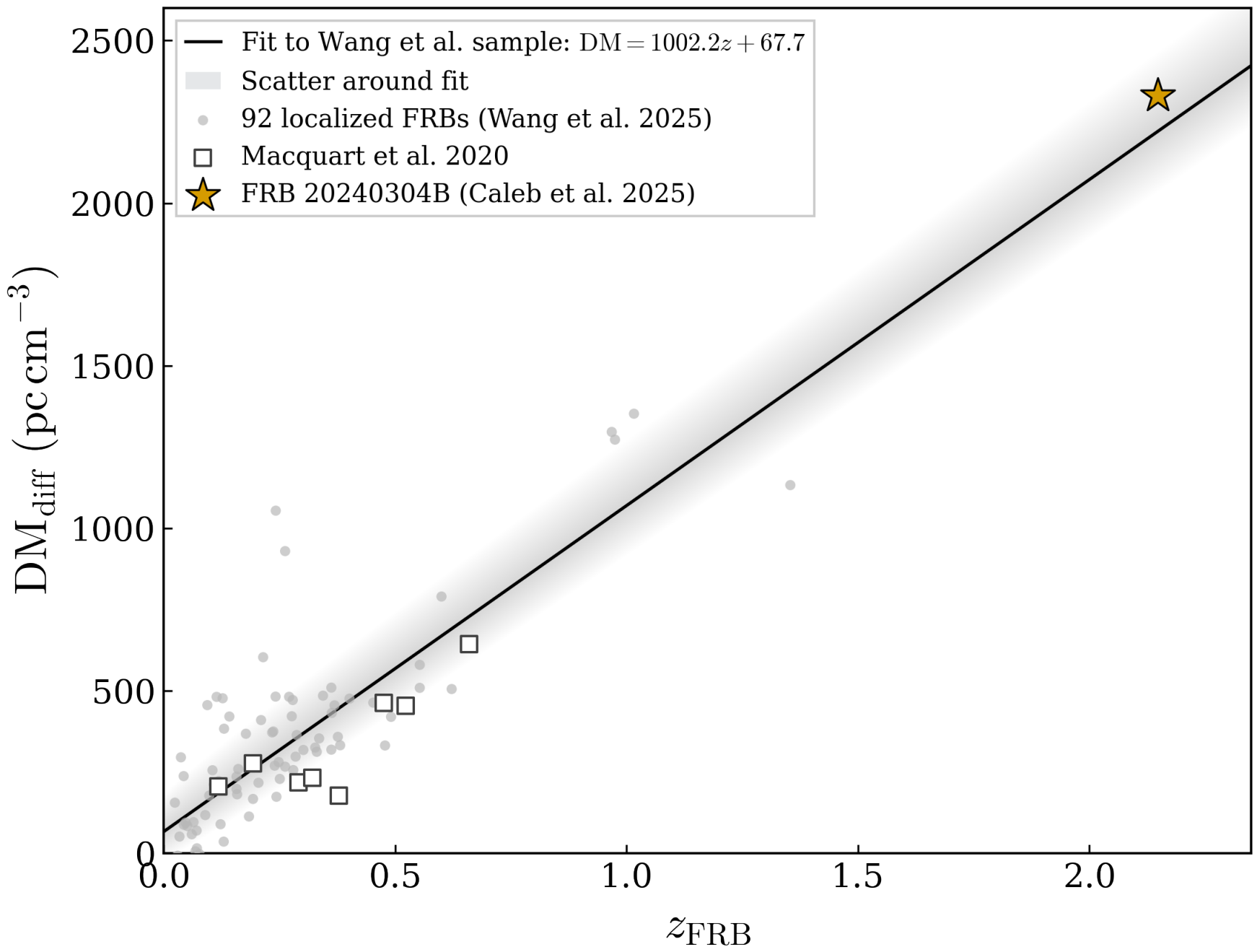

FIGURE 5.1: DM–$z$ relation of localized FRBs. Gray circles represent the sample of 92 well-localized FRBs compiled by Wang et al. (2025), while open squares denote the original benchmark sample from Macquart et al. (2020). The solid black line shows the best-fit relation to the Wang et al. sample, $\mathrm{DM} = 1002.2\,z + 67.7$, and the shaded region illustrates the scatter around this fit. The gold star marks FRB 20240304B, recently discovered by MeerTRAP and discussed by Caleb et al. (2025), as an extreme high-redshift system ($z = 2.148$). This source highlights the potential of future FRB observations to extend DM-based baryon studies to earlier cosmic epochs.

- **Cosmic-web baryons beyond halo-centered analyses.** Much of the current FRB DM interpretation is still organized around discrete structures such as host haloes, intervening haloes, and galaxy groups. A major next step is to extend this framework to the lower-density baryons that occupy filaments, sheets, and the outskirts of correlated structure. Large localized FRB samples from future facilities such as DSA-2000 and BURSTT should make it possible to move beyond measuring only the mean DM–$z$ relation and instead constrain the variance, topology, and environmental dependence of the cosmic-web contribution (Hallinan et al., 2019, Lin et al., 2022, Lee et al., 2022, Khrykin et al., 2024b, Zhang et al., 2025). In practice, this will require joint analyses with galaxy redshift surveys, X-ray/tSZ maps, and hydrodynamical simulations, so that FRBs can be used not merely as path-integrated baryon counters, but as statistical tracers of where diffuse baryons reside between and around haloes.

- **High-redshift FRBs and the peak-feedback era.** The simulation results in Part I indicate that FRB DM is sensitive to how baryons are redistributed by halo growth and AGN feedback. This sensitivity should become even more informative at intermediate and high redshift, when cosmic star formation and black-hole activity are closer to their peak and the partition of gas between galaxies, haloes, and the diffuse IGM is evolving rapidly. As localized FRBs extend to higher redshift, they may therefore provide a new way to test how feedback reshapes gaseous haloes across cosmic time, especially when combined with simulation suites that explicitly compare different feedback prescriptions (Kim et al., 2014, Roca-Fàbrega et al., 2024, Granizo et al., 2026). In this sense, FRBs may eventually constrain not only the baryon census at a given epoch, but also the redshift evolution of baryon cycling itself.

- **FRBs as probes of reionization.** A particularly exciting long-term prospect is the possibility of using sufficiently distant FRBs to probe the ionization history of the Universe. In principle, if sensitive surveys and deep optical/infrared follow-up can deliver a sample of well-localized high-$z$ bursts, the integrated DM signal could constrain both the timing of reionization and the degree of patchiness in the ionized IGM (Inoue, 2004, Deng & Zhang, 2014, Bhandari & Flynn, 2021). The same logic may also extend to the later thermal history of the diffuse IGM, including the effects of helium reionization that modulate the free-electron budget at intermediate redshift (Haardt & Madau, 2012). Realizing this goal will be challenging because host and source contributions become increasingly uncertain for individual events, but the payoff would be considerable: FRBs could provide a complementary probe of reionization that is sensitive directly to free electrons rather than to emission or absorption from selected luminous tracers.

- **Magnetized baryons and RM tomography.** This thesis has focused primarily on DM, but the next generation of FRB constraints should come from interpreting DM and RM together. Multi-frequency polarization data can help separate dense but weakly magnetized plasma from lower-density regions with strong coherent fields, thereby offering leverage on the magnetization of the Milky Way halo, the CGM/ICM, and source-local environments. Repeating FRBs such as FRB 20121102 already demonstrate that local RM can be large, variable, and astrophysically diagnostic (Michilli et al., 2018). As samples improve, the combination of RM, DM, scattering, and foreground-galaxy information may enable a form of magneto-ionic tomography in which FRBs constrain not only where baryons are, but also how magnetic fields are distributed across haloes, cluster outskirts, and compact-source environments.

- **Completing the Milky Way–host–foreground bookkeeping.** One of the clearest lessons from the unified decomposition in Equation (5.1) is that the remaining systematic floor in FRB cosmology may come from imperfect subtraction of nearby contributions rather than from the high-redshift IGM itself. The Milky Way halo, Local Group medium, local-volume structures, host galaxy, and foreground haloes all need to be modeled with higher fidelity and with realistic sightline-to-sightline variance (Yamasaki & Totani, 2020, Faerman et al., 2017, Huang et al., 2025). Progress on this front should come from combining Galactic electron-density models, constrained simulations of the nearby Universe, and better empirical host-DM priors calibrated by high-resolution galaxy simulations such as AGORA and related zoom-in studies (Kim et al., 2014, Roca-Fàbrega et al., 2024, Granizo et al., 2026). In practical terms, the goal is to turn the current bookkeeping identity into an operational inference framework in which each term is assigned a physically motivated prior rather than absorbed into an undifferentiated nuisance term.

- **Source-region DM/RM evolution beyond one-dimensional SNR models.** On the source side, the most immediate theoretical extension of Part II is to move beyond spherically symmetric ejecta models and explore genuinely multidimensional local environments. Young remnants are likely to be shaped by asymmetric explosions, anisotropic circumstellar mass loss, clumping, Rayleigh–Taylor mixing, and line-of-sight variations through the circumstellar medium; these effects can introduce substantial burst-to-burst or epoch-to-epoch differences even at fixed source age. In addition, future models should couple thermal plasma evolution to cosmic-ray acceleration and magnetic-field amplification at the shock, so that DM, RM, synchrotron emission, and free–free transparency are predicted self-consistently rather than in isolation (Piro, 2016, Yang & Zhang, 2017, Piro & Gaensler, 2018, Niu et al., 2025, Pandhi et al., 2026). Such developments are especially important for repeaters associated with persistent radio sources, because hybrid SNR+MWN scenarios may naturally require both secular ejecta evolution and compact magnetized nebular variability (Dai et al., 2017, Margalit & Metzger, 2018, Yang & Dai, 2019, Michilli et al., 2018). More broadly, once three-dimensional CSM structure is taken seriously, FRB DM evolution will become not only an age diagnostic, but also a probe of the progenitor's final mass-loss geometry and local shock physics.

# Appendix A

# Empirical Scatter Envelope in the DM–$z$ Relation

This appendix describes the construction of the empirical DM–$z$ relation and the associated scatter envelope shown in Figure 1.5. The purpose of this procedure is to provide a visual representation of the dispersion of the current FRB sample, rather than a rigorous statistical or physical model of the DM variance.

We first fit all localized FRBs in the sample with a linear relation

$$\mathrm{DM}_{\mathrm{fit}}(z) = az + b, \tag{A.1}$$

where the coefficients $a$ and $b$ are obtained from a least-squares fit to the full dataset. The residual for each source is then defined as

$$\Delta\mathrm{DM}_i = \mathrm{DM}_{\mathrm{cosmic},i} - (az_i + b). \tag{A.2}$$

From these residuals, we estimate a global standard deviation

$$\sigma_{\mathrm{res}} = \mathrm{std}(\Delta\mathrm{DM}_i), \tag{A.3}$$

which characterizes the overall dispersion of the sample relative to the best-fit relation.

To qualitatively capture the increasing spread toward higher redshift, we introduce an empirical redshift-dependent scatter amplitude

$$\sigma(z) = \sigma_{\mathrm{res}} \left(A + F\sqrt{z}\right), \tag{A.4}$$

where $A$ and $F$ are dimensionless coefficients controlling the baseline scatter and its redshift dependence, respectively. In this work, we adopt $A = 0.65$ and $F = 0.55$ as empirical values chosen to provide a visually reasonable representation of the data dispersion.

This parameterization is qualitatively reminiscent of the framework in which the dispersion of DM is related to large-scale structure and baryon inhomogeneity (e.g., McQuinn 2014, Macquart et al. 2020), where the fractional scatter may encode information about baryon distribution and feedback processes. In particular, the parameter $F$ can be interpreted as a phenomenological measure of the growth of line-of-sight variance with redshift, which may be linked to feedback-induced baryon redistribution.

However, we emphasize that the present choice of $A$ and $F$ is purely empirical and not derived from a formal statistical fit or physical model. A rigorous interpretation of these parameters in terms of baryon physics or feedback strength is beyond the scope of this work, and will be discussed in Section 3.6.

The upper and lower bounds of the shaded region are defined as

$$\mathrm{DM}_{\mathrm{upper}}(z) = \mathrm{DM}_{\mathrm{fit}}(z) + \sigma(z), \tag{A.5}$$

$$\mathrm{DM}_{\mathrm{lower}}(z) = \max\left[\mathrm{DM}_{\mathrm{fit}}(z) - 0.75\,\sigma(z),\, 0\right]. \tag{A.6}$$

This construction leads to a mildly asymmetric envelope, with a broader extension toward higher DM values. The asymmetry is introduced both to avoid unphysical negative values at low redshift and to qualitatively reflect the skewed distribution of observed DM values, where high-DM excursions are more common due to contributions from dense intervening structures.

We emphasize that this scatter envelope is empirical and intended for visualization only. It does not represent a statistically rigorous confidence interval, nor a physically motivated prediction of $\sigma_{\mathrm{DM}}(z)$. A more rigorous treatment would require either redshift-binned estimates of $\sigma_{\mathrm{DM}}(z)$ or direct modeling of line-of-sight variance from cosmological simulations.

# Appendix B

# Additional Methods, Tests, and Derivations for Part I (Cosmic Web)

## B.1 Impact of Resolution and Box Size on Density Profiles

Figure B.1 show the median density profiles from the L500 simulation for various mass ranges, similar to Figure 3.4.

Figure B.2 compares the density profiles of gas and dark matter across simulations with different box sizes and resolutions. Panels (a) and (b) illustrate the fiducial L25, L50, L100 simulations, while panels (c) and (d) display the NoBH L50, L100, L500 simulations. This comparison highlights the effects of box size and resolution on the density profiles of halos within similar mass ranges.

In panels (a) and (v), the gas and dark matter profiles in the L25 and L50 simulations exhibit slight deviations from the L100 results. For dark matter, the L25 and L50 results are relatively close and slightly higher than the L100 simulation. This discrepancy likely arises from sample variance due to the smaller box sizes, where fewer massive halos are available for statistical averaging. For gas, the profiles show distinct trends across mass ranges. In low-mass halos ($10^{10.5}\,M_\odot < M_H < 10^{12.5}\,M_\odot$), the L25 and L50 simulations yield similar results, deviating from the L100. However, for halos in the $10^{12.5}\,M_\odot < M_H < 10^{13.5}\,M_\odot$ range, the profiles near the halo center differ significantly across all box sizes, highlighting the crucial role of AGN feedback in this mass regime. In the highest mass range ($M_H > 10^{13.5}\,M_\odot$), the L25 and L50 profiles remain closer, with larger deviations observed for the L100 simulation.

In panels (c)) and (d), the NoBH simulations show different trends. For dark matter, the L50 and L500 profiles are more aligned, while the L100 profiles deviate slightly. This consistency between L50 and L500 simulations suggests that sample variance in the L100 box may contribute to the differences. For gas, the L500 simulation exhibits a steep rise in the central density, distinct from the L50 and L100 results. This steep rise is attributed to the lower resolution of the L500 simulation, where the gravitational softening length ($16\,h^{-1}$ ckpc) coincides with the scale of the steep density increase. The limited resolution smooths gas distributions and artificially concentrates gas particles near the halo center, leading to an overestimation of central densities.

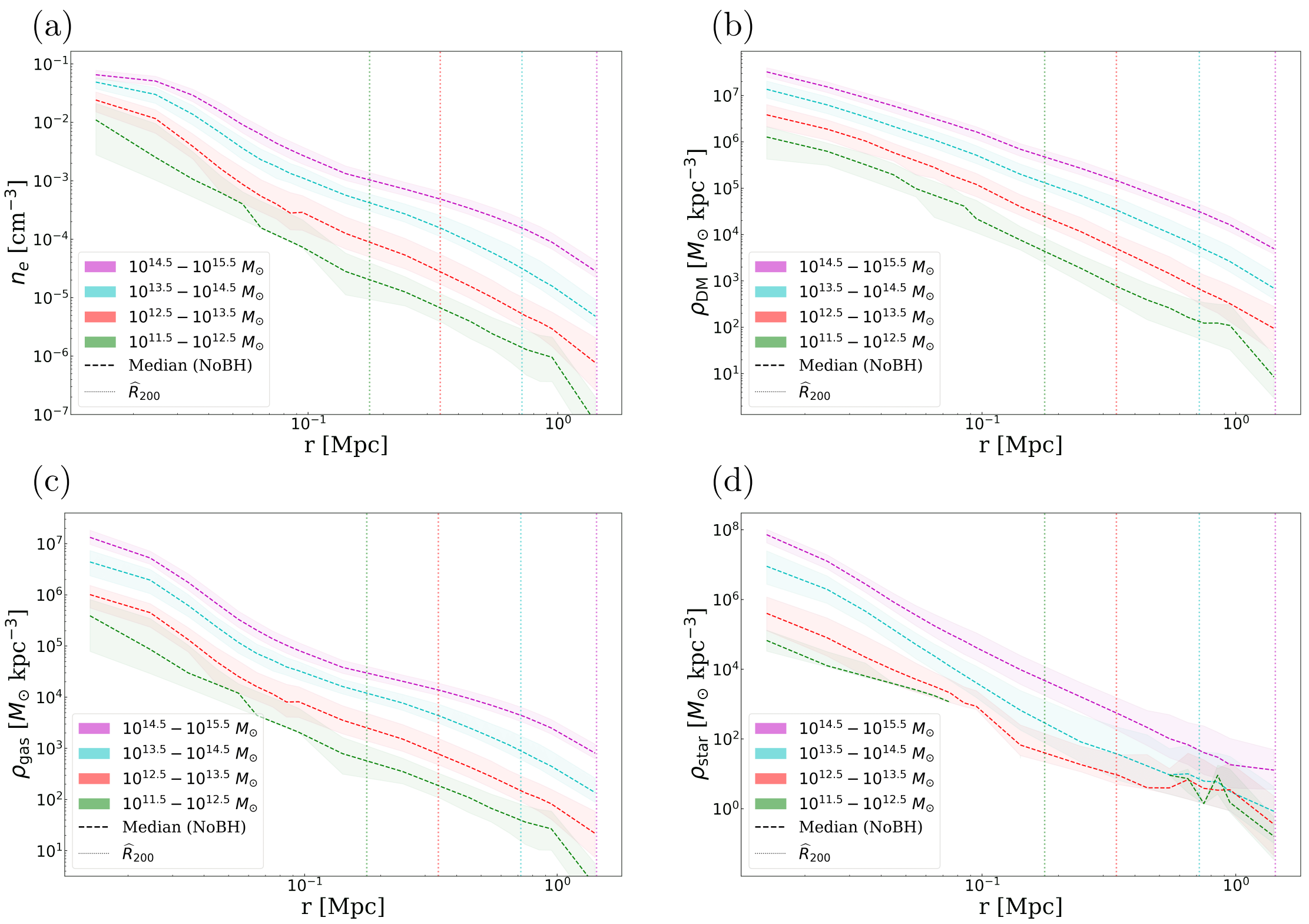

FIGURE B.1: Median density profiles for halos in the L500NoBH simulation, categorized by mass ranges: electron number density $n_e$ (panel a), dark matter density $\rho_{\rm DM}$ (panel b), gas density $\rho_{\rm gas}$ (panel c), and stellar mass density $\rho_{\rm star}$ (panel d). The shaded regions indicate $1\sigma$ uncertainties for 500 halos in each mass range. The vertical dotted lines represent the median $\widehat{R}_{200}$ of halos in each mass range.

The differences in gas and dark matter profiles highlight the impact of resolution and box size on simulations. Gas profiles are more sensitive to resolution due to the interplay of gravitational forces, hydrodynamics, and cooling processes. In contrast, dark matter profiles, driven solely by gravitational interactions, remain relatively robust against resolution changes. The findings highlight the necessity of using higher-resolution simulations for accurately studying baryonic processes, particularly in large-scale simulations with low mass resolution.

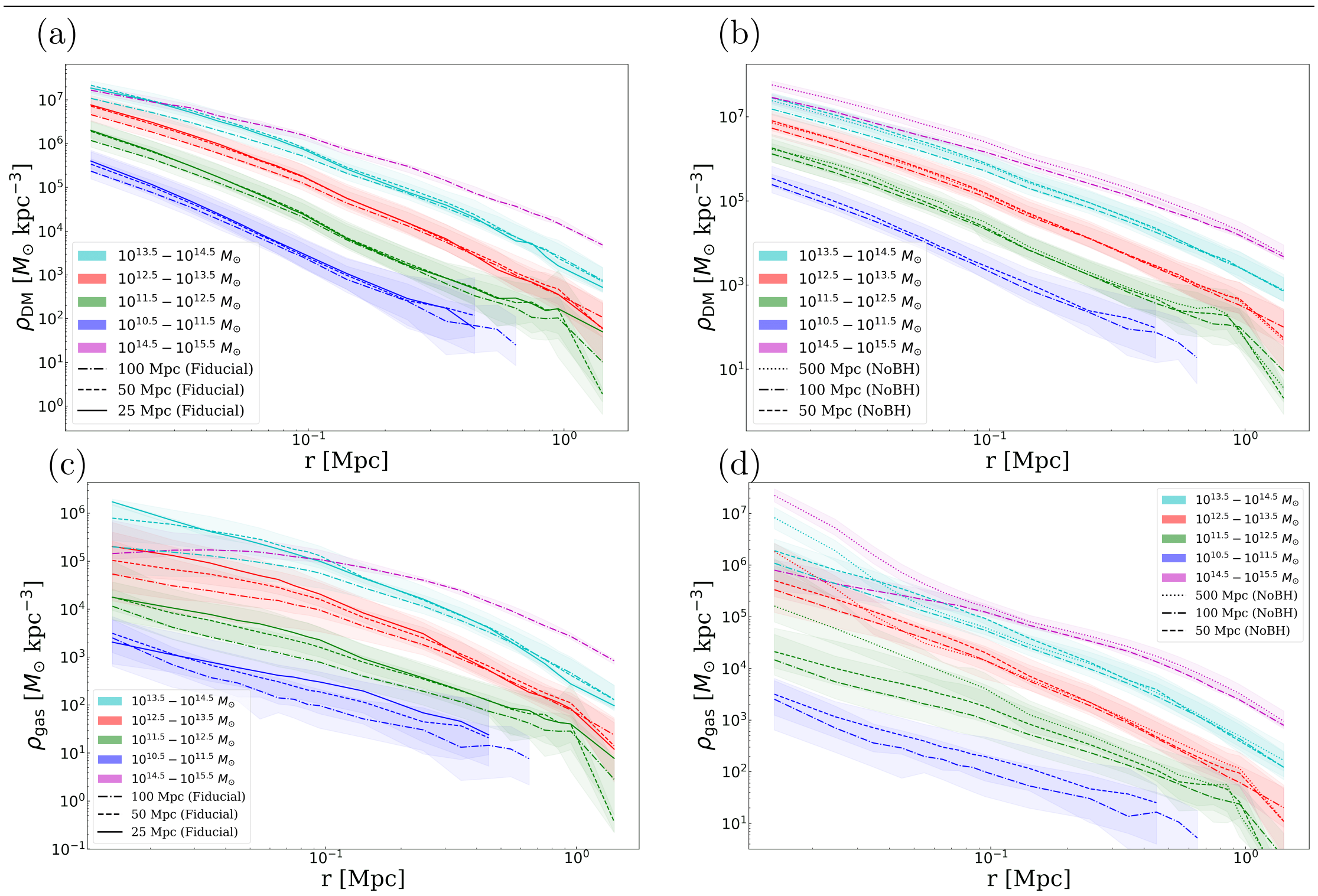

FIGURE B.2: Comparison of density profiles for halos in different box sizes (25, 50, 100, and 500 Mpc). The left column compares fiducial simulations in 25, 50, and 100 Mpc boxes, and the right column compares NoBH simulations in 50, 100, and 500 Mpc boxes.

## B.2 Implications of Random Sightline Placement

In our large-scale light cone simulation, FRBs are modeled as occurring along random sight lines through the cosmic web, without explicitly fixing their origin or termination within halos. This assumption allows for efficient statistical sampling but does not guarantee a "halo-to-halo" configuration. As a result, individual sight lines may start or end inside, near, or outside of halos.

In the real Universe, FRBs originate in galaxies and terminate in the MW. Our use of random sight lines may therefore lead to a modest underestimation of total DM, compared to an idealized configuration in which all sight lines pass through both a HH and the MW halo. This effect is partially mitigated by averaging over a large number of sight lines, which collectively sample diverse environments.

To assess this potential offset, we refer to our dedicated HH analysis (see Section 2.4, Figure 3.20e, f), where the local DM contribution from FRBs varies by environment:

- Galaxy Cluster (central): ∼1477–2003 pc cm$^{-3}$ (median: 1670.16)
- MW-like Galaxy (central): ∼83–813 pc cm$^{-3}$ (median: 163.53)
- MW-like Galaxy (off-center): ∼6.38–28.42 pc cm$^{-3}$ (median: 11.59)
- Dwarf Galaxy (central): ∼2.3–9.6 pc cm$^{-3}$ (median: 4.13)

- Dwarf Galaxy (off-center): $\sim$0.01–0.02 pc cm$^{-3}$

These values estimate the systematic DM contribution omitted by random-LOS sampling. Accordingly, we caution that our DM results represent only the cosmic contribution outside the MW and host galaxies, which are analyzed separately in Section 3.1.

## B.3 Host Galaxy DM Correction

In this work, we aim to compare the simulated diffuse intergalactic medium (IGM) dispersion measure with observations of localized fast radio bursts (FRBs). For this purpose, we use the latest FRB sample compiled by Sharma et al. (2026), which provides measurements of the extragalactic dispersion measure, $\mathrm{DM_{exgal}}$, after subtracting the Milky Way contribution.

Since $\mathrm{DM_{exgal}}$ still contains contributions from the host galaxy, a direct comparison with the simulated diffuse component is not appropriate. Therefore, for the comparison presented in Figure 3.2, we estimate a host-corrected dispersion measure, denoted as $\mathrm{DM_{diff}}$, by statistically subtracting the host galaxy contribution.

Following Sharma et al. (2026), we assume that the rest-frame host galaxy DM follows a log-normal distribution (Eq. 3.1), characterized by a mean value $\langle \mathrm{DM_{host}} \rangle \approx 128.6\,\mathrm{pc\,cm^{-3}}$ and a variance $\sigma^2_{\mathrm{DM_{host}}} \approx (150\,\mathrm{pc\,cm^{-3}})^2$. To properly account for the statistical properties of this distribution, we convert these values into the corresponding log-space parameters and perform a Monte Carlo sampling by drawing one realization of $\mathrm{DM_{host,rest}}$ for each FRB.

The corresponding observed-frame host contribution is then obtained as

$$\mathrm{DM_{host,obs}} = \frac{\mathrm{DM_{host,rest}}}{1+z}. \tag{B.1}$$

The diffuse component used in Figure 3.2 is thus estimated as

$$\mathrm{DM_{diff}} = \mathrm{DM_{exgal}} - \mathrm{DM_{host,obs}}. \tag{B.2}$$

To avoid unphysical values introduced by statistical fluctuations, negative $\mathrm{DM_{diff}}$ values are clipped to zero. We note that this statistical correction introduces additional scatter and potential bias, particularly at low redshift where the host contribution can be comparable to or exceed $\mathrm{DM_{exgal}}$.

To illustrate the impact of the host-galaxy correction, Figure B.3 show a comparison between the original extragalactic dispersion measures, $\mathrm{DM_{exgal}}$, and the host-corrected diffuse component, $\mathrm{DM_{diff}}$, for the Fiducial and NoBH simulations, respectively. The host subtraction systematically shifts the observed FRB sample toward lower DM values, improving the consistency with the simulated diffuse IGM component while preserving the overall scatter.

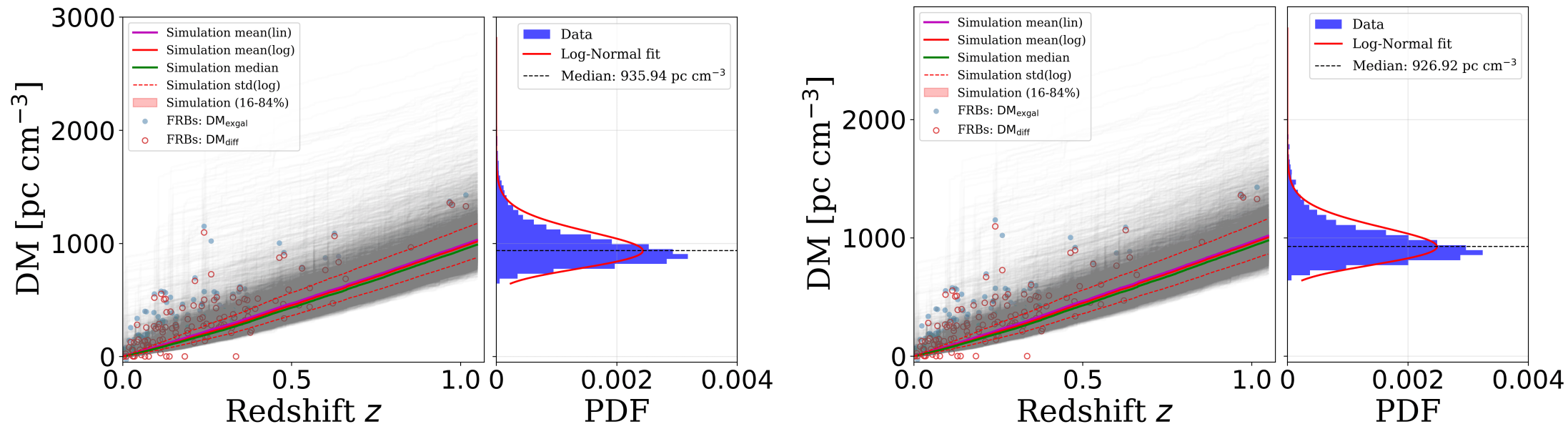

FIGURE B.3: Comparison between the observed extragalactic dispersion measure ($DM_{exgal}$; blue points) and the host-corrected diffuse component ($DM_{diff}$; red open circles), overlaid on the simulated DM–$z$ relation. Left panel: Fiducial simulation including AGN feedback. Right panel: NoBH simulation without AGN feedback. The background gray lines represent individual simulated sightlines, while the colored curves indicate the mean and median trends. The shaded region denotes the 16–84% scatter. The right-hand subpanels show the corresponding DM distributions at $z = 1$, with log-normal fits overlaid. The host subtraction systematically shifts the observed FRB sample toward lower DM values, improving the consistency with the simulated diffuse IGM component.

## B.4 Further Clarification on the Definition and Interpretation of $f_{\mathrm{diff}}$

In Section 2.3.3, we defined the diffuse baryon fraction, $f_{\mathrm{diff}}$, as the sum of those in the IGM and halos. However, both components are redshift-dependent, and thus $f_{\mathrm{diff}}$ evolves with redshift as $f_{\mathrm{diff}}(z) = f_{\mathrm{IGM}}(z) + f_{\mathrm{Halos}}(z)$ (see Figure 3.18). This relation can be rigorously obtained from cosmological simulations. However, the value of $f_{\mathrm{diff,\,obs}}$ constrained observationally via Equation 2.45 represents a distinct concept. It reflects the integrated contribution of diffuse gas and halos along the entire sightline up to a given redshift $z$, weighted by the DM contribution of intersected structures.

More precisely, for an FRB located at redshift $z = z_i$ providing a sightline $i$, the observed fraction can be written as

$$f^{(i)}_{\mathrm{diff,\,obs}}(z < z_i) = f^{(i)}_{\mathrm{IGM}}(z < z_i) + f^{(i)}_{\mathrm{Halos}}(z < z_i), \tag{B.3}$$

which depends not only on the redshift integration but also on the intervening halo distribution along the LoS. When we take a median or logarithmic mean over many FRBs, this yields a statistical measure $\langle f_{\mathrm{diff,\,obs}}(< z)\rangle$ that reflects both cosmic structure and observational selection effects. To achieve this goal, one would need a well-defined sample of localized FRBs with good redshift measurements.

Furthermore, Figure B.4 illustrates the composition of cosmic baryons at $z = 0$ in our CROCODILE simulation in terms of both physical phases and large-scale structures. Baryons are primarily composed of gas, with only minor contributions from stars and black holes.

In the *phase-based* classification, gas is divided using a temperature threshold of $10^4$ K to separate *cold* and *hot* components, and a baryon overdensity threshold of $\log_{10}(\rho_{\mathrm{gas}}/\bar{\rho}_b) = 3.5$ to distinguish *diffuse* from *dense* phases. In contrast, the *structure-based* classification is based on the CGM subtraction method (Section 3.4.2.2), in which gas within halos is excluded based on a cutoff radius $R_{\rm cut}$.

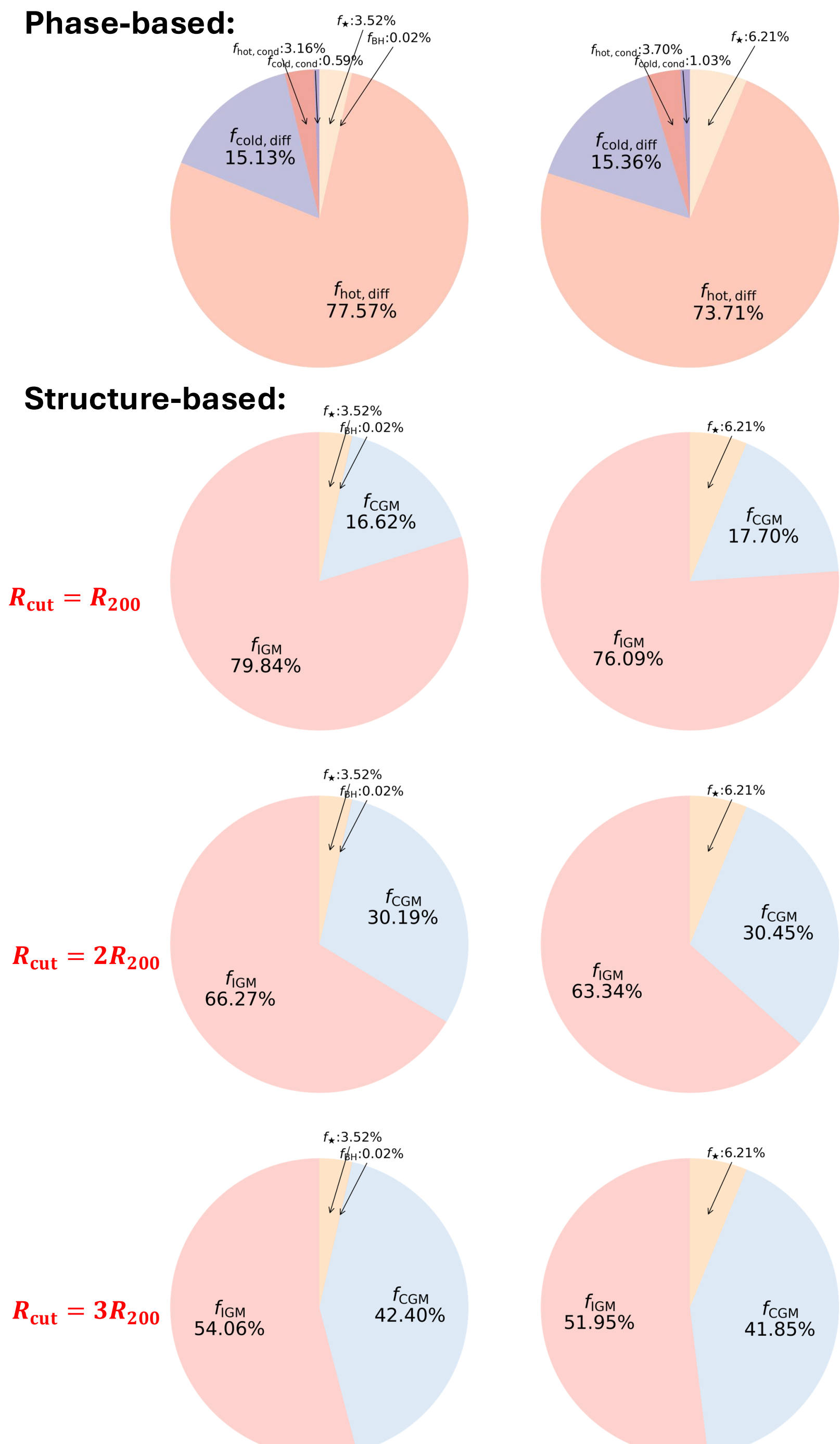

FIGURE B.4: Pie chart comparison of baryon composition at $z = 0$ under two classification schemes: gas phase-based (top row) and structure-based definitions (bottom three rows). The top row shows the distribution between cold/hot IGM/ phases. The remaining panels show baryon fractions under different CGM definitions based on $R_{\rm cut} = R_{200}, 2R_{200}, 3R_{200}$ for both fiducial (left) and NoBH (right) simulations.

This leads to two distinct definitions of the diffuse baryon fraction:

$$\begin{aligned} f_{\rm diff,\,phase} &= f_{\rm cold,\,diff} + f_{\rm hot,\,diff} \\ &= 1 - f_{\rm cold,\,dense} - f_{\rm hot,\,dense} - f_\star - f_{\rm BH}, \end{aligned} \tag{B.4}$$

$$\begin{aligned} f_{\rm diff,\,struct} &= f_{\rm CGM} + f_{\rm IGM} \\ &= 1 - f_\star - f_{\rm BH}. \end{aligned} \tag{B.5}$$

Each component can be measured directly from simulations, but cannot be independently constrained from FRB DM observations. For $f_{\rm diff,\,struct}$, we do not distinguish between the ISM and intra-cluster medium (ICM), as our CGM subtraction method does not depend on halo mass or internal structure — these components are effectively included in the CGM term.

By comparing fiducial and NoBH simulations, we find that the fiducial model yields more hot and diffuse gas in the phase-based decomposition, reflecting AGN-driven heating and gas ejection. Similarly, the structure-based decomposition shows a larger IGM fraction and a smaller CGM fraction in the fiducial case, indicating baryon redistribution due to AGN feedback. Additionally, the stellar mass fraction is clearly lower in the fiducial model, confirming suppression of star formation by AGN.

We find $f_{\rm diff,\,phase} = f_{\rm cold,\,diff} + f_{\rm hot,\,diff} \approx 93\%$ at $z = 0$, in agreement with estimates from Connor et al. (2024). Assuming that the IGM is primarily composed of hot diffuse gas (with $f_{\rm hot,\,diff} \approx 78\%$ in fiducial and 74% in NoBH), our values closely match those from the structure-based method with $R_{\rm cut} = R_{200}$, where $f_{\rm IGM} \approx 80\%$ (fiducial) and 76% (NoBH) at $z = 0$. As $R_{\rm cut}$ increases, the balance between CGM and IGM varies significantly, reaffirming the sensitivity of these components to the adopted halo boundary.

The quantity $\langle f_{\rm diff,\,obs}(< z)\rangle$ represents a redshift-integrated and sightline-averaged observable derived from many FRB sight lines. *It is not a snapshot of the instantaneous baryon fraction at a given redshift, but rather a statistical measure shaped by both the redshift integration and the distribution of intervening halos along the LoS.* For example, to interpret an FRB observed at $z = 1$, one must sample many sight-lines in a light-cone data up to $z = 1$, effectively stacking the contributions from all structures within that redshift range. As such, $\langle f_{\rm diff,\,obs}(< z)\rangle$ can be understood as a weighted average over the redshift-dependent curves of (instantaneous or differential) $f_{\rm CGM}(z)$ and $f_{\rm IGM}(z)$ shown in Figure B.6 (Appendix B.5).

While the IGM contribution dominates due to its nearlyCha universal presence in all directions, the halo contribution varies significantly depending on redshift and sightline geometry. At low redshift, many FRBs may not intersect any foreground halos, leading to an underestimated $\langle f_{\rm diff,\,obs}\rangle$, as seen in the behavior of the Macquart relation fit toward $z = 0$ in Figure 3.17b and Table B.3 ($f_{\rm diff}(z \to 0)$). At higher redshifts, although the number of intersecting halos increases, their relative DM contributions decrease due to smaller physical sizes and increased redshift dilution. Thus, $\langle f_{\rm diff,\,obs}\rangle$ encodes both the underlying cosmic gas distribution and observational sample bias effects.

## B.5 Redshift Evolution of Baryons in Different Cosmological Components

In this appendix, we present a comprehensive view of the redshift evolution of baryonic mass fractions across different cosmological components and classification schemes.

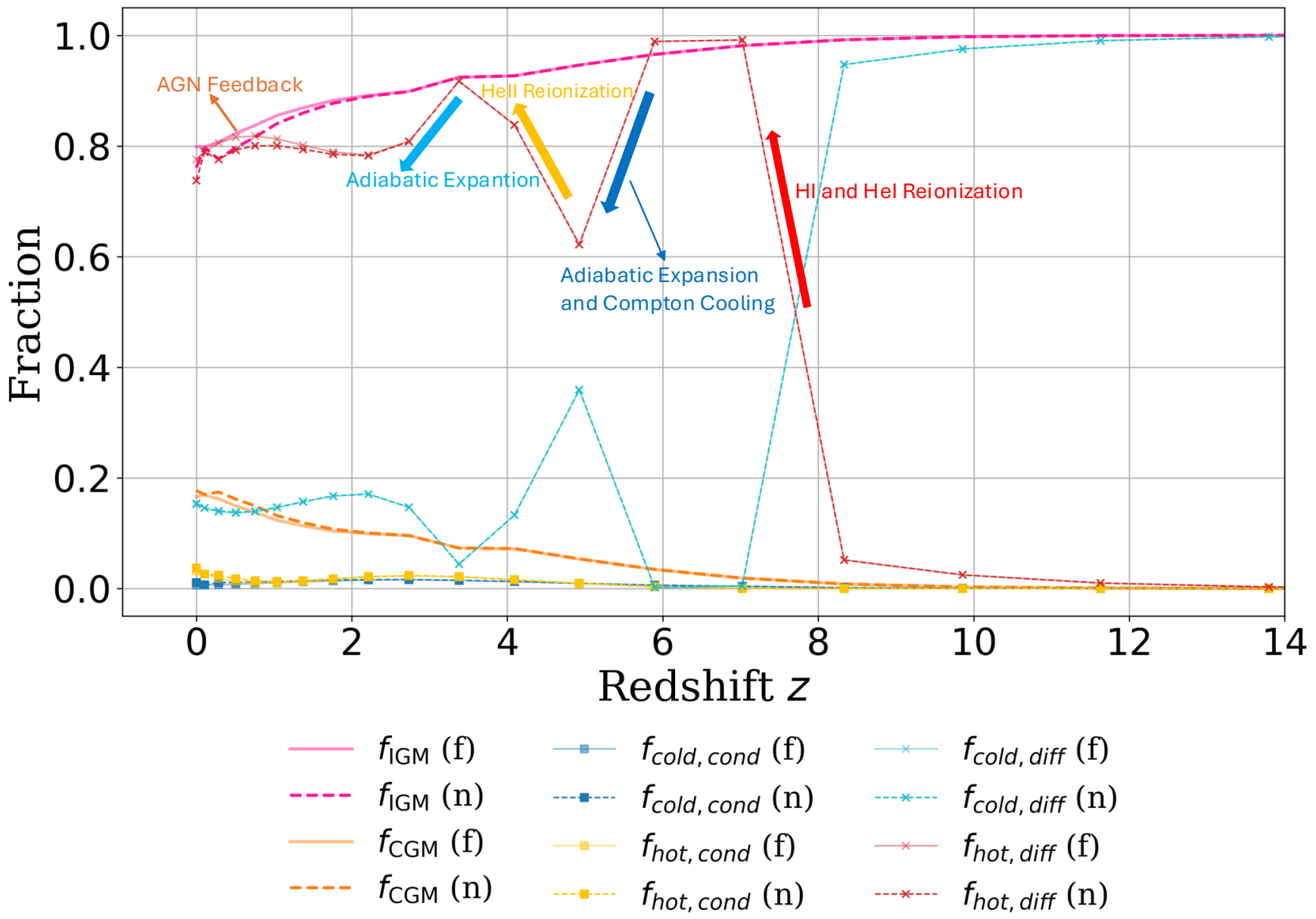

FIGURE B.5: Redshift evolution of baryonic mass fractions under two classification schemes using a CGM cutoff of $R_{\rm cut} = R_{200}$. Solid and dashed lines represent results from the fiducial and NoBH models, respectively. For the structure-based classification, the IGM and CGM are shown as pink and orange lines (without markers). For the phase-based classification, square markers denote condensed phases and x symbols denote diffuse phases. Cold and hot components are colored using cool and warm tones, respectively: deep blue for $f_{\rm cold,\,cond}$, yellow for $f_{\rm hot,\,cond}$, cyan-green for $f_{\rm cold,\,IGM}$, and bright red for $f_{\rm hot,\,IGM}$.

*IGM–CGM evolution:* Figure B.5 shows the evolution of gas baryon fractions under both phase-based and structure-based classifications. The corresponding values of $f_{\rm IGM}$ and $f_{\rm CGM}$ are listed in Tables B.1 and B.2, respectively. With a CGM cutoff of $R_{\rm cut} = R_{200}$, we find that $f_{\rm IGM}$ decreases from nearly unity at high redshift to $\lesssim 0.8$ at $z = 0$, while $f_{\rm CGM}$ increases from zero to $\lesssim 0.2$. This reflects the gradual incorporation of intergalactic gas into halos. The impact of AGN feedback on baryon redistribution appears mainly below $z \lesssim 2$ and is relatively modest in our CROCODILE simulations.

*Early phase evolution ($z > 8$):* At early times, $f_{\rm cold,diff}$ dominates due to the absence of collapsed halo structures and the overall low temperature of the gas. As redshift decreases below $z \sim 13$, $f_{\rm cold,diff}$ begins to decline slowly, while $f_{\rm hot,diff}$ shows a gradual increase. This transition reflects mild heating caused by adiabatic compression in regions of growing density perturbations, although the overall composition of the baryon phase remains largely unchanged during this epoch.

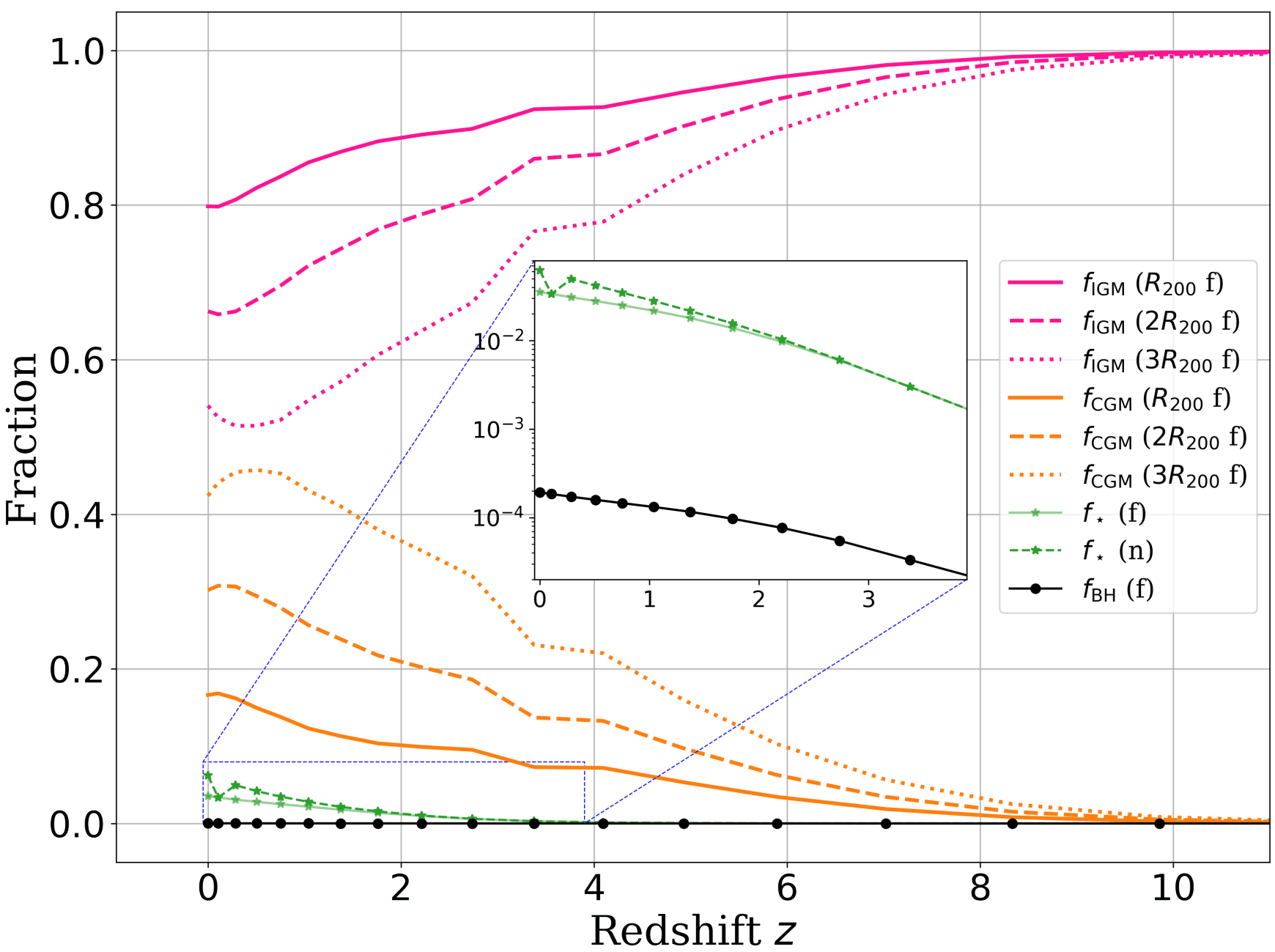

FIGURE B.6: Redshift evolution of baryon mass fractions in various cosmological components. The curves for $f_{\mathrm{CGM}}$ (orange) and $f_{\mathrm{IGM}}$ (pink) are plotted for different $R_{\mathrm{cut}}$ values: solid ($R_{200}$), dashed ($2R_{200}$), and dotted ($3R_{200}$). The evolution of $f_{\star}$ (green, five-pointed star markers) and $f_{\mathrm{BH}}$ (black, solid circles) is also shown for both fiducial and NoBH models. The inset panel zooms in on the redshift range $z < 4$ to highlight the detailed evolution of the stellar and black hole mass fractions.

*Reionization ($z \sim 8 \to 6$):* Reionization starts at $z \sim 8$, causing a rapid and nearly complete conversion of cold diffuse gas ($f_{\mathrm{cold,diff}}$) into hot diffuse gas ($f_{\mathrm{hot,diff}}$) by $z \sim 7$. After this point, both components stabilize, indicating the end of reionization and the establishment of a fully ionized diffuse medium.

*Thermal evolution ($z \approx 6$ to 2):* After hydrogen reionization completes around $z \sim 6$, the diffuse IGM cools through adiabatic expansion and Compton scattering off the CMB, leading to a decline in $f_{\mathrm{hot,diff}}$ and a corresponding rise in $f_{\mathrm{cold,diff}}$. From $z \sim 5$ to $z \sim 3.2$, HeII reionization driven by quasars gradually reheats the IGM (whose effect is included via the uniform UV background radiation), causing $f_{\mathrm{hot,diff}}$ to increase again. Below $z \sim 3.2$, after the completion of HeII reionization, the IGM enters another cooling phase due to adiabatic expansion. As a result, $f_{\mathrm{hot,diff}}$ begins to decline while $f_{\mathrm{cold,diff}}$ correspondingly rises. We also note that this thermal transition is sensitive to the adopted temperature threshold: lowering the cutoff from 10000 K to 8000 K significantly suppresses the apparent oscillatory feature.

*AGN feedback at low redshift ($z < 2$):* In this phase, AGN feedback becomes increasingly important and begins to heat and expel gas from condensed structures back into the diffuse phase. This effect is reflected by comparing the fiducial and NoBH simulations: in the fiducial run, both $f_{\mathrm{hot,cond}}$ and $f_{\mathrm{cold,cond}}$ show lower values, while $f_{\mathrm{hot,diff}}$ increases correspondingly. Thus, the apparent transition is not primarily a result of redshift evolution, but rather a manifestation of AGN-driven redistribution of baryons.

*Structure-based view:* Figure B.6 highlights how the evolution of $f_{\rm CGM}$ and $f_{\rm IGM}$ depends on $R_{\rm cut}$. For $R_{\rm cut} = 3R_{200}$, both fractions tend toward $\sim 0.5$ at $z = 0$. This emphasizes that the observationally inferred $\langle f_{\rm diff,\,obs} \rangle$ from FRBs is not a snapshot of the instantaneous IGM fraction at redshift $z$, but rather a LoS integrated average over all contributing structures at $z' < z$. At low redshift, many FRBs intersect no halos at all, resulting in a dominance of the IGM contribution to DM. The inset panel further shows that AGN feedback suppresses stellar growth in the fiducial model, while black hole growth remains limited throughout cosmic time.

TABLE B.1: Fractional baryon budget in the IGM for the Fiducial and NoBH runs at different redshifts and radial cuts.

| | $R_{200}$ | | $2R_{200}$ | | $3R_{200}$ | |
|---|---|---|---|---|---|---|
| $z$ | Fiducial | NoBH | Fiducial | NoBH | Fiducial | NoBH |
| 11.63 | 0.999 | 0.999 | 0.999 | 0.999 | 0.998 | 0.998 |
| 9.86 | 0.997 | 0.997 | 0.995 | 0.995 | 0.992 | 0.992 |
| 8.33 | 0.992 | 0.992 | 0.985 | 0.985 | 0.975 | 0.975 |
| 7.02 | 0.981 | 0.981 | 0.966 | 0.966 | 0.943 | 0.943 |
| 5.90 | 0.965 | 0.965 | 0.937 | 0.937 | 0.897 | 0.897 |
| 4.93 | 0.946 | 0.946 | 0.902 | 0.902 | 0.840 | 0.840 |
| 4.10 | 0.927 | 0.927 | 0.866 | 0.866 | 0.778 | 0.778 |
| 3.38 | 0.924 | 0.924 | 0.860 | 0.860 | 0.766 | 0.766 |
| 2.74 | 0.899 | 0.898 | 0.808 | 0.808 | 0.674 | 0.674 |
| 2.21 | 0.891 | 0.890 | 0.788 | 0.788 | 0.637 | 0.637 |
| 1.76 | 0.883 | 0.877 | 0.769 | 0.766 | 0.606 | 0.604 |
| 1.37 | 0.869 | 0.860 | 0.743 | 0.738 | 0.572 | 0.570 |
| 1.04 | 0.855 | 0.841 | 0.722 | 0.712 | 0.547 | 0.543 |
| 0.75 | 0.837 | 0.816 | 0.696 | 0.682 | 0.522 | 0.514 |
| 0.51 | 0.823 | 0.797 | 0.678 | 0.660 | 0.514 | 0.504 |
| 0.29 | 0.807 | 0.776 | 0.662 | 0.642 | 0.514 | 0.502 |
| 0.11 | 0.798 | 0.796 | 0.659 | 0.656 | 0.525 | 0.521 |
| 0.00 | 0.798 | 0.761 | 0.663 | 0.633 | 0.541 | 0.519 |

TABLE B.2: Fractional baryon budget in the CGM for the Fiducial and NoBH runs at different redshifts and radial cuts.

| | $R_{200}$ | | $2R_{200}$ | | $3R_{200}$ | |
|---|---|---|---|---|---|---|
| $z$ | Fiducial | NoBH | Fiducial | NoBH | Fiducial | NoBH |
| 11.63 | 0.001 | 0.001 | 0.001 | 0.001 | 0.002 | 0.002 |
| 9.86 | 0.003 | 0.003 | 0.005 | 0.005 | 0.008 | 0.008 |
| 8.33 | 0.008 | 0.008 | 0.015 | 0.015 | 0.025 | 0.025 |
| 7.02 | 0.019 | 0.019 | 0.034 | 0.034 | 0.056 | 0.056 |
| 5.90 | 0.034 | 0.034 | 0.063 | 0.063 | 0.103 | 0.103 |
| 4.93 | 0.053 | 0.053 | 0.097 | 0.097 | 0.160 | 0.160 |
| 4.10 | 0.072 | 0.072 | 0.133 | 0.133 | 0.220 | 0.220 |
| 3.38 | 0.073 | 0.073 | 0.137 | 0.137 | 0.231 | 0.231 |
| 2.74 | 0.095 | 0.096 | 0.186 | 0.186 | 0.320 | 0.320 |
| 2.21 | 0.099 | 0.100 | 0.202 | 0.202 | 0.353 | 0.352 |
| 1.76 | 0.104 | 0.107 | 0.217 | 0.219 | 0.380 | 0.380 |
| 1.37 | 0.113 | 0.119 | 0.239 | 0.240 | 0.410 | 0.409 |
| 1.04 | 0.123 | 0.131 | 0.257 | 0.260 | 0.431 | 0.429 |
| 0.75 | 0.138 | 0.149 | 0.279 | 0.284 | 0.453 | 0.451 |
| 0.51 | 0.149 | 0.162 | 0.294 | 0.299 | 0.457 | 0.454 |
| 0.29 | 0.162 | 0.174 | 0.307 | 0.309 | 0.455 | 0.448 |
| 0.11 | 0.168 | 0.170 | 0.308 | 0.311 | 0.441 | 0.445 |
| 0.00 | 0.166 | 0.177 | 0.302 | 0.304 | 0.424 | 0.418 |

## B.6 Origin of FRBs

The formation mechanism of FRBs remains an active field of research, with magnetars being the leading candidate. This theory is supported by observations of the magnetar SGR 1935+2154, which is associated with a similar burst, FRB 200428 (Bochenek et al., 2020). However, other formation channels may contribute to the overall FRB population, albeit to a lesser extent.

## B.7 Fitting Parameters for the Models

Table B.3 summarizes the best-fit parameters for the C-Exp, CPL-Exp, and DPL-Exp models. In the same appendix section, Table B.4 presents the MCMC-derived mNFW fitting results for the dark-matter halo and gas density profiles across the adopted halo-mass bins.

TABLE B.3: Fitting parameters and $\chi^2$ values for C-Exp, CPL-Exp, and DPL-Exp.

| Model | Case | $\kappa$ | $\tau$ | $\zeta$ | $\gamma$ | $\chi^2$/DOF |
|---|---|---|---|---|---|---|
| | $f_{\rm IGM}$: | $\kappa$ | $\tau$ | - | - | $\chi^2$/DOF |
| | Fiducial (FoF) | $0.27^{+0.02}_{-0.02}$ | $5.44^{+0.56}_{-0.51}$ | - | - | 0.142 |
| | NoBH (FoF) | $0.28^{+0.02}_{-0.02}$ | $5.03^{+0.54}_{-0.39}$ | - | - | 0.127 |
| | $R_{\rm cut} = R_{200}$ (Fiducial) | $0.30^{+0.01}_{-0.01}$ | $5.37^{+0.21}_{-0.24}$ | - | - | 0.017 |
| | $R_{\rm cut} = R_{200}$ (NoBH) | $0.33^{+0.02}_{-0.02}$ | $4.51^{+0.28}_{-0.28}$ | - | - | 0.037 |
| C-Exp | $R_{\rm cut} = 2R_{200}$ (Fiducial) | $0.27^{+0.02}_{-0.01}$ | $3.76^{+0.20}_{-0.30}$ | - | - | 0.107 |
| | $R_{\rm cut} = 2R_{200}$ (NoBH) | $0.28^{+0.01}_{-0.01}$ | $3.42^{+0.17}_{-0.23}$ | - | - | 0.106 |
| | $R_{\rm cut} = 3R_{200}$ (Fiducial) | $0.24^{+0.02}_{-0.02}$ | $2.67^{+0.34}_{-0.33}$ | - | - | 1.371 |
| | $R_{\rm cut} = 3R_{200}$ (NoBH) | $0.24^{+0.02}_{-0.01}$ | $2.53^{+0.24}_{-0.33}$ | - | - | 1.284 |
| | $f_{\rm diff}$ (Fiducial) | $0.87^{+0.05}_{-0.09}$ | $1.32^{+0.23}_{-0.10}$ | - | - | 0.057 |
| | $f_{\rm diff}$ (NoBH) | $0.90^{+0.11}_{-0.14}$ | $1.22^{+0.36}_{-0.16}$ | - | - | 0.129 |
| | $f_{\rm diff}(z \to 0)$ (Fiducial) | | 0.685 | | | |
| | $f_{\rm diff}(z \to 0)$ (NoBH) | | 0.666 | | | |
| | $f_{\rm IGM}$: | $\kappa$ | $\tau$ | $\zeta$ | $\gamma$ | $\chi^2$/DOF |
| | Fiducial (FoF) | $0.51^{+0.48}_{-0.51}$ | $3.74^{+14.91}_{-1.86}$ | $0.78^{+1.68}_{-0.23}$ | - | 0.143 |
| | NoBH (FoF) | $0.62^{+0.37}_{-0.61}$ | $3.01^{+10.30}_{-1.31}$ | $0.72^{+1.24}_{-0.18}$ | - | 0.123 |
| | $R_{\rm cut} = R_{200}$ (Fiducial) | $0.05^{+0.11}_{-0.05}$ | $9.56^{+7.33}_{-2.74}$ | $1.56^{+0.95}_{-0.35}$ | - | 0.016 |
| | $R_{\rm cut} = R_{200}$ (NoBH) | $0.25^{+0.25}_{-0.20}$ | $5.07^{+3.69}_{-1.62}$ | $1.09^{+0.51}_{-0.23}$ | - | 0.039 |
| CPL-Exp | $R_{\rm cut} = 2R_{200}$ (Fiducial) | $1.00^{+0.03}_{-0.07} \times 10^{-3}$ | $13.61^{+0.25}_{-0.32}$ | $2.66^{+0.02}_{-0.02}$ | - | 0.034 |
| | $R_{\rm cut} = 2R_{200}$ (NoBH) | $1.00^{+0.05}_{-0.01} \times 10^{-3}$ | $12.86^{+0.27}_{-2.97}$ | $2.71^{+0.01}_{-0.49}$ | - | 0.049 |
| | $R_{\rm cut} = 3R_{200}$ (Fiducial) | $1.00^{+0.01}_{-0.05} \times 10^{-3}$ | $10.88^{+0.48}_{-0.54}$ | $2.72^{+0.04}_{-0.04}$ | - | 0.651 |
| | $R_{\rm cut} = 3R_{200}$ (NoBH) | $1.00^{+0.05}_{-0.07} \times 10^{-3}$ | $10.58^{+0.039}_{-0.47}$ | $2.74^{+0.04}_{-0.03}$ | - | 0.560 |
| | $f_{\rm diff}$ (Fiducial) | $1.83^{+0.12}_{-0.16}$ | $0.24^{+0.19}_{-0.23}$ | $0.37^{+0.11}_{-0.11}$ | - | 0.014 |
| | $f_{\rm diff}$ (NoBH) | $1.86^{+0.05}_{-0.04}$ | $0.10^{+0.05}_{-0.10}$ | $0.31^{+0.03}_{-0.05}$ | - | 0.015 |
| | $f_{\rm diff}(z \to 0)$ (Fiducial) | | 0.655 | | | |
| | $f_{\rm diff}(z \to 0)$ (NoBH) | | 0.604 | | | |
| | $f_{\rm IGM}$: | $\kappa$ | $\tau$ | $\zeta$ | $\gamma$ | $\chi^2$/DOF |
| | Fiducial (FoF) | $8.89^{+2.21}_{-8.80} \times 10^{-7}$ | $18.94^{+5.11}_{-1.50}$ | $4.84^{+0.12}_{-0.34}$ | $0.67^{+4.34}_{-3.84}$ | 0.100 |
| | NoBH (FoF) | $8.31^{+7.20}_{-2.05} \times 10^{-7}$ | $18.22^{+3.47}_{-1.16}$ | $4.92^{+0.08}_{-0.25}$ | $0.66^{+4.09}_{-3.83}$ | 0.083 |
| | $R_{\rm cut} = R_{200}$ (Fiducial) | $8.99^{+4.27}_{-5.12} \times 10^{-7}$ | $25.32^{+1.38}_{-2.72}$ | $4.45^{+0.10}_{-0.19}$ | $0.87^{+3.68}_{-3.40}$ | 0.015 |
| | $R_{\rm cut} = R_{200}$ (NoBH) | $1.51^{+3.20}_{-6.22} \times 10^{-6}$ | $20.60^{+1.32}_{-1.66}$ | $4.55^{+0.18}_{-0.36}$ | $0.68^{+3.69}_{-3.39}$ | 0.040 |
| DPL-Exp | $R_{\rm cut} = 2R_{200}$ (Fiducial) | $3.18^{+1.88}_{-2.42} \times 10^{-6}$ | $23.29^{+2.43}_{-12.80}$ | $4.04^{+0.42}_{-2.32}$ | $1.03^{+3.43}_{-0.68}$ | 0.030 |
| | $R_{\rm cut} = 2R_{200}$ (NoBH) | $3.20^{+9.71}_{-3.19} \times 10^{-4}$ | $14.55^{+7.47}_{-8.19}$ | $3.01^{+1.54}_{-1.74}$ | $1.00^{+3.54}_{-0.27}$ | 0.052 |
| | $R_{\rm cut} = 3R_{200}$ (Fiducial) | $0.18^{+0.11}_{-0.12}$ | $8.18^{+3.74}_{-1.38}$ | $0.67^{+0.38}_{-0.19}$ | $2.03^{+0.99}_{-1.56}$ | 0.127 |
| | $R_{\rm cut} = 3R_{200}$ (NoBH) | $0.16^{+0.12}_{-0.12}$ | $7.28^{+3.79}_{-1.36}$ | $0.74^{+0.47}_{-0.21}$ | $1.89^{+0.968}_{-1.37}$ | 0.123 |
| | $f_{\rm diff}$ (Fiducial) | $1.97^{+0.00}_{-1.97}$ | $1.73^{+44.10}_{-0.10} \times 10^{-8}$ | $0.03^{+4.97}_{-0.01}$ | $9.82^{+4.82}_{-9.80}$ | 0.016 |
| | $f_{\rm diff}$ (NoBH) | $1.92^{+0.01}_{-1.32}$ | $1.12^{+15.91}_{-1.03} \times 10^{-8}$ | $0.03^{+4.97}_{-0.01}$ | $9.30^{+4.41}_{-9.80}$ | 0.016 |
| | $f_{\rm diff}(z \to 0)$ (Fiducial) | | 0.706 | | | |
| | $f_{\rm diff}(z \to 0)$ (NoBH) | | 0.684 | | | |

TABLE B.4: mNFW fitting parameters for Dark Matter and Gas Density profiles across halo mass bins.

| Profile | Parameter | Mass Range ($M_\odot$) | | | | |
|---|---|---|---|---|---|---|
| | | $10^{10.5}$–$10^{11.5}$ | $10^{11.5}$–$10^{12.5}$ | $10^{12.5}$–$10^{13.5}$ | $10^{13.5}$–$10^{14.5}$ | $10^{14.5}$–$10^{15.5}$ |
| Dark Matter | $y_0$ | $4.89^{+3.48}_{-3.30}$ | $5.18^{+3.28}_{-3.45}$ | $5.04^{+3.34}_{-3.31}$ | $4.18^{+3.76}_{-2.96}$ | $4.40^{+3.64}_{-3.02}$ |
| | $\alpha$ | $-1.06^{+3.04}_{-0.46}$ | $-0.77^{+2.81}_{-0.48}$ | $-0.29^{+2.71}_{-0.74}$ | $-0.31^{+1.06}_{-0.50}$ | $0.01^{+1.50}_{-0.66}$ |
| | $C$ | $9.99^{+6.87}_{-6.80}$ | $9.71^{+7.00}_{-6.78}$ | $9.91^{+6.86}_{-6.76}$ | $11.40^{+5.93}_{-6.88}$ | $11.38^{+5.99}_{-6.91}$ |
| | $\beta$ | $0.06^{+0.68}_{-1.34}$ | $-0.01^{+0.67}_{-1.28}$ | $0.34^{+0.82}_{-1.35}$ | $0.33^{+1.19}_{-1.47}$ | $-0.48^{+1.10}_{-1.10}$ |
| Gas | $y_0$ | $4.83^{+3.50}_{-3.39}$ | $5.41^{+3.15}_{-3.51}$ | $5.40^{+3.20}_{-3.40}$ | $4.52^{+3.59}_{-2.91}$ | $4.58^{+3.57}_{-3.07}$ |
| | $\alpha$ | $-1.29^{+3.69}_{-1.62}$ | $-0.85^{+0.31}_{-0.27}$ | $-0.33^{+0.48}_{-0.31}$ | $0.52^{+1.08}_{-0.64}$ | $1.05^{+1.63}_{-0.95}$ |
| | $C$ | $10.11^{+6.75}_{-6.67}$ | $9.35^{+7.23}_{-6.64}$ | $9.29^{+7.22}_{-6.68}$ | $11.33^{+5.95}_{-6.63}$ | $11.56^{+5.78}_{-6.49}$ |
| | $\beta$ | $0.85^{+0.78}_{-1.48}$ | $-0.99^{+0.77}_{-0.69}$ | $-0.77^{+0.88}_{-0.82}$ | $0.35^{+1.14}_{-1.45}$ | $-0.66^{+1.32}_{-0.91}$ |
| | $f_{\rm gas}$ | $0.034^{+0.019}_{-0.012}$ | $0.035^{+0.019}_{-0.012}$ | $0.082^{+0.049}_{-0.030}$ | $0.142^{+0.081}_{-0.052}$ | $0.155^{+0.091}_{-0.055}$ |

# Appendix C

# Additional Methods, Tests, and Derivations for Part II (SNR)

## C.1 Gaunt Factor for Free–Free Absorption

The free–free absorption coefficient used in Eq. (6) depends on the velocity–averaged Gaunt factor $g_{\rm ff}(\nu, T_{\rm e}, Z)$, which corrects the classical bremsstrahlung cross section for quantum and kinematic effects. The asymptotic behavior of the Gaunt factor was derived by Novikov & Thorne (1973) and later summarized in Fig. 5.2 of Rybicki & Lightman (1979). Its dependence can be expressed in terms of two dimensionless parameters:

$$u = \frac{h\nu}{k_{\rm B}T_{\rm e}}, \qquad \eta = \frac{k_{\rm B}T_{\rm e}}{Z^2 R_{\rm y}}, \tag{A1}$$

where $R_{\rm y} = 13.6\,{\rm eV}$ is the Rydberg energy.

The Gaunt factor admits three analytic limits:

**(1) Large–angle region** The transition to large–angle Rutherford scattering occurs when the impact parameter reaches the regime where

$$1 \gtrsim u \gtrsim \eta^{1/2}. \tag{C.1}$$

In this regime the Gaunt factor is of order unity,

$$g_{\rm ff}^{\rm (LA)} \simeq 1. \tag{A2}$$

**(2) Small–angle classical region ($u \ll 1$ and $\eta < 1$)**

$$g_{\rm ff}^{\rm (cl)} = \frac{\sqrt{3}}{\pi} \ln\left[\frac{1}{2\,\xi^{5/2}}\left(\frac{k_{\rm B}T_{\rm e}}{Z^2 R_{\rm y}}\right)^{1/2}\left(\frac{k_{\rm B}T_{\rm e}}{h\nu}\right)\right], \tag{A3}$$

where $\xi = e^{\gamma} \simeq 1.781$ and $\gamma$ is Euler's constant.

**(3) Small–angle U.P. region ($u \ll 1$ and $\eta > 1$)**

$$g_{\rm ff}^{(\rm UP)} = \frac{\sqrt{3}}{\pi} \ln\left[\frac{1}{\xi^2}\left(\frac{k_{\rm B}T_{\rm e}}{h\nu}\right)\right]. \tag{A4}$$

Because the SNR spans a broad range of temperatures ($T_{\rm e} \sim 10^4$–$10^7$ K) and the FRB frequencies lie in the GHz band, different radial layers naturally fall into different asymptotic regimes. To ensure numerical smoothness, we evaluate the Gaunt factor using a blended interpolation across the boundaries $u = \eta^{1/2}$ and $\eta = 1$. Throughout this work, we compute $g_{\rm ff}$ at $\nu = 10^9$ Hz.

## C.2 Characteristic Scales and Self-Similar Constants in the Wind Case

In this appendix, we summarize the characteristic scales and dimensionless constants adopted in the self-similar driven wave (SSDW) solution for a supernova remnant (SNR) expanding into a wind-like circumstellar medium (CSM) with density profile $\rho_{\rm w} \propto r^{-2}$ ($s = 2$).

### C.2.1 Characteristic Radius and Timescale

For a wind environment characterized by $K \equiv \dot{M}/(4\pi v_{\rm w})$, we define the normalized parameter $K_{13} \equiv K/(10^{13}\,{\rm g\,cm^{-1}})$. Following the standard SSDW scaling (e.g., Tang & Chevalier 2017; Zhao et al. 2021), the characteristic radius and timescale are given by

$$R_{\rm ch} \simeq 6.58 \times 10^2\,{\rm pc}\; M_1 \left(\frac{5.1}{K_{13}}\right), \tag{C.2}$$

and

$$t_{\rm ch} \simeq 2.86 \times 10^5\,{\rm yr}\; E_{51}^{-1/2}\, M_1^{3/2} \left(\frac{5.1}{K_{13}}\right), \tag{C.3}$$

The numerical factor 5.1 arises from the fiducial normalization $\dot{M} = 10^{-5}\,M_\odot\,{\rm yr^{-1}}$ and $v_w = 10\,{\rm km\,s^{-1}}$, for which $K \simeq 5.1 \times 10^{13}\,{\rm g\,cm^{-1}}$. These characteristic scales are used to normalize the self-similar evolution of the forward shock, reverse shock, and contact discontinuity.

### C.2.2 Contact Discontinuity and the Constant $\zeta_c$

We follow the standard parametrization of the ejecta and ambient-medium density profiles (e.g., TM99), writing

$$\rho(r,t) = \begin{cases} \rho_{\rm ej}(r,t), & r \le R_{\rm ej}(t), \\ \rho_{\rm a}(r), & r > R_{\rm ej}(t), \end{cases} \tag{C.4}$$

where the ambient medium is described by a power law

$$\rho_{\rm a}(r) = \eta_s\, r^{-s}, \tag{C.5}$$

and the freely expanding ejecta take the form

$$\rho_{\rm ej}(r,t) = \frac{M_{\rm ej}}{R_{\rm ej}^3(t)}\, f\!\left(\frac{r}{R_{\rm ej}(t)}\right). \tag{C.6}$$

Here $R_{\rm ej}(t)$ is the outer radius of the ejecta and $w \equiv r/R_{\rm ej}$. The ejecta structure function $f(w)$ is specified by a flat inner core and a power-law outer envelope,

$$f(w) = \begin{cases} f_0, & 0 \le w \le w_{\rm core}, \\ f_0\left(\dfrac{w_{\rm core}}{w}\right)^n, & w_{\rm core} \le w \le 1, \end{cases} \qquad (n > 5), \tag{C.7}$$

where $w_{\rm core}$ denotes the fractional core radius and $n$ is the outer power-law index of the ejecta. The normalization constant $f_0$ is fixed by the mass conservation condition $\int_0^1 4\pi w^2 f(w)\, dw = 1$, yielding

$$f_0 = \frac{3}{4\pi w_{\rm core}^n}\left[\frac{1 - n/3}{1 - (n/3)\, w_{\rm core}^{3-n}}\right]. \tag{C.8}$$

The radius of the contact discontinuity can be written as (Tang & Chevalier, 2017)

$$R_c^*(t^*) = \left[(\lambda_c\, t^*)^{-\tilde{\alpha}} + \left(c\, t^{*\beta}\right)^{-\tilde{\alpha}}\right]^{-1/\tilde{\alpha}}, \tag{C.9}$$

where $R_c^* \equiv R_c/R_{\rm ch}$ and $t^* \equiv t/t_{\rm ch}$. Here $\tilde{\alpha}$ is the interpolation exponent in the fitting formula of Tang & Chevalier (2017) and is unrelated to the DM time-scaling index $\alpha$ (defined in the main text via DM $\propto t^{-\alpha}$). This expression interpolates between the early-time free-expansion behavior $R_c^* \simeq \lambda_c t^*$ and the self-similar ejecta-dominated regime $R_c^* \simeq c\, t^{*\beta}$, with $\beta = (n-3)/(n-s)$. In the SSDW limit ($t \ll t_{\rm ch}$), $R_c$ reduce to

$$R_c(t) \simeq \zeta_c\, R_{\rm ch}\left(\frac{t}{t_{\rm ch}}\right)^{\frac{n-3}{n-s}}, \tag{C.10}$$

where $s = 2$ for the wind case. The dimensionless normalization constant $\zeta_c$ encapsulates the detailed structure of the self-similar solution and is given by

$$\zeta_c = \left(A\, f_0\, w_{\rm core}^n\, \lambda_c^{n-3}\right)^{\frac{1}{n-s}}. \tag{C.11}$$

where the constant $A$ is a dimensionless coefficient that depends on $(n,s)$, with values tabulated in Chevalier (1982). In this work, we adopt $A = 0.067$, corresponding to $n = 10$ and $s = 2$. The parameter $\lambda_c$ is a dimensionless constant related to the density profile of the ejecta and can be expressed as

$$\lambda_c^2 = 2\, w_{\rm core}^{-2}\left(\frac{5-n}{3-n}\right)\left(\frac{w_{\rm core}^{n-3} - n/3}{w_{\rm core}^{n-5} - n/5}\right). \tag{C.12}$$

## C.3 Analytic-Model Comparison Using Simulated Mean Molecular Weights

In these appendix figures, the analytic prescriptions are evaluated using the composition-dependent quantities extracted from the simulations. For the unshocked ejecta and unshocked CSM we adopt $\mu_a$, under the assumption that each atom in the unshocked ejecta contributes only one free electron (hence $\mu_e = \mu_a$), while for the shocked region we use $\mu_e$ and $\mu_m$ computed from the self-consistent ionization evolution. In the PG18 wind case, the time-dependent $\mu_e(t)$ in the shocked region introduces an early-time variation that is qualitatively consistent with the NEI features seen in our shocked-region simulations. By contrast, the PG18 shocked-ionized case shows a nearly negligible NEI imprint because the DM evolution depends primarily on the ratio $\mu_m/\mu_e$ (see Eq. 4.13); this ratio varies less than $\mu_e$ or $\mu_m$ individually, thereby suppressing the apparent ionization-driven modulation in DM($t$).

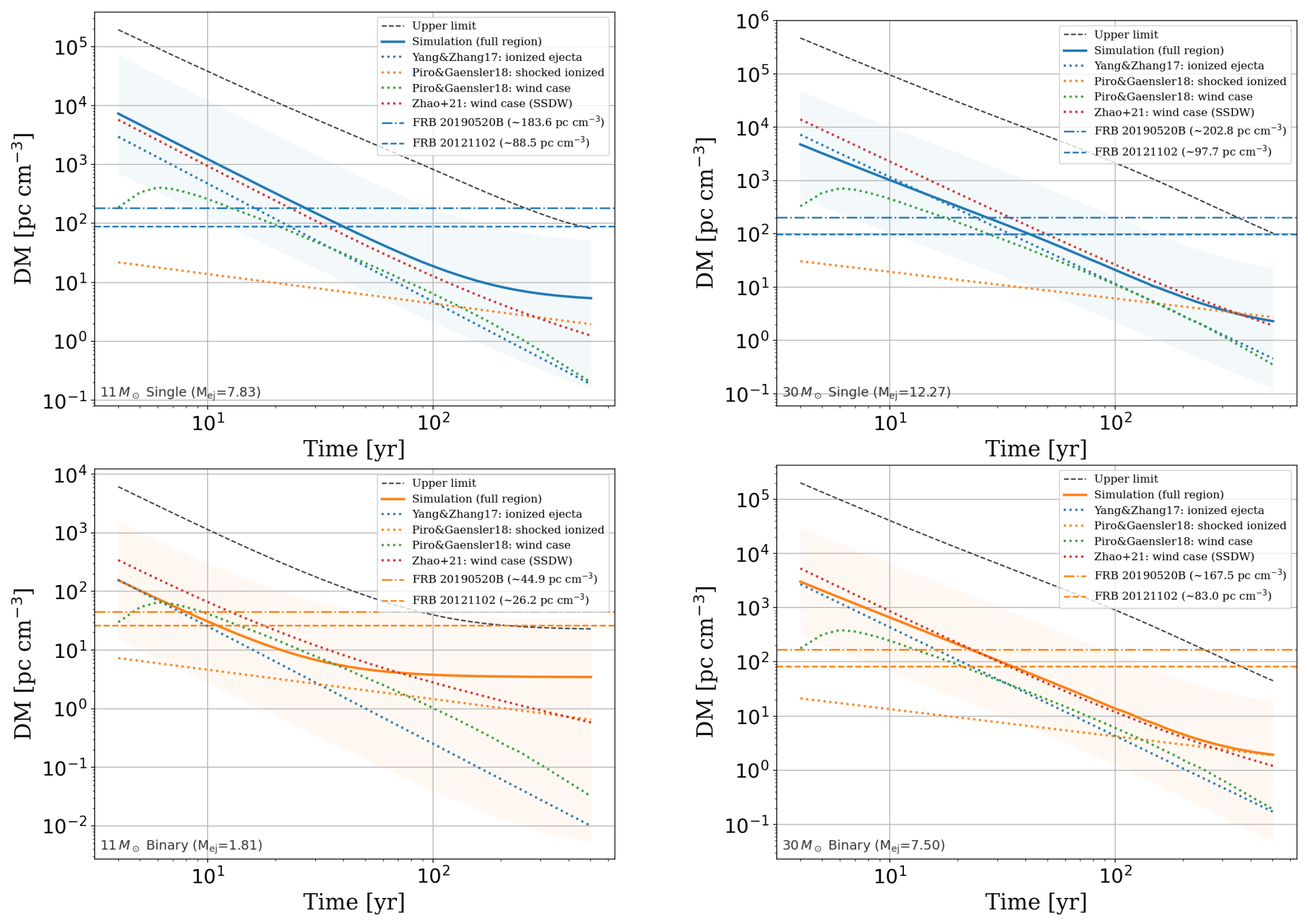

FIGURE C.1: Same format as Fig. 4.11, but the analytic prescriptions adopt the simulated mean molecular weights ($\mu_a$, $\mu_e$, $\mu_m$) for each model, rather than the fiducial values used in Section 4.7.4.

## C.4 Occurrence-Time and Curvature Table for FRB 20121102

In this appendix table, $t_{\rm occur}$ and the DM-derivative quantities are evaluated at the epoch where $\mathrm{dDM/dt} = -3.93\,\mathrm{pc\,cm^{-3}\,yr^{-1}}$, corresponding to the second-stage DM evolution of FRB 20121102.

TABLE C.1: Same format as Table 4.3, but for FRB 20121102. Here $t_{\rm occur}$ and DM-derivative quantities are defined at $\mathrm{dDM}/\mathrm{d}t = -3.93\,\mathrm{pc\,cm^{-3}\,yr^{-1}}$ (the second-stage DM evolution of FRB 20121102).

| Ionization config. | Quantity | $11\,M_\odot$ SS | $11\,M_\odot$ BS | $30\,M_\odot$ SS | $30\,M_\odot$ BS |
|---|---|---|---|---|---|
| FF: $\chi_{e,\rm unej}=1$, $\chi_{e,\rm ISM}=1$ | $t_{\rm esc}$ (yr) | 32 | 6 | 62 | 46 |
| | $t_{\rm occur}$ (yr) | 87 | 24 | 98 | 84 |
| | $\mathrm{d^2DM}/\mathrm{d}t^2$ ($\mathrm{pc\,cm^{-3}\,yr^{-2}}$) | 0.21 | 0.50 | 0.15 | 0.10 |
| | $\lvert \mathrm{d^2DM}/\mathrm{d}t^2\rvert_{\rm est}$ ($\mathrm{pc\,cm^{-3}\,yr^{-2}}$) | 0.09 | 0.33 | 0.07 | 0.09 |
| HF: $\chi_{e,\rm unej}=0.1$, $\chi_{e,\rm ISM}=1$ | $t_{\rm esc}$ (yr) | 13 | 0 | 17 | 13 |
| | $t_{\rm occur}$ (yr) | 40 | 11 | 42 | 35 |
| | $\mathrm{d^2DM}/\mathrm{d}t^2$ ($\mathrm{pc\,cm^{-3}\,yr^{-2}}$) | 0.28 | 1.07 | 0.24 | 0.29 |
| | $\lvert \mathrm{d^2DM}/\mathrm{d}t^2\rvert_{\rm est}$ ($\mathrm{pc\,cm^{-3}\,yr^{-2}}$) | 0.19 | 0.69 | 0.17 | 0.21 |
| MF: $\chi_{e,\rm unej}=0.01$, $\chi_{e,\rm ISM}=1$ | $t_{\rm esc}$ (yr) | 5 | 0 | 4 | 0 |
| | $t_{\rm occur}$ (yr) | 19 | 5 | 17 | 15 |
| | $\mathrm{d^2DM}/\mathrm{d}t^2$ ($\mathrm{pc\,cm^{-3}\,yr^{-2}}$) | 0.60 | 1.43 | 0.65 | 0.67 |
| | $\lvert \mathrm{d^2DM}/\mathrm{d}t^2\rvert_{\rm est}$ ($\mathrm{pc\,cm^{-3}\,yr^{-2}}$) | 0.40 | 1.56 | 0.45 | 0.46 |
| FH: $\chi_{e,\rm unej}=1$, $\chi_{e,\rm ISM}=0.1$ | $t_{\rm esc}$ (yr) | 32 | 6 | 62 | 46 |
| | $t_{\rm occur}$ (yr) | 87 | 24 | 98 | 84 |
| | $\mathrm{d^2DM}/\mathrm{d}t^2$ ($\mathrm{pc\,cm^{-3}\,yr^{-2}}$) | 0.21 | 0.50 | 0.15 | 0.10 |
| | $\lvert \mathrm{d^2DM}/\mathrm{d}t^2\rvert_{\rm est}$ ($\mathrm{pc\,cm^{-3}\,yr^{-2}}$) | 0.09 | 0.33 | 0.07 | 0.09 |
| HH: $\chi_{e,\rm unej}=0.1$, $\chi_{e,\rm ISM}=0.1$ | $t_{\rm esc}$ (yr) | 13 | 0 | 17 | 13 |
| | $t_{\rm occur}$ (yr) | 40 | 11 | 42 | 35 |
| | $\mathrm{d^2DM}/\mathrm{d}t^2$ ($\mathrm{pc\,cm^{-3}\,yr^{-2}}$) | 0.22 | 1.08 | 0.24 | 0.29 |
| | $\lvert \mathrm{d^2DM}/\mathrm{d}t^2\rvert_{\rm est}$ ($\mathrm{pc\,cm^{-3}\,yr^{-2}}$) | 0.20 | 0.69 | 0.17 | 0.21 |
| MH: $\chi_{e,\rm unej}=0.01$, $\chi_{e,\rm ISM}=0.1$ | $t_{\rm esc}$ (yr) | 5 | 0 | 4 | 0 |
| | $t_{\rm occur}$ (yr) | 19 | 5 | 17 | 15 |
| | $\mathrm{d^2DM}/\mathrm{d}t^2$ ($\mathrm{pc\,cm^{-3}\,yr^{-2}}$) | 0.60 | 1.43 | 0.65 | 0.67 |
| | $\lvert \mathrm{d^2DM}/\mathrm{d}t^2\rvert_{\rm est}$ ($\mathrm{pc\,cm^{-3}\,yr^{-2}}$) | 0.40 | 1.56 | 0.45 | 0.46 |

# Data and Code Availability

The stellar evolution inputs used in this thesis are based on publicly available MESA models. The corresponding progenitor data are available via Zenodo:

https://doi.org/10.5281/zenodo.5929871 (Farmer et al., 2023).

The data products associated with the supernova-remnant (SNR) studies presented in this thesis, including the time evolution of DM, RM, free–free optical depth, and the numerical datasets used to generate the figures, are publicly available through Zenodo:

https://doi.org/10.5281/zenodo.20486710

The corresponding source code used to generate these data products is available at:

https://github.com/zhaojosephzhang/SNR_FRB

In addition, the ARCOS (Analysis Repository for Cosmological FRB Studies in Simulations) software framework developed for FRB cosmology and simulation-based studies using the CROCODILE simulations has been publicly archived through Zenodo:

https://doi.org/10.5281/zenodo.20487842

The source code is available at:

https://github.com/zhaojosephzhang/ARCOS

Due to the large volume of the underlying CROCODILE/GADGET simulation outputs and the environment-specific dependencies of the simulation workflows, the raw simulation data are not publicly distributed.

All external observational datasets used in this thesis are publicly available and properly cited in the reference list.

Access to non-public simulation data may be provided upon reasonable request, subject to data ownership policies, collaboration agreements, and storage limitations.